\documentclass[12pt]{article}

\usepackage[normalem]{ulem}
\usepackage{amsthm}
\usepackage{bm}
\usepackage{mathtools}
\usepackage{thmtools}
\declaretheoremstyle[headfont=\normalfont]{normalhead}
\usepackage{color}
\usepackage{graphicx}
\usepackage{morefloats}
\usepackage[hidelinks]{hyperref}
\usepackage{amssymb}
\usepackage{textcomp}
\usepackage{url}
\usepackage{float}
\usepackage{appendix}
\usepackage{authblk}
\usepackage{caption}
\usepackage{booktabs}
\usepackage[shortlabels]{enumitem}
\usepackage{setspace}

\usepackage{algpseudocode}
\usepackage{algorithm}

\usepackage{mathrsfs}
\usepackage{MnSymbol}

\usepackage[
backend=biber,
giveninits=true,
uniquename=false,
date=year,
style=apa,
uniquelist=false,
sorting=nty,
url=false,
isbn=false,
doi=false,
eprint=false,
maxbibnames=5,
minbibnames=1,
maxcitenames=1,
mincitenames=1
]{biblatex}

\DeclareNameAlias{sortname}{family-given}

\newcommand{\norm}[1]{\lVert #1 \lVert}

\newcommand\C{\mathcal{C}}
\newcommand\E{\mathbb{E}}
\newcommand\curl{\left\{}
\newcommand\curr{\right\}}

\newcommand\bl{\left(}
\newcommand\br{\right)}

\newcommand{\calC}{\mathcal{C}}

\newcommand{\calF}{\mathcal{F}}

\newcommand{\calM}{\mathcal{M}}

\newcommand{\calT}{\mathcal{T}}

\newcommand{\ra}{\rightarrow}

\newcommand{\Var}{\operatorname{Var}}
\newcommand{\Cov}{\operatorname{Cov}}
\newcommand{\Corr}{\operatorname{Corr}}

\newcommand{\op}{\operatorname{op}}

\renewcommand\epsilon{\varepsilon}

\newcommand{\Exp}{\operatorname{Exp}}

\newcommand{\M}{\calM}

\newcommand{\sigmax}{\Sigma_X}
\newcommand{\sigmay}{\Sigma_{Y}}
\newcommand{\sigmaxy}{\Sigma_{XY}}

\newcommand{\rone}{{\mathbb{R}}}
\newcommand{\Ex}[1]{\mathbb{E} \left [  #1 \right ]}

\newcommand{\sigmaxhat}{\hat{\Sigma}_{X}}
\newcommand{\sigmayhat}{\hat{\Sigma}_{Y}}
\newcommand{\sigmaxyhat}{\hat{\Sigma}_{XY}}

\newcommand{\Tsc}{\mathcal{T}}
\newcommand{\txm}{T_x\M}
\newcommand{\Log}{\operatorname{Log}}
\newcommand{\Logy}{\operatorname{Log}_{\mu}y}

\newcommand{\Logyi}{\operatorname{Log}_{\hat \mu}y_i}
\newcommand{\Logyhat}{\operatorname{Log}_{\hat \mu}y_1}

\newcommand{\Jmu}{L^2(T\mu)}
\newcommand{\Jmuhat}{L^2(T\hat\mu)}
\newcommand{\dcorr}{{d^{(\text{corr})}}}

\newcommand{\hil}{\mathbb{H}}
\newcommand{\T}{\top}
\newcommand{\sumj}{\sum_{j=1}^\infty}
\newcommand{\ubar}{\overline{\mathcal{U}}}
\newcommand{\vbar}{\overline{\mathcal{V}}}
\newcommand{\clim}{\overline{\operatorname{Im} \bl \mathscr{K}_1 \br}}

\newcommand{\inner}[2]{ \langle #1, #2 \rangle }
\newcommand{\innerdouble}[2] {\llangle #1, #2 \rrangle}
\newcommand{\maximize}[2]{\underset{#2}{\operatorname{maximize}\ }#1}
\newcommand{\minimize}[2]{\underset{#2}{\operatorname{minimize}\ } #1}
\newcommand{\supp}[2]{\underset{#2}{\operatorname{sup}\ }#1}

\newcommand{\argmax}[2]{\underset{#2}{\operatorname{arg\,max\,}}#1}
\newcommand{\argmin}[2]{\underset{#2}{\operatorname{arg\,min\,}}#1}

\newcommand{\bhat}{\hat{B}}
\newcommand{\rootnfrac}{\frac{1}{\sqrt{N}}}
\newcommand{\nfrac}{{\frac{1}{N}}}
\newcommand{\rate}{\bl \frac{d}{N}\log \bl p\eta^{-1} \br \br}
\newcommand{\etathat}{\hat{\tilde{\eta_k}}}
\newcommand{\etat}{\tilde{\eta_k}}
\newcommand{\gammaterm}{\operatorname{min}\bl\gamma_{k-1}^2-\gamma_k^2,\gamma_k^2-\gamma_{k+1}^2\br}
\newcommand{\gammainvhat}{\hat \gamma_k^{-1}}
\newcommand{\gammainv}{\gamma_k^{-1}}

\newcommand{\thetagamma}{\delta_{\Gamma}}
\newcommand{\gammamuhat}{\Gamma_{\hat{\mu},\mu}}
\newcommand{\psihat}{\hat{\psi}}
\newcommand{\psitil}{\tilde{\psi}}
\newcommand{\psibar}{\bar{\psi}}
\newcommand{\sigmaz}{\Sigma_Z}
\newcommand{\phidif}{\munorm{\hat{\phi}_j \thetagamma \phi_j}}

\newcommand{\xtest}{X_{\text{test}}}
\newcommand{\ytest}{y_{\text{test}}}
\newcommand{\Ytest}{Y_{\text{test}}}
\newcommand{\ztest}{Z_{\text{test}}}
\newcommand{\xtests}{X_{\text{test},S}}

\newcommand{\infnorm}[1]{\left \| #1 \right \|_{\infty}}
\newcommand{\twonorm}[1]{\left \| #1 \right \|_2}
\newcommand{\onenorm}[1]{\left \| #1 \right \|_1}
\newcommand{\Fnorm}[1]{\left \| #1 \right \|_F}
\newcommand{\twoinfnorm}[1]{\left \| #1 \right \|_{2,\infty}}
\newcommand{\onetwo}[1]{\left \| #1 \right \|_{\ell_1,\ell_2}}
\newcommand{\maxnorm}[1]{\left \| #1 \right \|_{\operatorname{max}}}
\newcommand{\subgaussnorm}[1]{\left \| #1 \right \|_{\psi_2}}
\newcommand{\subexpnorm}[1]{\left \| #1 \right \|_{\psi_1}}
\newcommand{\munorm}[1]{\left \| #1 \right \|_{\mu}}
\newcommand{\muhatnorm}[1]{\left \| #1 \right \|_{\hat \mu}}

\newtheoremstyle{mydef}
{\topsep}{\topsep}%
{}{}%
{\itshape}{}
{\newline}
{%
  \rule{\textwidth}{0.0pt}\\*%
  \thmname{#1}~\thmnumber{#2}\thmnote{\-\ #3}.\\*[-1.5ex]%
  \rule{\textwidth}{0.0pt}}%

\begin{document}

\def\spacingset#1{\renewcommand{\baselinestretch}%
{#1}\small\normalsize} \spacingset{1}

\newtheorem{conjecture}{Conjecture}
\newtheorem{proposition}{Proposition}
\newtheorem{theorem}{Theorem}[section]
\newtheorem{informaltheorem}{Theorem (Informal)}[section]
\newtheorem{question}{Question}
\newtheorem{remark}{Remark}
\newtheorem{proposal}{Proposal}
\newtheorem{lemma}{Lemma}[section]
\newtheorem{corollary}{Corollary}[section]
\newtheorem{observation}{Observation}[section]
\newtheorem{assumption}{Assumption}[section]
\newtheorem{definition}{Definition}[section]

\author[1]{James Buenfil}
\author[2]{Eardi Lila}

\affil[1]{Department of Statistics, University of Washington, Seattle}
\affil[2]{Department of Biostatistics, University of Washington, Seattle}

\date{}

\title{Asymmetric canonical correlation analysis of Riemannian and high-dimensional data}
\maketitle
\begin{abstract}
\hskip -.2in
\noindent In this paper, we introduce a novel statistical model for the integrative analysis of Riemannian-valued functional data and high-dimensional data. We apply this model to explore the dependence structure between each subject's dynamic functional connectivity -- represented by a temporally indexed collection of positive definite covariance matrices -- and high-dimensional data representing lifestyle, demographic, and psychometric measures. Specifically, we employ a reformulation of canonical correlation analysis that enables efficient control of the complexity of the functional canonical directions using tangent space sieve approximations. Additionally, we enforce an interpretable group structure on the high-dimensional canonical directions via a sparsity-promoting penalty. The proposed method shows improved empirical performance over alternative approaches and comes with theoretical guarantees. Its application to data from the Human Connectome Project reveals a dominant mode of covariation between dynamic functional connectivity and lifestyle, demographic, and psychometric measures. This mode aligns with results from static connectivity studies but reveals a unique temporal non-stationary pattern that such studies fail to capture.
\end{abstract}


\begin{refsection}
\section{Introduction}\label{sec:introduction}
One of the primary goals of large-scale neuroimaging studies, such as the Human Connectome Project, ABCD, and the UK Biobank, is to understand the relationship between complex neuroimaging traits and non-imaging high-dimensional variables, including cognitive abilities, neurodegenerative conditions, mental health disorders, psychometric test scores, and other external factors \parencite{zhu2023statistical}. In the context of functional connectivity studies, such complex imaging data are typically networks that are derived from fMRI data and are characterized by a single covariance matrix that captures the temporal correlation between the fMRI signals of different brain regions. For instance, \textcite{xia2018linked} study correlation patterns between functional connectivity and psychiatric symptoms. Other studies, such as \textcite{smith2015positivenegative} and \textcite{liu2022improved}, investigate the relationship between functional connectivity and behavioral and demographic measures. 

Traditional analyses often view brain functional networks as static. Yet, there is growing evidence that these networks are inherently dynamic and exhibit significant temporal fluctuations \parencite{hutchison2013dynamic}, which appear to be linked to various aspects of human behavior \parencite{liegeois2019resting}. Therefore, they are best represented by a time-indexed collection of covariances, that is, a Riemannian manifold-valued function where the manifold consists of the space of symmetric positive definite (SPD) matrices. 

This work seeks to identify joint variation between these functional dynamic networks and multivariate variables, such as lifestyle, demographic, and psychometric measures. To this purpose, we develop a novel asymmetric canonical correlation analysis model that allows us to explore the underlying relationships between two data views: Riemannian manifold-valued functional data and high-dimensional variables. We refer to this setting as \textit{asymmetric} due to the different nature of the data views, which require different approaches to address their complexity. While our motivation stems from dynamic functional connectivity, the proposed method is general and can be applied to a variety of other settings.

Numerous models have been developed to model manifold-valued functional data, \parencites[see, e.g., ][]{pigoli2014distances}{dai2018principal}{ masarotto2019procrustes}{lin2019intrinsic}{ dubey2020functional,dubey2021modeling}{zhang2020mixedeffect}{zhou2021dynamic}{ghodrati2022distributionondistribution}{ghosal2023frechet}{stocker2023functional}, which can be more broadly viewed as object data \parencite{marron2021object} -- a generalization of functional data \parencites{ramsay2015functional}{hsing2015theoretical}{kokoszka2017introduction}. Regression models for manifold-valued data with low-dimensional predictors have been proposed in \textcites{petersen2019frechet}{zhao2021covariate}{zhou2023functional}. See also \textcite{petersen2022modeling} for a review. Nonetheless, models that facilitate the integration of manifold-valued functional data with high-dimensional variables have not been extensively explored. 

Canonical correlation analysis (CCA) is one of the principal tools for data integration \parencites{hotelling1936relations}{uurtio2018tutorial}{zhuang2020technical}{yang2021survey} and can be used to identify shared structure between two low-dimensional sets of variables by seeking linear combinations of these sets -- with weights referred to as canonical vectors -- that exhibit maximum correlation. CCA methods that go beyond low-dimensional data have largely focused on the symmetric setting, where both data views have the same structure or form. Extensions of CCA to high-dimensional data have been proposed, for instance, in \textcites{witten2009penalized}{lin2013group}{chen2013sparse}{gaynanova2016simultaneous}{gao2017sparse}{yoon2020sparse}{wang2021eigenvectorbased}. The setting of functional data has been considered in \textcites{he2010functional}{shin2015canonical}{huang2015functional} and that of more complex imaging data in \textcites{cho2022tangent}{liu2021noneuclidean}. CCA between data on Riemannian manifolds has been considered in \textcite{kim2014canonical}. Inferential aspects have been explored in \textcites{yang2015independence}{mckeague2022significance}{kessler2023computational}. Methods that estimate both shared and individual structure have been proposed in \textcites{lock2013joint}{ feng2018anglebased}{carmichael2020learning, shu2020dcca}{yuan2022doublematched}, and their connection to CCA has been studied in \textcite{murden2022interpretive}. 

Yet, despite the large body of literature on CCA and its extensions, existing approaches are not able to effectively estimate common structure between Riemannian manifold-valued functional data and high-dimensional multivariate variables, and more broadly, between imaging and high-dimensional data. To bridge this gap, we propose a model that leverages a regression-based characterization of CCA, which allows us to incorporate appropriate notions of complexity for the functional and high-dimensional canonical directions. Specifically, our approach takes advantage of the inherent smoothness and geometric nature of the functional data, employing tangent space approximations based on a data-driven function basis computed using the Riemannian Functional Principal Components Analysis (RFPCA) framework \parencites{dai2018principal}{lin2019intrinsic}{shao2022intrinsic}. Moreover, it tackles the high dimensionality of the multivariate data by imposing sparsity. It therefore performs feature selection, resulting in models that are more interpretable and mitigate overfitting issues. In the motivating application, this will result in the identification of a small and interpretable set of multivariate variables linked to specific functional dynamic connectivity patterns. On the other hand, the tangent-space representation ensures that the estimated functional canonical directions remain constrained to the non-linear space to which the data belong, i.e., the space of SPD matrices.

The asymmetric setting considered in this work is of interest not just for its potential applications but also methodologically, as it has some distinct features that are not found in the purely sparse or functional settings. Specifically, we show that if the functional data can be efficiently represented using a finite subspace, the proposed method can consistently estimate the high-dimensional canonical vectors without requiring the direct estimation of the precision matrix of the high-dimensional data -- a notoriously difficult problem and typically solvable only under specific structural assumptions \parencite{cai2016estimating}. This feature renders the proposed methodology novel even in the simpler setting of classical functional and high-dimensional data integration.

In addition to accommodating manifold-valued functional data and high-dimensional data, our proposed method has several other desirable properties in comparison to existing CCA models, which we highlight below:
\begin{enumerate}
\item It can estimate multiple canonical directions simultaneously, without requiring iterative deflation strategies and leveraging shared sparsity structure across canonical vectors.
\item It is computationally efficient, with its complexity essentially reducing to solving a regularized multivariate linear regression problem.
\item It does not require a consistent estimator for the precision matrix of the high-dimensional data.
\item The canonical vectors satisfy the correct orthogonality conditions, ensuring that the proposed approach is invariant to data rescaling, while simultaneously enforcing an interpretable sparsity structure on the high-dimensional canonical vectors.
\item When the number of observations is larger than the dimension of the high-dimensional data and no sparsity is imposed, our approach reduces to classical multivariate CCA.
\end{enumerate}

The rest of the paper is organized as follows. In Section~\ref{sec:model}, we introduce the proposed asymmetric CCA model. In Section \ref{sec:estimation}, we introduce the associated estimator and in Section \ref{sec:theory}, we explore its theoretical properties. In Section \ref{sec:application}, we apply our method to data from the Human Connectome Project to study dynamic functional connectivity. Simulation studies, proofs, and more technical details are left to the supplementary materials.

\section{Model}\label{sec:model}

\subsection{Elements of Riemannian geometry}\label{frechet_mean_section}
Let $\M$ be an $M$-dimensional Riemannian manifold and let $\txm$ denote the tangent space at a point $x \in \M$ equipped with Riemannian metric $\langle \cdot , \cdot \rangle_{x}$. Moreover, for any $x \in \M$, denote the exponential map by $\Exp_x: U \rightarrow \M$, where $U \subset \txm$ is an open set containing the origin that guarantees that this map is a bijection onto its range $\operatorname{Im}\left(\operatorname{Exp}_{x} \right)$. The logarithmic map at $x$, denoted by $\operatorname{Log}_x: \text{Im}\left (\Exp_x \right ) \rightarrow U$ is the inverse of the exponential map $\Exp_x$. We denote by $d_{\M}(\cdot,\cdot)$ the Riemannian distance function on $\M$, which generalizes the Euclidean distance to manifolds. We refer to \textcites{Lee2012}{lee2018introduction} for an introduction to the differential geometric concepts used in this work.

In our final application, $\M$ will represent the non-Euclidean manifold of SPD matrices equipped with the affine-invariant metric \parencites[see, e.g., ][]{fletcher2007riemannian}{pennec2019riemannian}. This is particularly well-suited to studying covariance matrices thanks to its natural affine-invariant property: given two random vectors $X,Y \in \rone^p$, let $\sigmax$ and $\sigmay$ denote their covariance matrices; for any rotation matrix $R \in \rone^{p \times p}$, the geodesic distance on the manifold is invariant to the rotation of $X$ and $Y$ by $R$, i.e. $d_{\M}\bl \sigmax,\sigmay \br =  d_{\M}\bl \Sigma_{RX},\Sigma_{RY} \br $.

The affine-invariant metric at $P \in \M$ between $W,Z \in T_P \M$ is defined as $\inner{W}{Z}_{\M} = \operatorname{tr}\bl P^{-1}WP^{-1}Z\br$. Let $\operatorname{exp}$ and $\operatorname{log}$ denote the matrix exponential and logarithm, defined here on the sets of symmetric matrices and positive definite matrices, respectively, and let $\norm{\cdot}_F$ denote the Frobenius norm of a matrix. Then, the affine-invariant Riemannian distance is defined as $d_{\M}(P, Q) = \|\text{log}(P^{-1/2} Q P^{-1/2})\|_F$. The logarithmic map $\operatorname{Log}_P(Q) = P^{1/2}\operatorname{log}\bl P^{-1/2}QP^{-1/2} \br P^{1/2}$ will allow us to compute unconstrained tangent space representations of our data, i.e., symmetric matrices. Roughly speaking, the tangent space representations allow us to apply simple Euclidean mathematical operations without breaking the geometry of the space of SPD matrices and the exponential map $\Exp_P(W) = P^{1/2}\operatorname{exp}\bl P^{-1/2}WP^{-1/2} \br P^{1/2}$ will allow us to map tangent space elements back to the manifold $\M$. In this case, the exponential and logarithmic maps $\operatorname{Exp}$ and $\operatorname{Log}$ are global bijections between $\M$ and the space of symmetric matrices. Alternative metrics that accommodate positive semi-definite covariances have been defined, for instance, in \textcites{dryden2009noneuclidean}{ pigoli2014distances}{masarotto2019procrustes}.

Next, we present the mathematical tools necessary to model Riemannian-valued functions. Let $\Tsc$ be a compact subset of $\mathbb{R}$ and let $\mu: \Tsc \rightarrow \M$ be a sufficiently smooth curve on $\mathcal{M}$. A vector field $V$ along $\mu$ is a map from $\mathcal{T}$ to the tangent bundle $T \mathcal{M}$ such that $V(t) \in T_{\mu(t)} \mathcal{M}$ for all $t \in \mathcal{T}$. The collection of vector fields $V$ along $\mu$ defines a vector space. Define $\Jmu$ to be the space of square integrable vector fields $V$ along $\mu$ equipped with inner product $\innerdouble{U}{V}_{\mu}:=\int_{\mathcal{T}}\langle V(t), U(t)\rangle_{\mu(t)} \mathrm{d}t$ and induced norm defined by $\|\cdot\|^2_\mu = \innerdouble{\cdot}{\cdot}_{\mu}$, where $U$ and $V$ are both vector fields along $\mu$. Then, $\Jmu$ is a separable Hilbert space \parencite{lin2019intrinsic}.

For a curve $\mu$ and Riemannian-valued function $y: \Tsc \to \M$, we denote as $\Logy$ the function $t \mapsto \Log_{\mu(t)} y(t)$. Similarly, for a vector field $V$ along $\mu$, we denote as $\Exp_{\mu}V$ the function $t \mapsto \Exp_{\mu(t)} V(t)$. In our setting, $y$ will be random, and $\mu$ will represent the mean of $y$. Under appropriate assumptions, the vector field ${\Logy}$ along $\mu$ will be a random element of $\Jmu$, which intuitively represents a linearized and centered version of $y$. Indeed, if $\M$ is a Euclidean space $\M = \rone^d$, then $\Log_{\mu(t)} y(t) = y(t) - \mu(t)$ for every $t \in \Tsc$. 

Later, we will need to compare vector fields along different curves $\mu$ and $\hat{\mu}$. To this purpose, following \textcite{lin2019intrinsic}, we introduce the parallel transport operator. We denote the parallel transport operator on $\M$ along geodesics as $\mathcal{P}_{x,p}: T_x \M \to T_p \M$. A fundamental property of this operator is that it preserves inner products of tangent vectors, i.e., for any $u,v \in T_x \M$, $\inner{u}{v}_x = \inner{\mathcal{P}_{x,p}u}{\mathcal{P}_{x,p}v}_p$. We can then define parallel transport for vector fields $U,V$ along curves $f,h: \Tsc \to \M$. Specifically, given $U \in L^2(Tf)$ and $V \in L^2(Th)$, we define $\Gamma_{f,h}U \in L^2\bl T h \br $ as the map $t \mapsto \mathcal{P}_{f(t),h(t)}U(t)$. Therefore, $\Gamma_{f,h}$ can be viewed as a map from $L^2(Tf)$ to $L^2(Th)$. Therefore, while $U$ and $V$ cannot be `compared' directly since for every $t$, $U(t)$ and $V(t)$ may belong to different tangent spaces, we can compare $\Gamma_{f,h}U$ and $V$, since they are both elements of $L^2(Th)$. In particular, $\norm{\Gamma_{f,h}U - V}_h$ quantitatively describes the difference between $U$ and $V$.
We refer to Proposition 2 of \textcite{lin2019intrinsic} for additional properties of the parallel transport operator.
\subsection{Modeling Riemannian-valued data}\label{sec:data}

Let $(y,X)$ be a pair of random variables, where $X \in \rone^p$ is a zero-mean random vector, with covariance $\sigmax \in \rone^{p \times p}$, representing the high-dimensional multivariate variables, and the process $y$ is a Riemannian-valued random process with continuous sample paths.
We assume that $\forall x \in \M, \forall t \in \Tsc$ we have $\E \left [ d_{\M}^2(y(t), x) \right ] < \infty$.

Next, we define the Fr\'echet mean of the process $y$ on $\M$ as
$$
\mu(t)=\underset{x \in \mathcal{M}}{\arg \min \text{ }} \E \left [ d_{\M}^2(y(t), x) \right ].
$$
We assume that the Fr\'echet mean $\mu(t)$ exists and is unique for every $t \in \Tsc$, and $\mu$ is a continuous function. For more details on the Fr\'echet mean, see \textcite{bhattacharya2003large}. Following \textcite{lin2019intrinsic}, we assume
\begin{equation*}
    \operatorname{Pr}\left\{\textrm{For all } t \in \mathcal{T}: y(t) \in \operatorname{Im}\left(\operatorname{Exp}_{\mu(t)}\right)\right\}=1,
\end{equation*}
which ensures that $\text{Log}_{\mu(t)} y(t)$ is defined almost surely for all $t \in \Tsc$. This condition is superfluous for many common manifolds, for example whenever $\operatorname{Exp}_x$ is surjective onto $\M$ for all $x \in \M$. For instance, this holds on the manifold of SPD matrices equipped with the affine-invariant metric.

Let the tensor product $U \otimes V: \Jmu \to \Jmu$,  between $U,V \in \Jmu$,  be defined as $\bl U \otimes V \br \bl W \br = \innerdouble{U}{W}_{\mu}V$ for all $W \in \Jmu$. If $\E \left [ \norm{\Logy}_{\mu}^2 \right ] < \infty$, then the covariance function $\C$ of $\Logy$ is defined as $\C = \E \left [ \Logy \otimes \Logy \right ]$ and is nonnegative and trace class. Therefore, it admits the eigendecomposition 
\begin{equation} \label{eq:operator_eigendecomposition}
    \C = \sumj \omega_j \phi_j \otimes \phi_j,
\end{equation} 
with $\omega_j$ a sequence of real numbers converging to $0$, and $\phi_j \in \Jmu$ satisfying $\innerdouble{\phi_j}{\phi_k}_{\mu} = \delta_{jk}$, where $\delta_{jk} = 1$ if $j=k$, and $0$ otherwise. The functions $\{\phi_j\}$ are called the population loading functions, or population principal components, of $\Logy$. Moreover, with probability one, we have that the process $\Logy$ admits a Principal Component expansion
\begin{equation*}
    \label{eq-logy}
    \Logy = \sumj Y_j \phi_j,
\end{equation*} 
where $Y_j = \innerdouble{\phi_j}{\Logy}_{\mu}$ are pairwise uncorrelated random variables, and satisfy $\E \left [ Y_j \right ] = 0$ and $\Var \bl  Y_j \br = \omega_j$. The variables $Y_j$ are called the population principal scores. For further details on the principal component basis and eigendecomposition of the $\mathcal{C}$, see Lemma \ref{closure_image_facts_thm} in the supplementary materials.

\subsection{Asymmetric Riemannian CCA}
\label{Asymmetric_CCA_section}
\begin{figure}[!htb]
\includegraphics[width=1.0\textwidth]{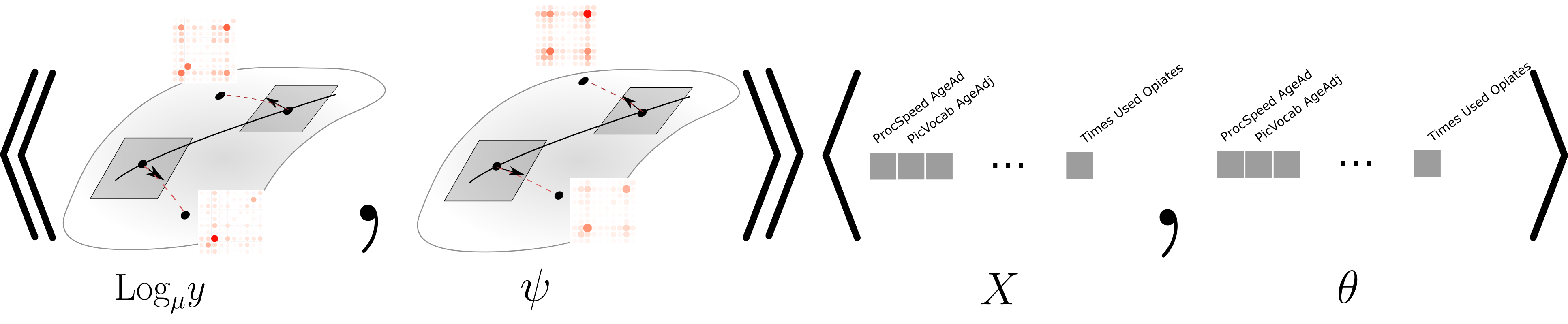}
\centering
\caption{In this figure, we illustrate the process of projecting the Riemannian-valued functional data and the high-dimensional data to define maximally correlated variables. We leverage tools from differential geometry to compute linear tangent representations $\text{Log}_{\mu} y$ of the temporally-indexed Riemannian-valued data $y$, which are equipped with a notion of inner product $\llangle \cdot, \cdot \rrangle_\mu$, that is, a projection operator. For the multivariate data, we use the conventional notion of projection, i.e., the Euclidean inner product. We therefore seek $\psi$ and $\theta$ whose respective data projections define maximally correlated variables.
}
\label{fig:model}
\end{figure}

In this section, we introduce our proposed asymmetric CCA model, which can be naturally formalized by mirroring the multivariate and functional versions \parencite{he2010functional} of the problem. We define the first canonical direction pair $(\psi_1,\theta_1)$ as a solution, if one exists, to the following problem
\begin{equation}\label{eq:cca_intuition}
    \maximize{\textrm{ Corr}^2 \left (\innerdouble{\Logy}{\psi}_{\mu},\inner{X}{\theta}\right )}{\psi \in \Jmu, \textrm{ } \theta \in \rone^p},
\end{equation}
where $\inner{\cdot}{\cdot}$ is the Euclidean inner product in $\rone^p$. Analogously, we can define the subsequent pairs $(\psi_k,\theta_k)$ to maximize the same objective function, with the condition that each pair is orthogonal to the previous ones, namely, $\innerdouble{\psi_k}{\calC \psi_{k'}}_{\mu} = \delta_{kk'}$ and $\theta_{k}^{\T} \sigmax \theta_{k'} = \delta_{kk'}$.
When they exist, we refer to $\psi_k$ as the $k$th canonical function, and to $\theta_k$ as the $k$th canonical vector. 
Given the canonical function $\psi_k \in \Jmu$, we can map it back to the original space via the exponential map. This procedure is illustrated in Figure~\ref{fig:model}.

While equation \eqref{eq:cca_intuition} provides an intuitive formulation of the canonical correlation problem, it has been noted in \textcite{cupidon_properties_2008} that the maximum of this problem may not be attained by any $\psi \in \Jmu$, $\theta \in \rone^p$. Despite this, the problem can still be reformulated with respect to the pair of canonical variables $\bl U,V \br = \bl \innerdouble{\Logy}{\psi}_{\mu},\inner{X}{\theta} \br $, resulting in the following maximization problem:
\begin{equation}\label{eq:cca_infinite}
\normalfont
    \maximize{\textrm{ Corr}^2 \left ( U,V \right )}{U \in \overline{\mathcal{U}}, \textrm{ } V \in \overline{\mathcal{V}}},
\end{equation}
where $\mathcal{U} = \{\innerdouble{\Logy}{\psi}_{\mu} : \textrm{ }\psi \in \Jmu \}$, $\mathcal{V} = \{\inner{X}{\theta} : \textrm{ }\theta \in \rone^p \}$, and $\overline{\mathcal{U}}$ and $\overline{\mathcal{V}}$ are appropriate closures of $\mathcal{U}$ and $\mathcal{V}$, respectively. This guarantees that an optimal canonical variable pair $(U,V)$ does exist. We emphasize that they \textit{cannot} necessarily be written in terms of the canonical directions, i.e., it does not necessarily hold that $U \in \ubar$ can be written as $U = \innerdouble{\Logy}{\psi}_{\mu}$ for some $\psi \in \Jmu$. To simplify the exposition, we defer the details of this formulation to Section \ref{appendixcca} of the supplementary materials (also see Remark \ref{remark:no_canonical_vectors}). In Theorem \ref{thm:consistency_of_scores_main}, we show that despite the nonexistence of population canonical directions for functional data, the \textit{canonical variables} corresponding to estimated canonical directions are consistent for $U$ and $V$. To investigate the theoretical properties of the \textit{canonical directions}, and in particular, to ensure that their population counterparts are well-defined, we make the following assumption. 
\begin{assumption} \label{remark:finite_dim_assumption}
There exists a complete orthonormal system $\{ \zeta_i \}_{i = 1}^{\infty}$ for $\Jmu$, and a set of indices $I \subset  \left \{1 , 2,\ldots \right \}$, with finite cardinality $\left | I \right | \equiv \dcorr \leq p$, such that
\begin{align*}
&\operatorname{Corr}\bl X_k,\innerdouble{\Logy}{\zeta_j}_{\mu} \br = 0, \qquad k=1, \ldots p, \forall j \in I^c,\\
&\operatorname{Corr}\bl\innerdouble{\Logy}{\zeta_i}_{\mu}, \innerdouble{\Logy}{\zeta_j}_{\mu} \br = 0, \qquad \forall i \in I, \forall j \in I^c,
\end{align*}
where $I^c$ denotes the complement of $I$ in $\{1 , 2,\ldots \}$.
\end{assumption}

Intuitively, Assumption~\ref{remark:finite_dim_assumption} implies that there are only a finite number of basis elements $\{\zeta_i\}_{i \in I}$ that capture the correlation between $X$ and $\Logy$ through the scores $\left \{ \innerdouble{\Logy}{\zeta_i}_{\mu} \right \}_{i \in I}$.
Crucially, through this assumption, we do not limit the dimensionality of the functional data. For more details on this assumption, see also Section \ref{subsec:reduction_to_finite_dim_problem} of the supplementary materials. Let $\{ \zeta_j \}_{j = 1}^{\infty}$ be an orthonormal system for $\Jmu$ satisfying Assumption \ref{remark:finite_dim_assumption}, and reorder the $\{ \zeta_j \}_{j = 1}^{\infty}$ so that $I = \left \{1 ,\ldots \dcorr \right \}$. Define $Y_j \equiv \innerdouble{\Logy}{\zeta_j}$ for $j=1,\ldots \dcorr$, so that the $\{Y_j\}$ are random variables with $\Var \bl Y_j \br < \infty$.
Note that in practice, to define the orthonormal system $\{\zeta_j \}_{j = 1}^{\infty}$, we could employ the principal components analysis in Section~\ref{sec:data}, or alternatively, we could design its basis functions to capture specific features of interest. Next, define $Y = (Y_1, \ldots, Y_\dcorr)$ and let $\sigmay$ be the $\dcorr \times \dcorr$ covariance of $Y$. Let $\norm{\cdot}_2$ denote the Euclidean 2-norm of a vector in $\rone^{\dcorr}$. Without loss of generality, we suppose that $X$ and $Y$ are mean $0$. Under Assumption \ref{remark:finite_dim_assumption}, the following theorem states that the canonical correlation problem in equation \eqref{eq:cca_infinite} is equivalent to solving a suitably formulated finite-dimensional regression problem.

\begin{theorem}
\label{multivariate-cca-theorem}
Under Assumption \ref{remark:finite_dim_assumption}, the CCA model in equation (\ref{eq:cca_intuition}) admits at most $\dcorr$ nontrivial canonical variable pairs $\{(U_k,V_k)\}$, and each pair $(U_k,V_k)$ can be written in terms of the associated canonical directions: $U_k = \innerdouble{\Logy}{\psi_k}_{\mu}$ and $V_k = \inner{X}{\theta_k}$ for some $\psi_k \in \Jmu$ and $\theta_k \in \rone^p$. Additionally, suppose $\sigmax \in \rone^{p \times p}$ and $\sigmay \in \rone^{\dcorr \times \dcorr}$ are invertible. Let $B$ be the solution to the multivariate least-squares problem
\begin{equation}\label{eq:opt}
    \minimize{ \E \left [ \|\sigmay^{-1/2}Y-B^{\T}X \|_2^2 \right ]}{B \in \rone^{p \times \dcorr}},
\end{equation}
and let
\begin{equation}
    B^{\T}\sigmax B = \tilde{H} D^2 \tilde{H}^{\T}
    \label{eq:eig}
\end{equation}
be an eigendecomposition of $B^{\T}\sigmax B$. Define
\begin{align}
    T = B \tilde{H} D^{-1} \in \rone^{p \times \dcorr},\\ 
    H = \sigmay^{-1/2}\tilde{H} \in \rone^{\dcorr \times \dcorr}.
    \label{eq:TH}
\end{align}
Then, the $k$th column of $H$, $\eta_k$, characterizes the $k$th canonical function $\psi_k$ through $\psi_k = \sum_{j=1}^\dcorr \eta_{kj} \zeta_j$, and the $k$th column of $T$ is the $k$th high-dimensional canonical vector $\theta_k$. Moreover, the optimum values attained by the maximization problem in equation \eqref{eq:cca_infinite} are the diagonal entries of $D^2$, which we denote by $\gamma_1^2, \ldots \gamma_\dcorr^2$.
\end{theorem}

The proof of Theorem \ref{multivariate-cca-theorem} can be found in Section~\ref{subsec:proof_of_main_thm} of the supplementary materials. This suggests a novel methodology for deriving estimates of the canonical functions $\{\psi_k\}$ and the canonical vectors $\{\theta_k\}$. This entails defining a subspace, spanned by $\{ \zeta_j \}_{j = 1}^{\dcorr}$, onto which the tangent space representations of the functional data are projected. Subsequently, the canonical functions and vectors can be characterized by the equations \eqref{eq:opt}-\eqref{eq:TH}, using empirical estimates in place of the theoretical population values. Note that in practice, we need to choose the dimension of this subspace by choosing a number of $\zeta_j$ to use, which we denote by $d$. In Theorem \ref{thm:consistency_of_scores_main} we show that for consistency of the canonical directions, all that is required is choosing $d \geq \dcorr$; we do not need to choose $d$ exactly equal to $\dcorr$.

Crucially, as opposed to other methods in the literature \parencites[see, e.g.,][]{chen2013sparse}{gao2017sparse}, the proposed model circumvents the direct estimation of $\Sigma_X^{-1}$, i.e., the precision matrix of the variable $X$, which is a notoriously difficult problem in high dimensions as it can be estimated only under restrictive structural assumptions. Our strategy yields interpretable results by enforcing sparsity directly on the canonical directions $\{\theta_k\}$ through an additional penalty term on the estimate of $B$. The complexity of the functional canonical direction is controlled by projecting the functional data on a finite-dimensional subspace. Such an approach leverages the smooth nature of the functional data (and its tangent space representation) --- which is reflected in the eigenvalues of the covariance rapidly decaying to zero --- suggesting that such a projection can serve as an efficient and interpretable approximation.

\subsubsection{On the existence of canonical directions and connections with partial least-squares}
In Theorem \ref{thm:consistency_of_scores_main} we show that even without Assumption \ref{remark:finite_dim_assumption}, given a new independent sample $(\xtest,\ytest)$, if we use our canonical direction estimates $\hat \psi_k$ and $\hat \theta_k$ to form estimated scores $\bl \hat U_k, \hat V_k \br \equiv \bl \innerdouble{\Log_{\hat \mu}\ytest}{\hat \psi_k}_{\hat \mu},\langle \hat \theta_k, \xtest \rangle\br$, where $\hat \mu$ has also been estimated from the data, then $(\hat U,\hat V)$ converge to the true scores $(U,V)$ given a sufficiently large $N$.
However, it should be noted that similar approximation results do not exist in general for the population canonical directions $\psi_k$ and $\theta_k$, which, as shown in \textcite{cupidon_properties_2008}, may not even be well defined. In other words, a maximizer of the population problem in equation (\ref{eq:cca_intuition}) may not be attained in $\Jmu$. This highlights that it is crucial for practitioners to be cautious in their interpretation of results derived from a CCA model. While the estimated canonical variables can generally be regarded as estimates of the population counterparts, the canonical directions should only be interpreted as estimates of the population canonical directions under a $d$-dimensional subspace approximation of the process and not necessarily of the underlying infinite-dimensional process, which, as mentioned earlier, may not even be well-defined. Table~\ref{dcorr_table} summarizes the conditions under which the quantities of interest are well-defined.
\begin{table}[h]
\centering
\begin{tabular}{|l|l|l|}
\hline
     & $\dcorr < \infty$ & $\dcorr = \infty$ \\ \hline
Canonical directions $(\theta_k,\psi_k)$   &   $\checkmark$  &  $\times$     \\ \hline
Canonical variables $(U_k,V_k)$         &      $\checkmark$  &   $\checkmark$  \\ \hline
\end{tabular}
\caption{Existence of canonical directions and variables, depending on $\dcorr$ in Assumption \ref{remark:finite_dim_assumption}.}\label{dcorr_table}
\end{table}

We also remark on how the situation changes when considering the partial least squares problem, as opposed to the canonical correlation problem. It is straightforward to show that by replacing equation \eqref{eq:opt} with 
\begin{equation}\label{eq:opt_pls}
    \minimize{ \E \left [ \|Y-B^{\T}X \|_2^2 \right ]}{B \in \rone^{p \times d}},
\end{equation}
namely, by omitting the standardization of the random coefficients in the basis expansion, the proposed reformulation defines a partial least squares model (see, e.g., \textcite{boulesteix2007partial}). For this model, the population canonical directions are more broadly defined, and it can be shown that the $d$-dimensional approximation of the functional data results in asymptotically negligible errors for both the canonical variables and directions. Intuitively, by comparing equations (\ref{eq:opt}) and (\ref{eq:opt_pls}), one can see that the functional data enter the partial least squares model via the unnormalized coefficients $Y$ of the first $d$ basis functions, leading to negligible `residual information' as $d$ increases, due to the compactness of the covariance of the functional data. On the other hand, in the CCA model, the functional data are incorporated through the normalized coefficients $\sigmay^{-1/2}Y$. This normalization step prevents the residual information from becoming negligible and may cause the canonical directions to diverge as $d$ increases.

In this work, we focus on CCA due to its ability to detect components of $y$ that are small in magnitude but correlated with $X$, as opposed to partial least squares models, which are sensitive to the scale of the signal. Such a feature is particularly critical in neuroimaging applications \parencite{wang2020finding}.

\section{Estimation}\label{sec:estimation}
Suppose we are given $N$ observations 
\begin{equation*}
(y_i,X_i), \quad i=1, \ldots, N,
\end{equation*}
each being a realization of the pair $(y,X)$. We propose the following estimation procedure, outlined in four steps.

\begin{enumerate}
\item[] \noindent \textbf{Step A: RFPCA}

\noindent We first compute the sample version of the Fr\'echet mean, defined as
\begin{equation*}
\hat{\mu}(t) = \underset{x \in \mathcal{M}}{\arg \min \text{ }} 
\frac{1}{N} \sum_{i=1}^N d_{\mathcal{M}}^2\left(y_i(t), x\right).
\end{equation*}
We then estimate the tangent space representations of the functional data observations using $\Logyi \in L^2\bl T\hat{\mu} \br$.
Next, we define an orthonormal basis for the tangent space representations using the RFPCA framework proposed in \textcites{dai2017optimal}{lin2019intrinsic}{shao2022intrinsic}
to estimate a data-driven basis $\{ \hat{\phi}_j \}_{j=1}^d$. Note that $d$ is not necessarily equal to $\dcorr$ from Section \ref{Asymmetric_CCA_section}. Specifically, we estimate the tangent-space covariance operator $\calC$ using the sample covariance function $\hat \calC = \frac{1}{N} \sum_{i=1}^N \Logyi \otimes \Logyi$. Each population loading function $\phi_j$ and associated eigenvalue $\omega_j$ can be estimated using the eigenfunction $\hat \phi_j$ and eigenvalue $\hat \omega_j$ of $\hat \calC$. The empirical Principal Component expansion of $\{\Logyi\}$ is then given by
\begin{equation*}
    \Logyi = \sum_{j=1}^d \hat Y_{ij} \hat \phi_j,
\end{equation*} 
where $\hat Y_{ij} = \innerdouble{\hat \phi_j}{\Logyi}_{\hat \mu}$ are the PC scores. Here we assume that the rank of the Principal Component expansion $d$ is such that $d < \operatorname{min}(p,N)$. For completeness, in Section~\ref{sec:apdx:irfpca} of the supplementary materials,
we provide a detailed description of the RFPCA algorithm, including a computationally efficient explicit basis construction for the space of SPD matrices equipped with the affine invariant metric.

\item[] \noindent \textbf{Step B: Regularized regression}

\noindent Next, we use the scores $\hat Y_{ij}$ to represent the manifold-valued functional data and estimate the canonical directions leveraging the characterization in Theorem~\ref{multivariate-cca-theorem}. We let $\mathbb{X} \in \mathbb{R}^{N \times p}$ and $\hat{\mathbb{Y}} \in \mathbb{R}^{N \times d}$ denote the data matrices $(X_{ij})_{ij}$ and $(\hat Y_{ij})_{ij}$, respectively, where our notation emphasizes that the entries of $\hat{\mathbb{Y}}$ are estimates.

Define $\sigmayhat = \frac{1}{N}\hat{\mathbb{Y}}^T \hat{\mathbb{Y}}$ and $\sigmaxhat = \frac{1}{N}\mathbb{X}^T\mathbb{X}$. We estimate the matrix $B$ in equation \eqref{eq:opt} using $\hat B$, which is derived by solving the following group lasso problem:
\begin{equation}\label{eq:grp_lasso}
\hat B = \argmin{ \frac{2}{N}\left \|  \hat{\mathbb{Y}} \sigmayhat^{-1/2} - \mathbb{X}B  \right \|_F^2 + \lambda \onetwo{B}}{B \in \mathbb{R}^{p \times d}},
\end{equation}
where $\| \cdot\|_F$ denotes the Frobenius norm and $\onetwo{B} = \sum_{i=1}^p \norm{b_i}_2$ is a group lasso penalty. Here, $b_i$ refers to the $i$th row of $B$. Note that the first term of the minimization problem in equation~(\ref{eq:grp_lasso}) is an empirical approximation of that in equation~\eqref{eq:opt}, where the data matrices $\mathbb{X}$ and $\hat{\mathbb{Y}}$ replace the random variables $X$ and $Y$, respectively.

\item[] \noindent \textbf{Step C: Eigenanalysis}

\noindent Given $\sigmaxhat^{1/2}\hat{B}$, we then compute its right singular vectors $\hat{\tilde{H}} \in \rone^{d \times d}$ and singular values matrix $\hat{D} \in \rone^{d \times d}$, that is,
\begin{equation*}
\hat{B}^T \sigmaxhat \hat{B} = \hat{\tilde{H}} \hat{D}^2 \hat{\tilde{H}}^T.
\end{equation*}

\item[] \noindent \textbf{Step D: Estimates computation}

\noindent We define
\begin{align}\label{eq:hatTH}
    \hat{T} = \hat{B} \hat{\tilde{H}} \hat{D}^{-1},\\
    \hat{H} = \sigmayhat^{-1/2}\hat{\tilde{H}}, \label{eq:hatTH2}
\end{align}
where $\hat{T} \in \rone^{p \times d}$ and $\hat{H} \in \rone^{d \times d}$ are estimates of $T$ and $H$, respectively. Then, $\hat{T} = \left[ \hat \theta_1, \ldots, \hat \theta_d\right]$ is a matrix whose columns $\hat{\theta}_k$ are the estimates of $\theta_k$, and $\hat{H} = \left[\hat \eta_1, \ldots, \hat \eta_d \right]$ is a matrix whose columns $\hat{\eta}_k$ are the estimates of $\eta_k$. The estimated canonical functions are therefore given by $\hat{\psi}_k = \sum_{j=1}^d \hat{\eta}_{kj} \hat{\phi}_j$, for $k = 1, \ldots d$, resulting in the estimated canonical functions and vectors $(\hat{\psi}_k, \hat \theta_k)$.
\end{enumerate}

\begin{algorithm}[hbt!]
\caption{Asymmetric Sparse-Functional CCA}\label{alg:asymmetric_sparse_fun_cca}
\hspace*{\algorithmicindent} \textbf{Input:} Pairs $(y_i, X_i)_{i = 1, \ldots N}$ of manifold-valued functional data and high-dimensional data; number of principal components $d$ chosen for the manifold-valued functional data.
\begin{algorithmic}
\State 1. Obtain $\hat{\phi}_j$, $\hat{\omega}_j$, for $j = 1, \ldots d$, and $\mathbb{Y}$ applying Intrinsic RFPCA to $(y_i)_{i = 1, \ldots N}$.
\State 2. Compute $\sigmayhat = \operatorname{diag}\bl \hat{\omega}_j \br $ and $\sigmaxhat = \frac{1}{N}\mathbb{ X}^{\T} \mathbb{ X}$.
\State 3. Compute $\hat{B}$ solving the group lasso problem in equation \eqref{eq:grp_lasso} using the \texttt{glmnet} package \parencite{glmnet}.
\State 4. Compute $\hat{H} = \left[\hat \eta_1, \ldots, \hat \eta_d \right]$ and $\hat{T} = \left[ \hat \theta_1, \ldots, \hat \theta_d\right]$ in equations \eqref{eq:hatTH} and \eqref{eq:hatTH2}.
\State 5. Compute the estimated canonical functions $\hat{\psi}_k = \sum_{j=1}^d \hat{\eta}_{kj} \hat{\phi}_j$ for $k = 1, \ldots d$.
\State 6. Return $\{\hat{\theta}_k\}_{k=1}^{d}$, the estimated canonical vectors associated with $X$, and $\{\hat{\psi}_k\}_{k=1}^{d}$, the estimated canonical functions associated with $y$.
\end{algorithmic}
\end{algorithm}

The sparsity-promoting regularization norm employed in equation \eqref{eq:grp_lasso} encourages entire rows of the matrix $B$ to be set to zero. From the equation $\hat{T} = \hat{B}\hat{\tilde{H}}\hat{D}^{-1}$, it follows that the corresponding rows of $\hat{T}$ will also be zero. This yields canonical vectors $\{\theta_k\}$ with a group sparsity structure, meaning they share identical sparsity patterns.

The main steps of the estimation procedure are summarized in Algorithm~\ref{alg:asymmetric_sparse_fun_cca}, which we refer to as \textit{asymmetric sparse-functional CCA}.

\subsection{Selection of hyperparameters}
In the RFPCA step of Algorithm~\ref{alg:asymmetric_sparse_fun_cca}, it is necessary to select the number of principal components $d$, which needs to be less than or equal to both $N$ and $p$. Additionally, the regularization parameter $\lambda$ needs to be chosen in the regularized regression step. We recommend using cross-validation to select these parameters. Specifically, for each choice of $d$, the optimal $\lambda$ in the regression step can be selected via cross-validation, as implemented in the \texttt{glmnet} package. Then, we select the value of \( d \) that yields the largest out-of-sample (or cross-validated) canonical correlations. For instance, if the user is interested in finding the top $k$ canonical directions, then we recommend using the sum of the first $k$ estimates of out-of-sample canonical correlations.

If the user believes that the functional data has finite rank, a hypothesis testing approach can be used to obtain an upper bound on $d$, for example by adapting the method proposed in \textcite{charkaborty2022testing}.

\subsection{Special instances}\label{sec:special_inst}
To demonstrate the versatility of our model, we present a few special cases. Although some of these settings are simpler than the motivating neuroimaging application, the proposed method still provides an innovative approach to analyzing such data. 

\begin{itemize}

\item In situations where $y_i \in \mathcal{M}$, meaning our imaging data are manifold-valued observations without a temporal dimension, Algorithm~\ref{alg:asymmetric_sparse_fun_cca} can be adapted by using tangent-space PCA \parencite{marron2021object} rather than RFPCA, similar to the setting considered in \textcite{kim2014canonical}. This model is especially useful for studying static connectivity networks. 

\item When the imaging data take the form of classical functional data, that is $y_i(t) \in \mathcal{M} \subset \mathbb{R}$ for all $t \in \calT$, one can apply Algorithm~\ref{alg:asymmetric_sparse_fun_cca} by replacing RFPCA with classical FPCA \parencites{ramsay2015functional}{ yao2005functional}. In addition, when $y_i(t) \in \mathcal{M} \subset \mathbb{R}^d$, multivariate FPCA can be employed \parencite{happ2018multivariate}.
\end{itemize} 

\begin{algorithm}[!hbt]
\caption{Asymmetric Sparse CCA}\label{alg:asymmetric_sparse_cca}
\hspace*{\algorithmicindent} \textbf{Input:} Pairs $(Y_i, X_i)_{i = 1, \ldots N}$ of low- and high-dimensional data. Let $\mathbb{Y} = (Y_{ij})_{ij}$ and $\mathbb{X} = (X_{ij})_{ij}$.
\begin{algorithmic}
\State 1. Compute $\sigmayhat = \frac{1}{N}\mathbb{Y}^{\T} \mathbb{Y}$ and $\sigmaxhat = \frac{1}{N}\mathbb{ X}^{\T} \mathbb{ X}$.
\State 2. Compute $\hat{B}$ solving the group lasso problem 
\begin{equation}\label{eq:grp_lasso_mult}
\hat B = \argmin{ \frac{2}{N}\left \|  \mathbb{Y} \sigmayhat^{-1/2} - \mathbb{X}B  \right \|_F^2 + \lambda \onetwo{B}}{B \in \mathbb{R}^{p \times d}}
\end{equation}
using the \texttt{glmnet} package \parencite{glmnet}.
\State 3. Compute $\hat{H} = \left[\hat \eta_1, \ldots, \hat \eta_d \right]$ and $\hat{T} = \left[ \hat \theta_1, \ldots, \hat \theta_d\right]$ in equations \eqref{eq:hatTH} and \eqref{eq:hatTH2}.
\State 4. Return $\{\hat{\eta}_k\}_{k=1}^{d}$, the estimated canonical vectors associated with $\{Y_i\}_{i=1}^N$, and $\{\hat{\theta}_k\}_{k=1}^{d}$, the estimated canonical vectors associated with $\{X_i\}_{i=1}^N$.
\end{algorithmic}
\end{algorithm}

Central to the proposed methodology is a CCA model for pairs of observations $(Y_i, X_i)$, where $Y_i \in \rone^d$, $X_i \in \rone^p$, $d \ll N$, and the covariance of $Y_i$ is full-rank. In the imaging setting, we use a dimension reduction model to compute the low-dimensional component. However, this setting may also be of independent interest and plays a crucial role in the development of the theoretical results. Therefore, we outline the algorithm for this particular setting in Algorithm~\ref{alg:asymmetric_sparse_cca}, and we refer to it as \textit{asymmetric sparse CCA}.

In this special case, our approach is closely related to the Eigenvector-CCA model proposed in \textcite{wang2021eigenvectorbased}. Yet, notable differences exist between the two approaches. For example, we ensure that the estimated canonical vectors satisfy the correct orthogonality conditions $\hat{H}^{\T}\sigmayhat \hat{H} = I_d$ and $\hat{T}^{\T}\sigmaxhat \hat{T} = I_d$. Furthermore, our proposed model does not rely on the assumption that the data have been generated from a regression model.

\section{Theory}
\label{sec:theory}

Here, we investigate the convergence properties of the proposed estimators. We first study the asymptotic properties of the asymmetric sparse CCA model outlined in Algorithm~\ref{alg:asymmetric_sparse_cca}, which sets the stage for studying the asymptotic convergence properties of the asymmetric Sparse-Functional CCA model outlined in Algorithm~\ref{alg:asymmetric_sparse_fun_cca}.

\subsection{Estimation error rates for Asymmetric Sparse CCA}
In this section, we state error bounds for the asymmetric sparse CCA model outlined in Algorithm~\ref{alg:asymmetric_sparse_cca}. We assume $N$ observations $Y_i \in \rone^d$ and $X_i \in \rone^p$ are independent copies of the random variables $Y$ and $X$, respectively. We denote with $\gamma_k$ the $k$th canonical correlation attained in the population version of the problem and recall that $T = \left [ \theta_1, \ldots \theta_d \right ] \in \rone^{p \times d}$. Moreover, we denote with $K = \operatorname{max} \left \{i \in \{ 1, \ldots d \}: \gamma_i > 0 \right \}$ the number of nontrivial canonical vectors. To simplify the notation, we use the conventions $\gamma_{d+1}^2=-\infty$ and $\gamma_0^2=\infty$. We use $\operatorname{cond}\bl A \br = \twonorm{A}\twonorm{A^{-1}}$ to denote the condition number of an invertible matrix $A$, and $\twonorm{A}$ to denote the operator norm of $A$, or equivalently, the square root of the largest eigenvalue of $A^{\T}A$. The norm $\twoinfnorm{A}$ denotes the maximum Euclidean norm of the rows of $A$, and $\onetwo{A} = \sum_{i=1}^p \norm{a_i}_2$, where $a_i$ is the $i$th row of $A$. The notation $a \lesssim b$ indicates inequality up to an absolute constant, i.e., there exists an absolute constant $C>0$ such that $a \leq Cb$. Next, we introduce the main assumptions.
\begin{assumption}\label{asm:dist_multivariate}
    The random variables $X$ and $Y$ are strict sub-Gaussian random vectors with invertible covariance matrices $\sigmax$ and $\sigmay$, respectively. Strict sub-Gaussian random vectors are introduced in Definition~\ref{def:strict_sub_gaussian} of the supplementary materials.
\end{assumption}

\begin{assumption}\label{asm:growth_multivariate}
  It holds that $d \leq p$, $d\log(p) = o(N)$, $\operatorname{cond}\bl \sigmay \br^2d = o(N)$, and $\gamma_1 > \ldots > \gamma_K$ are bounded from below and are distinct.  
\end{assumption}

\begin{assumption}
\label{asm:minor}
    The norms $\twoinfnorm{\sigmax},\onetwo{T}$ are bounded from above and are larger than $1$,
    $\twonorm{\sigmax^{-1}}, \twonorm{\sigmay^{-1}} \geq 1$, and $\hat{\eta}^{\T}\sigmayhat^{1/2} \sigmay^{1/2}\eta \geq 0$ for $k=1, \ldots K$.
\end{assumption}

The sub-Gaussian condition in Assumption \ref{asm:dist_multivariate} ensures that $X$ and $Y$ do not have heavy tails, allowing us to use standard concentration results for the estimation of $\sigmax$ and $\sigmay$. Strict sub-Gaussianity \parencite{kereta2021estimating} facilitates the proofs by allowing the sub-Gaussian norm of a random variable and its variance to be used interchangeably.

In Assumption \ref{asm:growth_multivariate}, the condition that $d \log(p) = o(N)$ allows $p$ to grow exponentially in $N/d$ (i.e., $p \lesssim e^{N/d}$) while still retaining consistency of the estimator for the canonical vectors. The critical component of the condition $\operatorname{cond}\bl \sigmay \br^2d = o(N)$ is that $d = o(N)$, which ensures that $\sigmay$ can be estimated at a sufficiently fast rate by its sample estimator $\sigmayhat$. The presence of $\operatorname{cond}\bl \sigmay \br^2$ allows us to show that $\twonorm{\sigmayhat} \lesssim \twonorm{\sigmay}$ and to ignore lower order terms of $\frac{d}{N}$, simplifying the theorem statement. We assume that the correlations $\gamma_1, \ldots \gamma_K$ are distinct in order to estimate each canonical vector separately instead of estimating entire subspaces.

Assumption \ref{asm:minor} is not essential, and mainly serves to simplify the statement of the theorem. Since the canonical vectors are defined only up to a sign, we use the condition $\hat{\eta}^{\T}\sigmayhat^{1/2} \sigmay^{1/2}\eta \geq 0$ to account for the sign ambiguity of the CCA solutions, allowing us to compare the estimates of the canonical vectors with their population counterparts through the differences $\twonorm{\theta_k-\hat{\theta}_k}$ and $\twonorm{\eta_k-\hat{\eta}_k}$.

\begin{theorem} \label{thm:main_theorem_multivariate_case}
 Suppose Assumptions \ref{asm:dist_multivariate}-\ref{asm:minor} hold. Fix $\alpha \in (0,1)$, and for some absolute constant $C>0$, define the regularization parameter in Algorithm~\ref{alg:asymmetric_sparse_cca} as $\lambda = C\sqrt{\frac{d}{N}\log (p\alpha^{-1})}$. Then, with probability $1-\alpha$, we have that, for $k = 1, \ldots K$,
 \begin{align}
        \twonorm{\theta_k-\hat \theta_k}^2 & \lesssim {\bl \frac{d}{N}\log \bl p\alpha^{-1} \br \br}^{1/2}\frac{\gamma_1^2\twoinfnorm{\sigmax}\onetwo{T}^2}{ {\min_{j \neq k} \left |{\gamma_k}^2 - {\gamma_j}^2 \right |^2}}\frac{ \twonorm{\sigmax^{-1}}}{\gamma_k^2},
\end{align}
\begin{align}  
        \twonorm{\eta_k - \hat{\eta}_k}^2 &\lesssim {\bl \frac{d}{N}\log \bl p\alpha^{-1} \br \br} ^{1/2} \frac{\gamma_1^2 \twoinfnorm{\sigmax} \onetwo{T}^2}{{\min_{j \neq k} \left |{\gamma_k}^2 - {\gamma_j}^2 \right |}^2}\twonorm{\sigmay^{-1}},
\end{align}
where $\theta_k$ and $\eta_k$ denote the high- and low-dimensional population canonical vectors, respectively.
\end{theorem}
The proof follows directly from Theorem \ref{thm:multivariate_canonical_vector_bound_probabilistic_slow} in the supplementary materials. We refer to this bound as a ``slow"-rate bound, as it makes fewer assumptions but results in slower convergence rates relative to the sample size $N$. Specifically, we make no sparsity assumptions on the high-dimensional canonical vectors. In Theorem \ref{thm:multivariate_canonical_vector_bound_probabilistic_fast} in the supplementary materials, we provide the ``fast"-rate bound, where under more restrictive assumptions, the term ${\bl \frac{d}{N}\log \bl p\alpha^{-1} \br \br} ^{1/2}$ is replaced by ${ \frac{d}{N}\log \bl p\alpha^{-1} \br}$, similar to what is observed in lasso regression problems \parencite{hastie2015statistical}. The proof of Theorem~\ref{thm:main_theorem_multivariate_case} hinges on two key components: firstly, deterministic group lasso bounds for in-sample prediction error \parencite{gaynanova2020prediction}, and secondly, the rates at which $\twoinfnorm{\Sigma_{XY} - \hat{\Sigma}_{XY}}$ and $\twonorm{B^{\T}(\sigmax - \sigmaxhat) B}$ converge to zero under the sub-Gaussian assumptions for $X$ and $Y$. Here, $\sigmaxhat$ and $\hat{\Sigma}_{XY}$ represent the sample covariance matrices. As an intermediate step in the proof, we show Theorem \ref{thm:operator_norm_bound} in the supplementary materials, which gives similar slow and fast rate bounds for the estimated canonical correlations $\hat{\gamma}_k$.

Under the stated assumptions, the canonical vector estimates are consistent. Moreover, our rates of convergence depend on the dimension of the high-dimensional data, $p$, only through $\log(p)$.
The bounds for the $k$th canonical directions depend on the nearest canonical correlation gaps, resembling those concerning the variance in the PCA literature.

We emphasize that our rates are dependent on $\sigmax$ only through $\twoinfnorm{\sigmax}$, and not $\twonorm{\sigmax}$. The norm $\twoinfnorm{\sigmax}$ can be much smaller than $\twonorm{\sigmax}$, particularly when many of the $X_j$'s are correlated with one another. This property highlights the robustness of the proposed methodology in the high-dimensional setting, where highly correlated covariates are commonplace.

We are able to establish our error bounds for each canonical vector $\theta_k$, $\eta_k$, independently, and these bounds depend on each other only through the norms of the canonical vectors $\onetwo{T}^2$, and through the neighboring canonical correlation gaps. It is also worth noting that the error associated with $\theta$ depends on $\sigmax^{-1}$ but not $\sigmay^{-1}$. Similarly, the error associated with $\eta$ depends on $\sigmay^{-1}$ but not $\sigmax^{-1}$. Hence, $Y$ can be poorly behaved without impacting the estimation of $\theta$, and vice-versa.

\subsection{Estimation error rates for canonical directions from Asymmetric Sparse-Functional CCA} \label{sec:main_manuscript_canonical_vectors_functional_CCA}

In this section, we investigate the asymptotic properties of our proposed estimators $\hat{\psi}_k$ and $\hat{\theta}_k$, outlined in Algorithm~\ref{alg:asymmetric_sparse_fun_cca}, for the canonical functions $\psi_k$ and canonical vectors $\theta_k$. In this setting, the observations are pairs of Riemannian-valued functional data $y_i \in \Jmu$ and high-dimensional multivariate data $X_i \in \rone^p$. Given the technical nature of many of the assumptions, we refer the reader to Assumptions \ref{assumption:manifold}-\ref{assumption:minor} in the supplementary materials for a complete list.

As in the multivariate case, we denote with $\gamma_k$ the $k$th canonical correlation attained in the population version of the problem, and we denote with $K = \operatorname{max} \left \{i \in \{ 1, \ldots \dcorr \}: \gamma_i > 0 \right \}$ the number of nontrivial canonical vectors. We again use the conventions $\gamma_{K+1}^2=-\infty$ and $\gamma_0^2=\infty$. Recall that we denote by $d$ the number of principal components we use in the estimation step and $p$ the dimension of the multivariate data. We denote by $\dcorr$ the dimensionality with which the finite-correlation Assumption \ref{remark:finite_dim_assumption} holds, with $\dcorr \leq d \leq p$. We suppose that the canonical vectors $\{\theta_k\}$ are $s$-sparse with a group sparsity pattern. We let $X_S$ denote the random vector where we omit covariates $\{X_j\}$ that do not contribute to the association structure with $Y$. For the high-dimensional terms to match the speed of convergence of the functional terms, we assume that $\sigmax^{1/2}$ satisfies the group restricted eigenvalue condition RE$(s,3,d)$, introduced in Definition~\ref{def:group_restricted_eigenvalue} in the supplementary materials, with parameter $\kappa = \kappa(s,d,\sigmax^{1/2})$, which yields `fast'-rate bounds.

We do not provide the `slow'-rate bounds as in Theorem \ref{thm:main_theorem_multivariate_case}. The terms resulting from using Intrinsic RFPCA and estimating the Frech\'et mean $\mu$ converge at a rate of $1/N$, while the terms resulting from solving the multivariate CCA problem would converge at a rate of $1/\sqrt{N}$; considering the `slow'-rate bound would amount to ignoring the contribution to the error from the functional estimation steps.
\begin{theorem}
    \label{thm:main_theorem_general_case}
For some absolute constant $C>0$, define the regularization parameter in Algorithm~\ref{alg:asymmetric_sparse_fun_cca} as $\lambda = C\sqrt{\frac{d}{N}\log (p)}$. Then, under Assumptions \ref{assumption:manifold}-\ref{assumption:minor}, for $k=1, \ldots K$, we have
    \begin{align}
        \norm{ \psi_k - \Gamma_{\hat{\mu},\mu}\hat \psi_k}_{\mu}^2 &= O_P \bl \frac{ d^2s \log(p)}{N} \frac{\munorm{\psi_k}^2 \kappa \twoinfnorm{\sigmax} }{\min_{j \neq k} \min \curl \left |{\gamma_k}^2 - {\gamma_j}^2 \right |, \left|{\gamma_k} - {\gamma_j} \right | \curr^2}\br,
    \end{align}
    \begin{align}
        \twonorm{\theta_k - \hat{\theta}_k}^2 &= O_P\bl \frac{ds \log(p)}{N }\frac{\twonorm{\Sigma_{X_S}^{-1}}^{1/2}+ \bl \frac{\gamma_1}{\gamma_k}\br^2 \twoinfnorm{\sigmax} \twonorm{\sigmax^{-1}}\kappa^2 }{\min_{j \neq k} \min \curl \left |{\gamma_k}^2 - {\gamma_j}^2 \right |, \left|{\gamma_k} - {\gamma_j} \right | \curr^2} \br,
    \end{align}
where we have omitted the terms $\Ex{\munorm{\Log_{\mu}y}^4}$ and $\Var \bl \innerdouble{\phi_j}{\Logy}_{\mu} \br$ for $j=1, \ldots d$.
\end{theorem}
The theorem presented here is a special case of Theorems~\ref{thm:canonical_function_error} and \ref{thm:canonical_vector_error} in the supplementary materials. As in Theorem~\ref{thm:main_theorem_multivariate_case}, the rate of convergence depends on $p$ only through the term $\log(p)$ and on $\sigmax$ only through $\twoinfnorm{\sigmax}$. Our rate also depends on the dimensionality of the reduced representation of the functional data, $d$, linearly and quadratically in the estimation of $\theta_k$ and $\psi_k$, respectively. The quadratic term $d^2$ is most likely not tight but arises from our choice to estimate each $\phi_j$ via $\hat{\phi}_j$, individually, rather than estimating subspaces. As in Theorem \ref{thm:main_theorem_multivariate_case}, the convergence rates depend on the neighboring correlation gaps.

It follows from Theorem \ref{thm:main_theorem_general_case} that if terms other than $d$, $s$, $p$, and $N$ are treated as constants, then, if $d^2s \log(p) = o(N)$, we have that $\hat \psi_k$ and $\hat \theta_k$ are consistent estimators for $\psi_k$ and $\theta_k$, respectively. Thus, for the proposed methodology, $p$ is allowed to grow exponentially with respect to $\frac{N}{d^2s}$ (i.e., $p \lesssim e^{\frac{N}{d^2s}}$) and consistency is retained.

\subsection{Estimation error rates for canonical variables from Asymmetric Sparse-Functional CCA}
In this section, we investigate the asymptotic properties of the canonical variables. In general, without an assumption such as Assumption \ref{remark:finite_dim_assumption}, the canonical directions $\psi_k$ and $\theta_k$ do not necessarily exist. However, even in the absence of such an assumption (equivalently, when $\dcorr = \infty$), using our proposed estimators $\hat \psi_k$ and $\hat \theta_k$ we still obtain a form of asymptotic consistency.

Recall that we denote by $d$ the number of principal components used to represent the functional data, and $p$ the dimension of the multivariate data. Given observations $\bl X_i,y_i\br_{i=1}^N$, we use Algorithm \ref{alg:asymmetric_sparse_fun_cca} to obtain the estimates $\hat \mu$, $\left \{ \hat \phi_j \right \}_{j=1}^d$, $\hat H = \left[\hat \eta_1, \ldots, \hat \eta_d \right]$, $\hat{T} = \left[ \hat \theta_1, \ldots, \hat \theta_d\right]$, and $\hat{\psi}_k = \sum_{j=1}^d \hat{\eta}_{kj} \hat{\phi}_j$ for $k = 1, \ldots d$.

We define the out-of-sample scores as follows. Let $\bl \xtest,\ytest \br$ be a new and independent data point drawn from the same distribution as the sample. We define
\begin{equation}
    \bl \hat U_k, \hat V_k \br \equiv \bl\innerdouble{\Log_{\hat \mu}\ytest}{\hat \psi_k}_{\hat \mu}, \inner{\xtest}{\hat \theta_k}\br,
\end{equation}
which represent the canonical scores obtained from the new data point using the canonical vectors estimated from the sample. Moreover, let $(U_k,V_k)$ denote the solution to the infinite-dimensional population problem \eqref{eq:cca_infinite}, defined with respect to $\ytest$ and $\xtest$.

Following Theorem 10.2.3 in \textcite{hsing2015theoretical}, we provide a probabilistic bound for $\E \left [ \bl U_k - \hat U_k\br^2 \left. \right \vert \bl X_i,y_i\br_{i=1}^N \right ]$ and $\E \left [ \bl V_k - \hat V_k\br^2 \left. \right \vert \bl X_i,y_i\br_{i=1}^N \right ]$ as the sample size $N$, the number of selected principal components $d$, and the dimension of the high-dimensional data $p$ go to infinity. We choose this notion of error, where we condition on the sample, because it allows us to derive a result that is comparable to our canonical vector consistency results, while also integrating out the randomness of $\bl\xtest,\ytest\br$.

We make the same assumptions as in Section \ref{sec:main_manuscript_canonical_vectors_functional_CCA}, with the important exception that we no longer make Assumption \ref{remark:finite_dim_assumption}, and therefore allow the case $\dcorr = +\infty$. In other words, we allow for all principal scores of the functional data $y$ to be correlated with the components of $X$. We denote with $\gamma_k^*$ the $k$th canonical correlation attained in the infinite-dimensional population version of the problem \eqref{eq:cca_infinite}, and we denote with $K = \operatorname{max} \left \{i \in \{ 1, \ldots p \}: \gamma_i^* > 0 \right \}$ the number of nontrivial canonical vectors. We use the conventions ${\gamma_{K+1}^*}^2=-\infty$ and ${\gamma_0^*}^2=\infty$.

In Theorem \ref{thm:consistency_of_scores_main}, which we are about to present, the operator $\mathscr{C}_{12}$ appears, which is a cross-covariance operator containing information about the correlation between $\Logy$ and $X$. We defer its definition to equation \eqref{eq:def_of_C_12} in \ref{sec:finite_infinite_CCA_theory}, due to its technical nature. We also employ its rank-$d$ principal component approximation, which we denote as $\mathscr{C}_{12}^{(d)}$. The norm of the difference $\norm{\mathscr{C}_{12} - \mathscr{C}_{12}^{(d)}}$ then represents how well our $d$-dimensional subspace captures the true correlation structure between the infinite-dimensional functional data $y$ and high-dimensional data $X$.

\begin{theorem}\label{thm:consistency_of_scores_main}
Let $(U_k,V_k)$ be the solution pair to the problem in the infinite-dimensional population version of the problem \eqref{eq:cca_infinite} between $\ytest$ and $\xtest$. Let $\bl \hat U_k, \hat V_k \br \equiv \bl \innerdouble{\Log_{\hat \mu}\ytest}{\hat \psi_k}_{\hat \mu},\inner{\xtest}{\hat \theta_k}\br$, where $\hat \theta_k$, $\hat \eta_k$ and $\hat \mu$ have been estimated via Algorithm \ref{alg:asymmetric_sparse_fun_cca}. For some absolute constant $C>0$, define the regularization parameter in Algorithm~\ref{alg:asymmetric_sparse_fun_cca} as $\lambda = C\sqrt{\frac{d}{N}\log (p)}$.
Then, under Assumptions \ref{assumption:manifold}-\ref{assumption:minor}, with the exception of Assumption \ref{remark:finite_dim_assumption}, for $k=1, \ldots K$, we have
\begin{align}
     &\max \left \{ \E \left [ \bl U_k - \hat U_k\br^2 \left. \right \vert \bl X_i,y_i\br_{i=1}^N \right ], \E \left [ \bl V_k - \hat V_k\br^2 \left. \right \vert \bl X_i,y_i\br_{i=1}^N \right ] \right \}
     \\
    &=O_P\bl\frac{{\gamma_1^*}^2 \norm{\mathscr{C}_{12} - \mathscr{C}_{12}^{(d)}}^2}{\min_{j \neq k} \left |{\gamma_k^*}^2 - {\gamma_j^*}^2 \right |} + \frac{ ds\log p}{N}\frac{ \twoinfnorm{\sigmax}\kappa }{\min_{j \neq k} \min \curl \left |{\gamma_k^*}^2 - {\gamma_j^*}^2 \right |, \left|{\gamma_k^*} - {\gamma_j^*} \right | \curr^2}  \br,
\end{align}
where we have omitted the terms $\Ex{\munorm{\Log_{\mu}y}^4}$ and $\Var \bl \innerdouble{\phi_j}{\Logy}_{\mu} \br$ for $j=1, \ldots d$.
\end{theorem}

The theorem presented here is a special case of Theorem~\ref{thm:consistency_of_scores} in the supplementary materials, whose proof can be found in Section~\ref{appendix:consistency_of_scores}. Here, we recover the classical trade-off in selecting $d$: increasing $d$ decreases the first term (a “bias” term arising from approximating the infinite-dimensional problem using a finite number of principal components), while increasing the second term (a “variance” term due to estimation error in solving the sample version of the finite-dimensional problem).

Comparing the variance term of these rates with the rates in Theorem~\ref{thm:main_theorem_general_case} for the canonical directions, we observe one primary difference. Focusing on the canonical direction $\theta_k$ for simplicity, we see that the $\Sigma_X^{-1/2}$ factors have disappeared. These factors are necessarily present in the canonical vector rates because the $\theta_k$ are derived by taking unit-norm vectors and scaling them by $\Sigma_X^{-1/2}$. The canonical variables, on the other hand, are defined from the canonical directions as $\theta_k^{\T} X$, where the $\Sigma_X^{-1/2}$ hidden in the definition of $\theta_k$ “cancels out” with the $\Sigma_X^{1/2}$ in $X$, removing the factor $\Sigma_X^{-1/2}$ from the canonical variable rates.

\section{Application to dynamic functional connectivity}
\label{sec:application}
\subsection{Data and preprocessing}
We analyze resting-state fMRI images from 1003 subjects in the Human Connectome Project dataset \parencite{vanessen2012human}. Throughout the duration of these 15-minute fMRI scans, participants were at rest and not engaging in any specific activities. Details on the acquisition process can be found in \textcites{glasser2013minimal}{smith2013restingstate}. The fMRI images have been pre-processed using the minimal pre-processing HCP pipeline \parencite{glasser2013minimal}, including spatial
artifact, distortion removal, and mapping onto a common reference template \parencite{smith2013restingstate}.

\begin{figure}[!htb]
\includegraphics[width=1.0\textwidth]{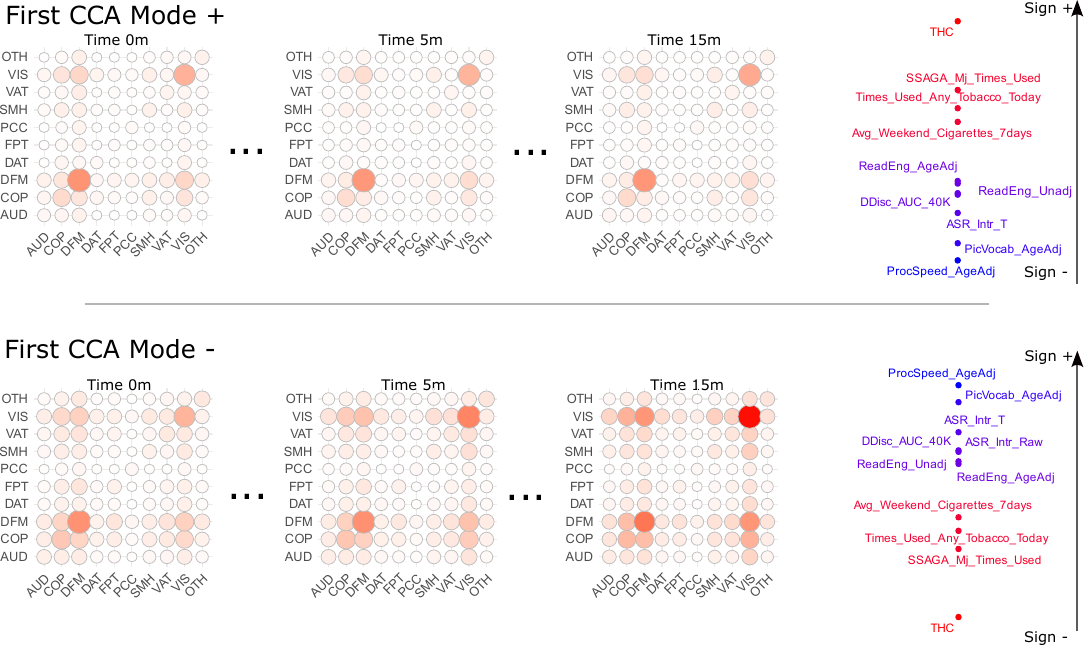}
\centering
\caption{This figure illustrates the first mode of covariation between dynamic connectivity and behavioral measures. On the top panel, we show $\left(\Exp_{\hat{\mu}}\left( -c \hat{\psi}_1 \right), -c \hat \theta_1 \right)$, which we refer to as `First CCA Mode +', on the bottom panel we show $\left(\Exp_{\hat{\mu}}\left( +c \hat{\psi}_1 \right), +c \hat \theta_1 \right)$, which we refer to as `First CCA Mode -'. These represent two extremities of the spectrum identified by the first mode of covariation. Within each panel, we show the canonical function of SPD covariances $\Exp_{\hat{\mu}}\left( \pm \hat{\psi}_1 \right)$ at three different times, and a subset of the selected entries of the canonical vector $\pm \hat \theta_1$. The depicted mode of covariation suggests that subjects with an increasing variance over time within the visual (VIS) and default mode (DFM) functional systems, as well as an increasing covariance between these systems, positively correlate with higher scores in `ProcSpeed\textunderscore AgeAdj' -- assessing processing speed -- and `PicVocab\textunderscore AgeAdj' -- evaluating language/vocabulary comprehension and negatively correlate with using cannabis and opiates (variables THC, SSAGA\textunderscore Mj\textunderscore Use, and SSAGA\textunderscore Times\textunderscore Used\textunderscore Opiates).
}
\label{fig:codimension}
\end{figure}

We define 360 spatially localized regions of interest (ROIs) using the multimodal parcellation proposed in \textcite{glasser2016multimodal}. These 360 regions are further aggregated into 10 distinct functional systems following the definition in \textcite{power2011functional}. These are the somatosensory/motor network (SMH), cingulo-opercular network (COP), auditory network (AUD), default mode network (DFM), visual network (VIS), frontoparietal network (FPT), salience network (SAL), ventral attention network (VAT), dorsal attention network (DAT), and a category for Other Regions (OTH), which includes areas that are not strictly classified within the aforementioned functional systems.

We partition the fMRI data into 20 time intervals of equal length. For each interval, we reduce the fMRI data to a `functional fingerprint' representation that is a $10 \times 10$ SPD covariance that captures the temporal correlation between the fMRI signals of different functional systems within a specific time interval. These matrices are denoted as $y_i(t_j)$ where $i = 1,\ldots, N = 1003$ represents the subject and $j = 1,\ldots,20$ denotes the time interval. We conducted sensitivity analysis by repeating the analysis with 10 and 40 time intervals of equal length. The results are virtually identical, as expected, since the time dynamics of the estimated mode of covariation, shown in Figure~\ref{fig:codimension_details}, does not seem to be constrained by the number of time points.

In addition, an extensive set of 150 subject traits of lifestyle, demographic, and psychometric measures are also provided for the same cohort of 1003 subjects. We denote these by $X_i$, with $i = 1,\ldots, N = 1003$. To account for potential confounding factors, we regressed out of the 150 variables nine confounders identified in \textcite{smith2015positivenegative}, and the squares of the continuous ones, using multivariable linear regression.

\subsection{Analysis}
We apply Algorithm~\ref{alg:asymmetric_sparse_fun_cca} to the pairs $\left( y_i(\cdot), X_i \right)$. Specifically, we model the SPD-valued functional data $\{y_i(\cdot)\}$ using the affine-invariant Riemannian metric. The choice of this metric is primarily driven by its ability to avoid the swelling effect \parencite{dryden2009noneuclidean}. The affine-invariant metric has been shown to be effective in prediction tasks \parencites{barachant2013classification}{ pervaiz2020optimising}.

The Fr\'echet mean $\hat \mu$ and tangent space representations $\{\Logyi\}$ are computed. See Section~\ref{sec:special_inst} for details. Both the hyperparameters $\lambda$ and $d$, the number of PCs used to reduce the dimension of the SPD-valued functional data, are chosen by cross-validation. Specifically, for every candidate $d$, the parameter $\lambda$ is chosen to minimize the cross-validated prediction error of the regression model in equation \eqref{eq:grp_lasso}, while $d$ is chosen by examining the scree plot of the cross-validated canonical correlations. We chose the smallest $d$ for which the cross-validated correlations appear to level off, that is, $d = 12$. The outlined procedure results in a set of $K$ estimated canonical directions $\left( \hat{\psi}_k, \hat{\theta}_k \right)_{k=1}^K$, where $\{\hat{\theta}_k\}$ are the canonical vectors associated with $\{X_i\}$, and $\{\hat{\psi}_k\}$ are the (tangent-space) representations of the canonical functions associated with $\{y_i\}$. After inspection of the cross-validated correlations and their associated variance, we decided to retain only the first pair of canonical directions. 

\subsection{Results and Discussion}

\begin{figure}[!htb]
\includegraphics[width=1.0\textwidth]{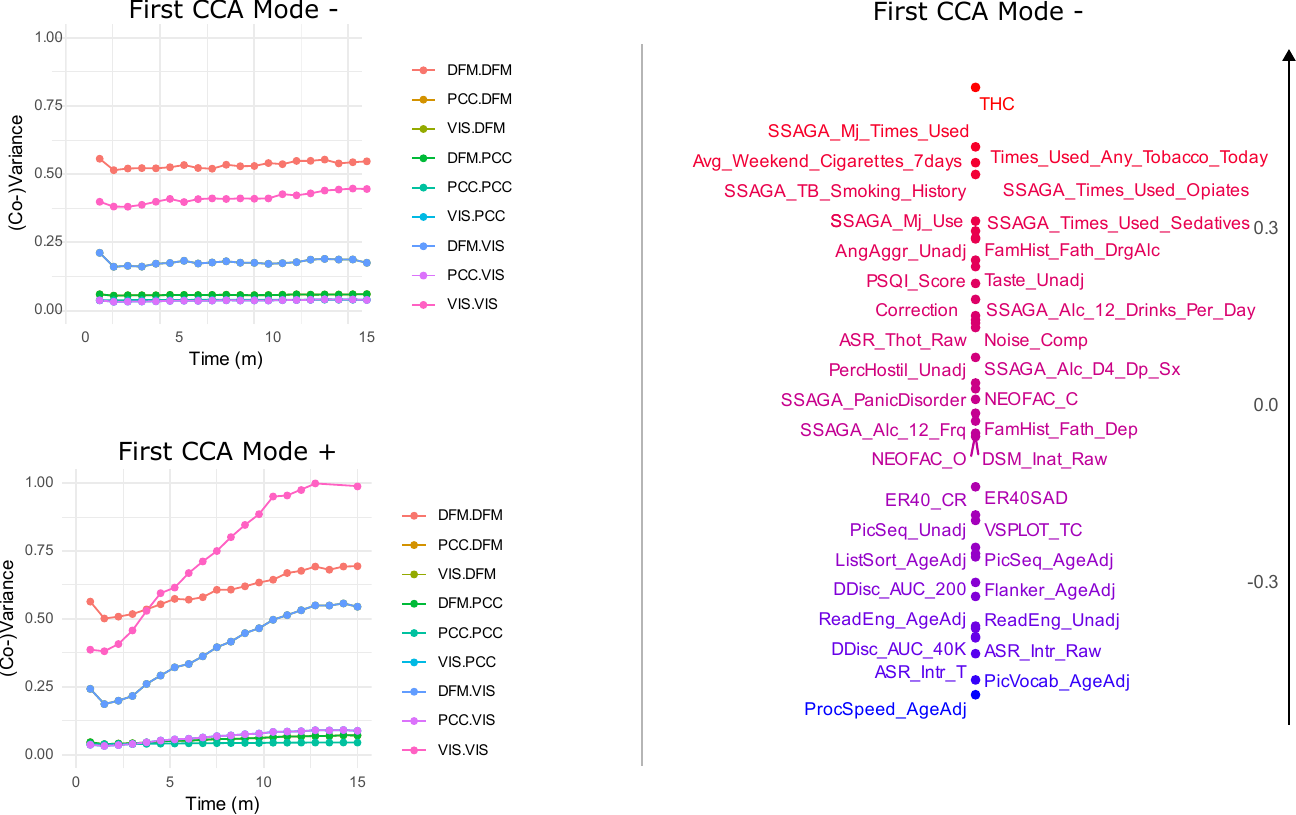}
\centering
\caption{On the left panel, for both `First CCA Mode -' and `First CCA Mode +', we show the temporal dynamics of selected entries of the dynamic mode of connectivity shown in Figure~\ref{fig:codimension}. Notably, some of these, e.g., the DFM-PCC covariance, remain stationary for both `First CCA Mode +' and `First CCA Mode -', while others, e.g., the DFM-VIS covariance, have markedly different patterns. On the right panel, we show a complete list of the 39 variables, of the canonical vector $\pm \hat \theta_1$, selected by the proposed model out of an initial set of 150, along with their relative importance.}
\label{fig:codimension_details}
\end{figure}

In Figure~\ref{fig:codimension}, we display the first canonical direction $\left( \hat{\psi}_1, \hat{\theta}_1 \right)$ by plotting 
\begin{equation*}
\left(\Exp_{\hat{\mu}}\left( -c \hat{\psi}_1 \right), -c \hat \theta_1 \right), \qquad
\left(\Exp_{\hat{\mu}}\left( +c \hat{\psi}_1 \right), +c \hat \theta_1 \right),
\end{equation*}
for a fixed positive constant $c$. In the figure, we refer to $\left(\Exp_{\hat{\mu}}\left( -c \hat{\psi}_1 \right), -c \hat \theta_1 \right)$ as `First CCA Mode -' and to $\left(\Exp_{\hat{\mu}}\left( +c \hat{\psi}_1 \right), +c \hat \theta_1 \right)$ as `First CCA Mode +'. Intuitively, these represent the two extremities of the first mode of covariation between functional dynamic connectivity and lifestyle, demographic, and psychometric measures. The exponential map $\Exp_{\hat{\mu}}(\cdot)$ allows us to map the canonical function back to the manifold of SPD-valued functions. 

The estimated first pair of canonical directions appears to link subjects with increasing variance over time within the visual and default mode functional systems, and increasing covariance over time between these functional systems, to `positive' lifestyle, demographic, and psychometric measures, such as better `ProcSpeed\textunderscore AgeAdj' score, which tests the `speed of processing' and better `PicVocab\textunderscore AgeAdj' score, which tests the ability to match an audio recording of a word to the most closely related picture. On the other hand, a more `stationary' connectivity pattern is associated with more `negative' lifestyle traits, such as a positive test for THC (THC), whether the subject has ever used cannabis (SSAGA\textunderscore Mj\textunderscore Use), and the number of times the subject has used opiates	(SSAGA\textunderscore Times\textunderscore Used\textunderscore Opiates). The cross-validated correlation of the identified mode of covariation is $0.075$, whereas its in-sample correlation is $0.259$. These values are relatively high compared to correlation analyses conducted in other studies using the same dataset \parencite{lin2020mapping}.

The multivariate component of the identified mode of covariation resembles the one found between static functional connectivity and lifestyle, demographic, and psychometric variables in \textcite{smith2015positivenegative}. However, as illustrated in Figure~\ref{fig:codimension_details}, our analysis reveals the non-stationary nature of this mode, with the latter portion of the scan emerging as the most informative in terms of functional connectivity. It is during this phase that the differences between the extremities of the mode of covariation become more evident.

It is possible that a latent variable linked to both the identified dynamic connectivity and the behavioral components of the first mode of covariation is responsible for the observed correlation between them. This variable potentially reflects the subjects' experience, such as growing impatience or distractions, during the 15-minute resting-state MRI session where they were instructed not to engage in specific tasks.  
Indeed, it appears that the `First CCA Mode -' subjects (who are more likely to test positive for THC and have used opiates) maintain a consistent `wandering mind', whereas the `First CCA +' subjects (who are likely to have better pattern completion skills and language/vocabulary comprehension) show a behavioral drift. This results in a progressive activation of the visual cortex and default mode network, and their cooperation, which might reflect a growing unease and consequent search for external stimuli.

\subsection{Comparison with other approaches}
We also explore replacing our proposed asymmetric sparse CCA model with standard CCA, incorporating a ridge penalty added to the covariance of the behavioural data, and with the PMA-sparse CCA model proposed in \textcite{witten2009penalized}. Cross-validating standard CCA yielded an estimated out-of-sample correlation for the first mode of covariation of $0.051$, while applying the PMA-sparse CCA model yielded a correlation of $0.027$. However, standard CCA does not perform selection of the important lifestyle, demographic, and psychometric measures, and did not show a particularly high level of consistency, both in the weights and signs assigned to the behavioural variables, compared to those obtained by our method. We believe this may be due to overfitting. On the other hand, the first mode of covariation computed using PMA-sparse CCA showed greater similarity to our mode of co-variation, although the out-of-sample correlation of the identified mode was lower than that of our proposed approach, indicating that the mode found might be suboptimal.

Further, we run an analogous analysis on static connectivity, that is, we compute the functional fingerprints using the whole time series (see also Section~ \ref{sec:special_inst} for methodological details). The results of this analysis are qualitatively similar to those obtained from the dynamic connectivity analysis. Specifically, a very similar behavioural mode of variation is identified, and this is associated with a static connectivity mode of variation that resembles the one in Figure~\ref{fig:codimension} at time 15m. However, completely removing the temporal component of the connectome does not provide insights into the possibility that the observed association reflects some subjects’ growing unease in the scanner. This highlights the importance of dynamic connectivity studies.

\section{Discussion and conclusions}
In this paper, we introduce a novel statistical model for identifying shared variation patterns between manifold-valued functional data and high-dimensional data. We refer to this setting as asymmetric due to the differing nature of the data. The proposed asymmetric CCA approach is designed to control the complexity of the canonical directions associated with the functional data by using Riemannian FPCA. This facilitates the identification of a lower-dimensional, smooth subspace onto which these data can be projected. Moreover, our approach controls the complexity of the high-dimensional canonical directions, which lack spatial structure, through a sparsity-promoting penalty that leads to the selection of the important variables. As opposed to other methods in the literature, this is achieved without requiring the estimation of the precision matrix of the high-dimensional data, which is, in general, prohibitive.

We apply asymmetric CCA to explore the association structure between resting-state dynamic functional connectivity, represented as time-indexed covariance matrices, and high-dimensional behavioral, lifestyle, and demographic features. Our analysis reveals a non-stationary pattern in functional connectivity, indicating that the usual assumption of temporal stationarity may not hold, even in resting-state studies. While this work focuses on an application in dynamic connectivity, the proposed method can be easily adapted to accommodate different Riemannian structures and to employ different data representation models, paving the way for several future extensions.

Yet, our approach has some limitations. For example, the dimension reduction step is unsupervised, which may result in the loss of small signals that are highly correlated with the high-dimensional data. In future work, we hope to explore supervised extensions of our method to address this limitation. Moreover, our theoretical analysis does not directly generalize to the setting of sparsely observed Riemannian data, as the downstream analysis of the CCA estimators requires proving the convergence of the Intrinsic RFPCA estimators in expectation, rather than in probability. While we establish these results for the fully observed case, extending the proof to the sparsely observed setting is non-trivial.

\setcounter{maxnames}{10}
\printbibliography
\end{refsection}

\newpage
\appendix
\appendixpage
\setcounter{page}{1}
\setcounter{equation}{0}
\renewcommand{\theequation}{S\arabic{equation}}

\begin{refsection}
The supplementary materials are organized as follows. In Section \ref{sec:simulations}, we study the empirical performance of the proposed method by means of simulation studies. In Section~\ref{appendixcca}, we formalize the CCA problem for random elements of Hilbert spaces and prove Theorem~\ref{multivariate-cca-theorem}. In Section~\ref{sec:proof_asymmetric_sparse_cca}, we present intermediate results and associated proofs, and conclude with the proof of Theorem \ref{thm:main_theorem_multivariate_case}, our theoretical result on the asymptotic errors made by the canonical direction estimates of Asymmetric Sparse CCA. In Section~\ref{appendix:proof_of_canonical_function_section}, we prove Theorem \ref{thm:main_theorem_general_case}, our theoretical result on the asymptotic errors made by the canonical direction estimates of Asymmetric Sparse-Functional CCA. In Section \ref{appendix:consistency_of_scores}, we prove Theorem \ref{thm:consistency_of_scores_main}, our theoretical result on the asymptotic errors made by the canonical variable estimates of Asymmetric Sparse-Functional CCA, in the absence of a finite-dimensional correlation structure assumption. In Section \ref{subsec:identities}, we present several norm and matrix identities that are utilized throughout the supplementary materials. Finally, in Section~\ref{sec:apdx:irfpca}, we provide additional details on the Intrinsic RFPCA algorithm \parencite{lin2019intrinsic}, which is used in the proposed Algorithm \ref{alg:asymmetric_sparse_fun_cca}. We also present an explicit basis construction for the space of symmetric positive definite matrices equipped with the affine invariant metric, providing a computational speed-up compared to the Gram-Schmidt procedure proposed in \textcite{lin2019intrinsic}.

\section{Simulations}
\label{sec:simulations}
We perform numerical experiments to investigate the finite sample performance of the proposed approach. First, we describe the data generation process. Then, we discuss the metrics utilized to evaluate the methods' performance. Lastly, we introduce the alternative approaches for comparison with our method and comment on the results.

\subsection{Data generation}
Recall that $y:\Tsc \ra \M$ is a random Riemannian process, $X \in \rone^p$ is a high-dimensional random vector, and $\mu:\Tsc \ra \M$ is a fixed smooth curve on $\M$ modeling the population mean of $y$. Here, we fix $p=200$ and choose $\M$ to be the manifold of $m \times m$ SPD matrices, with $m=3$. We let the time domain of $y$ be $\Tsc = [-1,1]$. In the following, we aim to generate realizations of $(y,X)$ according to a model that ensures that the $K$ population canonical vectors and canonical functions are prespecified vectors $\{ \theta_k \}_{k=1}^K \subset \rone^p$ and functions $\{\psi_k\}_{k=1}^K \subset \Jmu$, respectively, with $K=2$. We apply our proposed method and alternative approaches to this data to estimate the canonical vectors and functions, and then compare these estimates with the prespecified population quantities.

The procedure to generate the data is as follows. Take a random vector $Y \in \rone^d$, a set of vectors $ \{ \eta_k \}_k \subset \rone^{d}$, with $d=3$, and an orthonormal basis $\{\phi_j\} \subset \Jmu$. Moreover, define $\Logy = \sum_{j=1}^d Y_j \phi_{j}$ and $y = \Exp_{\mu} \bl \Logy \br$. It follows from Theorem \ref{multivariate-cca-theorem} that if the multivariate data $(Y,X)$ have population canonical vector $(\eta_k,\theta_k)$ then the functional/multivariate data $(y,X)$ will have population canonical pairs $(\psi_k,\theta_k)$, with $\psi_k = \sum_{j=1}^d \eta_{kj} \phi_j$. Additionally, we impose a group-sparse structure on the canonical vectors $\{\theta_k\}$. To replicate a realistic setting, we add an extra mode of variation $W\phi_{d+1}$ to $\Logy$, with $W$ a random variable that is independent of $X$ and $Y$, and with $\Var\bl W \br = 1/2$. This aims to contaminate the observations without affecting the canonical functions and vectors.

To generate the multivariate data $(Y,X)$ given prespecified canonical pairs $ \{(\eta_k,\theta_k) \}$, we use the model introduced in \textcite{chen2013sparse}. As is stated in Proposition 2.1 in \textcite{chen2013sparse}, this enables us to choose the canonical directions $ \{(\eta_k,\theta_k)\}$ and correlations $\{ \gamma_k\}$ for $k =1, \ldots K$ freely, while retaining the flexibility to specify $\Sigma_X \in \rone^{p \times p}$ and $\Sigma_Y \in \rone^{d \times d}$. We define $(Y,X)$ as
\begin{equation}\label{eq:sim_cca_mult}
    \begin{pmatrix}
    Y\\
    X
    \end{pmatrix}
    \sim \mathcal{N} \left (0,
    \begin{pmatrix}
    \sigmay\quad \Sigma_{YX}\\
    \Sigma_{XY} \quad \sigmax
    \end{pmatrix} \right ),
\end{equation}
where $\mathcal{N}$ denotes the multivariate normal distribution and $\Sigma_{YX} =\Sigma_Y\left(\sum_{k=1}^K \gamma_k \eta_k \theta_k^{T}\right) \Sigma_X$. It is easy to show that the population canonical vectors of $(Y,X)$ are $(\eta_k, \theta_k)$ with correlations $\gamma_k$, for $k=1,\ldots K$. The set of canonical vectors $\{\eta_k\}$ is defined by generating $K$ orthogonal random vectors, which are then normalized to satisfy the constraint $\eta_k^{\T}\sigmay\eta_j = \delta_{kj}$. Similarly, the canonical vectors $\{\theta_k \}$ are randomly generated and constrained to satisfy the condition $\theta_k^{\T}\sigmax \theta_j = \delta_{kj}$. The group sparsity assumption is enforced by ensuring that only $k_1=20$ elements of each canonical vector (the same elements across all vectors) are non-zero. Additionally, the variables $X_j$ corresponding to the non-zero components of $\theta_j$ have marginal covariance matrix $\Sigma_{X_S} = \text{diag}( \underbrace{2, \ldots 2}_{10}, \underbrace{1, \ldots 1}_{10} ) \in \rone^{k_1 \times k_1}$. The covariance $\sigmax$ is then defined as
\begin{equation}
\sigmax = \begin{pmatrix}
     \Sigma_{X_S} \quad 0\\
    0 \quad I_{p-k_1}
    \end{pmatrix}.
\end{equation}
The covariance $\Sigma_Y$ is set to be diagonal with diagonal values being $3,2,1$. The true canonical correlations are chosen to be $\gamma_1 = .95$ and $\gamma_2 = .6$.

We let the mean curve $\mu$ at each $t \in \calT$ be a SPD matrix $\mu(t) \in \rone^{m \times m}$. We set $\mu(0)$ by randomly generating its eigenvectors and setting the associated eigenvalues equal to $\left(1,2,3\right)$. The mean $\mu(t)$ at the other time-points $t \in \calT$ is generated by applying a time-variant rotation to the eigenvectors of $\mu(0)$. We choose each principal component $\phi_j$ to take the form $\phi_j(t) = E(t)P(t)$, for $j = 1, \ldots d$, where $E$ is chosen at random from a set of orthogonal basis vectors for $\Jmu$, and $P$ is chosen at random from a basis of orthogonal polynomials on $[-1,1]$. This ensures that $\{\phi_j\}$ are orthogonal to one another as elements of $\Jmu$.

In our experiments, we generate $N$ i.i.d. pairs $(Y_i,X_i)$ from the multivariate CCA model in equation (\ref{eq:sim_cca_mult}), for different choices of $N$. Next, we generate $y_i$ via $\Exp_{\mu} \bl \sum_{j=1}^{d} Y_{ij} \phi_{j} \br$ and evaluate it at $L=50$ locations $t_l \in [-1,1]$, yielding $y_i(t_l)$ for $i=1, \ldots N$ and $l=1,\ldots L$. The observations 
\begin{equation*}
\left(\{y_i(t_l)\}_{l}, X_i \right)_i, \qquad i=1,\ldots,N,
\end{equation*}
are used to estimate the canonical vectors and functions and compare different approaches.

\subsection{Metrics}
We use the following metrics to compare the estimated accuracy of the models considered.

\begin{enumerate}[A.]
\item \textbf{Normalized Euclidean error for the canonical vector} \begin{equation*}
    \twonorm{\theta_1/\twonorm{\theta_1}-\hat\theta_1/\twonorm{\hat\theta_1}}
\end{equation*}
This is a natural metric for evaluating the estimation accuracy.

\item \textbf{F1-score for the canonical vector}
\begin{equation*}
    2\cdot \frac{P \cdot R}{P + R},
\end{equation*} 
where $P = \frac{TP}{TP + FP}$ is the precision, $R = \frac{TP}{TP + FN}$ is the recall, $TP$ is the number of true positives, $FP$ the number of false positives, $FN$ the number of false negatives.\\

\item \textbf{$L^2$ Parallel transport error for the canonical function} 
\begin{equation*}
    \norm{\Gamma_{\hat{\mu},\mu}\hat{\psi}_1 - \psi_1}_{\mu}
\end{equation*}
This metric allows us to use the $\Jmu$ norm to compare the estimates to the true population analog, by parallel transporting $\hat{\psi}_1 \in \Jmuhat$ and defining $\Gamma_{\hat{\mu},\mu}\hat{\psi}_1 \in \Jmu$.

\item \textbf{Tangent Correlation}\\
Using a large test set $\{ \tilde{y}_i,\tilde{x}_i \}$, generated from the same distribution as the training data, we compute the sample correlation as follows:

\begin{equation*}
    \Corr\bl \bl \innerdouble{\Log_{\mu}\tilde{y}_i}{\Gamma_{\hat{\mu},\mu}\hat{\psi}_1}_{\mu}\br_i, \bl \tilde{x}_i^{\T}\hat{\theta}_1 \br_i \br.
\end{equation*}
We refer to this metric as the `Tangent' correlation, as it respects the manifold structure of the data.
\item \textbf{Euclidean Correlation}\\
Using a large test set $\{ \tilde{y}_i,\tilde{x}_i \}$, generated from the same distribution as the training data, we compute the sample correlation as follows:
\begin{equation*}
    \Corr\bl \bl \operatorname{vec}\bl\tilde{y}_i\br^{\T}\operatorname{vec}\bl\hat{\psi}_1\br\br_i, \bl \tilde{x}_i^{\T}\hat{\theta}_1 \br_i \br.
\end{equation*}
We refer to this metric as the `Euclidean' correlation, as it ignores the manifold structure of the data.
\end{enumerate}
\subsection{Approaches for comparison}
We compare 4 different approaches, detailed below.
\begin{enumerate}
    \item \textbf{Proposed approach.} We apply Algorithm \ref{alg:asymmetric_sparse_fun_cca} without modifications. We use cross-validation to choose the regularization parameter $\lambda$ in the group-lasso regression step as implemented by the \texttt{glmnet} package \parencite{glmnet}.

\item \textbf{Sparse PCA-based approach:} Intrinsic RFPCA + sparse PCA + classical CCA. We use Intrinsic RFPCA, as in our approach, to reduce the dimensionality of the functional data. We use sparse PCA, using the \texttt{elasticnet} R package \parencite{zou2006sparse}, to reduce the dimensionality of the multivariate data. Then, we use the estimated PCA scores as input for classical multivariate CCA. We provide sparse PCA with the exact number of principal components that are correlated with the functional data, i.e., $k_1=20$, and restrict the number of non-zero principal loadings per principal component to be $2$.

\item \textbf{Sparse CCA-based approach:} Intrinsic RFPCA + sparse CCA. We again use Intrinsic RFPCA to reduce the dimension of the functional data. Next, we use the Penalized Matrix Analysis (PMA) approach to sparse CCA proposed in \textcite{witten2009penalized} to compute canonical pairs between the PC scores from Intrinsic RFPCA and the high-dimensional data. The PMA approach to sparse CCA assumes that the covariance matrices of the data are the identity matrices, giving it a slight disadvantage. We choose the amount of penalization for $\theta_1$ using the suggested permutation-type approach \parencite{witten2009penalized}, and choose the penalization parameter for $\eta_1$ to induce virtually no penalization.

\item \textbf{Multivariate FPCA-based approach:} Multivariate FPCA + Asymmetric sparse CCA in Algorithm~\ref{alg:asymmetric_sparse_cca}. This approach is analogous to the one proposed, except that the Intrinsic RFPCA step is replaced by multivariate FPCA \parencite{happ2018multivariate}. Therefore, it disregards the SPD manifold structure of the data. Specifically, it transforms each SPD matrix into a vector extracting the lower triangular part of the matrix. Then it applies multivariate FPCA to the resulting vector-valued functions.
\end{enumerate}

We have chosen these alternative approaches in order to dissect specific components of the CCA problem. Specifically, approach 2 isolates the effect of selecting important features and identifying correlated components in two separate stages, and approach 3 isolates the effect of not taking advantage of the group sparsity structure in the canonical vectors and making restrictive assumptions on the covariance of the high-dimensional data. Approach 4 isolates the effect of treating manifold data as if it were Euclidean. Note that Approach 4 is technically solving a different canonical correlation problem than approaches 1-3 as it aims to maximize the Euclidean correlation rather than the tangent space correlation. For this reason, the underlying population canonical vectors and functions differ from those in the proposed model. Therefore, we only use metric E when evaluating the performance of approach 4.

Moreover, depending on the choice of $\sigmax$, either the PMA sparse CCA approach or the sparse PCA approach is at a disadvantage. Assuming $\sigmax$ to be the identity matrix meets the assumptions of PMA sparse CCA, but renders dimension reduction through sparse PCA less effective. Conversely, choosing $\sigmax$ not to be the identity matrix benefits sparse PCA at the expense of the PMA sparse CCA approach.

\subsection{Results and Discussion}
In our experiments, we set $p = 200$ and vary $N$. For each value of $N$, we run 15 trials. We provide the Intrinsic RFPCA model with the true rank, indicating the number of functional principal components associated with the variable $X$, that is, $d=3$.
\begin{figure}[ht]
\includegraphics[width=1.0\textwidth]{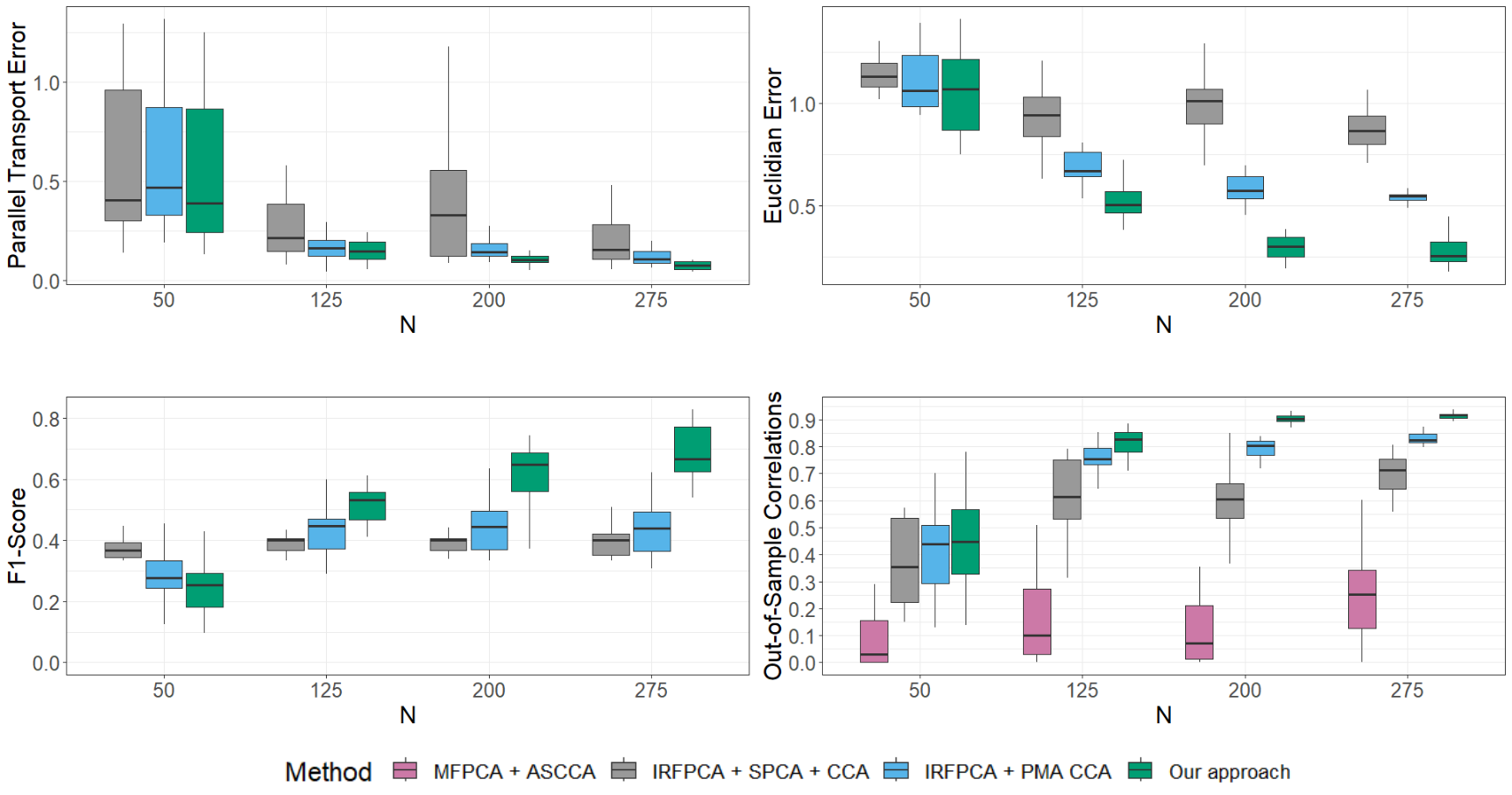}
\centering
\caption{(Top left): Performance evaluation using metric A, which measures the normalized Euclidean error in the first high-dimensional canonical vector, on approaches 1-3. (Top right): Performance evaluation using metric C, which is the parallel transport error in the first canonical function, on approaches 1-3. (Bottom left): Performance evaluation using metric B, the F1-score of the first estimated high-dimensional canonical vector compared to the associated population vector, on approaches 1-3. (Bottom right): Performance evaluation using out-of-sample correlations. We use out-of-sample tangent correlation (metric D) for approaches 1-3, and out-of-sample Euclidean correlation (metric E) for approach 4.}
\label{fig:all_errors}
\end{figure}

In Figure~\ref{fig:all_errors}, we present the performance of approaches 1-3 measured using all defined metrics A-E. As previously mentioned, Approach 4 is assessed using only metric E due to its differing underlying model. In the high-dimensional setting, where $N = 50$ and $p=200$, all four approaches showed similar performance across all metrics. This setting likely identifies the detectability limits of CCA methods. However, when provided with more samples, our approach quickly outperforms the other approaches. Differences in performance were more notable in the estimation of the Euclidean error for the canonical vectors, F1-score, and out-of-sample correlation. This suggests that the most challenging aspect of the setting considered is estimating the canonical vectors as opposed to the canonical functions. This can be explained by the similar modeling strategies adopted for the functional data. Approach 4, while able to find correlated components in the data according to the Euclidean notion of correlation (E), suffered from bias due to treating the functional data as Euclidean.

For approach 2, the differences in performance can be explained by its two-step strategy that involves first selecting the important features and reducing the dimension of the multivariate data, followed by identifying correlated components. Specifically, the sparse PCA step is based solely on the variance structure of $X$, and not on its correlation with the functional data. In our simulation, the variables of $X$ correlated with $Y$ have the same or smaller variance than those not correlated with $Y$. As a result, sparse PCA, which is unsupervised, struggles to tease them apart. 

\section{Canonical Correlation Analysis of Random Elements of Hilbert Spaces}
\label{appendixcca}
In this section, we provide a more rigorous formalization of the CCA model. We mirror the development of \textcite{hsing2015theoretical} and \textcite{huang2015functional}, but provide a less technical presentation by emphasizing the role of the canonical variables rather than the canonical vectors when formulating the general CCA problem. For an introduction to Hilbert space concepts and random elements taking values in Hilbert spaces, we refer to \textcite{hsing2015theoretical}. For an introduction to classical CCA, we refer to \textcite{uurtio2018tutorial}.

In Section \ref{subsec:problem_statement_CCA_general}, we define the infinite-dimensional version of the CCA problem, and establish the existence of solutions in our asymmetric setting (Theorem \ref{theorem_problem}). In Section \ref{subsec:preliminary_results}, we state preliminary definitions and results for the subsequent sections. In Section \ref{subsec:reduction_to_finite_dim_problem}, we state the necessary assumption (Assumption \ref{assumption_cor}) to reduce the infinite-dimensional CCA problem to a finite-dimensional CCA problem (Theorem \ref{theorem_problem_finite}). In Section \ref{sec:finite_infinite_CCA_theory}, we study the difference between the canonical variable solutions of the finite- and infinite-dimensional CCA problems (Theorem \ref{thm:finite_infinite_score}). In Section \ref{sec:apdx:proofs_of_CCA_formalization_theorems}, we prove the results of the section and additionally prove Theorem \ref{multivariate-cca-theorem}.

\subsection{Problem Statement} \label{subsec:problem_statement_CCA_general}
Let $\chi_1$ and $\chi_2$ be measurable functions from a probability space $\bl \Omega,\calF,\mathbb{P} \br$ to separable Hilbert spaces $\hil_1$ and $\hil_2$, respectively \parencite[See Section 7.2.~of][]{hsing2015theoretical}. Here, $\hil_1$ and $\hil_2$ are arbitrary Hilbert spaces, but throughout the paper they correspond to $\hil_1 = \Jmu$ and $\hil_2 = \rone^p$, and similarly $\chi_1$ and $\chi_2$ correspond to $\chi_1 = \Logy$ and $\chi_2 = X$. Hilbert space inner products are denoted by $\langle \cdot, \cdot \rangle$, with associated norms $\norm{\cdot}$. The specific choice of the norm or inner product will be clear from the context. We assume that $\Ex{\norm{\chi_i}^2} < \infty$ so that the mean and covariance of $\chi_i$ are well-defined for $i=1,2$. The mean element of $\chi_i$ is defined as $h_i \equiv \Ex{\chi_i} \in \hil_i$, and for simplicity, we assume $h_i = 0$  for $i = 1,2$.

A seemingly natural way to formalize the canonical correlation problem for the infinite-dimensional case, which is analogous to the finite-dimensional case, is
\begin{equation}
    \maximize{\textrm{ Corr}^2 \left (\inner{\chi_1}{f},\inner{\chi_2}{g} \right )}{f \in \hil_1, \textrm{ } g \in \hil_2}
\end{equation}
where $\textrm{Corr}$ is the usual correlation defined between two finite-variance real-valued random variables defined on $\bl \Omega,\calF,\mathbb{P} \br$. If they exist, the solution $(f,g)$ would be the first canonical vector pair.
Equivalently, we can write this problem in terms of the canonical variables $U,V$ as
\begin{equation}
    \maximize{\textrm{ Corr}^2 \left ( U,V \right )}{U \in \mathcal{U}, \textrm{ } V \in \mathcal{V}}
\end{equation}
where $\mathcal{U} = \left \{\inner{\chi_1}{f} : \textrm{ }f \in \hil_1 \right \}$, $\mathcal{V} = \left \{\inner{\chi_2}{g} : \textrm{ }g \in \hil_2  \right \}$.
However, the maximum of this problem may not be attained by any $U \in \mathcal{U}$, $V \in \mathcal{V}$ \parencite{cupidon_properties_2008}. 

It turns out we can amend this by simply taking the closures of $\mathcal{U}$ and $\mathcal{V}$. Let $\mathbb{L}^2 \bl \Omega,\calF,\mathbb{P} \br$ denote the Hilbert space of square-integrable random variables on $\Omega$ with inner product $\langle U,V \rangle = \Cov(U,V)$ for $U,V \in \mathbb{L}^2 \bl \Omega,\calF,\mathbb{P} \br$. Note that $U$ being square-integrable means that $\Var(U) < \infty$. If we replace
$\mathcal{U}$, $\mathcal{V}$ in the problem above with their closures as subsets of $\mathbb{L}^2 \bl \Omega,\calF,\mathbb{P} \br$, denoted $\ubar$, $\vbar$, (for further discussion of $\ubar$, $\vbar$, see the discussion following Example 7.6.5 of \textcite{hsing2015theoretical}) then it can be shown that the maximum \textit{will} be attained for some $U \in \ubar$, $V \in \vbar$, provided that a certain linear operator is assumed to be compact. In our setting, this compactness condition holds because we use $\hil_2 = \rone^p$, a finite-dimensional space. The result can be found in Theorem 10.1.2 of \textcite{hsing2015theoretical}, which we restate in our context here.

The following result establishes the existence of solutions to the general CCA problem in our asymmetric setting where $\textrm{dim}(\hil_1) = \infty$ but $\textrm{dim}(\hil_2) = p < \infty$, and that there are at most $p$ nontrivial solutions.
\begin{theorem}
    \label{theorem_problem}
    If $\textrm{dim}(\hil_1) = \infty$ and $\textrm{dim}(\hil_2) = p < \infty$, then there exists $U_1 \in \ubar$ and $V_1 \in \vbar$ which are solutions to $$\supp{\operatorname{ Corr}^2 \left ( U,V \right )}{U \in \ubar, \textrm{ } V \in \vbar},$$
with $\Var(U_1) = \Var(V_1) = 1$. For $k = 2, \ldots p$, there exists $U_k \in \ubar$ and $V_k \in \vbar$, which are solutions to 

$$\underset{\substack{U \in \ubar: \operatorname{Cov}\left ( U, U_i\right)=0, i=1, \ldots, k-1 \\
    V \in \vbar: \operatorname{Cov}\left ( V, V_i\right)=0, i=1, \ldots, k-1 }}{\operatorname{sup}} \operatorname{ Corr}^2 \left ( U,V \right ),$$
with $\Var(U_k) = \Var(V_k) = 1$.
Moreover, for all $U \in \ubar$ and $V \in \vbar$ which are uncorrelated with $U_1,\ldots U_p$, $V_1, \ldots V_p$, respectively, the pair $(U,V)$ is a trivial solution, that is,
$$\underset{\substack{U \in \ubar: \operatorname{Cov}\left ( U, U_i\right)=0, i=1, \ldots, k \\
    V \in \vbar: \operatorname{Cov}\left ( V, V_i\right)=0, i=1, \ldots, k }}{\operatorname{sup}} \operatorname{ Corr}^2 \left ( U,V \right ) = 0.$$
The pairs $\{(U_k,V_k)\}_{k=1}^p$ are called the canonical variable pairs. We refer to the problem of finding $\{(U_k,V_k)\}_{k=1}^p$ as the population canonical correlation problem.
\end{theorem}

\begin{remark}
    Intuitively, we must cast the problem in terms of the canonical variables $U \in \ubar$, $V \in \vbar$ rather than canonical vectors $f \in \hil_1$, $g \in \hil_2$ because $\mathcal{U} = \left \{\inner{\chi_1}{f} : \textrm{ }f \in \hil_1 \right \}$ and $\mathcal{V} = \left \{\inner{\chi_2}{g} : \textrm{ }g \in \hil_2  \right \}$ are not large enough for the supremum of the CCA problem to be attained. To emphasize this point, given an optimal $U \in \ubar$ of the form $U = \lim_{j \to \infty} \inner{\chi_1}{f_j}$ for some sequence $\bl f_j \br_j \in \hil_1$, then $U$ can not necessarily be written as an inner product $U = \inner{f}{\chi_1}$, because $\bl f_j \br_j$ may not even be a Cauchy sequence in $\hil_1$, and thus not converge in $\hil_1$. For further intuition on this property of the problem, see Remark \ref{remark:no_canonical_vectors}.
\end{remark}
In the next section, we introduce an assumption that allows us to make the infinite-dimensional CCA problem finite-dimensional, and furthermore formulate the CCA problem in terms of the canonical vectors rather than the canonical variables. From now on, we assume $\textrm{dim}(\hil_2) = p < \infty$ as in Theorem \ref{theorem_problem}.
\subsection{Background}
\label{subsec:preliminary_results}

We begin with preliminary definitions and properties of the random element $\chi_1$. In particular, we introduce the covariance operator $\mathscr{K}_1$. The eigenvectors of $\mathscr{K}_1$, also known as the principal components of $\chi_1$, are fundamental to our approach for two reasons: they provide a data-driven subspace for projecting $\chi_1$ and their properties simplify our proofs. 

Given that $\E \left [ \norm{\chi_1}^2 \right ] < \infty$ and $\chi_1$ has mean $0$, the covariance operator of $\chi_1$ is well-defined as $\mathscr{K}_1 \equiv \E \left [ \chi_1 \otimes \chi_1 \right ]$. Here, the tensor product $f \otimes g: \hil \to \hil$ between $f,g \in \hil$ for a Hilbert space $\hil$, is defined as $\bl f \otimes g \br \bl h \br = \inner{f}{h}g$ for all $h \in \hil$.

In the lemma below, we collect the properties of $\chi_1$ and $\mathscr{K}_1$ used in what follows. An orthonormal sequence of elements $\left \{ e_j\right \}_{j=1}^{\infty}$ of a Hilbert space $\hil$ such that $\overline{\text{span}\{e_j\}} = \hil$ is referred to as a complete orthonormal system (CONS) for $\hil$. For convenience, we state this result supposing that $\mathscr{K}_1$ has infinitely many eigenvalues, but an analogous result holds if $\mathscr{K}_1$ has only finitely many eigenvalues.
\begin{lemma}
    \label{closure_image_facts_thm}
    Let $\operatorname{Im} \bl \mathscr{K}_1 \br$ denote the image of $ \mathscr{K}_1$ and $\overline{\operatorname{Im} \bl \mathscr{K}_1 \br} \subseteq \hil_1$ denote its closure in $\hil_1$. Then, the following statements hold.
    \begin{enumerate}
        \item With probability 1, $\chi_1 \in \overline{\operatorname{Im} \bl \mathscr{K}_1 \br}$, and for any $f \in \operatorname{Im} \bl \mathscr{K}_1 \br^\perp$, $\inner{f}{\chi_1} = 0$
        \item $\mathscr{K}_1$ has the eigendecomposition $\mathscr{K}_1 = \sum_{j=1}^{\infty} \omega_j e_j \otimes e_j$, where $e_j \in \hil_1$ for $j=1, \ldots \infty$ and $\omega_1 \geq \omega_2 \geq \ldots > 0$. $\{ e_j \}_{j = 1}^{\infty}$ forms a CONS of $\clim$, and $\omega_j \rightarrow 0$ as $j \rightarrow \infty$. We refer to $\{(e_j,\omega_j)\}_{j=1}^\infty$ as the eigensystem of $\mathscr{K}_1$, with eigenvectors $e_j$ and eigenvalues $\omega_j$. The $e_j$ are also referred to as principal components of $\chi_1$.
        \item With probability $1$, $\chi_1 = \sum_{j=1}^{\infty}\inner{\chi_1}{e_j}e_j$. We refer to the $\{\left\langle\chi_1, e_j\right\rangle\}_{j=1}^{\infty}$ as the principal scores, and they are uncorrelated random variables with $\Ex{\inner{\chi_1}{e_j}} = 0$ and $\Var\bl \inner{\chi_1}{e_j}\br = \omega_j$.
        \item $\ubar = \left \{\sum_{j=1}^\infty a_j \inner{\chi_1}{e_j}:\sumj \omega_j a_j^2 < \infty \right \}$ and $\{\left\langle\chi_1, e_j\right\rangle/\omega_j^{1/2}\}_{j=1}^\infty$ forms a CONS for $\ubar$.
    \end{enumerate}
\end{lemma}
\begin{remark}
    Item 1 elucidates the role of $\clim$ as the subspace of $\hil_1$ where $\chi_1$ resides. Item 2 shows that the eigenvectors $\{e_j\}$ are a CONS for $\clim$, which implies Item 3, that the principal scores $ \{ \inner{\chi_1}{e_j} \}$ characterize $\chi_1$. Item 4 establishes that the set of potential canonical variables, $\ubar$, is equivalent to the set of linear combinations of the principal scores with finite variance.
\end{remark}
\begin{remark} \label{remark:no_canonical_vectors}
    Item 4 sheds light on why it is necessary to take the closure of $\mathcal{U}$ in order to guarantee a solution to the infinite-dimensional CCA problem in Theorem \ref{theorem_problem}, and further on why the canonical vectors are not defined in the infinite-dimensional CCA problem.
    
    From Items 1 and 2 it follows that
    \begin{equation} \label{eq:U_set_equation}
        \mathcal{U} \equiv \left \{ \inner{f}{\chi_1}: f \in \hil_1 \right \} = \left \{ \sumj a_j \inner{e_j}{\chi_1} : \sumj a_j^2 < \infty \right \}.
    \end{equation}
    For an element $\inner{f}{\chi_1} \in \mathcal{U}$ defined by $f = \sumj a_je_j \in \hil_1$, since $\Var\bl \inner{e_j}{\chi_1} \br = \omega_j$, we have that $\Var \bl \inner{f}{\chi_1} \br = \sumj a_j^2 \omega_j$. Since the $\omega_j$ decay to $0$ as $j$ approaches infinite, there are sequences $\left \{ a_j \right \}$ that do not satisfy $\sumj a_j^2 < \infty$ but do satisfy $\sumj a_j^2 \omega_j < \infty$, in other words, there are finite variance linear combinations of the $\inner{e_j}{\chi_1}$ that are not contained in $\mathcal{U}$. These are the elements of $\ubar$ that are missing in $\mathcal{U}$:
    \begin{equation}
        \mathcal{U} = \left \{ \sumj a_j \inner{e_j}{\chi_1} : \sumj a_j^2 < \infty \right \} \subset \ubar =   \left \{ \sumj a_j \inner{e_j}{\chi_1} : \sumj a_j^2\omega_j < \infty \right \}.
    \end{equation}
    If the canonical variable solution to the infinite-dimensional CCA problem \eqref{eq:infinite_dim_canonical_variable_problem} belongs to $\mathcal{U}$ but not $\ubar$, then it has no corresponding canonical vector.
\end{remark}
For an arbitrary CONS $\left \{ e_j \right \}_{j=1}^{\infty}$ for $\clim$, the associated scores $\left \{ \inner{e_j}{\chi_1} \right \}$ may not be orthogonal in $\ubar$, but as the following result shows, the associated scores still span $\ubar$. This is the property that allows us to not rely on the principal component basis and instead use an arbitrary CONS for $\hil_1$ in Assumption \ref{assumption_cor}.
\begin{lemma}
\label{eigenvector_theorem}
For any complete orthonormal system $\left \{ e_j \right \}_{j=1}^{\infty}$ of $\overline{\operatorname{Im} \bl \mathscr{K}_1 \br}$, $\ubar = \overline{\operatorname{span} \left \{ \inner{e_j}{\chi_1} \right \} }$. Thus, any element of $U \in \ubar$ can be written as $\sumj a_j \inner{\chi_1}{e_j}$ where the $a_j$ are such that $\Var \bl U \br < \infty$.
\end{lemma}
\subsection{Reduction to a finite-dimensional problem}
\label{subsec:reduction_to_finite_dim_problem}
We can now introduce Assumption~\ref{assumption_cor} on the correlation structure between $\chi_1$ and $\chi_2$.

\begin{assumption}
    \label{assumption_cor}
    There exists a complete orthonormal system $\{ e_j \}_{j = 1}^{\infty}$ for $\hil_1$ and a set of indices $I \subset  \left \{1 , 2,\ldots \right \}$, with finite cardinality $\left | I \right | = d$, such that
\begin{align}
&\operatorname{Corr}\bl V,\inner{\chi_1}{e_j}\br = 0, \qquad \forall  V \in \vbar, j \in I^c,\label{align:first_item}\\
&\operatorname{Corr}\bl\inner{\chi_1}{e_i}, \inner{\chi_1}{e_j} \br = 0, \qquad \forall i \in I, \forall j \in I^c, \label{align:second_item}
\end{align}
where $I^c$ denotes the complement of $I$ in $ \mathbb{N} = \{1 , 2,\ldots \}$.
\end{assumption}
\begin{remark}
    The complete orthonormal system $\left \{ e_j \right \}_j$ is not required to be the principal component basis.
    In the case when $\hil_2 = \rone^p$ and $\chi_2 = X$, equation \eqref{align:first_item} can be rewritten as
    \begin{equation}
        \operatorname{Corr}\bl X_i,\inner{\chi_1}{e_j} \br = 0, \qquad \forall  i = 1, \ldots p, j \in I^c.
    \end{equation}
    Intuitively, this assumption states that all elements $\psi$ of $\hil_1$ whose projections $\inner{\chi_1}{\psi}$ are correlated with $X$ belong to a $d$-dimensional subspace.
\end{remark}
\begin{remark}
    Assumption \ref{assumption_cor} is weaker than the assumption that $\chi_1$ admits a finite-dimensional representation $\chi_1 = \sum_{j=1}^d \inner{\chi_1}{e_j}e_j$, for a set of vectors $\left \{ e_j \right \}_{j=1}^d \subset \hil_1$. To see this, we first note that the elements $\left \{ e_j \right \}_{j=1}^d \subset \hil_1$ are orthonormal. Then, we complete $\left \{ e_j \right \}_{j=1}^d \subset \hil_1$ to form a CONS $\{ e_j \}_{j = 1}^{\infty}$ for $\hil_1$, and take $I = \{1, \ldots d \}$. Given that the elements $\{e_j\}$ are orthonormal, we have that $\inner{\chi_1}{e_j} = 0$ for all $j \in I^c$, with probability $1$. Hence, conditions \eqref{align:first_item} and \eqref{align:second_item} are satisfied.
\end{remark}
Making this assumption enables us to reduce the infinite-dimensional CCA problem to a finite-dimensional CCA problem; moreover, it allows us to formulate the CCA problem in terms of the population quantities of interest, the canonical vectors, rather than the canonical variables (Theorem \ref{theorem_problem_finite}). Theorem \ref{theorem_problem_finite} can be viewed as a generalization of Theorem 1 of \textcite{krzysko2013canonical}. There, it is  assumed that $\chi_1$ has a finite-dimensional representation, whereas here we make the weaker Assumption~\ref{assumption_cor}.
\begin{theorem}
    \label{theorem_problem_finite}
    Reorder the complete orthonormal system $\{ e_j \}_{j = 1}^{\infty}$ for $\hil_1$ in Assumption~\ref{assumption_cor} so that $I = \left \{1 ,\ldots d \right \}$. Then, under Assumption~\ref{assumption_cor}, the solution to the population canonical correlation problem in Theorem~\ref{theorem_problem} is found for a $U \in \ubar_d = \left \{ \sum_{i =1}^d a_i \inner{\chi_1}{e_i}: a_i \in \rone \right \}$.\\
    Moreover, when $\hil_2 = \rone^p$ and $\chi_2 = X$, the problem is equivalent to the following multivariate (finite-dimensional) canonical correlation problem, where $Y$ is the $d$-dimensional random vector such that $Y_j= \inner{\chi_1}{e_j}$, the $j$th score associated with $e_j$, for $j = 1, \ldots d$:
    \begin{align} \label{eq:canonical_pair}
        (a_1,b_1) &= \argmax{\operatorname{Corr}^2\bl a^{\T}Y,b^{\T}X \br}{a \in \rone^d, b \in \rone^p, \Var(a^{\T}Y) = \Var(b^{\T}X) = 1},\\
        (a_k,b_k) &= \argmax{\operatorname{Corr}^2\bl a^{\T}Y,b^{\T}X \br}{\substack{a \in \rone^d, b \in \rone^p, \Var(a^{\T}Y) = \Var(b^{\T}X) = 1\\ \Cov \left ( a^{\T}Y, a_i^{\T}Y\right)=0, i=1, \ldots, k \\
    \Cov \left ( b^{\T}X, b_i^{\T}X\right)=0, i=1, \ldots, k}}, \qquad k = 2, \ldots min(p,d).
    \end{align}
    We call the pair $\bl \sum_{j=1}^d a_{kj} e_j,b_k \br$ the $k$th canonical pair, since $a^{\T}Y = \inner{\sum_{j=1}^d a_{kj} e_j}{\chi_1}$, and $b^{\T}X = \inner{b}{\chi_2}$, where $a_{kj}$ is the $j$th entry of $a_k$, for $k = 1, \ldots\operatorname{min}(p,d)$.
\end{theorem}
This result is central to the proof of Theorem \ref{multivariate-cca-theorem}.
\subsection{Error between finite- and infinite-dimensional problems}\label{sec:finite_infinite_CCA_theory}
In this section, we analyze the error between the finite- and infinite-dimensional CCA problems. In particular, we quantify the error between the canonical variables obtained from solving these problems. We continue to write as $(U_k,V_k)$ the canonical variable solutions to the infinite-dimensional CCA problem in Theorem \ref{theorem_problem}, while we denote as $(U_k^{(d)},V_k^{(d)})$ the canonical variable solutions to the finite-dimensional CCA problem in Theorem \ref{theorem_problem_finite} where we have used a $d$-dimensional Assumption \ref{assumption_cor}. Thus, by definition, $U_k^{(d)} = \eta_k^{\T}Y$ and $V_k^{(d)} = \theta_k^{\T}X$.

The relationship between these problems is well-understood, and we refer to Chapter 10 of \textcite{hsing2015theoretical} for further background. We now introduce the notation and machinery necessary to write down the error between $U_k$ and $U^{(d)}_k$, and $V_k$ and $V^{(d)}_k$, provided in Theorem \ref{thm:finite_infinite_score}. This result is used to show Theorem \ref{thm:consistency_of_scores}.

For the remainder of this section, we suppose that the complete orthonormal system $\{ e_j \}_{j = 1}^{\infty}$ for $\hil_1$ in Assumption \ref{assumption_cor} are the principal components of $\chi_1$ (extended from a CONS for $\clim$).

\subsubsection{Background}
The idea central to deriving our bound is that there are spaces $\mathbb{G}_1 \subseteq \hil_1$ and $\mathbb{G}_2 \subseteq \hil_2$ that are congruent to $\ubar$ and $\vbar$, respectively, so that the infinite-dimensional CCA problem in Theorem \ref{theorem_problem} can be written over $\mathbb{G}_1$ and $\mathbb{G}_2$ instead of $\ubar$ and $\vbar$, and subsequently in terms of an operator, $\mathscr{C}_{12}$, between $\mathbb{G}_1$ and $\mathbb{G}_2$. The infinite-dimensional CCA solution is then derived from the singular vector decomposition of $\mathscr{C}_{12}$, while the solution to the finite-dimensional CCA problem is derived from the singular vector decomposition of a principal component-approximation of $\mathscr{C}_{12}$, $\mathscr{C}_{12}^{(d)}$. The error between $\mathscr{C}_{12}$ and $\mathscr{C}_{12}^{(d)}$ quantifies the error between the canonical variable solutions for the two problems.

Define 
\begin{equation}
    \mathbb{G}_1 \equiv \text{Im} \bl \mathscr{K}_1^{1/2} \br = \left \{ \sumj \omega_j^{1/2} a_j e_j : \sumj a_j^2 < \infty \right \},
\end{equation}
where $(\omega_j,e_j)$ are the eigenvector-eigenvalue pairs of $\mathscr{K}_1$. We define a new inner product on $\mathbb{G}_1 \subseteq \hil_1$, which is not the one inherited from $\hil_1$. For $f,g \in \mathbb{G}_1$ with representations $f = \sumj \omega_j^{1/2} a_j e_j$ and $g = \sumj \omega_j^{1/2} b_j e_j$, we define $\inner{f}{g}_{\mathbb{G}_1} = \sumj a_j b_j$. It is straightforward to verify that $\mathbb{G}_1$ with this inner product is a Hilbert space with a CONS $\left \{\tilde{e}_{1j} \right \}_{j=1}^\infty$ where $\tilde{e}_{1j}\equiv \omega_j^{1/2}e_j$.

We have the following congruency between $\ubar$ and $\mathbb{G}_1$. We say that two Hilbert spaces, $\hil_1$ and $\hil_2$ with norms $\norm{\cdot}_1$ and $\norm{\cdot}_2$, respectively, are congruent if there exists a bijective function $\pi:\hil_1 \rightarrow \hil_2$ such that $\norm{f-g}_1 = \norm{\pi(f)-\pi(g)}_2$ for all $f,g \in \hil_1$.
\begin{lemma}
    $\mathbb{G}_1$ is congruent to $\ubar$, with $\pi:\mathbb{G}_1 \rightarrow \ubar$ defined as the map which takes $f \in \mathbb{G}_1$ with representation $f = \sumj a_j \tilde{e}_{1j}$ to its image $\pi(f) \in \ubar$:
    \begin{equation}
        \pi(f) = \sumj a_j\frac{\inner{\chi_1}{e_j}}{\omega_j^{1/2}}.
    \end{equation}
\end{lemma}
The proof of this result follows directly from Theorem 2.4.16 of \textcite{hsing2015theoretical}, recalling that $\left \{ {\inner{\chi_1}{e_j}}/{\omega_j^{1/2}} \right \}_{j=1}^\infty$ form a CONS for $\ubar$ by Item 4 of Lemma \ref{closure_image_facts_thm}.

We define $\mathbb{G}_2$ and $\pi_2: \mathbb{G}_2 \rightarrow \vbar$ analogously based on the covariance operator for $\chi_2$ for clarity of our presentation, but note that since $\hil_2$ is assumed to be finite-dimensional, $\mathbb{G}_2$ is also finite-dimensional, with $\text{dim} \bl \mathbb{G}_2 \br = p$. We similarly define a CONS $\left \{\tilde{e}_{2j} \right \}_{j=1}^p$ for $\mathbb{G}_2$.

We can now write the problem over $U,V$, which reads as
\begin{equation}\label{eq:infinite_dim_canonical_variable_problem}
\maximize{\operatorname{ Corr}^2 \left ( U,V \right )}{\substack{U \in \ubar, \textrm{ } V \in \vbar,\\
\Var(U) = \Var(V) = 1}},
\end{equation}
in terms of elements of $\mathbb{G}_1$ and $\mathbb{G}_2$ instead:
\begin{equation} \label{eq:variable_problem}
\maximize{\operatorname{ Corr}^2 \left ( \pi_1(f),\pi_2(g) \right )}{\substack{f \in \mathbb{G}_1, \textrm{ } g \in \mathbb{G}_2,\\
\Var(\pi_1(f)) = \Var(\pi_2(g)) = 1}}.
\end{equation}
We define an operator $\mathscr{C}_{12}$ between $\mathbb{G}_2$ and $\mathbb{G}_1$ such that the following property holds. For any $f \in \mathbb{G}_1$, $g \in \mathbb{G}_2$, we have
\begin{equation} \label{eq:def_of_C_12}
    \Cov \bl \pi_1(f), \pi_2(g) \br = \inner{f}{\mathscr{C}_{12}g}_{\mathbb{G}_1}.
\end{equation}
This operator exists, is bounded, and has $\norm{\mathscr{C}_{12}} \leq 1$ by Theorem 10.1.1 of \textcite{hsing2015theoretical}. We denote by $\mathscr{C}_{21}: \mathbb{G}_1 \rightarrow \mathbb{G}_2$ the adjoint of $\mathscr{C}_{12}$, which satisfies $\Cov \bl \pi_1(f), \pi_2(g) \br = \inner{\mathscr{C}_{21}f}{g}_{\mathbb{G}_2}$.

In our setting where $\mathbb{G}_2$ is finite-dimensional, $\mathscr{C}_{12}$ is immediately compact, so that we can solve \eqref{eq:variable_problem} in terms of the singular vector decomposition of $\mathscr{C}_{12}$. In fact, taking the singular vector decomposition of $\mathscr{C}_{12}$ and applying equation \eqref{eq:def_of_C_12} comprise the proof of Theorem \ref{theorem_problem}. We note that if $\mathbb{G}_2$ were infinite-dimensional, we would need to assume compactness of $\mathscr{C}_{12}$.

\subsubsection{Equivalence of finite-dimensional canonical variables}
Before stating our error bound, we need one more result on an equivalent definition of the canonical variables $U_k^{(d)}$ and $V_k^{(d)}$. We additionally introduce a principal component approximation of $\mathscr{C}_{12}$.

Since we have CONS $\left \{\tilde{e}_{1i} \right \}_{i=1}^\infty$ and $\left \{\tilde{e}_{2j} \right \}_{j=1}^p$ for $\mathbb{G}_1$ and $\mathbb{G}_2$, respectively, we can write $\mathscr{C}_{12}$ as
\begin{equation}
    \mathscr{C}_{12} = \sum_{j=1}^p \sum_{i=1}^\infty \inner{\tilde{e}_{2j}}{\mathscr{C}_{12} \tilde{e}_{1i}}_{\mathbb{G}_1}.
\end{equation}
We then define its finite-dimensional approximation by the first $d$ principal components of $\mathscr{K}_1$ as
\begin{equation}
    \mathscr{C}_{12}^{(d)} = \sum_{j=1}^p \sum_{i=1}^d \inner{\tilde{e}_{2j}}{\mathscr{C}_{12} \tilde{e}_{1i}}_{\mathbb{G}_1}.
\end{equation}
We define $\mathscr{C}_{21}^{(d)}$ in an analogous way.

\begin{lemma} \label{lemma:equivalence_of_score_definitions}
     Recall $\bl U_k^{(d)},V_k^{(d)} \br$, the canonical variables obtained by solving the finite-dimensional CCA problem in Theorem \ref{theorem_problem_finite} obtained by making a $d$-dimensional Assumption \ref{assumption_cor}:
     \begin{equation}\label{eq:finite_dim_score_problem}
        \bl U_k^{(d)},V_k^{(d)} \br = \argmax{\Corr^2 \bl U,V \br}{\substack{U \in \ubar_d, V \in \vbar,\\ \Var (U) = \Var (V) = 1}}.
    \end{equation}
    Let $\bl f_k, g_k \br$ be the $k$th pair of right and left singular vectors of $\mathscr{C}_{12}^{(d)}$, respectively. Let $\bl \tilde{U}_k^{(d)},\tilde{V}_k^{(d)} \br$ be defined by
    \begin{align}
        \tilde{U}_k^{(d)} &= \pi_1\bl f_k \br,\\
        \tilde{V}_k^{(d)} &= \pi_2\bl g_k \br.
    \end{align}
    Then, $U_k^{(d)} = \tilde{U}_k^{(d)}$ and $\tilde{V}_k^{(d)} = V_k^{(d)}$.
\end{lemma}
\begin{remark}
    In this paper, we have defined $\bl U_k^{(d)},V_k^{(d)} \br$ directly as the solution to a finite-dimensional CCA problem. In \textcite{hsing2015theoretical}, finite-dimensional canonical variables  are studied and defined as $\bl \tilde{U}_k^{(d)},\tilde{V}_k^{(d)} \br$. This result shows that these two definitions are equivalent.
\end{remark}

\subsubsection{Error bound}
We can finally state the following result about the error in making a $d$-dimensional Assumption \ref{assumption_cor}. The proof follows directly from Theorem 5.2.2 and the proof of equation (10.19) in Theorem 10.2.3 in \textcite{hsing2015theoretical}, the triangle inequality, and Lemma \ref{lemma:equivalence_of_score_definitions}.
\begin{theorem} \label{thm:finite_infinite_score}
     Let $(U_k,V_k)$ be the infinite-dimensional canonical variables obtained by solving the problem in Theorem \ref{theorem_problem}. Let $\bl U_k^{(d)},V_k^{(d)} \br$ be the canonical variables obtained by solving the finite-dimensional CCA problem in Theorem \ref{theorem_problem_finite} obtained by making a $d$-dimensional Assumption \ref{assumption_cor} with the principal components of $\chi_1$. Then for $k \leq d$, as $d \rightarrow \infty$, we have
    \begin{equation}
        \max \left \{ \Ex{\bl  U_k - U_k^{(d)}  \br^2}, \Ex{\bl  V_k - V_k^{(d)}  \br^2} \right \} \lesssim \frac{{\gamma_1^*}^2 \norm{\mathscr{C}_{12} - \mathscr{C}_{12}^{(d)}}^2}{\min_{j \neq k} \left |{\gamma_k^*}^2 - {\gamma_j^*}^2 \right |},
    \end{equation}
     where $\gamma_i^*$ denote the infinite-dimensional canonical correlations, i.e. the singular values of $\mathscr{C}_{12}$.
\end{theorem}
This result is used in the proof of Theorem \ref{thm:consistency_of_scores}.
\subsection{Proofs}
\label{sec:apdx:proofs_of_CCA_formalization_theorems}

\noindent \textbf{Proof of Lemma \ref{closure_image_facts_thm}:}\\
The first item is part 3 of Theorem 7.2.5 of \textcite{hsing2015theoretical}. The second and third items are Theorem 7.2.6 and Theorem 7.2.7 of \textcite{hsing2015theoretical}, respectively.
For the fourth item, equation \eqref{eq:U_set_equation} gives
\begin{equation}
    \mathcal{U} \equiv \left \{ \inner{f}{\chi_1}: f \in \hil_1 \right \} = \left \{ \sumj a_j \inner{e_j}{\chi_1} : \sumj a_j^2 < \infty \right \}.
\end{equation}
From this, it is clear that $\mathcal{U} \subseteq \left \{\sum_{j=1}^\infty a_j \inner{\chi_1}{e_j}:\sumj \omega_j a_j^2 < \infty \right \}$, because $\sum_{j=1}^\infty a_j^2$ implies $\sum_{j=1}^\infty  \omega_j a_j^2$. We also have that
\begin{equation}
    \overline{\text{span}\bl \left \{ \right \inner{\chi_1}{e_j} \}_{j=1}^\infty \br} = \left \{\sum_{j=1}^\infty a_j \inner{\chi_1}{e_j}:\sumj \omega_j a_j^2 < \infty \right \},
\end{equation}
from the definition of the norm in $\mathbb{L}^2\bl \Omega, \mathcal{F},\mathbb{P} \br$ and because we can calculate the norm squared as $\Var \bl \sum_{j=1}^\infty a_j \inner{\chi_1}{e_j} \br = \sumj \omega_j a_j^2$ by the continuity of the inner product. In particular, this set is closed, so $\mathcal{U} \subseteq \left \{\sum_{j=1}^\infty a_j \inner{\chi_1}{e_j}:\sumj \omega_j a_j^2 < \infty \right \}$ implies $\overline{\mathcal{U}} \subseteq \left \{\sum_{j=1}^\infty a_j \inner{\chi_1}{e_j}:\sumj \omega_j a_j^2 < \infty \right \}$. To show the reverse inclusion, it suffices to show that ${\text{span}\bl \left \{ \right \inner{\chi_1}{e_j} \}_{j=1}^\infty \br} \subseteq \mathcal{U}$, which is clear from equation \eqref{eq:U_set_equation}.

That $\{\left\langle\chi_1, e_j\right\rangle/\omega_j^{1/2}\}_{j=1}^\infty$ forms a CONS for $\ubar$ follows from $\ubar = \overline{\text{span}\bl \left \{ \right \inner{\chi_1}{e_j} \}_{j=1}^\infty \br}$ as was just shown, and normalizing the already orthogonal $\inner{\chi_1}{e_j}$ to have norm $1$. \hfill \qedsymbol\\
    
\noindent \textbf{Proof of Lemma \ref{eigenvector_theorem}:}\\
We begin by employing the fourth item of Lemma \ref{closure_image_facts_thm}, which states that $\{\inner{\chi_1}{e_j}/\omega_j^{1/2}\}$ is a CONS for $\ubar$, where the $e_j$ are the eigenvectors of $\mathscr{K}_1$, and $\omega_j$ are the corresponding eigenvalues. Therefore, given an arbitrary CONS for $\clim$, $\{ f_j \}_{j=1, \ldots \infty}$, to complete the proof it suffices to show that $\overline{\text{span} \left \{ \inner{f_j}{\chi_1} \right \} } = \overline{\text{span} \left \{ \inner{e_j}{\chi_1} \right \} }$.

To show the $\subseteq$ direction, it suffices to show that $\inner{f_k}{\chi_1} \in \overline{\text{span} \left \{ \inner{e_j}{\chi_1} \right \} }$, for every $k$, by the definitions of closure and span of a set of vectors. Since the functions $\{e_j\}$ form a CONS for $\clim$, there exists a sequence $\bl a_j \br_{j=1}^{\infty}$ of scalars such that $f_k = \sumj a_j e_j$. Therefore, $\inner{f_k}{\chi_1} = \sumj a_j \inner{\chi_1}{e_j}$ by continuity of the inner product on $\hil_1$, and we have $\inner{f_k}{\chi_1} \in \overline{\text{span} \left \{ \inner{e_j}{\chi_1} \right \} }$.

To show the $\supseteq$ direction, we must show that $\inner{e_k}{\chi_1} \in \overline{\text{span} \left \{ \inner{f_j}{\chi_1} \right \} }$ for every $k$, which follows by similar arguments. \hfill \qedsymbol\\

\noindent \textbf{Proof of Theorem \ref{theorem_problem_finite}:}\\
We prove the statement for the first canonical pair; the proof for the remaining canonical pairs follows from a similar argument. Let $(U,V)$ be the first canonical pair of the population canonical correlation problem in Theorem~\ref{theorem_problem}. We consider the CONS $\{ e_j \}_{j = 1}^{\infty}$ for $\hil_1$ from Assumption \ref{assumption_cor}, and reorder its elements so that $I = \left \{1, \ldots d \right \}$. By Lemma \ref{eigenvector_theorem}, we write $U \in \ubar$ as $U = \sumj a_j \inner{\chi_1}{e_j}$. We will show that under Assumption \ref{assumption_cor}, we can find a $Q = \sumj q_j \inner{\chi_1}{e_j} \in \ubar$, with $q_j = 0$ for $j>d$, that attains the same maximum value as $U$. Thus we will have $Q \in \ubar_d \equiv \left \{ \sum_{i=1}^d a_i \inner{\chi_1}{e_i}: a_i \in \rone \right \}$, completing the proof of the first statement of the Theorem. For random variables with variance $1$, such as $U$ and $V$, we have that $\Cov(U,V) = \operatorname{Corr}(U,V)$. We use these interchangeably throughout the proof.

If the optimum value of $\operatorname{Corr}^2(U,V)$ is $0$, then we can select any $Q$ with $q_j = 0$ for $j>d$ and $\Var(Q) = 1$. Therefore, from now on, we focus on the case where $\operatorname{Corr}^2(U,V) \neq 0$.
Let 
     \begin{equation}
        U = \sum_{j=1}^d a_j \inner{\chi_1}{e_j} + \sum_{j=d+1}^\infty a_j \inner{\chi_1}{e_j} \equiv W + Z.
     \end{equation}
Then, from continuity of the inner product and Assumption \ref{assumption_cor}, it follows that $\operatorname{Cov}(Z,V) = 0$ and $\operatorname{Cov}(W,Z) = 0$, from conditions \eqref{align:first_item} and \eqref{align:second_item} respectively.
Before constructing $Q$, we note that the variance of $W$ must be less than or equal to $1$. To see this, we use
\begin{equation}
    1 = \Var(U) = \Var(W +Z) = \Var(W) + 2\Cov(W,Z) + \Var(Z) = \Var(W) + \Var(Z), 
\end{equation}
since $\Cov(W,Z) = 0$. Then, $\Var(W) \leq 1$ since both $\Var(W)$ and $\Var(Z)$ are positive and sum to $1$.

Now, we construct a canonical variable $Q$ with the desired property. The optimal value of the CCA population problem in Theorem~\ref{theorem_problem} under assumption \ref{assumption_cor} is 
    \begin{align}
        \operatorname{Corr}^2(U,V) &= \operatorname{Cov}^2(W+Z,V)\\
        &= \bl \operatorname{Cov}(W,V) + \operatorname{Cov}(Z,V) \br^2\\
        &= \operatorname{Cov}^2(W,V)\\
        &= \operatorname{Corr}^2(W,V)\Var(W). \label{optimum}
    \end{align}
Having established $\Var(W) \leq 1$, there are three cases, either $\Var(W) =0$, $0 <\Var(W) < 1$, or $\Var(W) = 1$. In the case $\Var(W) = 1$, we take $Q = W$, and using equation \eqref{optimum}, we see that the pair $(W,V)$ attains the same maximum correlation as $(U,V)$. This completes the proof as $W$ is of the desired form.
Now, we will show that the other two cases, $\Var(W) =0$, $0 <\Var(W) < 1$, are not possible. $\Var(W) = 0$ cannot hold since, by equation \eqref{optimum}, we would have $\Cov(U,V) = 0$, which we have already ruled out. Assume towards a contradiction that $0 < \Var(W) < 1$, let $c = \frac{1}{\Var(W)^{1/2}}>1$, and define $Q = cW$. Then, we have that $\Var(Q) = c^2\Var(W) = 1$, and
    \begin{equation}
         \operatorname{Corr}^2(U,V) = \operatorname{Corr}^2\bl W,V \br \Var(W) < \operatorname{Corr}^2 \bl Q,V \br,
    \end{equation}
by equation \eqref{optimum}, $\operatorname{Corr}^2 \bl W,V \br = \operatorname{Corr}^2 \bl Q,V \br$, and $\Var(W) < 1$. However, this is a contradiction as it would imply that the pair $(Q,V)$ attains a larger value of the objective than $(U,V)$. This completes the proof of the first statement.

Having established the existence of a solution of the stated form for $Q$, that we are able to reformulate the CCA problem in terms of the finite-dimensional vectors $\{a_k\}_{k}$ rather than $U \in \ubar$ follows from the definition of $\ubar_d$ and the bilinearity of the inner product $\inner{\cdot}{\cdot}$ on $\hil_1$. In the case that $\hil_2 = \rone^p$ and $\chi_2 = X$, we are able to reformulate the problem in terms of the finite-dimensional vectors $\{b_k\}_k$ rather than $V \in \vbar$ due to the following argument. We have $\mathcal{V} \equiv \{\inner{\chi_2}{g} : \textrm{ }g \in \hil_2 \} = \operatorname{span} \left \{ \inner{\chi_2}{e_j}, j = 1,\ldots p \right \}$ (where the $e_j$ here are the standard unit vectors for $\rone^p$) is isomorphic to $\rone^p$, which is complete. Thus, $\{\inner{\chi_2}{g} : \textrm{ }g \in \hil_2 \}$ is complete, so its completion in $\mathbb{L}^2 \bl \Omega,\calF,\mathbb{P} \br$ is itself, i.e.  $ \vbar = \mathcal{V}$. Therefore, $\vbar = \operatorname{span} \left \{ \inner{\chi_2}{e_j}, j = 1,\ldots p \right \} = \{g^{\T}X : g \in \rone^p \}$, i.e. the set of linear combinations of $X_1, \ldots X_p$. 

The number of nontrivial canonical variables has changed from $p$ in Theorem~\ref{theorem_problem} to $\operatorname{min}(p,d)$.  This is because, in a finite-dimensional CCA problem concerning random vectors of dimensions $p$ and $d$, the smaller of the two dimensions is the upper limit for the number of nontrivial canonical variables \parencite{uurtio2018tutorial}. This completes the proof.
\hfill \qedsymbol \\

\noindent \textbf{Proof of Lemma \ref{lemma:equivalence_of_score_definitions}:}\\
Using the congruency of $\ubar$ with $\mathbb{G}_1$ and $\vbar$ with $\mathbb{G}_2$, the optimization problem in \eqref{eq:finite_dim_score_problem} can be rewritten as
\begin{equation}
    \maximize{\Corr^2 \bl \pi_1(f),\pi_2(g) \br}{\substack{f \in \text{span}\bl \tilde{e}_{11},\ldots \tilde{e}_{1d} \br, g \in \mathbb{G}_2,\\
    \Var \bl \pi_1(f) \br = \Var \bl \pi_2(g) \br = 1}}.
\end{equation}
Using the definition of $\mathscr{C}_{12}$, we have
\begin{equation}
    \Corr^2 \bl \pi_1(f),\pi_2(g) \br = \inner{\mathscr{C}_{21}f}{g}_{\mathbb{G}_2}^2.
\end{equation}
Because $f \in \text{span}\bl \tilde{e}_{11},\ldots \tilde{e}_{1d} \br$, we obtain $\inner{\mathscr{C}_{21}f}{g}_{\mathbb{G}_2} = \inner{\mathscr{C}_{21}^{(d)}f}{g}_{\mathbb{G}_2} = \inner{f}{\mathscr{C}_{12}^{(d)} g}_{\mathbb{G}_1}$.
Therefore, the maximum of this problem is determined by the singular vectors of $\mathscr{C}_{12}^{(d)}$, and we see that ${U}_k^{(d)} = \pi_1\bl f_k \br$ and ${V}_k^{(d)} = \pi_2\bl g_k \br$. \hfill \qedsymbol
\subsection{Proof of Theorem \ref{multivariate-cca-theorem}}\label{subsec:proof_of_main_thm}
Given that Assumption \ref{remark:finite_dim_assumption} is equivalent to Assumption \ref{assumption_cor}, by applying Theorem \ref{theorem_problem_finite}, we readily derive the first part of the theorem. This establishes that there are at most $\dcorr$ nontrivial canonical variable pairs $(U_k,V_k)$. Moreover, each pair $(U_k,V_k)$ can be written in terms of the canonical directions $U_k = \innerdouble{\Logy}{\psi_k}_{\mu}$ and $V_k = X^{\T}\theta_k$, for some $\psi_k \in \Jmu$ and $\theta_k \in \rone^p$. Additionally, $(\psi_k,\theta_k) = \bl \sum_{j=1}^d a_{kj} \phi_j,b_k \br$, where the pairs $(a_k,b_k)$ are defined in Theorem~\ref{theorem_problem_finite} as the solution to a multivariate CCA problem, and the functions $\{\phi_j\}$ form the CONS for $\Jmu$, defined in Section~\ref{Asymmetric_CCA_section}.

It remains to be shown that the solutions $(a_k,b_k)$ to the multivariate CCA problem can be characterized by the equations \eqref{eq:opt}-\eqref{eq:TH}. We focus on the finite-dimensional optimization problem, in equation \eqref{eq:canonical_pair}, that defines the first canonical pair. This is equivalent to
$$\supp{a_1^{\T}\Sigma_{XY}b_1}{a_1^{\T}\Sigma_Xa_1 = 1 = b_1^{\T}\sigmay b_1}.$$
Now using the assumption that $\Sigma_X$ and $\Sigma_Y$ are invertible, we make a change of variables $\tilde{a}_1 = \Sigma_X^{1/2}a_1$, $\tilde{b}_1 = \Sigma_Y^{1/2}b_1$ and obtain the equivalent problem
$$\supp{\tilde{a}_1^{\T}\Sigma_X^{-1/2}\Sigma_{XY}\sigmay^{-1/2}\tilde{b}_1}{\tilde{a}_1^{\T}\tilde{a}_1 = 1 = \tilde{b}_1^{\T}\tilde{b}_1}$$
Let $U\Gamma V^{\T} = \Sigma_X^{-1/2}\Sigma_{XY}\sigmay^{-1/2}$ be a singular value decomposition of $\Sigma_X^{-1/2}\Sigma_{XY}\sigmay^{-1/2}$, where $U \in \rone^{p \times \dcorr}$, $\Gamma \in \rone^{\dcorr \times \dcorr}$, $V \in \rone^{\dcorr \times \dcorr}$, $U^{\T}U = I_\dcorr = V^{\T}V$, and where $\Gamma$ is a diagonal matrix with the diagonal elements $\gamma_1, \ldots \gamma_\dcorr$, in descending order. Note that $p \geq \dcorr$. Then it follows from standard properties of the SVD that the first columns of $U$ and $V$, denoted as $u_1$ and $v_1$ respectively, are the solutions to the above problem, i.e. $(\tilde{a}_1,\tilde{b}_1) = (u_1,v_1)$.
Similarly, it can be shown that the $k$th columns of $U$ and $V$, $u_k$ and $v_k$ respectively, are such that $(\tilde{a}_k,\tilde{b}_k) = (u_k,v_k)$, and that the optimal correlations are the singular values $\gamma_1, \ldots \gamma_\dcorr$. Undoing the change of variables, it can be seen that the solutions to the original problems in equation \eqref{eq:canonical_pair} are the pairs formed by the $k$th columns of the matrices $\Sigma_X^{-1/2}U$ and $\sigmay^{-1/2}V$. The associated squared correlations are the diagonal entries of $\Gamma^2$.

Now let $B$ be the solution to the optimization problem
\begin{equation}
    \minimize{ \textrm{ }\E \left [ \|\sigmay^{-1/2}Y-B^{\T}X \|_2^2 \right ]}{B \in \rone^{p \times \dcorr}}.
\end{equation}
It is straightforward to show that $B = \sigmax^{-1}\sigmaxy\sigmay^{-1/2}$. Therefore, we have
\begin{equation}
    \Sigma_X^{1/2}B = U \Gamma V^{\T},
\end{equation}
\begin{equation}
    B^{\T}\Sigma_XB = V\Gamma^2V^{\T}
\end{equation}
and
\begin{equation}
    BV\Gamma^{-1} = \Sigma_X^{-1/2}U.
    \label{sparsitypls}
\end{equation}
Identifying $\tilde{H}$, $H$, and $T$ in equations \eqref{eq:opt}-\eqref{eq:TH} with $V$, $\sigmay^{-1/2}V$, and $\Sigma_X^{-1/2}U$, respectively, completes the proof.

\section{Asymmetric Sparse CCA: Proof of Theorem~\ref{thm:main_theorem_multivariate_case}}\label{sec:proof_asymmetric_sparse_cca}

\subsection{Notation}
\label{appendix:notation}
For a vector $x \in \rone^p$ with entries $\{x_j\}$ we define its infinity norm $\infnorm{x} = \operatorname{max}_j(|x_j|)$, its Euclidean norm $\twonorm{x} = \sqrt{\sum_{j=1}^p x_j^2}$, and its $\ell_1$ norm $\onenorm{x} = \sum_{j=1}^p |x_j|$. For a matrix $A \in \rone^{p \times d}$ with singular values $\sigma_1, \ldots \sigma_d$, its operator norm is $\twonorm{A} = \operatorname{max}_i(|\sigma_i|)$. To denote the $i$th row of the matrix $A$, we use $A_i$, and for the entry in the $i$th row and $j$th column, we use $a_{ij}$. We define the matrix norms $\Fnorm{A} = \left( \sum_{i=1}^p \sum_{j=1}^d a_{ij}^2 \right)^{1/2}$, $\onetwo{A}=\sum_{i=1}^p \twonorm{A_i}$, and $\maxnorm{A}=\max_{(i,j)} |a_{i,j}|$.

Given the normed spaces $(\rone^d,\norm{\cdot}_{\alpha})$ and $(\rone^p,\norm{\cdot}_{\beta})$, and a matrix $A \in \rone^{p \times d}$, we define the matrix norm induced by $\norm{\cdot}_{\alpha}$ and $\norm{\cdot}_{\beta}$ as \begin{equation}
    \norm{A}_{\alpha,\beta} = \supp{\norm{Ax}_{\beta}}{\norm{x}_{\alpha}=1}.
\end{equation}
For additional properties of the matrix norms used throughout the paper, we refer to Section \ref{subsec:identities}.
We use the notation $x \lesssim y$ for $x,y \in \rone$ to indicate that $x \leq C y$, with $C$ some positive absolute constant.

\subsection{Sub-Gaussian random vectors}
Now we briefly define sub-Gaussian random vectors and state basic properties that we use in the proofs. We refer the reader to \textcite{vershynin2018high} for a more comprehensive introduction to sub-Gaussian random variables and vectors. 

A random variable $X$ is sub-Gaussian if, for some constant $C>0$, it satisfies 
\begin{equation}
\mathbb{P}\{|X| \geq t\} \leq 2 \exp \left(-t^2 / C\right) \quad \text { for all } t \geq 0.
\end{equation}
The sub-Gaussian norm of $X$ is defined as 
\begin{equation}
\|X\|_{\psi_2}=\inf \left\{t>0: \mathbb{E} \exp \left(X^2 / t^2\right) \leq 2\right\}.
\end{equation}
A random vector $X \in \rone^p$ is called sub-Gaussian if $\inner{X}{x}$ is sub-Gaussian for all $x \in \rone^p$. The sub-Gaussian norm of $X$ is defined as
\begin{equation}
\|X\|_{\psi_2}=\sup _{x \in \rone^p: \twonorm{x} = 1}\|\langle X, x\rangle\|_{\psi_2}.
\end{equation}
From its definition, it is clear that $\subgaussnorm{X_i} \leq \subgaussnorm{X}$, where $X_i$ is the $i$th element of $X$. To simplify our analysis, we will also assume that sub-Gaussian vectors $X$ satisfy the variance-proxy condition defined below.
\begin{definition} \label{assumption:subgauss_proxy}
A sub-Gaussian random vector $X$ satisfies the variance-proxy condition if there exists a constant $K_X$ such that for any $x \in \rone^p$, $\subgaussnorm{\inner{X}{x}} \leq K_X \Var\bl\inner{X}{x}\br^{1/2}$.
\end{definition}
Intuitively, this condition implies that the sub-Gaussian norms of the one-dimensional marginals of $X$ can be used as proxies for their standard deviations. Note that the reverse inequality $\Var\bl\inner{X}{x}\br^{1/2} \leq K \subgaussnorm{\inner{X}{x}}$ for $K = \sqrt{2}$ is always satisfied when $X$ has mean $0$ (Proposition 2.5.2. (ii) of \textcite{vershynin2018high}). Moreover, for a Gaussian random vector $X$, this proxy assumption holds with $K_X = 1$. If $X$ is a zero-mean sub-Gaussian random vector that satisfies the variance-proxy condition and has covariance matrix $\Sigma_X$, it follows from the definition above that $\subgaussnorm{X} \leq K_X \twonorm{\Sigma_X}^{1/2}$. Additionally, it is straightforward to show that $\operatorname{max}_i(\subgaussnorm{X_i}) \leq K_X \twoinfnorm{\Sigma_X}^{1/2}$, where $X_i$ is the $i$th entry of $X$. The proxy assumption allows us to compare sub-Gaussian norms of vectors to one another through their variances.

Throughout our proofs, we assume that the variance-proxy condition applies to the random vectors $X$, $B^{\T}X$, $Y$, $\sigmay^{-1/2}Y$, and $\sigmay^{-1}Y$. To simplify our assumptions, for the main theorems in this section, we conveniently assume that $X$ and $Y$ are strict sub-Gaussians, as defined in \parencite{kereta2021estimating}:
\begin{definition} \label{def:strict_sub_gaussian}
    A sub-Gaussian random vector $X$ is called strict sub-Gaussian if there exists a constant $K_X$ such that for any matrix $U \in \rone^{k \times p}$, the following inequality is satisfied: 
\begin{equation}
     \subgaussnorm{UX} \leq K_X \twonorm{\Sigma_{UX}}^{1/2}.
\end{equation}
\end{definition}

\subsection{Proof of Theorem \ref{thm:main_theorem_multivariate_case}}

Recall that the matrices $\mathbb{X} \in \mathbb{R}^{N \times p}$ and $\mathbb{Y} \in \mathbb{R}^{N \times d}$ consist of $N$ samples of the random vectors $X \in \mathbb{R}^p$ and $Y \in \mathbb{R}^d$, respectively. We assume that $\sigmax$ and $\sigmay$ are invertible, and without loss of generality, we assume that $X$ and $Y$ have mean $0$. 

To estimate $\sigmax$ and $\sigmay$, we use their respective sample covariance estimates $\sigmayhat = \mathbb{Y}^{\T}\mathbb{Y}/N$ and $\sigmaxhat = \mathbb{X}^{\T}\mathbb{X}/N$. Define $B = \sigmax^{-1}\sigmaxy\sigmay^{-1/2}$, let $\bhat$ be the solution to the sample group lasso problem \eqref{eq:grp_lasso}, and let $\lambda$ be the associated penalization constant. In the setting of Theorem~\ref{multivariate-cca-theorem}, if we define $\tilde H$ by the eigendecomposition $B^{\T}\sigmax B \equiv \tilde{H} D^2 \tilde{H}^{\T}$, then by letting 
\begin{align}
    T = B \tilde{H} D^{-1} \in \rone^{p \times d},\\ 
    H = \sigmay^{-1/2}\tilde{H} \in \rone^{d \times d},
\end{align}
it follows that the $k$th column of $H$, $\eta_k$, is the $k$th canonical vector associated with $Y$, and the $k$th column of $T$, $\theta_k$, is the $k$th canonical vector associated with $X$. Moreover, the diagonal entries of $D^2$ are the squared population canonical correlations $\gamma_1 > \ldots > \gamma_d$, which we assume are distinct. This allows us to focus on estimating individual canonical vectors rather than subspaces spanned by canonical vectors sharing identical correlations. 

We denote the columns of $\tilde H$ as $\tilde \eta_k$ and denote by $\{\hat \theta_k\}$ and $\{\hat \eta_k\}$ the estimates of the canonical vectors, and by $\{\hat{\gamma}_k\}$ the estimated canonical correlations, that is, the diagonal entries of $\hat{D}$. Note that by definition, the squared population correlations $\gamma_1^2 \ldots \gamma_d^2$ are the eigenvalues of $B^{\T}\sigmax B$ and the estimated squared correlations $\hat{\gamma}_1^2, \ldots \hat{\gamma}_d^2$ are the eigenvalues of $\bhat^T\sigmaxhat\bhat$. In the remainder of this section, we derive bounds on the estimation error for the canonical correlations, quantified by $| \gamma_k^2 - \hat{\gamma}_k^2|$, and the canonical vectors, quantified by $\twonorm{\eta_k - \hat{\eta}_k}^2$ and $\twonorm{\theta_k - \hat{\theta}_k}^2$.

\subsubsection{Deterministic bounds}\label{appendix:det_bounds_mult}
We begin by presenting our deterministic results. To establish fast-rate bounds, we use the Group restricted eigenvalue condition, analogously to the lasso regression problem \parencite{hastie2015statistical} and similar to \textcite{gaynanova2020prediction} in the context of penalized optimal scoring.

\begin{definition}[Group restricted eigenvalue condition] \label{def:group_restricted_eigenvalue}
    A matrix $Q \in \rone^{q \times p}$ satisfies the Group restricted eigenvalue condition RE$(s,c,d)$ with parameter $\kappa$ if for all sets $S \subset \{1, \ldots p\}$ with $|S| \leq s$, we have that, for all $ A \in \rone^{p \times d}$ such that $\onetwo{A_{\bar S}} \leq c \onetwo{A_S}$,\begin{equation}
    \norm{QA}_F \geq \frac{\norm{A_S}_F^2}{\kappa}.
\end{equation}
Here, $|S|$ denotes the cardinality of $S$, and $\bar S = \{1, \ldots p\} \backslash S$.
\end{definition}

The following lemma establishes a deterministic bound for the $2$-norm of the difference between the linear operators $B^{\T}\sigmax B$ and $\bhat^T\sigmaxhat\bhat$. In turn, this quantity will be used to bound the errors $|\gamma_k^2- \hat{\gamma}_k^2|$, $\twonorm{\eta_k - \hat{\eta}_k}^2$ and $\twonorm{\theta_k - \hat{\theta}_k}^2$.
\begin{lemma}
The following inequality holds:
\label{lemma:deterministic_bound_1}
\begin{align*}
    \twonorm{\bhat^T\sigmaxhat\bhat - B^{\T}\sigmax B} & \leq \rootnfrac\twonorm{\mathbb{X}B} \rootnfrac \Fnorm{\mathbb{X}(\bhat - B)} + \twonorm{B^{\T}(\sigmax - \sigmaxhat)B} + \gamma_1\twonorm{(\bhat-B)^{\T}\sigmax^{1/2}}\\
    & + \nfrac \Fnorm{\mathbb{X}(\bhat - B)}^2 + \onetwo{\bhat-B}\twoinfnorm{(\sigmax-\sigmaxhat)B}.
\end{align*}
\end{lemma}
In the equation above, the first-order terms appear on the first line, while the second-order terms appear on the second line of the equation. In this section, wherever possible, we will keep the convention.

Let $E = \mathbb{Y}\sigmayhat^{-1/2}-\mathbb{X}B$. Next, we derive `slow'- and `fast'-rate deterministic bounds.

\begin{lemma}
    \label{lemma:fast-slow-rate-appendix1}
    If $\lambda \geq \frac{2}{N}\twoinfnorm{\mathbb{X}^{\T}E}$, then the following slow-rate bound holds:
    \begin{align*}
        \twonorm{\bhat^T\sigmaxhat\bhat - B^{\T}\sigmax B} & \lesssim \rootnfrac \twonorm{\mathbb{X}B} \sqrt{\lambda} \onetwo{B}^{1/2} + \twonorm{B^{\T}(\sigmax - \sigmaxhat)B}\\
        &+ \gamma_1 \onetwo{B}^{1/2} \bl \lambda \onetwo{B} + \maxnorm{\sigmaxhat - \sigmax} \br^{1/2}\\
         &+ \lambda \onetwo{B} + \onetwo{B}\twoinfnorm{(\sigmax-\sigmaxhat)B}.
    \end{align*}
    If, additionally, $B$ has at most $s$ non-zero rows, and $\rootnfrac \mathbb{X}$ satisfies the Group restricted eigenvalue condition
    RE$(s,3,d)$ with parameter $\kappa_X$, then the following fast-rate bound holds:
    \begin{align*}
        \twonorm{\bhat^T\sigmaxhat\bhat - B^{\T}\sigmax B}& \lesssim \rootnfrac \twonorm{\mathbb{X}B} \kappa_X^{1/2}s^{1/2}\lambda + \twonorm{B^{\T}(\sigmax - \sigmaxhat)B} + \gamma_1 \twonorm{\sigmax}^{1/2} \kappa_X s^{1/2} \lambda\\
        & + \kappa_X s \lambda^2 + \twoinfnorm{(\sigmax-\sigmaxhat)B} \kappa_X s \lambda.
    \end{align*}
\end{lemma}
Note that in the fast-rate bound $\sqrt{\lambda}$ and $\onetwo{B}$ are replaced with $\lambda$ and $\kappa_X s$, respectively. Next, we derive a bound for $\frac{1}{N}\twoinfnorm{\mathbb{X}^{\T}E}$.
\begin{lemma}
The following inequality holds:
    \label{lemma:lambda_0_bound}
    \begin{align*}
        \frac{1}{N}\twoinfnorm{\mathbb{X}^{\T}E} &\leq \twoinfnorm{\bl \sigmaxyhat - \sigmaxy \br\sigmay^{-1/2}} + \twoinfnorm{\sigmaxy \bl \sigmayhat^{-1/2} - \sigmay^{-1/2} \br} + \twoinfnorm{\bl \sigmax - \sigmaxhat \br B}\\
        &+ \twoinfnorm{\bl \sigmaxyhat - \sigmaxy \br \bl \sigmayhat^{-1/2} - \sigmay^{-1/2} \br}.
    \end{align*}
\end{lemma}
Denote the right-hand side of the equation in Lemma~\ref{lemma:lambda_0_bound} as $\lambda_0$. Given that $\frac{2}{N}\twoinfnorm{\mathbb{X}^{\T}E} \leq 2\lambda_0$, choosing $\lambda \geq 2\lambda_0$ ensures that $\lambda \geq \frac{2}{N}\twoinfnorm{\mathbb{X}^{\T}E}$. Thus, we can replace the assumption $\lambda \geq \frac{2}{N}\twoinfnorm{\mathbb{X}^{\T}E}$ with the assumption $\lambda \geq 2 \lambda_0$. Later, we will establish a high-probability bound for $\lambda_0$.

Due to the fact that $\twoinfnorm{\bl \sigmax - \sigmaxhat \br B} \leq \lambda_0$, we obtain the following simplification of Lemma \ref{lemma:fast-slow-rate-appendix1}, where the fourth and fifth terms are combined.
\begin{lemma}
\label{lemma:fast-slow-rate-appendix2}
    If $\lambda \geq 2\lambda_0$, then the following slow-rate bound holds:
    \begin{align*}
        \twonorm{\bhat^T\sigmaxhat\bhat - B^{\T}\sigmax B} & \lesssim \rootnfrac \twonorm{\mathbb{X}B} \sqrt{\lambda} \onetwo{B}^{1/2} + \twonorm{B^{\T}(\sigmax - \sigmaxhat)B}\\ 
        &+ \gamma_1 \onetwo{B}^{1/2} \bl \lambda \onetwo{B} + \maxnorm{\sigmaxhat - \sigmax} \br^{1/2}\\
        & + \lambda \onetwo{B}.
    \end{align*}
    If, additionally, $B$ has at most $s$ non-zero rows, and $\rootnfrac \mathbb{X}$ satisfies the Group restricted eigenvalue condition
    RE$(s,3,d)$ with parameter $\kappa_X$, then the following fast-rate bound holds:
    \begin{align*}
        \twonorm{\bhat^T\sigmaxhat\bhat - B^{\T}\sigmax B} &\lesssim \rootnfrac \twonorm{\mathbb{X}B} \kappa_X^{1/2}s^{1/2}\lambda + \twonorm{B^{\T}(\sigmax - \sigmaxhat)B}\\
        &+ \gamma_1 \kappa_X^{1/2}s^{1/2}\bl 1 + \kappa_Xs\maxnorm{\sigmaxhat - \sigmax} \br^{1/2} \lambda\\
        & + \kappa_X s \lambda^2.
    \end{align*}
\end{lemma}

Lemma~\ref{lemma:fast-slow-rate-appendix2} shows that the rate of convergence will ultimately be determined by $\lambda_0$, $\twonorm{B^{\T}(\sigmax - \sigmaxhat)B}$, and $\maxnorm{\sigmaxhat - \sigmax}$.

\subsubsection{Probabilistic bounds}\label{appendix:prob_bounds_mult}
From now on, we assume that $X$ and $Y$ are sub-Gaussian random vectors and that the variance-proxy condition in Definition~\ref{assumption:subgauss_proxy} holds for the random vectors $X$, $B^{\T}X$, $Y$, $\Sigma_Y^{-1/2}Y$, and $\sigmay^{-1}Y$. We will repeatedly use the union bound and omit for simplicity the absolute constants arising from its applications.

First, we present an intermediary result that will be used to derive a probabilistic upper bound for $\lambda_0$.
\begin{lemma}
    \label{lemma:prob_bound_for_lambda_0}
    Let $X \in \rone^p$ and $Z \in \rone^d$ be zero-mean random vectors with covariance matrices $\sigmax$ and $\Sigma_Z$ and cross-covariance matrix $\Sigma_{XZ}$. Assume the entries of $X$ and $Z$ are sub-Gaussian random variables with norms $\subgaussnorm{X_i} = g_i$ and $\subgaussnorm{Z_j} = h_i$, for $i= 1, \ldots p$ and $j = 1, \ldots d$. Let $g = \operatorname{max}(g_i)$ and $h = \operatorname{max}(h_j)$. Let $\mathbb{X} \in \rone^{N \times p}$ and $\mathbb{Z} \in \rone^{N \times d}$ be data matrices such that the pairs of rows $\{(\mathbb{X}_i^{\T},\mathbb{Z}_i^{\T})\}$ are independent samples from the joint distribution $(X,Z)$. 
    If $d \leq p$ and $\operatorname{log}(p) = o(N)$, then for any fixed $\eta \in (0,1)$, with probability at least $1-\eta$,
    \begin{equation}
        \twoinfnorm{\Sigma_{XZ} - \frac{1}{N}\mathbb{X}^{\T}\mathbb{Z}} \lesssim g h \sqrt{\frac{d}{N}\operatorname{log}(p\eta^{-1})}.
    \end{equation}
\end{lemma}
\begin{remark}
    In Lemma~\ref{lemma:prob_bound_for_lambda_0}, it is stated that for a fixed $\eta$, if $\lim \bl \operatorname{log}(p)/N \br = 0$ as $p$ and $N$ go to infinity, then, eventually, the stated bound holds.
\end{remark}
Next, we derive probabilistic upper bounds for $\lambda_0$, bounding the terms in Lemma~\ref{lemma:lambda_0_bound}.
\begin{lemma}
    \label{lemma:prob_bound_for_lambda_0_2}
    If $d \leq p$ and $\operatorname{log}(p) = o(N)$, then for any fixed $\eta \in (0,1)$, with probability $1-\eta$,
    \begin{equation}
        \twoinfnorm{\bl \sigmaxyhat - \sigmaxy \br\sigmay^{-1/2}} \lesssim \operatorname{max}_i(\subgaussnorm{X_i}) \sqrt{\frac{d}{N}\operatorname{log}(p\eta^{-1})}
    \end{equation}
    and
    \begin{equation}
        \twoinfnorm{\bl \sigmax - \sigmaxhat \br B} \lesssim \operatorname{max}_i(\subgaussnorm{X_i}) \gamma_1 \sqrt{\frac{d}{N}\operatorname{log}(p\eta^{-1})}.
    \end{equation}
    Moreover, if $d = o(N)$, then
    \begin{equation}
        \twoinfnorm{\sigmaxy \bl \sigmayhat^{-1/2} - \sigmay^{-1/2} \br} \lesssim \twoinfnorm{\sigmax}^{1/2}\gamma_1 \sqrt{\frac{d + \operatorname{log}(\eta^{-1})}{N}}
    \end{equation}
    and
    \begin{equation}
        \twoinfnorm{\bl \sigmaxyhat - \sigmaxy \br \bl \sigmayhat^{-1/2} - \sigmay^{-1/2} \br} \lesssim \twoinfnorm{\bl \sigmaxyhat - \sigmaxy \br\sigmay^{-1/2}}.  
    \end{equation}
\end{lemma}

\begin{remark}
    \label{remark:lambda_0_prob_bound_subgauss_bound}
    As noted in Section~\ref{appendix:notation}, $\operatorname{max}_i(\subgaussnorm{X_i}) \lesssim \twoinfnorm{\sigmax}^{1/2}$. 
\end{remark}

Using Lemmas~\ref{lemma:lambda_0_bound} and \ref{lemma:prob_bound_for_lambda_0_2}, and Remark~\ref{remark:lambda_0_prob_bound_subgauss_bound}, it straightforward to derive the following result.
\begin{lemma}
\label{lemma:lambda_0_final_bound}
    If $d \leq p$, $\operatorname{log}(p) = o(N)$, and $d = o(N)$, then for any fixed $\eta \in (0,1)$, with probability $1-\eta$,
    \begin{equation}
        \lambda_0 \lesssim  \twoinfnorm{\sigmax}^{1/2} \sqrt{\frac{d}{N}\operatorname{log}(p\eta^{-1})}.
    \end{equation}
\end{lemma}

Next, we establish bounds on the other terms appearing in \ref{lemma:fast-slow-rate-appendix2}.
\begin{lemma}
\label{lemma:other_terms}
     If $\operatorname{log}(p) = o(N)$, then for any fixed $\eta \in (0,1)$, with probability $1-\eta$,
     \begin{equation}
         \maxnorm{\sigmax - \sigmaxhat} \lesssim \operatorname{max}(\subgaussnorm{X_i}^2) \sqrt{\frac{\operatorname{log}\bl p\eta^{-1} \br}{N}}.
     \end{equation}
     If $d = o(N)$, then for any fixed $\eta \in (0,1)$, with probability $1-\eta$,
     \begin{equation}
          \twonorm{B^{\T}(\sigmax - \sigmaxhat)B} \lesssim \gamma_1^2 \sqrt{\frac{d + \operatorname{log}\bl \eta^{-1} \br}{N}},
     \end{equation}
     and
     \begin{equation}
         \rootnfrac \twonorm{\mathbb{X}B} \lesssim \gamma_1.
     \end{equation}
\end{lemma}

Before presenting our final bounds, we establish that the group-restricted eigenvalue condition holds for the design matrix $\frac{1}{\sqrt{N}}\mathbb{X}$, with high probability, assuming that the same condition holds for $\sigmax^{1/2}$.
\begin{lemma}
\label{lemma:probabilistic_group_restricted_eigenvalue_condition}
    Suppose $\sigmax^{1/2}$ satisfies the group restricted eigenvalue condition RE$(s,3,d)$ with parameter $\kappa = \kappa(s,d,\sigmax^{1/2})$. If $ \operatorname{max}_{i=1,\ldots p}(\subgaussnorm{X_i}^4) \kappa^2 s^2\log\bl p\br = o(N)$ and $s^2\operatorname{log}(p) = o(N)$, then for any fixed $\eta$, with probability $1-\eta$, $\frac{1}{\sqrt{N}}\mathbb{X}$ satisfies the group restricted eigenvalue condition RE$(s,3,d)$ with parameter $\kappa_X$, where
    \begin{equation}
        0 < \kappa_X \leq 2 \kappa.
    \end{equation}
\end{lemma}

Next, we state our probabilistic bound for $\twonorm{\bhat^T\sigmaxhat\bhat - B^{\T}\sigmax B}$. The proof of the slow-rate bound follows straightforwardly from Lemmas \ref{lemma:fast-slow-rate-appendix2}, \ref{lemma:lambda_0_final_bound} and \ref{lemma:other_terms}. The proof of the fast-rate bound follows similarly from Lemmas \ref{lemma:fast-slow-rate-appendix2}, \ref{lemma:lambda_0_final_bound} and \ref{lemma:other_terms}, with the addition of Lemma~\ref{lemma:probabilistic_group_restricted_eigenvalue_condition}.

\begin{theorem}
\label{thm:operator_norm_bound}
Assume $X$ and $Y$ are sub-Gaussian random vectors and that $X$, $B^{\T}X$, $\sigmay^{-1/2}Y$ satisfy the variance-proxy condition \ref{assumption:subgauss_proxy}. Moreover, assume that $d \leq p$, $\log(p) = o(N)$, and $d = o(N)$. Fix $\eta \in (0,1)$, and for some absolute constant $C>0$, let $\lambda = C \twoinfnorm{\sigmax}^{1/2}\sqrt{\frac{d}{N}\log (p\eta^{-1})}$. Then, with probability $1-\eta$, the following slow-rate bound holds:
\begin{align*}
    &\twonorm{\bhat^T\sigmaxhat\bhat - B^{\T}\sigmax B} \lesssim\\
    & \bl \frac{d}{N}\log(p\eta^{-1})\br^{1/4} \left [\gamma_1^2 + \twoinfnorm{\sigmax}^{1/2} \onetwo{B} + \gamma_1 \twoinfnorm{\sigmax}^{1/4}\onetwo{B}^{1/2} \bl 1 +
    \onetwo{B}^{1/2} + \twoinfnorm{\sigmax}^{1/4} \br \right].
\end{align*}
Under the additional assumption that $B \in \rone^{p \times d}$ has at most $s$ nonzero rows, $\sigmax^{1/2}$ satisfies the group restricted eigenvalue condition RE$(s,3,d)$ with parameter $\kappa = \kappa(s,d,\sigmax^{1/2})$, $s^2\log(p) = o(N)$, and $\twoinfnorm{\sigmax}^2\kappa^2s^2\log(p) = o(N)$, then the following fast-rate bound holds:
\begin{equation}
    \twonorm{\bhat^T\sigmaxhat\bhat - B^{\T}\sigmax B} \lesssim \bl \frac{d}{N}\log\bl p\eta^{-1}\br \br^{1/2} \left [ \gamma_1\bl \gamma_1 + \kappa^{1/2}s^{1/2}\twoinfnorm{\sigmax}^{1/2} \br \right].
\end{equation}
\end{theorem}

\begin{corollary}
    \label{cor:simplified_operator_norm_bound}
    In the setting of Theorem~\ref{thm:operator_norm_bound}, under the additional assumption that $\twoinfnorm{\sigmax},\onetwo{B} \geq 1$, then the expression of the slow-rate bound simplifies as follows: 
\begin{equation}
    \twonorm{\bhat^T\sigmaxhat\bhat - B^{\T}\sigmax B} \lesssim \bl \frac{d}{N}\log\bl p\eta^{-1}\br \br^{1/4} \left [ \gamma_1 \twoinfnorm{\sigmax}^{1/2} \onetwo{B} \right].
\end{equation}
Under the additional assumption that $\twoinfnorm{\sigmax},\kappa \geq 1$, then the expression of the fast-rate bound simplifies as follows: 
\begin{equation}
    \twonorm{\bhat^T\sigmaxhat\bhat - B^{\T}\sigmax B} \lesssim \bl \frac{d}{N}\log\bl p\eta^{-1}\br \br^{1/2} \left [ \gamma_1 \twoinfnorm{\sigmax}^{1/2}s^{1/2}\kappa^{1/2} \right].
\end{equation}
\end{corollary}
\begin{remark}
    The eigenvalues of the matrices $\bhat^T\sigmaxhat\bhat$ and $ B^{\T}\sigmax B$ are $\{ \hat{\gamma}_k^2\}$ and $\{\gamma_k^2 \}$ respectively. Then by Weyl's inequality (\textcite{bhatia2013matrix} Corollary III.2.6.), because we have bounded the operator norm of the difference between these two matrices, we immediately obtain bounds on the estimation error of $\gamma_k^2 $ by $\hat{\gamma}_k^2$ in Algorithm~\ref{alg:asymmetric_sparse_cca}.
\end{remark}
To establish bounds for $\twonorm{\theta_k-\hat \theta_k}$ and $\twonorm{\eta_k - \hat \eta_k}$, we first introduce a few supporting lemmas.

\begin{lemma}
    \label{lemma:B_minus_B_bound}
    Under the slow-rate bound assumptions stated in Corollary~\ref{cor:simplified_operator_norm_bound}, for any fixed $\eta \in (0,1)$, with probability at least $1-\eta$, we have
    \begin{equation}
        \twonorm{B-\hat B} \lesssim \bl \frac{d}{N}\log\bl p\eta^{-1}\br \br^{1/4} \twonorm{\sigmax^{-1/2}}\onetwo{B}\twoinfnorm{\sigmax}^{1/2}.
    \end{equation}
    Under the fast-rate bound assumptions stated in Corollary~\ref{cor:simplified_operator_norm_bound}, for any fixed $\eta \in (0,1)$, with probability at least $1-\eta$, we have
    \begin{equation}
        \twonorm{B-\hat B} \lesssim \bl \frac{d}{N}\log\bl p\eta^{-1}\br \br^{1/2} \kappa s^{1/2} \twoinfnorm{\sigmax}^{1/2}.
    \end{equation}
\end{lemma}

\begin{lemma}
    \label{lemma:y_covariance_bounds}
    Suppose that $Y \in \rone^d$ is a sub-Gaussian vector, $d = o(N)$ and that $Y$ satisfies the variance-proxy condition. Then, for any fixed $\eta \in (0,1)$, with probability $1-\eta$, we have
    \begin{equation}
        \twonorm{\sigmay - \sigmayhat} \lesssim \twonorm{\sigmay} \sqrt{\frac{d\log \bl \eta^{-1}\br}{N}}.
    \end{equation}
    Additionally, suppose that $\sigmay^{-1}Y$ satisfies the variance-proxy condition, and $\twonorm{\sigmay}^2d = o(N)$. Then for fixed $\eta \in (0,1)$, with probability $1-\eta$, we have
    \begin{equation}
        \twonorm{\sigmay^{-1/2} - \sigmayhat^{-1/2}} \lesssim \twonorm{\sigmay^{1/2}}\twonorm{\sigmay^{-1/2}}^2 \sqrt{\frac{d\log \bl \eta^{-1}\br}{N}}.
    \end{equation}
\end{lemma}

Studying the theoretical properties of CCA through the lens of regression, using the matrix $B$, has been convenient thus far. However, for our final results, we bound $B =\sigmax^{-1/2}\tilde{T}D\tilde{H} = T D \tilde{H}$ in terms of quantities that are more directly related to the CCA problem. Using identity \ref{item:one_two_norm_inequality} in Section~\ref{subsec:identities}, along with the standard properties of the $2$-norm, and noting that $\tilde{T}$, $\tilde{H}$ are orthogonal matrices and $D$ is a diagonal matrix with diagonal values no greater than $1$, we observe that
\begin{align}
    \onetwo{B} \leq &\onetwo{T} \leq \onetwo{\sigmax^{-1/2}},\\
    \twonorm{B} \leq &\twonorm{T} \leq \twonorm{\sigmax^{-1/2}}.
\end{align}
Hence, in Corollary~\ref{cor:simplified_operator_norm_bound}, we can replace the assumption that $\onetwo{B} \geq 1$ with the assumption that $\onetwo{T} \geq 1$.

Next, we state our probabilistic bounds on the estimated canonical vectors. We denote with $K = \operatorname{max} \left \{i \in \{ 1, \ldots d \}: \gamma_i > 0 \right \}$ the number of nontrivial canonical vectors. Moreover, to simplify the notation, we use the conventions $\gamma_{K+1}^2=-\infty$ and $\gamma_0^2=\infty$.

\begin{theorem}
    \label{thm:multivariate_canonical_vector_bound_probabilistic_slow}
    Under the slow-rate bound assumptions stated in Corollary~\ref{cor:simplified_operator_norm_bound} and assuming that the canonical correlations $\gamma_1, \ldots \gamma_K$ are bounded from below, $Y$ and $\sigmay^{-1}Y$ satisfy the variance-proxy condition, and $\hat{\eta}^{\T}\sigmayhat^{1/2} \sigmay^{1/2}\eta \geq 0$, for $k=1, \ldots K$.
    
    \noindent If $d\log\bl p \br \twoinfnorm{\sigmax}^2\onetwo{T}^4 = o(N)$, then, for any fixed $\eta \in (0,1)$, with probability $1-\eta$,
    \begin{equation}
        \twonorm{\theta_k-\hat \theta_k} \lesssim \rate^{1/4}\frac{\gamma_1 \twoinfnorm{\sigmax}^{1/2}\onetwo{T} \twonorm{\sigmax^{-1/2}}}{\gamma_k \gammaterm}, \qquad k=1, \ldots K.
    \end{equation}
    If $\twonorm{\sigmay}^2d = o(N)$ and $\twonorm{\sigmay}^2\twonorm{\sigmay^{-1}}^2d = o(N)$, then, for any fixed $\eta \in (0,1)$, with probability $1-\eta$,
\begin{equation}
    \twonorm{\eta_k - \hat{\eta}_k} \lesssim \rate ^{1/4} \frac{\gamma_1 \twoinfnorm{\sigmax}^{1/2} \onetwo{T}\twonorm{\sigmay^{-1/2}}}{\gammaterm}, \qquad k=1, \ldots K.
\end{equation}
\end{theorem}
\begin{theorem}
    \label{thm:multivariate_canonical_vector_bound_probabilistic_fast}
    Under the fast-rate bound assumptions stated in Corollary~\ref{cor:simplified_operator_norm_bound} and assuming that the canonical correlations $\gamma_1, \ldots \gamma_K$ are bounded from below, $Y$ and $\sigmay^{-1}Y$ satisfy the variance-proxy condition, and $\hat{\eta}_k^{\T}\sigmayhat^{1/2} \sigmay^{1/2}\eta_k \geq 0$, for $k=1, \ldots K$.
    
    \noindent If $d\log\bl p \br \twoinfnorm{\sigmax}s\kappa = o(N)$, then for any fixed $\eta \in (0,1)$, with probability $1-\eta$,
    \begin{equation}
        \twonorm{\theta_k-\hat \theta_k} \lesssim \rate^{1/2}\frac{\gamma_1 \twoinfnorm{\sigmax}^{1/2} \twonorm{T}s^{1/2}\kappa  }{\gamma_k\gammaterm}, \qquad k=1, \ldots K.
    \end{equation}
    If $\twonorm{\sigmay}^2d = o(N)$, then for any fixed $\eta \in (0,1)$, with probability $1-\eta$,
\begin{equation}
    \twonorm{\eta_k - \hat{\eta}_k} \lesssim \rate^{1/2} \twonorm{\sigmay^{-1/2}}\operatorname{max}\left \{ \frac{\gamma_1 \twoinfnorm{\sigmax}^{1/2} s^{1/2}\kappa^{1/2}}{\gammaterm}, \twonorm{\sigmay^{1/2}}\twonorm{\sigmay^{-1/2}} \right \},
\end{equation}
with $k=1, \ldots K$.
\end{theorem}

\subsection{Proofs for the deterministic bounds in Section \ref{appendix:det_bounds_mult} and for the probabilistic bounds in Section~\ref{appendix:prob_bounds_mult}}
\label{sec:apdx:proofs_for_multivariate_CCA_proof_section}

\noindent \textbf{Proof of Lemma \ref{lemma:deterministic_bound_1}:}\\
The triangle inequality is used repeatedly without comment. By adding and subtracting $\bhat^{\T}\sigmax B$, we have
\begin{equation}
    \twonorm{\bhat^T\sigmaxhat\bhat - B^{\T}\sigmax B} \leq \twonorm{\bhat^T(\sigmaxhat \bhat - \sigmax B)} + \twonorm{(\bhat -B)^{\T}\sigmax B}.
\end{equation}
Since 
\begin{equation}
    \sigmaxhat \bhat - \sigmax B = \sigmaxhat(\bhat-B) + (\sigmaxhat - \sigmax)B
\end{equation}
by adding and subtracting $\sigmaxhat B$, we deduce that
\begin{equation}
    \twonorm{\bhat^T\sigmaxhat\bhat - B^{\T}\sigmax B} \leq \underbrace{\twonorm{\bhat^{\T}\sigmaxhat(\bhat-B)}}_{\text{Term I}} + \underbrace{\twonorm{\bhat^{\T}(\sigmaxhat - \sigmax) B}}_{\text{Term II}} + \underbrace{\twonorm{(\bhat -B)^{\T}\sigmax B}}_{\text{Term III}}.
\end{equation}
We bound each term individually. Recall that $\sigmaxhat = \frac{1}{N}\mathbb{X}^{\T}\mathbb{X}$.\\
Term I: We have
\begin{equation}
    \twonorm{\bhat^{\T}\sigmaxhat(\bhat-B)} \leq \rootnfrac\twonorm{\mathbb{X}\bhat} \Fnorm{\mathbb{X}(\bhat-B)}\rootnfrac
\end{equation}
using $\twonorm{AB} \leq \twonorm{A}\twonorm{B} \leq \twonorm{A}\Fnorm{B}$.
Since \begin{equation}
    \twonorm{\mathbb{X}\bhat} \leq \twonorm{\mathbb{X}B} + \twonorm{\mathbb{X}(\bhat-B)},
\end{equation}
we have
\begin{equation}
    \twonorm{\bhat^{\T}\sigmaxhat(\bhat-B)} \leq \rootnfrac\twonorm{\mathbb{X}B} \Fnorm{\mathbb{X}(\bhat-B)}\rootnfrac + \nfrac \Fnorm{\mathbb{X}(\bhat-B)}^2.
\end{equation}
Term II: We have
\begin{align}
    \twonorm{\bhat^{\T}(\sigmaxhat - \sigmax) B} & \leq \twonorm{B^{\T}(\sigmaxhat - \sigmax) B} + \twonorm{(\bhat-B)^{\T}(\sigmaxhat - \sigmax) B}\\
    & \leq  \twonorm{B^{\T}(\sigmaxhat - \sigmax) B} + \onetwo{\bhat-B}\twoinfnorm{(\sigmaxhat - \sigmax) B}
\end{align}
using $\twonorm{A^{\T}B} \leq \onetwo{A} \twoinfnorm{B}$.\\
Term 3: We have
\begin{equation}
    \twonorm{(\bhat -B)^{\T}\sigmax B} \leq \twonorm{(\bhat-B)^{\T}\sigmax^{1/2}} \twonorm{\sigmax^{1/2}B} \leq \gamma_1\twonorm{(\bhat-B)^{\T}\sigmax^{1/2}}
\end{equation}
since $\sigmax^{1/2}B = \tilde{T}D \tilde{H}$ where $\tilde{T}$ and $\tilde{H}$ are orthogonal and $D$ is diagonal.

Combining these results, we obtain the statement of the lemma. $\hfill \qedsymbol$\\
\noindent \textbf{Proof of Lemma \ref{lemma:fast-slow-rate-appendix1}:}\\
For reference, Lemma \ref{lemma:deterministic_bound_1} gives the bound
\begin{align}
        \twonorm{\bhat^T\sigmaxhat\bhat - B^{\T}\sigmax B} &\lesssim \rootnfrac \twonorm{\mathbb{X}B} \sqrt{\lambda} \onetwo{B}^{1/2} + \twonorm{B^{\T}(\sigmax - \sigmaxhat)B}\\ 
        &+ \gamma_1 \onetwo{B}^{1/2} \bl \lambda \onetwo{B} + \maxnorm{\sigmaxhat - \sigmax} \br^{1/2}\\
         &+ \lambda \onetwo{B} + \onetwo{B}\twoinfnorm{(\sigmax-\sigmaxhat)B}.
\end{align}
\textit{Proof of slow-rate bound}:\\
Assuming that $\lambda \geq \frac{2}{N}\twoinfnorm{\mathbb{X}^{\T}E}$, then by Theorem 1 and Corollary 1 of \textcite{gaynanova2020prediction} we have
\begin{align}
     &\nfrac \Fnorm{\mathbb{X}(\bhat-B)}^2 \lesssim \lambda \onetwo{B},\\
    &\Fnorm{(\bhat-B)^{\T}\sigmax^{1/2}}^2 \lesssim \onetwo{B} \bl \lambda + \onetwo{B}\maxnorm{\sigmaxhat - \sigmax} \br,
\end{align}
and
\begin{equation}
    \onetwo{\bhat} \lesssim \onetwo{B}.
\end{equation}
This last equation implies that $\onetwo{\bhat - B} \lesssim \onetwo{B}$ by the triangle inequality.
Applying these bounds to the terms in Lemma \ref{lemma:deterministic_bound_1} establishes the slow-rate bound.\\
\textit{Proof of fast-rate bound}:\\
Assuming that $\lambda \geq \frac{2}{N}\twoinfnorm{\mathbb{X}^{\T}E}$, that $B$ has at most $s$ nonzero rows, and assuming the group restricted eigenvalue condition on $\rootnfrac \mathbb{X}$, then by Theorem 2 and Corollary 2 of \textcite{gaynanova2020prediction} we have
\begin{align}
     &\nfrac \Fnorm{\mathbb{X}(\bhat-B)}^2 \lesssim \kappa_X s \lambda^2 \\
    &\Fnorm{(\bhat-B)^{\T}\sigmax^{1/2}}^2 \lesssim \kappa_Xs\bl 1 + \kappa_Xs\maxnorm{\sigmaxhat - \sigmax} \br \lambda^2 \label{eq:sigmax_one_half_B_bound}
\end{align}
and
\begin{equation}
    \onetwo{\bhat - B} \lesssim \kappa_X s \lambda.
\end{equation}
Applying these bounds to the terms in Lemma \ref{lemma:deterministic_bound_1} establishes the fast-rate bound. $\hfill \qedsymbol$\\
\noindent \textbf{Proof of Lemma \ref{lemma:lambda_0_bound}:}\\
From the definition of $E$, $E = \mathbb{Y} \sigmayhat^{-1/2} - \mathbb{X}B$, we have
\begin{align}
\frac{1}{N}{\mathbb{X}^{\T}E} & = \nfrac \mathbb{X}^{\T} \mathbb{Y} \sigmayhat^{-1/2} - \nfrac \mathbb{X}^{\T}\mathbb{X}B \\
&= \sigmaxyhat \sigmayhat^{-1/2} - \sigmaxhat B\\
    &= \bl \sigmaxyhat \sigmayhat^{-1/2} - \sigmaxy \sigmay^{-1/2} \br + \bl \sigmaxy \sigmay^{-1/2}  - \sigmaxhat B \br. \label{eq:add_subtract}
\end{align}
Now considering the first term in equation \eqref{eq:add_subtract}, by adding and subtracting $\sigmaxy \sigmayhat^{-1/2}$, and subsequently adding and subtracting $\sigmay^{-1/2}$ to $\sigmayhat^{-1/2}$, we have
\begin{align}
     \sigmaxyhat \sigmayhat^{-1/2} - \sigmaxy \sigmay^{-1/2} & =  \bl \sigmaxyhat - \sigmaxy \br \sigmayhat^{-1/2} + \sigmaxy \bl \sigmayhat^{-1/2} - \sigmay^{-1/2} \br \\
    &= \bl \sigmaxyhat - \sigmaxy \br \bl \sigmayhat^{-1/2} - \sigmay^{-1/2} \br +\bl \sigmaxyhat - \sigmaxy \br \sigmay^{-1/2} + \sigmaxy \bl \sigmayhat^{-1/2} - \sigmay^{-1/2} \br.
\end{align}
For the second term, we have
\begin{align}
    \sigmaxy \sigmay^{-1/2}  - \sigmaxhat B &= \sigmax \sigmax^{-1}\sigmaxy\sigmay^{-1/2} - \sigmaxhat B\\
    &= \bl \sigmax - \sigmaxhat \br B.
\end{align}
Combining these equalities and using the triangle inequality completes the proof.  $\hfill \qedsymbol$\\
\noindent \textbf{Proof of Lemma \ref{lemma:prob_bound_for_lambda_0}:}\\
The proof is adapted from Lemma 7 of \textcite{gaynanova2020prediction} but considers cross-covariance matrices rather than the covariance matrices.
Let $X_{ij}$ denote entry $(i,j)$ of $\mathbb{X}$ and $Z_{ij}$ denote entry $(i,j)$ of $\mathbb{ Z}$. Then
\begin{equation}
    \nfrac \bl \mathbb{ X}^{\T} \mathbb{ Z} \br_{kl} = \nfrac \sum_{i=1}^N X_{ik}Z_{il}, \qquad k=1, \ldots p, \text{ }l = 1, \ldots d.
\end{equation}
Let $\Sigma_{XZ}$ be the cross-covariance matrix of $X$ and $Z$ with $(k,l)$ entry equal to $\sigma_{kl} = \mathbb{ E}\left [ X_{ik} Z_{il} \right ]$. Then, $ X_{ik} Z_{il} - \sigma_{kl}$ are each mean $0$ subexponential random variables, since $$\subexpnorm{ X_{ik} Z_{il}} \leq \subgaussnorm{X_{ik}}\subgaussnorm{Z_{il}} = g_k h_l \leq gh$$
by Lemma 2.7.7.~of \textcite{vershynin2018high}, and because
$$\subexpnorm{ X_{ik} Z_{il} - \sigma_{kl}} \lesssim \subexpnorm{ X_{ik} Z_{il}}
$$
by Exercise 2.7.10.~of \textcite{vershynin2018high}. Thus, $\bl \Sigma_{XZ} -  \nfrac \mathbb{ X}^{\T} \mathbb{ Z} \br_{kl}$ is a sum of independently and identically distributed (i.i.d.) subexponential random variables, since each for fixed $k$ and $l$, the $X_{ik} Z_{il} - \sigma_{kl}$ are mean $0$ subexponential i.i.d. random variables over $i =1 , \ldots N$.

By Corollary 2.8.3.~of \textcite{vershynin2018high}, (Berstein's inequality), for each $k$ and $l$, and for every $t > 0$,
\begin{equation}
    P \bl \left | \bl \Sigma_{XZ} -  \nfrac \mathbb{ X}^{\T} \mathbb{ Z} \br_{kl} \right | \geq t \br \leq 2 \operatorname{exp}\left [ -c N \operatorname{min}\bl\frac{t^2}{K^2},\frac{t}{K} \br \right],
\end{equation}
where $K = Cgh$, for absolute constants $c$ and $C$. Applying a union bound, we have
\begin{equation}
    P \bl \maxnorm{\Sigma_{XZ} -  \nfrac \mathbb{ X}^{\T} \mathbb{ Z}} \geq t \br \leq 2 d p \operatorname{exp}\left [ -c N \operatorname{min}\bl\frac{t^2}{K^2},\frac{t}{K} \br \right].
\end{equation}
When $t \leq K$, we have
\begin{equation} \label{eq:rhs_probability}
    P \bl \maxnorm{\Sigma_{XZ} -  \nfrac \mathbb{ X}^{\T} \mathbb{ Z}} \geq t \br \leq 2 d p \operatorname{exp}\left [ -c N \operatorname{min}\frac{t^2}{K^2} \right].
\end{equation}
since $t^2/K^2 \leq t/K$ if and only if $t \leq K$.
Letting the right-hand side of equation \eqref{eq:rhs_probability} be denoted as $\eta$, we solve for $t$ in terms of $\eta$ to obtain
\begin{equation}
    t = \sqrt{\operatorname{log}\bl 2dp\eta^{-1} \br \frac{K^2}{cN}},
\end{equation}
and so that, for $\eta \in (0,1)$, if $\sqrt{\operatorname{log}\bl 2dp\eta^{-1} \br \frac{K^2}{cN}} \leq Cgh$, then with probability at least $1-\eta$ we have
\begin{equation}
    \maxnorm{\Sigma_{XZ} -  \nfrac \mathbb{ X}^{\T} \mathbb{ Z} } \lesssim gh \sqrt{\operatorname{log}\bl 2dp\eta^{-1} \br \frac{1}{N}}.
\end{equation}
$\operatorname{log}\bl 2dp\eta^{-1} \br \leq \operatorname{log}(2) + 2\operatorname{log}\bl p\eta^{-1} \br $ since $d \leq p$, and because we suppose $\operatorname{log}(p) = o(N)$,
it follows that for fixed $\eta \in (0,1)$, with probability $1-\eta$,
\begin{equation}
    \maxnorm{\Sigma_{XZ} -  \nfrac \mathbb{ X}^{\T} \mathbb{ Z} } \lesssim gh \sqrt{\operatorname{log}\bl p\eta^{-1} \br \frac{1}{N}}.
\end{equation}
Using $\twoinfnorm{A} \leq \sqrt{d} \maxnorm{A}$ for any $A \in \rone^{p \times d}$ completes the proof. $\hfill \qedsymbol$\\
\noindent \textbf{Proof of Lemma \ref{lemma:prob_bound_for_lambda_0_2}:}\\
To establish
\begin{equation} \label{eq:to_show1}
    \twoinfnorm{\bl \sigmaxyhat - \sigmaxy \br\sigmay^{-1/2}} \lesssim \operatorname{max}_i(\subgaussnorm{X_i}) \sqrt{\frac{d}{N}\operatorname{log}(p\eta^{-1})},
\end{equation}
we can use Lemma \ref{lemma:prob_bound_for_lambda_0} where $X = X$ and $Z= \sigmay^{-1/2}Y$, since $\sigmaxy \sigmay^{-1/2} = \Sigma_{X,\sigmay^{-1/2}Y}$. Then, the sub-Gaussian norms are $g = \operatorname{max}_i(\subgaussnorm{X_i})$ and $h = \operatorname{max}_i \bl \subgaussnorm{\bl\sigmay^{-1/2}Y\br_i}\br$,
where $\bl\sigmay^{-1/2}Y\br_i$ is the $i$th entry of $\sigmay^{-1/2}Y\in \rone^d$. Using Definition \ref{assumption:subgauss_proxy} for $\sigmay^{-1/2}Y$, we have
\begin{equation}
    h \leq K_{\sigmay^{-1/2}Y} \twonorm{\sigmay^{-1/2} \sigmay \sigmay^{-1/2}}^{1/2} = K_{\sigmay^{-1/2}Y}.
\end{equation}
Treating $K_{\sigmay^{-1/2}Y}$ as an absolute constant establishes equation \eqref{eq:to_show1}.

Establishing
\begin{equation} \label{eq:to_show2}
    \twoinfnorm{\bl \sigmax - \sigmaxhat \br B} \lesssim \operatorname{max}_i(\subgaussnorm{X_i}) \gamma_1 \sqrt{\frac{d}{N}\operatorname{log}(p\eta^{-1})}
\end{equation}
follows an identical argument except that we let $Z = B^{\T}X$, so $$h \leq K_{B^{\T}X}\twonorm{B^{\T}\sigmax B}^{1/2}$$ in the final step. The identity $\twonorm{B^{\T}\sigmax B} = \gamma_1^2$ establishes equation \eqref{eq:to_show2}.

To deduce that
\begin{equation} \label{eq:main_eq_to_deduce}
    \twoinfnorm{\sigmaxy \bl \sigmayhat^{-1/2} - \sigmay^{-1/2} \br} \lesssim \twoinfnorm{\sigmax}^{1/2}\gamma_1 \sqrt{\frac{d + \operatorname{log}(\eta^{-1})}{N}},
\end{equation}
we begin by using $\twoinfnorm{AB} \leq \twoinfnorm{A}\twonorm{B}$ and $\sigmay^{-1/2} \sigmay^{1/2} = I_d$ to obtain
\begin{equation}
    \twoinfnorm{\sigmaxy \bl \sigmayhat^{-1/2} - \sigmay^{-1/2} \br}  \leq \underbrace{\twoinfnorm{\sigmaxy \sigmay^{-1/2}}}_{\text{Term I}}\underbrace{\twonorm{ \sigmay^{1/2}\bl \sigmayhat^{-1/2} - \sigmay^{-1/2} \br}}_{\text{Term II}}.
\end{equation}
Considering Term I, we have
\begin{align}
    \twoinfnorm{\sigmaxy \sigmay^{-1/2}} & = \twoinfnorm{\sigmax^{1/2}\sigmax^{-1/2}\sigmaxy \sigmay^{-1/2}}\\
    &\leq \twoinfnorm{\sigmax^{1/2}} \twonorm{\sigmax^{1/2}B}\\
    &= \twoinfnorm{\sigmax^{1/2}} \gamma_1.
\end{align}
In the below, we are able to bound Term II without incurring unnecessary factors of $\twonorm{\sigmay^{-1/2}}$ using the results of \textcite{kereta2021estimating} which pertain to precision matrix estimation along subspaces. The main idea is to bound $\twonorm{ \sigmay^{1/2}\bl \sigmayhat^{-1/2} - \sigmay^{-1/2} \br}$ in terms of $\twonorm{\sigmay^{1/2}\bl \sigmayhat^{-1} - \sigmay^{-1}\br \sigmay^{1/2}}$, to which the results of \textcite{kereta2021estimating} can be applied. That $\twonorm{ \sigmay^{1/2}\bl \sigmayhat^{-1/2} - \sigmay^{-1/2} \br}$ is not simply equal to $\twonorm{\sigmay^{1/2}\bl \sigmayhat^{-1} - \sigmay^{-1}\br \sigmay^{1/2}}$ is due to $\sigmay$ and $\sigmayhat$ not necessarily commuting with one another. We begin with
\begin{equation}
    \twonorm{ \sigmay^{1/2}\bl \sigmayhat^{-1/2} - \sigmay^{-1/2} \br} = \twonorm{\sigmay^{1/2} \sigmayhat^{-1/2} - I}.
\end{equation}
Using identity \ref{item:product_SPD_matrices} in Section \ref{subsec:identities} and that both $\sigmay^{1/2}$ and $\sigmayhat^{1/2}$ are positive definite along with their inverses,
\begin{equation}
   \text{Term II} = \twonorm{\bl \sigmay^{1/2} \sigmayhat^{-1} \sigmay^{1/2}\br^{1/2} - I} = \twonorm{\bl \sigmay^{1/2}\sigmayhat^{-1}\sigmay^{1/2}\br^{1/2} - \bl \sigmay^{1/2}\sigmay^{-1}\sigmay^{1/2}\br^{1/2}}.
\end{equation}
Using $\twonorm{A^{1/2} - B^{1/2}} \leq \frac{1}{2}\operatorname{max}\bl \twonorm{A^{-1}}, \twonorm{B^{-1}} \br^{1/2} \twonorm{A - B}$ for positive definite matrices $A$ and $B$, we deduce that
\begin{align}
     \text{Term II} & \leq \frac{1}{2}\operatorname{max}\bl \twonorm{\bl \sigmay^{1/2}\sigmayhat^{-1}\sigmay^{1/2} \br^{-1}}, \twonorm{I_d^{-1}} \br^{1/2} \twonorm{\sigmay^{1/2}\sigmayhat^{-1}\sigmay^{1/2} - \sigmay^{1/2}\sigmay^{-1}\sigmay^{1/2}}\\
     &= \underbrace{\frac{1}{2}\operatorname{max}\bl \twonorm{\sigmay^{-1}\sigmayhat}, 1 \br^{1/2}}_{\text{Term II.I}} \underbrace{\twonorm{\sigmay^{1/2}\bl \sigmayhat^{-1} - \sigmay^{-1}\br \sigmay^{1/2}}}_\text{{Term II.I}}.
\end{align}
We bound $\twonorm{\sigmay^{-1}\sigmayhat}$ in Term II.I with
\begin{align}
    \twonorm{\sigmay^{-1}\sigmayhat} &\leq \twonorm{\sigmay^{-1}\bl \sigmayhat - \sigmay \br} + \twonorm{\sigmay^{-1}\sigmay}\\
    & = \twonorm{\sigmay^{-1/2}\bl \sigmayhat - \sigmay \br\sigmay^{-1/2}} + 1.
\end{align}
We apply a result concerning covariance estimation along subspaces, Lemma 2 of \textcite{kereta2021estimating}, to deduce that for fixed $\eta \in (0,1)$ with probability $1-\eta$,
\begin{equation}
    \twonorm{\sigmay^{-1/2}\bl \sigmayhat - \sigmay \br\sigmay^{-1/2}} \lesssim \subgaussnorm{\sigmay^{-1/2}Y}^2\operatorname{max} \bl \sqrt{\frac{2d + \operatorname{log}(\eta^{-1})}{N}},\frac{2d + \operatorname{log}(\eta^{-1})}{N} \br.
\end{equation}
From Definition \ref{assumption:subgauss_proxy} for $\sigmay^{-1/2}Y$ combined with the assumption that $d = o(N)$, we have that in the limit this term is bounded by $1$. Therefore, for fixed $\eta \in (0,1)$, with probability $1-\eta$, $\twonorm{\sigmay^{-1}\sigmayhat} \lesssim 1$. From this, $\text{Term II.I} \lesssim 1$ as well.\\

Having bounded $\twonorm{ \sigmay^{1/2}\bl \sigmayhat^{-1/2} - \sigmay^{-1/2} \br}$ in terms of $\text{Term II.II} = \twonorm{\sigmay^{1/2}\bl \sigmayhat^{-1} - \sigmay^{-1}\br \sigmay^{1/2}}$, we will bound the latter term. We use Theorem 10 of \textcite{kereta2021estimating} directly, implying that for fixed $\eta \in (0,1)$, if $N \gtrsim (d + \operatorname{log}(\eta^{-1})\subgaussnorm{\sigmay^{-1/2}Y}^4$, then with probability $1-\eta$,
\begin{equation}
    \twonorm{\sigmay^{1/2}\bl \sigmayhat^{-1} - \sigmay^{-1}\br \sigmay^{1/2}} \lesssim \subgaussnorm{\sigmay^{-1/2}Y}^2\sqrt{\frac{\operatorname{rank}\bl \sigmay \br + \operatorname{log}(\eta^{-1})}{N}}.
\end{equation}
By Definition \ref{assumption:subgauss_proxy} for $\sigmay^{-1/2}Y$, $\subgaussnorm{\sigmay^{-1/2}Y} \lesssim 1$, so that the assumption $d = o(N)$ ensures that for fixed $\eta$, $N \gtrsim \bl d + \operatorname{log}(\eta^{-1}) \br \subgaussnorm{\sigmay^{-1/2}Y}^4$ eventually.

With our bounds for both Term II.I and Term II.II, we deduce that for fixed $\eta \in (0,1)$, with probability $1-\eta$,
\begin{equation} \label{eq:term_II_bound_used_later}
    \text{Term II} = \twonorm{ \sigmay^{1/2}\bl \sigmayhat^{-1/2} - \sigmay^{-1/2} \br} \lesssim \sqrt{\frac{d + \operatorname{log}(\eta^{-1})}{N}}.
\end{equation}
Now having bounded both Term I and Term II, we finally establish that for fixed $\eta \in (0,1)$, if $d = o(N)$, with probability $1-\eta$,
\begin{equation}
    \label{equation:sigmayhat_inverse_one_half_bound}
    \twoinfnorm{\sigmaxy \bl \sigmayhat^{-1/2} - \sigmay^{-1/2} \br} \lesssim \twoinfnorm{\sigmax}^{1/2}\gamma_1 \sqrt{\frac{d + \operatorname{log}(\eta^{-1})}{N}},
\end{equation}
completing the proof of \eqref{eq:main_eq_to_deduce}.\\

To show
\begin{equation}
    \twoinfnorm{\bl \sigmaxyhat - \sigmaxy \br \bl \sigmayhat^{-1/2} - \sigmay^{-1/2} \br} \lesssim \twoinfnorm{\bl \sigmaxyhat - \sigmaxy \br\sigmay^{-1/2}},
\end{equation}
we begin with
\begin{equation} \label{eq:two_factors}
    \twoinfnorm{\bl \sigmaxyhat - \sigmaxy \br \bl \sigmayhat^{-1/2} - \sigmay^{-1/2} \br}  \leq \twoinfnorm{\bl \sigmaxyhat - \sigmaxy \br\sigmay^{-1/2}} \twonorm{ \sigmay^{1/2}\bl \sigmayhat^{-1/2} - \sigmay^{-1/2} \br},
\end{equation}
using $\sigmay^{-1/2} \sigmay^{1/2} = I_d$ and $\twoinfnorm{AB} \leq \twoinfnorm{A}\twonorm{B}$. From equation \eqref{eq:term_II_bound_used_later} we obtain that with probability $1-\eta$, the second factor in \eqref{eq:two_factors} is bounded by an absolute constant as $d = o(N)$, completing the proof. \hfill \qedsymbol\\

\noindent \textbf{Proof of Lemma \ref{lemma:other_terms}:}\\
That for fixed $\eta \in (0,1)$, if  $\operatorname{log}(p) = o(N)$, then with probability $1-\eta$,
\begin{equation}
    \maxnorm{\sigmax - \sigmaxhat} \lesssim \operatorname{max}(\subgaussnorm{X_i}^2) \sqrt{\frac{\operatorname{log}\bl p\eta^{-1} \br}{N}}
\end{equation}
follows from Lemma 7 of \textcite{gaynanova2020prediction}.

That for fixed $\eta \in (0,1)$, if  $d = o(N)$, then with probability $1-\eta$,
     \begin{equation}
        \label{equation:sigma_BTX_no_sparsity}
          \twonorm{B^{\T}(\sigmax - \sigmaxhat)B} \lesssim \gamma_1^2 \sqrt{\frac{\operatorname{rank}\bl B^{\T}\sigmax \br + \operatorname{log}\bl \eta^{-1} \br}{N}}
     \end{equation}
follows from Lemma 2 of \textcite{kereta2021estimating} in addition to Definition \ref{assumption:subgauss_proxy} applied to $B^{\T}X$, which implies that $\subgaussnorm{B^{\T}X} \leq K_{B^{\T}X}\twonorm{B^{\T}\sigmax B}^{1/2}$.
Using $\operatorname{rank}\bl AB \br \leq \operatorname{min}\bl \operatorname{rank} \bl A \br, \operatorname{rank} \bl B \br \br $, we have
\begin{equation}
    \operatorname{rank}\bl B^{\T}\sigmax \br = \operatorname{rank}\bl \sigmax B \br = \operatorname{rank} \bl \sigmaxy \sigmay^{-1/2} \br \leq d,
\end{equation}
which establishes the desired result.\\
To show that for fixed $\eta \in (0,1)$, if  $d = o(N)$, then with probability $1-\eta$,
     \begin{equation}
         \rootnfrac \twonorm{\mathbb{X}B} \lesssim \gamma_1,
     \end{equation}
we begin with \begin{equation}
    \rootnfrac \twonorm{\mathbb{X}B} = \rootnfrac \twonorm{B^{\T}\mathbb{X}^{\T}\mathbb{X}B}^{1/2} = \twonorm{B^{\T} \sigmaxhat B}^{1/2},
\end{equation}
which holds since $\twonorm{A} = \twonorm{A^{\T}A}^{1/2}$. Adding and subtracting $B^{\T} \sigmax B$ and using the triangle inequality, we obtain that for fixed $\eta \in (0,1)$, with probability $1- \eta$,
\begin{equation}
    \twonorm{B^{\T} \sigmaxhat B} \leq \twonorm{B^{\T} \sigmax B} + \twonorm{ B^{\T} \bl \sigmaxhat - \sigmax \br B} \lesssim \gamma_1^2.
\end{equation}
In the last inequality, we have used $d = o(N)$ and equation \eqref{equation:sigma_BTX_no_sparsity} to deduce that $\twonorm{ B^{\T} \bl \sigmaxhat - \sigmax \br B}$ becomes smaller that $\gamma_1^2$ eventually. This completes the proof. \hfill \qedsymbol\\

\noindent \textbf{Proof of Lemma \ref{lemma:probabilistic_group_restricted_eigenvalue_condition}:}\\
By Lemma 6 of \textcite{gaynanova2020prediction}, if suffices to show that under the condition $s^2\log(p) = o(N)$, that for fixed $\eta$, with probability $1-\eta$, we have $s\maxnorm{\sigmax -\sigmaxhat} \leq (32 \kappa)^{-1}$. By the first item of Lemma \ref{lemma:other_terms}, we then have that for fixed $\eta$ with probability $1-\eta$,
\begin{equation}
    \kappa s\maxnorm{\sigmax-\sigmaxhat} \lesssim \kappa s\operatorname{max}_i(\subgaussnorm{X_i}^2) \sqrt{\frac{\log\bl p \eta^{-1}\br}{N}}.
\end{equation}
It therefore suffices that $\operatorname{max}_i(\subgaussnorm{X_i}^4) \kappa^2 s^2\log\bl p\br = o(N)$, since then with probability $1-\eta$, $\kappa s\maxnorm{\sigmax-\sigmaxhat}$ is arbitrarily small. But this is the assumed condition, so the proof is complete. \hfill \qedsymbol\\

\noindent \textbf{Proof of Lemma \ref{lemma:B_minus_B_bound}:}\\
\textit{Proof of slow-rate bound}:\\
Under the slow-rate assumptions of Corollary \ref{cor:simplified_operator_norm_bound} and by Corollary 1 in \textcite{gaynanova2020prediction}, for fixed $\eta$, with probability $1-\eta$, we have
\begin{equation}
    \twonorm{B-\bhat}^2 \lesssim \twonorm{\sigmax^{-1}}\onetwo{B}\bl \lambda + \onetwo{B}\maxnorm{\sigmaxhat - \sigmax} \br.
\end{equation}
Then
\begin{equation}
    \lambda \lesssim \twoinfnorm{\sigmax}^{1/2}\rate^{1/2},
\end{equation}
and by Lemma \ref{lemma:other_terms},
\begin{equation}
    \maxnorm{\sigmax-\sigmaxhat} \lesssim \twoinfnorm{\sigmax}\rate^{1/2}.
\end{equation}
The last two equations bound $\twonorm{B - \bhat}$, and since $\twoinfnorm{\sigmax},\onetwo{B} \geq 1$, the proof of the slow-rate bound is complete.\\
\textit{Proof of fast-rate bound}:\\
Under the slow-rate assumptions of Corollary \ref{cor:simplified_operator_norm_bound} and  by Theorem 2 in \textcite{gaynanova2020prediction}, for fixed $\eta$, with probability $1-\eta$, we have
\begin{equation}
    \twonorm{B-\bhat} \lesssim \kappa s^{1/2} \lambda.
\end{equation}
Again using $\lambda \lesssim \twoinfnorm{\sigmax}^{1/2}\rate^{1/2}$, the result is shown. \hfill \qedsymbol\\

\noindent \textbf{Proof of Lemma \ref{lemma:y_covariance_bounds}:}\\
The proof of the first statement follows from Lemma 2 of \textcite{kereta2021estimating} and Definition \ref{assumption:subgauss_proxy} applied to $Y$.\\
To show the bound on $\twonorm{\sigmay^{-1/2} -\sigmayhat^{-1/2}}$, we begin by using identity \ref{item:MVT_square_root_matrices} in Section \ref{subsec:identities} applied to $\sigmay^{-1}$ and $\sigmayhat^{-1}$. We have
\begin{equation}
    \twonorm{\sigmay^{-1/2} -\sigmayhat^{-1/2}} \leq \frac{1}{2}\operatorname{max}\bl\twonorm{\sigmay},\twonorm{\sigmayhat}\br^{1/2}\twonorm{\sigmay^{-1} - \sigmayhat^{-1}}.
\end{equation}
To bound $\twonorm{\sigmay^{-1} - \sigmayhat^{-1}}$, we use Corollary 11 of \textcite{kereta2021estimating}: if $$\bl \operatorname{rank}\bl \sigmay \br + \log\bl \eta^{-1} \br \br \subgaussnorm{\sigmay^{1/2}Y}^4 \lesssim N$$ eventually, then with probability $1-\eta$,
\begin{equation}
    \twonorm{\sigmay^{-1} - \sigmayhat^{-1}} \lesssim \subgaussnorm{\sigmay^{-1}Y}^2 \sqrt{\frac{\operatorname{rank}\bl \sigmay\br + \log \bl \eta^{-1} \br}{N}}.
\end{equation}
The variance proxy condition on $\sigmay^{1/2}Y$ (Definition \ref{assumption:subgauss_proxy}) implies that $\subgaussnorm{\sigmay^{-1/2}Y}$ is bounded by an absolute constant. The condition $d = o(N)$ then ensures $\bl \operatorname{rank}\bl \sigmay \br + \log\bl \eta^{-1} \br \br \subgaussnorm{\sigmay^{1/2}Y}^4 \lesssim N$ eventually. The variance proxy condition on $\sigmay^{-1}Y$ implies $\subgaussnorm{\sigmay^{-1}Y} \lesssim \twonorm{\sigmay^{-1/2}}$.

To bound $\twonorm{\sigmayhat}$, we use the triangle inequality to obtain $\twonorm{\sigmayhat} \leq \twonorm{\sigmayhat - \sigmay} + \twonorm{\sigmay}$. Then, using the first statement of the lemma and with the additional assumption that $\twonorm{\sigmay}^2d = o(N)$, we have with probability $1-\eta$,
\begin{equation} \label{eq:simplifying_sigmayhat}
    \twonorm{\sigmayhat} \lesssim \twonorm{\sigmay}.
\end{equation}
Combining these results together establishes the statement of the lemma and the proof is complete. \hfill \qedsymbol\\

\noindent \textbf{Proof of Theorems \ref{thm:multivariate_canonical_vector_bound_probabilistic_slow} and \ref{thm:multivariate_canonical_vector_bound_probabilistic_fast}:}\\
\textit{Proof of bounds for $\theta$}:\\
By definition, we have that
\begin{align}
        &\theta_k = B \tilde \eta_k \gamma_k^{-1} \label{eq:theta_k_def}\\ 
        &\hat \theta_k = \hat B \hat{\tilde{\eta_k}} \hat \gamma_k^{-1}.\label{eq:theta_hat_k_def}
\end{align}
We bound $\twonorm{\theta_k - \hat{\theta}_k}$ by bounding all three of $\twonorm{B-\hat{B}}$, $\twonorm{ \tilde{\eta}_k - \hat{\tilde{\eta_k}}}$, and $\left | \gamma_k^{-1} - \hat \gamma_k^{-1} \right |$.
For ease of notation, we denote $\tilde \eta_k$ by $v$. We begin with
\begin{align}
    \twonorm{\theta_k - \hat\theta_k} &\leq \twonorm{Bv\gammainv - Bv\gammainvhat} + \twonorm{Bv\gammainvhat - \bhat \hat{v} \gammainvhat}\\
    &\leq \left | \gammainv - \gammainvhat \right | \twonorm{Bv} + \left | \gammainvhat \right |\twonorm{ B v - \bhat \hat{v}}. \label{eq:rhs_term}
\end{align}
Examining $\twonorm{ B v - \bhat \hat{v}}$ on the right-hand side of \eqref{eq:rhs_term}:
\begin{align}
    \twonorm{ B v - \bhat \hat{v}} & \leq \twonorm{Bv - B \hat{v}} + \twonorm{B\hat{v} - \bhat \hat{v}}\\
    &\leq \twonorm{B}\twonorm{v - \hat v} + \twonorm{B - \bhat} \twonorm{\hat{v}}.
\end{align}
Using \eqref{eq:rhs_term} and since $\twonorm{\hat v} = 1$, we have
\begin{equation}
\label{eq:theta_bound_partial_progress}
    \twonorm{\theta_k - \hat\theta_k} \leq \left | \gammainv - \gammainvhat \right | \twonorm{Bv} + \left | \gammainvhat \right | \bl \twonorm{B}\twonorm{v - \hat v} + \twonorm{B - \bhat} \br.
\end{equation}
where we note that we now have bounded $\twonorm{\theta_k - \hat{\theta}_k}$ in terms of $\twonorm{B-\hat{B}}$, $\twonorm{ \tilde{\eta}_k - \hat{\tilde{\eta_k}}}$, and $\left | \gamma_k^{-1} - \hat \gamma_k^{-1} \right |$.
To bound $\left | \gammainv - \gammainvhat \right | $, we can use identity \ref{item:difference_in_inverses_bound} from Section \ref{subsec:identities}, giving us that
\begin{equation} \label{eq:inverse_gamma_bound}
    \left | \gammainv - \gammainvhat \right | \leq \operatorname{min}\bl \gamma_k,\hat{\gamma}_k\br^{-3} \left | \gamma_k^2 - \hat{\gamma}_k^2 \right |.
\end{equation}
Bounding the two factors in equation \eqref{eq:inverse_gamma_bound} amounts to establishing that $\gamma_k$ is close to $\hat{\gamma}_k$. For this, we apply Weyl's inequality (\textcite{bhatia2013matrix} Corollary III.2.6.) to obtain
\begin{equation}
    \label{eq:weyl}
    \left | \gamma_k^2 - \hat{\gamma}_k^2 \right | \leq \twonorm{B^{\T}\sigmax B - \bhat^{\T}\sigmaxhat\bhat}.
\end{equation}
since $\gamma_k^2$ is the $k$th eigenvalue of $B^{\T}\sigmax B$, and $\hat{\gamma}_k^2$ is the $k$th eigenvalue of $\bhat^{\T}\sigmaxhat\bhat$. We then deduce that
\begin{equation} \label{eq:using_gamma_bounded_from_below}
    \gamma_k^2 - \twonorm{B^{\T}\sigmax B - \bhat^{\T}\sigmaxhat\bhat} \leq \hat{\gamma}_k^2.
\end{equation}
From equation \eqref{eq:using_gamma_bounded_from_below} we establish that $\frac{1}{2}\gamma_k^2 \leq \hat{\gamma}_k^2$ by using
\begin{equation}
    \twonorm{B^{\T}\sigmax B - \bhat^{\T}\sigmaxhat\bhat} \lesssim \frac{1}{2}\gamma_k^2,
\end{equation}
which holds asymptotically in both the fast and slow rate cases under our assumption that the $\gamma_k$ are bounded from below and from Corollary \ref{cor:simplified_operator_norm_bound}. From $\frac{1}{2}\gamma_k^2 \leq \hat{\gamma}_k^2$ we also deduce
\begin{equation}
    \operatorname{min}\bl \gamma_k,\hat{\gamma}_k\br^{-2} \lesssim \frac{1}{\gamma_k^2},
\end{equation}
and additionally that
\begin{equation}
    \frac{1}{\hat{\gamma}_k} \lesssim \frac{1}{\gamma_k}.
\end{equation}
We now use our results thus far regarding $\gamma_k$ and $\hat{\gamma_k}$ to obtain a simplified equation \eqref{eq:theta_bound_partial_progress}:
\begin{equation} \label{eq:simplified_theta_bound}
    \twonorm{\theta_k - \hat \theta_k} \lesssim \frac{1}{\gamma_k^3}\twonorm{B^{\T}\sigmax B - \bhat^{\T}\sigmaxhat\bhat}\twonorm{Bv} + \frac{1}{\gamma_k} \bl \twonorm{B}\twonorm{v-\hat v} + \twonorm{B-\bhat}\br
\end{equation}
To bound $\twonorm{v - \hat v}$, we can use the Davis-Kahan theorem (Corollary 3 of \textcite{yu2015useful}). Assuming that $\etat^{\T}\etathat \geq 0$, then
\begin{equation}
    \twonorm{\etat - \etathat} \leq \frac{2^{3/2}\twonorm{B^{\T}\sigmax B - \bhat^{\T}\sigmaxhat\bhat}}{\operatorname{min}\bl\gamma_{k-1}^2-\gamma_k^2,\gamma_k^2-\gamma_{k+1}^2\br},
\end{equation}
because $\etat$ is the $k$th eigenvector of $B^{\T}\sigmax B$, and $\etathat$ is the $k$th eigenvector of $\bhat^{\T}\sigmaxhat\bhat$. From this and equation \eqref{eq:simplified_theta_bound}, we obtain
\begin{equation}
    \twonorm{\theta_k - \hat \theta_k} \lesssim \frac{1}{\gamma_k^2}\twonorm{B^{\T}\sigmax B - \bhat^{\T}\sigmaxhat\bhat}\twonorm{\theta_k} + \frac{1}{\gamma_k} \bl \twonorm{B}\frac{\twonorm{B^{\T}\sigmax B - \bhat^{\T}\sigmaxhat\bhat}}{\gammaterm} + \twonorm{B-\bhat}\br,
\end{equation}
where we have also used the definition of $\theta_k$, $\theta_k = B v \gamma_k^{-1}$.
Rearranging this expression, we have
\begin{equation}
    \twonorm{\theta_k - \hat \theta_k} \lesssim \bl \frac{\twonorm{\theta_k}}{\gamma_k^2} + \frac{\twonorm{B}}{\gamma_k \gammaterm} \br \twonorm{B^{\T}\sigmax B - \bhat^{\T}\sigmaxhat\bhat} + \frac{\twonorm{B-\bhat}}{\gamma_k}. \label{eq:thm_C_2_theta_bound}
\end{equation}

Now we use our bounds for $ \twonorm{B^{\T}\sigmax B - \bhat^{\T}\sigmaxhat\bhat}$ and $\twonorm{B-\bhat}$ depending on if we are in the slow or fast rate case. In the slow rate case, we also use
\begin{equation}
    \twonorm{\theta_k} = \twonorm{\sigmax^{-1/2}\tilde{\theta_k}} \leq \twonorm{\sigmax^{-1/2}}\twonorm{\tilde{\theta_k}} = \twonorm{\sigmax^{-1/2}},
\end{equation}
where we have used the standard result of classical CCA that $\theta_k = \sigmax^{-1/2}\tilde{\theta_k}$ for some unit vector $\tilde{\theta_k}$ \parencite{uurtio2018tutorial}.

The proofs for both the fast and slow rate then follow directly from Lemma \ref{lemma:B_minus_B_bound}, Theorem \ref{cor:simplified_operator_norm_bound}, and rearranging of terms. This completes the proof of the bounds on $\twonorm{\theta_k - \hat{\theta}_k}$.\\
\textit{Proof of bounds for $\eta$}:\\
By definition, we have
\begin{align}
    \eta_k = \sigmay^{-1/2}\etat,\\
    \hat{\eta}_k = \sigmayhat^{-1/2}\etathat.
\end{align}
We bound $\twonorm{\eta_k - \hat{\eta}_k}$ with the triangle inequality:
\begin{align}
    \twonorm{\eta_k - \hat{\eta}_k} & \leq \twonorm{\sigmay^{-1/2}\etat - \sigmay^{-1/2}\etathat} + \twonorm{\sigmay^{-1/2}\etathat - \sigmayhat^{-1/2}\etathat}\\
    &\leq \twonorm{\sigmay^{-1/2}}\twonorm{\etat - \etathat} + \twonorm{\sigmay^{-1/2} - \sigmayhat^{-1/2}}\twonorm{\etathat}.
\end{align}
To simplify this expression, we can use $\twonorm{\etathat} =1$, the Davis-Kahan Theorem for $\twonorm{\etat - \etathat}$ as in the proof for the $\theta$ bounds, and the second statement of Lemma \ref{lemma:y_covariance_bounds}. 
For clarity, we state the assumptions required for these results: $d = o(N)$, $\twonorm{\sigmay}^2d = o(N)$, the variance proxy condition (Definition \ref{assumption:subgauss_proxy}) for $Y$ and $\sigmay^{-1}Y$, and $\etat^{\T}\etathat \geq 0$ for $k=1, \ldots K$. Then, for $\eta \in (0,1)$, we have
\begin{equation}\label{eq:fast_and_slow_ref}
    \twonorm{\eta_k - \hat{\eta}_k} \lesssim \twonorm{\sigmay^{-1/2}} \left [\frac{\twonorm{\bhat^T\sigmaxhat\bhat - B^{\T}\sigmax B}}{\gammaterm} + \twonorm{\sigmay}^{1/2}\twonorm{\sigmay^{-1/2}}\sqrt{\frac{d\log\bl \eta^{-1} \br}{N}} \right ].
\end{equation}
Now we apply Corollary \ref{cor:simplified_operator_norm_bound} to equation \eqref{eq:fast_and_slow_ref} under the fast and slow rate assumptions. In the slow rate case, we have
\begin{align*}
    \twonorm{\eta_k - \hat{\eta}_k} &\lesssim \twonorm{\sigmay^{-1/2}} \Bigg [ \frac{\gamma_1 \twoinfnorm{\sigmax}^{1/2} \onetwo{B}}{\gammaterm}\bl \frac{d}{N}\log\bl p\eta^{-1}\br \br^{1/4}\\ 
    &+ \twonorm{\sigmay}^{1/2}\twonorm{\sigmay^{-1/2}}\bl\frac{d\log\bl \eta^{-1} \br}{N}\br^{1/2}\Bigg ].
\end{align*}
Factoring out $\bl\frac{d}{N}\br^{1/4}$ we obtain
\begin{align*}
   \twonorm{\eta_k - \hat{\eta}_k} &\lesssim \twonorm{\sigmay^{-1/2}} \bl\frac{d}{N}\br^{1/4}\Bigg [ \frac{\gamma_1 \twoinfnorm{\sigmax}^{1/2} \onetwo{B}}{\gammaterm}\log\bl p\eta^{-1}\br ^{1/4}\\ 
   &+ \twonorm{\sigmay}^{1/2}\twonorm{\sigmay^{-1/2}}\log\bl \eta^{-1} \br^{1/2}\bl \frac{d}{N} \br^{1/4}\Bigg ].
\end{align*}
In the bracketed expression we are able to combine the first and second terms, since we assume $\gamma_1$ is bounded from below, using the additional assumption that $\twonorm{\sigmay}^2\twonorm{\sigmay^{-1}}^2d = o(N)$, and because the other terms in the first term are greater than or equal to $1$. Then, with probability $1-\eta$,
\begin{equation}
    \twonorm{\eta_k - \hat{\eta}_k} \lesssim  \frac{\gamma_1 \twoinfnorm{\sigmax}^{1/2} \onetwo{B}\twonorm{\sigmay^{-1/2}}}{\gammaterm} \bl\frac{d\log\bl p\eta^{-1}\br}{N}\br ^{1/4},
\end{equation}
completing the proof of the slow-rate bound for $\eta$.\\

In the fast-rate case, under the fast-rate bound assumptions of Corollary \ref{cor:simplified_operator_norm_bound} and applying Corollary \ref{cor:simplified_operator_norm_bound} to equation \eqref{eq:fast_and_slow_ref}, we establish
\begin{equation}
    \twonorm{\eta_k - \hat{\eta}_k} \lesssim  \twonorm{\sigmay^{-1/2}}\operatorname{max}\bl \frac{\gamma_1 \twoinfnorm{\sigmax}^{1/2} s^{1/2}\kappa^{1/2}}{\gammaterm}, \twonorm{\sigmay}^{1/2}\twonorm{\sigmay^{-1}}^{1/2}\br \bl\frac{d\log\bl p\eta^{-1}\br}{N}\br^{1/2},
\end{equation}
completing the proof of the fast-rate bound for $\eta$. \hfill \qedsymbol

\section{Asymmetric Sparse-Functional CCA: Proof of Theorem~\ref{thm:main_theorem_general_case}}
\label{appendix:proof_of_canonical_function_section}
In this section, we prove Theorem \ref{thm:main_theorem_general_case}. Recall that we use Assumption \ref{remark:finite_dim_assumption} where the $\{ \phi_j \}_{j = 1}^{\infty}$ are the principal components of $\Logy$, and without loss of generality suppose that $I = \left \{1 ,\ldots \dcorr \right \}$. We suppose that $d$, the number of principal components we use in practice, satisfies $d \geq \dcorr$. Define $Y_j \equiv \innerdouble{\Logy}{\phi_j}$ for $j=1,\ldots d$, so that the $\{Y_j\}$ are random variables with $\Var \bl Y_j \br \equiv \omega_j$. Let $Y = (Y_1,\ldots,Y_d)^{\T}$. Then, by Theorem \ref{multivariate-cca-theorem} the canonical pairs $\{(\psi_k,\theta_k)\}$ can be computed by solving a $\dcorr$-dimensional multivariate CCA problem between $X$ and $Y^{\dcorr} \equiv (Y_1,\ldots,Y_\dcorr)^{\T}$. It is straightforward to show that thanks to Assumption \ref{remark:finite_dim_assumption} and the properties of the singular value decomposition, we can equivalently solve a $d$-dimensional multivariate CCA problem between $X$ and $Y$:
\begin{equation} \label{eq:ideal}
    \maximize{\Corr^2\bl \eta_1^{\T}Y,\theta_1^{\T}X\br}{\Var\bl \eta_1^{\T}Y\br = \Var\bl\theta_1^{\T}X\br = 1},
\end{equation}
with subsequent canonical pairs defined analogously, where $Y$ has replaced $Y^{\dcorr}$. We denote with $K = \operatorname{max} \left \{i \in \{ 1, \ldots \dcorr \}: \gamma_i > 0 \right \}$ the number of nontrivial canonical vectors in this problem, and to simplify the notation, we use the conventions $\gamma_{K+1}^2=-\infty$ and $\gamma_0^2=\infty$. Here, $\eta_k \in \rone^d$ and $\theta_k \in \rone^p$ for $k=1, \ldots K$. The canonical pairs are then given by $\left(\psi_k = \sum_{j=1}^d\phi_j\eta_{kj},\theta_k \right)$, where $\eta_{kj} = 0$ for $j>\dcorr$. Recall that $\C \equiv \E \left [ \Logy \otimes \Logy \right ]$ admits the expansion $\C = \sumj \omega_j \phi_j \otimes \phi_j$, where $\{\phi_j\}$ are the eigenfunctions of $\C$ with associated eigenvalues $\{\omega_j\}$. We let $\gamma_1^2 \ldots \gamma_K^2$ denote the squared canonical correlations attained by the pairs $\bl \psi_1, \theta_1\br, \ldots \bl \psi_K, \theta_K\br$.

We denote by $\hat \psi_k$ and $\hat \theta_k$ the canonical vectors estimated using our proposed Algorithm~\ref{alg:asymmetric_sparse_fun_cca}. In practice, we are given a sample of $N$ independent pairs 
\begin{equation}
(y_i,X_i), \quad i=1, \ldots, N,
\end{equation}
where each pair $(y_i,X_i)$ is an independent observation of the pair $(y,X)$. Here, the functions $\{y_i\}$ are assumed to be fully observed on $\Tsc$. We denote $\tau \equiv |\Tsc|$, the length of the time interval of the functional data. We store the observations $\{X_i\}$ in a matrix $\mathbb{X} \in \rone^{N \times p}$.

In Algorithm~\ref{alg:asymmetric_sparse_fun_cca}, we estimate $\mu$ using the sample Fr\'echet mean, denoted as $\hat \mu$, and estimate the eigenfunctions $\{\phi_j\}$ using $\hat \phi_j$, which are the eigenfunctions of the sample covariance function $\hat \calC \equiv \frac{1}{N} \sum_{i=1}^N \Logyi \otimes \Logyi$. Hence, the functional data can be represented using the vector $Z \in \rone^d$, where its $j$th element is 
\begin{equation} \label{eq:def_of_Z}
    Z_j \equiv \innerdouble{\Logyhat}{\hat \phi_j}_{\hat \mu}.
\end{equation}
We note that in the definition of $Z$, both $\hat \mu$ and $\hat \phi_j$ depend on $y_1$ since their estimation depends on the full sample $y_1, \ldots y_N$. We also note that the distribution of $Z$ depends on the sample size $N$. In practice, we solve the following problem for the first pair of canonical variables:
\begin{equation} \label{eq:in_practice}
    \maximize{\Corr^2\bl a_1^{\T}Z,b_1^{\T}X\br}{\Var\bl a_1^{\T}Z\br = \Var\bl b_1^{\T}X\br = 1}.
\end{equation}
The subsequent canonical pairs can be defined analogously. We denote the solutions to these problems as $\bl a_1, b_1\br, \ldots \bl a_K, b_K\br$. We let $\tilde{\gamma}_1^2 \ldots \tilde{\gamma}_K^2$ denote the squared canonical correlations attained by the pairs $\bl a_1, b_1\br, \ldots \bl a_K, b_K\br$.
We expect that, under appropriate assumptions, ${a_k}$ and ${b_k}$ will closely approximate ${\eta_k}$ and ${\theta_k}$, respectively, provided that $\hat{\mu}$ and $\{\hat{\phi}_j\}$ closely approximates $\mu$ and $\{\phi_j\}$, respectively.

We assume that the canonical vectors $\{\theta_k\}$ are group $s$-sparse, and that the associated vectors $\{b_k\}$ are also group $s$-sparse. Let the support $S \subseteq \{ 1, \ldots p \}$ represent the indices of non-zero elements of $\theta_k$ or $b_k$, with cardinality $|S| \leq 2s$. This sparsity condition allows us to simplify equations \eqref{eq:ideal} and \eqref{eq:in_practice} by replacing $X$ with $X_S \in \rone^{|S|}$, the random vector consisting of only the entries $\{X_j:j \in S\}$. Moreover, $\theta_k$ and $b_k$ can be replaced with $\theta_{k,S}$ and $b_{k,S}$, respectively.

We begin by deriving an error bound for the estimation of $\psi_k$ using $\hat{\psi}_k$. Since the population quantity $\psi_k$ belongs to $\Jmu$ and our estimate $\hat \psi_k$ belongs to $\Jmuhat$, we use the parallel transport operator $\Gamma_{\mu,\hat \mu}$ to define estimation error, as proposed in \textcite{lin2019intrinsic}. For ease of notation, we denote $\Gamma_{f,g}U - V$ as $U \thetagamma V \in L^2\bl Tg \br$, for any vector fields $U \in L^2\bl T f \br$ and $V \in L^2\bl T g \br$.

\subsection{Bounding the canonical function error}
\label{sec:apdx:subsec:assumptions_general_case_CCA}

Next, we derive a bound for $\hat \psi_k \thetagamma \psi_k \in \Jmu$. Recall that, by definition, $\psi_k = \sum_{j=1}^d \phi_j \eta_{kj}$ and $\hat \psi_k = \sum_{j=1}^d \hat{\phi}_j \hat{\eta}_{kj}$. 
\begin{lemma} \label{lemma:start_of_proof}
    The following inequality holds:
    \begin{equation}
        \norm{\hat \psi_k \thetagamma \psi_k}_{\mu}^2 \lesssim \underbrace{\twonorm{\hat \eta_k - a_k}^2 \vphantom{\sumj}}_{\text{Term I}} + \underbrace{\twonorm{a_k - \eta_k}^2 \vphantom{\sumj}}_{\text{Term II}} + \underbrace{\bl \infnorm{\eta_k}\sum_{j=1}^d \munorm{\hat{\phi}_j \thetagamma \phi_j} \br^2}_{\text{Term III}}.
    \end{equation}
\end{lemma}
\begin{remark} \label{remark:expectations_necessary}
    Term II, that is $\twonorm{a_k - \eta_k}^2$, captures differences between the population CCA problem described in equations \eqref{eq:ideal} and that in \eqref{eq:in_practice}. The non-random nature of this term complicates the analysis, as it requires deriving bounds for the expectation rather than establishing probability bounds.
\end{remark}
The first term, $\twonorm{\hat{\eta}_k - a_k}$, will be bounded using our multivariate CCA arguments. The second and third terms will be bounded in the following section.

\subsubsection{Bounding Terms II and III of Lemma \ref{lemma:start_of_proof}}
Assuming that $\sigmay$ and $\sigmaz$ are invertible, and from the definitions of $a_k$ and $\eta_k$ as the solutions to the problems in equations \eqref{eq:in_practice} and \eqref{eq:ideal} respectively, we have that
\begin{align}
    a_k = \sigmaz^{-1/2}\tilde{a}_k,\\
    \eta_k = \sigmay^{-1/2}\tilde{\eta}_k,
\end{align}
where $\tilde{a}_k$ is the $k$th eigenvector (unit vector) of $A^{\T}A$ and $\tilde{\eta}$ is the $k$th eigenvector (unit vector) of $C^{\T}C$, where we have
\begin{align}
    A = \Sigma_{X_S}^{-1/2}\Sigma_{{X_S}Z}\sigmaz^{-1/2},\\
    C = \Sigma_{X_S}^{-1/2}\Sigma_{{X_S}Y}\sigmay^{-1/2}.
\end{align}
Applying inequality \ref{item:AaBb} in Section~\ref{subsec:identities}, we have that
\begin{equation}
    \twonorm{a_k-\eta_k}^2 \lesssim \twonorm{\sigmaz^{-1/2} - \sigmay^{-1/2}}^2 + \twonorm{\sigmay^{-1}}\twonorm{\tilde{a}_k - \tilde{\eta}_k}^2.
\end{equation}
Noting that $A^{\T}A$ has the same eigenvectors as $\left | A \right |$, where  $\left | A \right | \equiv \bl A^{\T}A \br^{1/2}$, we can apply the Davis-Kahan theorem (Corollary 3 of \textcite{yu2015useful}) and obtain
\begin{equation} \label{eq:a_minus_eta_tilde_bound}
    \twonorm{\tilde{a}_k-\tilde{\eta}_k} \lesssim \frac{\twonorm{\left |A \right| - \left |C \right |}}{\operatorname{min}\bl\gamma_{k-1}-\gamma_k,\gamma_k-\gamma_{k+1}\br}, \qquad k=1, \ldots K,
\end{equation}
where we assume $\tilde{a}_k^{\T}\tilde{\eta}_k \geq 0$ and we use the conventions $\gamma_{K+1}=-\infty$, and $\gamma_0=\infty$.
Next, note that
\begin{equation}
    \twonorm{\left |A \right| - \left |C \right |} \leq \Fnorm{\left |A \right| - \left |C \right |} \leq \sqrt{2}\Fnorm{A-C},
\end{equation}
where the second inequality is identity \ref{item:bhatia_VII_39} from Section \ref{subsec:identities}. We can then bound $\Fnorm{A-C}^2$ in terms of $\twonorm{\sigmaz^{-1/2}-\sigmay^{-1/2}}^2$ and $\mathbb{E}\left [\twonorm{Y-Z}^2 \right]$ using the following lemma.
\begin{lemma} \label{lemma:A_minus_C_Frobenious}
    Under the group $s$-sparsity assumptions on $b_k$ and $\theta_k$, we have that
    \begin{equation}
        \Fnorm{A-C}^2 \leq 2s \left [ \mathbb{E}\left [\twonorm{Z}^2\right ] \twonorm{\sigmaz^{-1/2}-\sigmay^{-1/2}}^2 + \twonorm{\sigmay^{-1}}\mathbb{E}\left [\twonorm{Y-Z}^2 \right] \right ].
    \end{equation}
\end{lemma}
Next, we bound $\twonorm{\sigmaz^{-1/2}-\sigmay^{-1/2}}^2$ in terms of $\mathbb{E}\left [\twonorm{Y-Z}^2 \right]$ using the following lemma.

\begin{lemma} \label{lemma:minus_square_root_difference_ZY}
It can be shown that
    \begin{equation}
    \twonorm{\sigmaz^{-1/2}-\sigmay^{-1/2}}^2 \lesssim \E \left [\twonorm{Z-Y}^2 \right] \operatorname{max}\bl \twonorm{\sigmay^{-1}},\twonorm{\sigmaz^{-1}}\br^3\operatorname{max}\bl \E \left [ \twonorm{Z}^2 \right ],\E \left [ \twonorm{Y}^2 \right] \br.
\end{equation}
Additionally, we have that
\begin{equation}
        \twonorm{\sigmay - \sigmaz}^2 \lesssim \operatorname{max}\bl \E \left [ \twonorm{Z}^2 \right ],\E \left [ \twonorm{Y}^2 \right] \br \E \left [\twonorm{Z-Y}^2 \right].
    \end{equation}
\end{lemma}

Combining the results of this section we obtain the following bound on $\twonorm{a_k - \eta_k}^2$ in terms of $\E \left [ \twonorm{Z-Y}^2 \right ]$.
\begin{lemma} \label{lemma:a_minus_eta_with_z_minus_y}
Under the group $s$-sparsity assumptions on $b_k$ and $\theta_k$, we have
    \begin{align*}
        \twonorm{a_k - \eta_k}^2 &\lesssim \Ex{\twonorm{Z-Y}^2}\\
        &\times \Bigg [\bl 1 + \frac{s\twonorm{\sigmay^{-1}}\Ex{\twonorm{Z}^2}}{\operatorname{min}\bl\gamma_{k-1}-\gamma_k,\gamma_k-\gamma_{k+1}\br^2}\br \operatorname{max}\bl \twonorm{\sigmay^{-1}},\twonorm{\sigmaz^{-1}} \br^3 \operatorname{max}\bl \Ex{\twonorm{Z}^2},\Ex{\twonorm{Y}^2} \br\\ 
        &+ \frac{s\twonorm{\sigmay^{-1}}^2}{\operatorname{min}\bl\gamma_{k-1}-\gamma_k,\gamma_k-\gamma_{k+1}\br^2} \Bigg]
    \end{align*}
    Additionally, if we assume that $\twonorm{\sigmaz^{-1}} \lesssim \twonorm{\sigmay^{-1}}$, $\E \left [\twonorm{Z}^2 \right ] \lesssim \E \left [\twonorm{Y}^2 \right ]$, and $\twonorm{\sigmay^{-1}},\E \left [\twonorm{Y}^2 \right ] \geq 1$, then the statement simplifies as follows:
    \begin{equation}
    \label{eq:Z_minus_Y_explanation1}
        \twonorm{a_k - \eta_k}^2 \lesssim \E \left [ \twonorm{Z-Y}^2 \right ]\frac{s \twonorm{\sigmay^{-1}}^4 E \left [\twonorm{Y}^2 \right]^2}{\operatorname{min}\bl\gamma_{k-1}-\gamma_k,\gamma_k-\gamma_{k+1}\br^2}.
    \end{equation}
\end{lemma}
\begin{remark}
In subsequent discussions, we will detail the conditions necessary for the additional assumptions stated here to hold.
\end{remark}

Having established a bound for $\twonorm{a_k - \eta_k}^2$ in terms of $\E \left [ \twonorm{Z-Y}^2 \right ]$, we now turn our attention to bounding $\E \left [ \twonorm{Z-Y}^2 \right ]$ using 'known' quantities. In the process, we will derive a bound for $\Ex{\munorm{\hat{\phi}_j \thetagamma \phi_j}}$, which will enable us to derive a probabilistic bound. This, in turn, will be used to bound Term III in Lemma \ref{lemma:start_of_proof}.

To establish a bound for $\E \left [ \twonorm{Z-Y}^2 \right ]$, we first begin by introducing a lemma to bound $\twonorm{Z-Y}^2$.
\begin{lemma} \label{lemma:z_minus_y}
It can be shown that
    \begin{equation} \label{eq:Z_minus_Y_explanation2}
        \twonorm{Z-Y}^2 \leq 2d \munorm{\Log_{\hat \mu} y_i \thetagamma \Log_{\mu}y_i}^2 + 2 \sum_{j=1}^d \munorm{\Log_{\mu}y_i}^2\phidif^2.
    \end{equation}
\end{lemma} 
Next, we aim to establish a bound for  $\phidif^2$. Consider an operator $\hat \C$ on $\Jmuhat$. We use parallel transport to define $\Phi \hat{\C}$ as the operator on $\Jmu$ such that $\Phi \hat{\C}\bl V \br = \Gamma_{\hat{\mu},\mu}\bl \hat{\C}\bl \Gamma_{{\mu},\hat{\mu}}V\br \br \in \Jmu$, for every $V \in \Jmu$. We also define the operator $\hat{\C}_{\mu} \equiv \frac{1}{N} \sum_{i=1}^N \Log_{\mu}y_i \otimes \Log_{\mu}y_i$ on $\Jmu$. Moreover, we use $\norm{\cdot}_{\op}$ to denote the operator norm on $\Jmu$.
\begin{lemma} \label{lemma:phi_dif}
    For any $j \geq 1$, we have that
    \begin{equation}
        \phidif^2 \lesssim \frac{\norm{\C - \hat \C}_{\op}^2 + \norm{\hat{\C}_{\mu} - \Phi \hat{\C}}_{\op}^2}{\operatorname{min}\bl\omega_{j-1}-\omega_j, \omega_j - \omega_{j+1}\br^2}.
    \end{equation}
\end{lemma}
We introduce the following lemma to bound the expectation of the terms in the previous lemma.
\begin{lemma} \label{lemma:operator_expectation_bounds}
    If $\Ex{\munorm{\Log_{\mu}y_1}^4} < \infty$, then
    \begin{equation}
        \Ex{\norm{\C - \hat \C}_{\op}^2} \leq \frac{1}{N}\Ex{\munorm{\Log_{\mu}y_1}^4}.
    \end{equation}
    Additionally, under the assumption that $\Ex{\munorm{\Log_{\hat \mu} y_i \thetagamma \Log_{\mu}y_i}^4}^{1/2} \lesssim \Ex{\munorm{\Log_{\hat \mu} y_i \thetagamma \Log_{\mu}y_i}^2}$, we have
    \begin{equation}
        \Ex{\norm{\hat{\C}_{\mu} - \Phi \hat{\C}}_{\op}^2} \lesssim \bl \Ex{\munorm{\Log_{\mu}y_i}^4}^{1/2} + \Ex{\munorm{\Log_{\hat \mu} y_i \thetagamma \Log_{\mu}y_i}^2} \br \Ex{\munorm{\Log_{\hat \mu} y_i \thetagamma \Log_{\mu}y_i}^2}.
    \end{equation}
\end{lemma}

The lemma above shows that, in order to bound $\Ex{\phidif^2}$, it is first necessary to bound $\Ex{\munorm{\Log_{\hat \mu} y_i \thetagamma \Log_{\mu}y_i}^2}$. To do this, we need the following more technical results. But first, we state some preliminary definitions. Let $T\M$ denote the tangent bundle of $\M$, and let $\nabla$ denote the Riemannian connection on $\M$. The next result, which is a mean value theorem for the parallel transport operation, is used in the proof of Lemma \ref{lemma:technical_2}.
\begin{lemma} \label{lemma:technical_1}
    For a smooth vector field $U: \M \ra T\M$, $x,y \in \M$, with minimizing geodesic $\gamma(t)$ between $x$ and $y$ (so that $\gamma(0) = x$ and $\gamma(d(x,y)) = y$), we have that
    \begin{equation}
        \norm{\mathcal{P}_{y,x}U(y) - U(x)}_x \leq d(y,x) \supp{\norm{\nabla_{\gamma'(c)}U(\gamma(c))}_{\gamma(c)}}{c \in \left [0,d(x,y) \right]}.
    \end{equation}
\end{lemma}
\begin{remark}
    When the vector field $U$ has bounded Hessian $H$, defined below, we can use this result to bound the parallel transport error by the geodesic distance $d(x,y)$ between the base points of the vector field.
\end{remark}
For $x \in \M$ and $t \in \Tsc$, define $f_t(x) \equiv \frac{1}{2}d^2\bl x, y_1(t) \br$. Let $H_t$ be the Riemannian Hessian of $f_t$, i.e. $H_t(x):T_x\M \ra T_x\M$ such that for all $x \in \M$, $t \in \Tsc$, and $v \in T_x\M$, $H_t(x)(v) = \nabla_v \operatorname{grad}f_t(x)$. For a mapping $A(x)$, which for each $x$ is an operator $A(x):T_x\M \ra T_x\M$, we define the operator norm at $x$ of $A$: $\norm{A}_{\op,x} \equiv \supp{\norm{A(x)(v)}_x}{v \in T_x\M,\norm{v}_x = 1}$. Recall that $\tau = |\Tsc|$ is the length of the time interval of the functional data.
\begin{lemma} \label{lemma:technical_2}
    Assume that
    \begin{enumerate}
        \item \label{item:hadamard_assumption} $\M$ is a complete, simply-connected Riemannian manifold with nonpositive sectional curvature.
        
        \item $\supp{\Ex{d\bl y_1(t),y_2(t) \br^3}}{t \in \Tsc} < \infty$.\label{item:expectation_finite_assumption} 
        \item $\supp{\norm{H_t(x)}_{\op,x}^2}{t \in \Tsc,x\in \M} \lesssim 1$ with probability $1$.\label{item:hessian_assumption_bound}
    \end{enumerate}
    Then,
    \begin{equation}
        \Ex{\munorm{\Log_{\hat \mu} y_i \thetagamma \Log_{\mu}y_i}^2} \lesssim \frac{\tau}{N}.
    \end{equation}
\end{lemma}
The next lemma combines the results of Lemmas \ref{lemma:phi_dif}, \ref{lemma:operator_expectation_bounds}, and \ref{lemma:technical_2}.
\begin{lemma} \label{lemma:expectation_phi_dif}
    Under the assumptions of Lemma \ref{lemma:technical_2}, and additionally assuming that\\ $\Ex{\munorm{\Log_{\hat \mu} y_i \thetagamma \Log_{\mu}y_i}^4}^{1/2} \lesssim \Ex{\munorm{\Log_{\hat \mu} y_i \thetagamma \Log_{\mu}y_i}^2}$, and  $\Ex{\munorm{\Log_{\mu}y_1}^4}\geq 1$, we have
    \begin{equation}
        \Ex{\phidif^2} \lesssim \frac{1}{N}\frac{\tau \Ex{\munorm{\Log_{\mu}y_1}^4}}{\operatorname{min}\bl\omega_{j-1}-\omega_j, \omega_j - \omega_{j+1}\br^2}.
    \end{equation}
\end{lemma}
From Lemma \ref{lemma:expectation_phi_dif} and Markov's inequality, we obtain the following inequality, which we can use to bound Term III in Lemma~\ref{lemma:start_of_proof}.
\begin{corollary}
    \label{cor:prob_bound_on_phi_dif}
    Under the assumptions of Lemma \ref{lemma:expectation_phi_dif}, we have that
    \begin{equation}
        \phidif^2 = O_P \bl \frac{1}{N}\frac{\tau \Ex{\munorm{\Log_{\mu}y_1}^4}}{\operatorname{min}\bl\omega_{j-1}-\omega_j, \omega_j - \omega_{j+1}\br^2} \br.
    \end{equation}
\end{corollary}
Now we can combine the results of Lemmas \ref{lemma:z_minus_y}, \ref{lemma:technical_2} and \ref{lemma:expectation_phi_dif} to obtain a bound on $\Ex{\norm{Z-Y}^2}$.
\begin{lemma} \label{lemma:z_minus_y_expectation_bound}
    Under the assumptions of Lemma \ref{lemma:expectation_phi_dif} and additionally assuming that\\
    $\Ex{\phidif^4}^{1/2} \lesssim \Ex{\phidif^2}$, we have that
    \begin{equation}
        \Ex{\twonorm{Z-Y}^2} \lesssim \frac{\tau d}{N}\Ex{\munorm{\Log_{\mu}y_1}^4}^{3/2} \underset{j=1,\ldots d}{\operatorname{max}}\bl \frac{1}{ \omega_j-\omega_{j+1}}\br^2.
    \end{equation}
\end{lemma}

The proof follows from applying the Cauchy-Schwarz inequality and collecting similar terms.

Now we can state the conditions under which the additional assumptions of Lemma \ref{lemma:a_minus_eta_with_z_minus_y} will hold. From now on, we keep tracking the terms $\tau$ and $\Ex{\munorm{\Log_{\mu}y_1}^4}$ in the error bounds, but we assume they are constant.
\begin{lemma} \label{lemma:conditions_for_assumptions_to_hold}
    We have that
    $\twonorm{\sigmay^{-1}} = \omega_d^{-1}$, $\twonorm{\sigmay} = \omega_1$, and $\Ex{\twonorm{Y}^2} = \sum_{j=1}^d\omega_j$. Additionally, if $\Ex{\twonorm{Y}^2},\twonorm{\sigmay^{-1}} \geq 1$, and
    \begin{equation} \label{eq:lemma_assumption}
        \frac{d}{N}\frac{\sum_{j=1}^d\omega_j}{\omega_d^2}\underset{j=1,\ldots d}{\operatorname{max}}\bl \frac{1}{ \omega_j-\omega_{j+1}}\br^2 = o(1),
    \end{equation}
    then $\Ex{\twonorm{Z-Y}^2} = o(1)$, $\twonorm{\sigmaz^{-1}} \lesssim \twonorm{\sigmay^{-1}}$, $\twonorm{\sigmaz} \lesssim \twonorm{\sigmay}$ and $\Ex{\twonorm{Z}^2} \lesssim \Ex{\twonorm{Y}^2}$.
\end{lemma}
Now we can combine Lemmas \ref{lemma:a_minus_eta_with_z_minus_y}, \ref{lemma:z_minus_y_expectation_bound}, and \ref{lemma:conditions_for_assumptions_to_hold} to obtain a final bound on $\twonorm{a_k - \eta_k}^2$.
\begin{theorem}
\label{thm:a-eta}
    Under the assumptions of Lemmas \ref{lemma:z_minus_y_expectation_bound} and  \ref{lemma:conditions_for_assumptions_to_hold}, we have that
    \begin{equation}
        \twonorm{a_k - \eta_k}^2 \lesssim \frac{\tau s d}{N}\underset{j=1,\ldots d}{\operatorname{max}}\bl \frac{1 }{ \omega_j-\omega_{j+1}}\br^2 \bl \frac{ \sum_{j=1}^d\omega_j}{\omega_d^2}\br^2 \frac{\Ex{\munorm{\Log_{\mu}y_1}^4}^{3/2}}{\operatorname{min}\bl\gamma_{k-1}-\gamma_k,\gamma_k-\gamma_{k+1}\br^2}.
    \end{equation}
\end{theorem}

Before we combine the bounds for the three components detailed in Lemma \ref{lemma:start_of_proof}, we show that $\tilde{\gamma}_k$ and $\gamma_k$ are asymptotically equivalent, which allows us to simplify the expression of our final bounds. Recall that the correlations $\{\tilde{\gamma}_k\}$ are defined using equation \eqref{eq:in_practice} as the canonical correlations between $X$ and $Z$, while the correlations $\{\gamma_k\}$ are the canonical correlations between $X$ and $Y$. To this end, we first state the following bound, which follows directly from Lemmas \ref{lemma:A_minus_C_Frobenious} and \ref{lemma:minus_square_root_difference_ZY}. This will also be used later to derive a bound for $\twonorm{\theta_k-\hat{\theta}_k}$.
\begin{lemma}\label{lemma:A_minus_C_bound}
    Under the assumptions of Lemma \ref{lemma:conditions_for_assumptions_to_hold}, we have that
    \begin{equation}
        \Fnorm{A-C}^2 \lesssim s \twonorm{\sigmay^{-1}}^3 \Ex{\twonorm{Y}^2}^2 \Ex{\twonorm{Z-Y}^2}.
    \end{equation}
\end{lemma}

We can prove the next lemma by using Lemmas~\ref{lemma:z_minus_y_expectation_bound}, \ref{lemma:conditions_for_assumptions_to_hold}, and \ref{lemma:A_minus_C_bound}.
\begin{lemma} \label{lemma:gamma_and_gamma_tilde_equivalent}
    Under the assumptions of Lemmas \ref{lemma:z_minus_y_expectation_bound} and \ref{lemma:conditions_for_assumptions_to_hold}, and further assuming that $\tau$ and $\Ex{\munorm{\Log_{\mu}y_1}^4}$ are absolute constants, that the canonical correlations $\{\gamma_k\}$ and $\{\tilde{\gamma}_k\}$ are bounded from below and that
    \begin{equation}
        \frac{ds}{N} \frac{\bl \sum_{j=1}^d\omega_j \br^2} {\omega_d^3}\underset{j=1,\ldots d}{\operatorname{max}}\bl \frac{1}{ \omega_j-\omega_{j+1}}\br^2 = o(1),
    \end{equation}
    then $\gamma_k$ and $\tilde{\gamma}_k$ are asymptotically equivalent, i.e., $\gamma_k \lesssim \tilde{\gamma}_k$ and $\tilde{\gamma}_k \lesssim \gamma_k$. Additionally, $\gamma_k^2 \lesssim \tilde{\gamma}_k^2$ and $\tilde{\gamma}_k^2 \lesssim \gamma_k^2$. 
\end{lemma}

Now we are now in a position to establish bounds for all three terms in Lemma \ref{lemma:start_of_proof}, using Theorem \ref{thm:multivariate_canonical_vector_bound_probabilistic_fast} (applied to $X$ and $Z$ and using the bound for $\eta$), Theorem \ref{thm:a-eta}, and Corollary \ref{cor:prob_bound_on_phi_dif} for Term I, II, and III, respectively. This will yield our final bound for the canonical functions $\norm{\hat \psi_k \thetagamma \psi_k}_{\mu}^2$, which is presented in Section~\ref{sec:final_rates_sparse_functional}.

\subsection{Bounding high-dimensional canonical vector error}
In this section, we derive a bound for the estimation error of the high-dimensional canonical vectors, $\twonorm{\theta_k- \hat{\theta}_k}$. Having already derived a bound for the canonical functions, the proof is straightforward. We start with an application of the triangle inequality. For $k=1, \ldots K$,
\begin{equation}
    \twonorm{\theta_k- \hat{\theta}_k}^2 \lesssim \twonorm{\theta_k - b_k}^2 + \twonorm{b_k - \hat{\theta}_k}^2,
\end{equation}
where $b_k$ is the high-dimensional canonical vector given by the solution to problem \eqref{eq:in_practice}. Assuming $s$-group sparsity for $\theta_k$ and $b_k$, we represent the associated vectors with non-zero entries as $\theta_{k,S}$ and $b_{k,S}$. Recall that these are at most $2s$-dimensional. By definition, we then have $\twonorm{\theta_k - b_k} =\twonorm{\theta_{k,S} - b_{k,S} }$, hence
\begin{equation} \label{eq:theta_minus_b_first_bound}
    \twonorm{\theta_k- \hat{\theta}_k}^2 \lesssim \twonorm{\theta_{k,S} - b_{k,S} }^2 + \twonorm{b_k - \hat{\theta}_k}^2.
\end{equation}
To bound the second term, we can use Theorem \ref{thm:multivariate_canonical_vector_bound_probabilistic_fast}, applied to the random vectors $X$ and $Z$. To bound the second term, we can make the following argument, which is similar to the one made to bound $\twonorm{\eta_k-a_k}$. Let
\begin{align}
    A = \Sigma_{X_S}^{-1/2}\Sigma_{{X_S}Z}\sigmaz^{-1/2},\\
    C = \Sigma_{X_S}^{-1/2}\Sigma_{{X_S}Y}\sigmay^{-1/2}.
\end{align}
Then, from classical CCA \parencite{uurtio2018tutorial},
we have that $\theta_{k,S} = \Sigma_{X_S}^{-1/2} \tilde{\theta}_{k,S}$, where $\tilde{\theta}_{k,S}$ is the $k$th eigenvector of $CC^{\T}$, and $b_k = \Sigma_{X_S}^{-1/2} \tilde{b}_{k,S}$, where $\tilde b_{k,S}$ is the $k$th eigenvector of $AA^{\T}$. Therefore,
\begin{align}
    \twonorm{\theta_{k,S} - b_{k,S} } &= \twonorm{\Sigma_{X_S}^{-1/2} \tilde{\theta}_{k,S} - \Sigma_{X_S}^{-1/2} \tilde{b}_{k,S}}\\
    & = \twonorm{\Sigma_{X_S}^{-1/2}\bl \tilde{\theta}_{k,S} - \tilde{b}_{k,S}\br}\\ 
    & \leq \twonorm{{\Sigma_{X_S}^{-1/2}}} \twonorm{\tilde{\theta}_{k,S} - \tilde{b}_{k,S}}. \label{eq:b_tilde_theta_tilde_term}
\end{align}
Since $A^{\T}A$ has the same eigenvectors as $\left | A \right | = \bl AA^{\T} \br^{1/2}$, we can then use the Davis-Kahan theorem \parencite{yu2015useful} to derive the following bound. If $\tilde{\theta}_{k,S}^{\T}\tilde{b}_{k,S} \geq 0$, then
\begin{equation}
    \twonorm{\tilde{\theta}_{k,S} - \tilde{b}_{k,S}} \lesssim \frac{\twonorm{\left |A^{\T} \right | - \left | C^{\T} \right |}}{\operatorname{min}\bl \gamma_{k-1}-\gamma_k,\gamma_k - \gamma_{k+1}\br}.
\end{equation}
The term $\twonorm{\left |A^{\T} \right| - \left |C^{\T} \right |}$ can be bounded as follows:
\begin{equation}
    \twonorm{\left |A^{\T} \right| - \left |C^{\T} \right |} \leq \Fnorm{\left |A^{\T} \right| - \left |C^{\T} \right |} \leq \sqrt{2}\Fnorm{A^{\T}-C^{\T}} = \sqrt{2}\Fnorm{A-C},
\end{equation}
where the second inequality is identity \ref{item:bhatia_VII_39} from Section \ref{subsec:identities}. Having established a bound for $\twonorm{\tilde{\theta}_{k,S} - \tilde{b}_{k,S}}$ in terms of $\Fnorm{A-C}$, we can apply similar arguments to those used to bound $\twonorm{a_k -\eta_k}$ in order to derive a bound for $\twonorm{\tilde{\theta}_{k,S} - \tilde{b}_{k,S}}$ in terms of $\Ex{\twonorm{Z-Y}^2}$. Combining the results of this section, using Lemmas \ref{lemma:z_minus_y_expectation_bound} and \ref{lemma:A_minus_C_bound}, we establish the following result. 
\begin{lemma} \label{lemma:theta_minus_b_bound}
    Under the assumptions of Lemmas \ref{lemma:z_minus_y_expectation_bound} and \ref{lemma:conditions_for_assumptions_to_hold}, and if $\tilde{\theta}_{k,S}^{\T}\tilde{b}_{k,S} \geq 0$, we have that
\begin{equation}
    \twonorm{\theta_{k,S} - b_{k,S} }^2 \lesssim \frac{\twonorm{{\Sigma_{X_S}^{-1/2}}}^2 s \twonorm{\sigmay^{-1}}^3 \Ex{\twonorm{Y}^2}^2}{{\operatorname{min}\bl \gamma_{k-1}-\gamma_k,\gamma_k - \gamma_{k+1}\br^2}} \frac{\tau d}{N}\Ex{\munorm{\Log_{\mu}y_1}^4}^{3/2} \underset{j=1,\ldots d}{\operatorname{max}}\bl \frac{1}{ \omega_j-\omega_{j+1}}\br^2.
\end{equation}
\end{lemma}

Next, by applying Theorem~\ref{thm:multivariate_canonical_vector_bound_probabilistic_fast}, Lemma \ref{lemma:theta_minus_b_bound}, and equation \eqref{eq:theta_minus_b_first_bound}, we can establish the final bound for the high-dimensional canonical vector error $\twonorm{\theta_k- \hat{\theta}_k}^2$, which is presented in Section~\ref{sec:final_rates_sparse_functional}.

\subsection{Final rates}\label{sec:final_rates_sparse_functional}
In this section, we present our final results. Recall that we denote as $K = \operatorname{max} \left \{i \in \{ 1, \ldots \dcorr \}: \gamma_i > 0 \right \}$ the number of nontrivial canonical vectors and we use the conventions $\gamma_{K+1}^2 = -\infty$ and $\gamma_0^2 = +\infty$. The random vector $Z$ was defined in equation \eqref{eq:def_of_Z} and represents a `sample' version of $Y$ where $\phi_j$ and $\mu$ are replaced by their finite-sample estimates.

For clarity, we first provide a comprehensive list of our assumptions. For the definitions of the quantities that appear below, please see the beginning of Section \ref{appendix:proof_of_canonical_function_section}.
\begin{assumption}[Manifold Properties]
    \begin{enumerate}
    \item[]
    \item The manifold $\M$ is a complete simply-connected Riemannian manifold with nonpositive sectional curvature.
    \item The curvature $\supp{\norm{H_t(x)}_{\op,x}^2}{t \in \Tsc,x\in \M}$ is bounded with probability $1$.
\end{enumerate}
\label{assumption:manifold}
\end{assumption}

\begin{assumption}[Distributional Assumptions]
\begin{enumerate}
    \item[] 
    \item The random vectors $X$ and $Z$ are strict sub-Gaussian random vectors (Definition \ref{def:strict_sub_gaussian}).\label{item:assumption:distributional1}
    \item The covariance matrices $\sigmax$, $\sigmay$, and $\sigmaz$ are invertible. \label{item:assumption:distributional2}
    \item The group $s$-sparsity assumption holds for $\{b_k\}$ and $\{\theta_k\}$. \label{item:assumption:distributional3}
    \item The matrix $\sigmax^{1/2}$ satisfies the group restricted eigenvalue condition RE$(s,3,d)$ (Definition \ref{def:group_restricted_eigenvalue}) with parameter $\kappa = \kappa(s,d,\sigmax^{1/2})$ .\label{item:assumption:distributional4}
    \item The functional data are such that $\supp{\Ex{d\bl y_1(t),y_2(t) \br^3}}{t \in \Tsc} < \infty$. \label{item:assumption:distributional5}
\end{enumerate}
\label{assumption:distributional}
\end{assumption}

\begin{assumption}[Rate Assumptions]
    \begin{enumerate}
    \item[] 
    \item There exists a complete orthonormal system for $\Jmu$, $\{ \phi_j \}_{j = 1}^{\infty}$, such that a $\dcorr$-dimensional Assumption \ref{remark:finite_dim_assumption} holds with $p\geq d \geq \dcorr$, where $d$ is the chosen number of principal components in Algorithm \ref{alg:asymmetric_sparse_fun_cca};  \label{item:assumption:rate1}
    \item $\operatorname{cond}\bl \sigmay \br^2d = o(N)$;\label{item:assumption:rate2}
    \item $\kappa^2s^2d\log(p) = o(N)$; \label{item:assumption:rate3}
    \item $ds\frac{\bl \sum_{j=1}^d\omega_j \br^2} {\omega_d^3}\underset{j=1,\ldots d}{\operatorname{max}}\bl \frac{1}{ \omega_j-\omega_{j+1}}\br^2 = o(N)$; \label{item:assumption:rate4}
    \item The correlations $\gamma_1, \ldots \gamma_K$ are bounded from below and distinct from one another, as well as $\tilde{\gamma}_1, \ldots \tilde{\gamma}_K$. \label{item:assumption:rate5}
    \item $\Ex{\phidif^4}^{1/2} \lesssim \Ex{\phidif^2}$; \label{item:assumption:rate6}
    \item $\Ex{\munorm{\Log_{\hat \mu} y_i \thetagamma \Log_{\mu}y_i}^4}^{1/2} \lesssim \Ex{\munorm{\Log_{\hat \mu} y_i \thetagamma \Log_{\mu}y_i}^2}$. \label{item:assumption:rate7}
\end{enumerate}
\label{assumption:rate}
\end{assumption}

\begin{assumption}[Minor Assumptions]
    \begin{enumerate}
    \item[]
    \item The quantities $\twoinfnorm{\sigmax},\onetwo{T}$ are bounded from above and are $\geq 1$. \label{item:assumption:minor1}
    \item The variables $\tau$ and $\Ex{\munorm{\Log_{\mu}y_1}^4}$ are constants.\label{item:assumption:minor2}
    \item The following quantities are larger than 1: $\kappa, \omega_1,\omega_d^{-1},\infnorm{\eta},\twonorm{\sigmax^{-1}}, \twonorm{\sigmaz^{-1}}$. \label{item:assumption:minor3}
    \item $a_k^{\T}\sigmaz^{1/2} \sigmay^{1/2}\eta_k \geq 0$ and $\hat{\eta}_k^{\T}\hat{\Sigma_Z}^{1/2}\sigmaz^{1/2}a_k\geq 0$ for $k=1, \ldots K$. \label{item:assumption:minor5}
\end{enumerate}
\label{assumption:minor}
\end{assumption}

Next, we present our main results.
\begin{theorem}[Canonical Function Error Bound]
    \label{thm:canonical_function_error}
    Under Assumptions \ref{assumption:manifold}-\ref{assumption:minor}, we have
    \begin{align}
        \norm{\hat \psi_k \thetagamma \psi_k}_{\mu}^2 &= O_P \bl \frac{d^2s \log(p)}{N} \br \frac{\tau}{\min_{j \neq k} \min \curl \left |{\gamma_k}^2 - {\gamma_j}^2 \right |, \left|{\gamma_k} - {\gamma_j} \right | \curr^2} \\
        &\cdot \bl \frac{\sum_{j=1}^d \omega_j}{\omega_d^2}\br^2 \underset{j=1,\ldots d}{\operatorname{max}}\bl \frac{1 }{ \omega_j-\omega_{j+1}}\br^2  \kappa \twoinfnorm{\sigmax} \infnorm{\eta}^2 \Ex{\munorm{\Log_{\mu} y}^4}^{3/2} ,
    \end{align}
with $k=1, \ldots K$. 
\end{theorem}

\begin{theorem}[Canonical Vector Error Bound]
\label{thm:canonical_vector_error}
    Under Assumptions \ref{assumption:manifold}-\ref{assumption:minor}, and additionally assuming that $\theta_{k,S}^{\T}\Sigma_{X_S}b_{k,S} \geq 0$, we have
    \begin{align}
        \twonorm{\theta_k - \hat{\theta}_k}^2 &= O_P\bl \frac{d s \log(p)}{N }\br\frac{\tau}{\min_{j \neq k} \min \curl \left |{\gamma_k}^2 - {\gamma_j}^2 \right |, \left|{\gamma_k} - {\gamma_j} \right | \curr^2}\\
        &\cdot \left [ \frac{\bl \sum_{j=1}^d\omega_j \br^2}{\omega_d^3} \underset{j=1,\ldots d}{\operatorname{max}}\bl \frac{1}{ \omega_j-\omega_{j+1}}\br^2 \Ex{\munorm{\Log_{\mu} y}^4}^{3/2}\twonorm{\Sigma_{X_S}^{-1/2}} + \bl \frac{\gamma_1}{\gamma_k}\br^2 \twoinfnorm{\sigmax}\kappa^2 \twonorm{\sigmax^{-1}}   \right ],
    \end{align}
with $k=1, \ldots K$.
\end{theorem}

Here we make a few remarks on the assumptions of Theorems \ref{thm:canonical_function_error} and \ref{thm:canonical_vector_error}. Some of the more technical assumptions arise from avoiding overly simplifying assumptions. For instance, with the exception of the curvature-related quantity $H_t(x)$, we do not assume that the random variables/functions are bounded. Furthermore, we avoid assuming Gaussianity of the random variables of interest and instead assume these are sub-Gaussians. As in \textcite{lin2019intrinsic}, we do not assume that the Frech\'et mean $\mu$ is known and instead estimate it using its sample version $\hat{\mu}$; this choice introduces significant complexity, making it necessary to use the parallel transport operator to compare estimates and estimands, which are defined in different tangent spaces. These challenges are further compounded by the dimensionality reduction step that takes place before CCA, which requires that we derive bounds in \textit{expectation} rather than in probability. Specifically, this requires showing that $\Ex{ d\bl \hat{\mu}(t),\mu(t) \br^2} \lesssim \frac{1}{N}$, i.e., equation (\ref{eq:frechet_mean_expectation_bound}),
using the results of \textcite{schotz2019convergence} as opposed to those of \textcite{lin2019intrinsic}. See also Remark \ref{remark:expectations_necessary}.

\noindent \textbf{Remarks on Assumption \ref{assumption:manifold}:}\\
\noindent Assumption \ref{assumption:manifold} is required to bound the term $\Ex{ d\bl \hat{\mu}(t),\mu(t) \br^2}$. See also Lemma \ref{lemma:technical_2}. Here, $H_t(x)$ is the Riemannian Hessian of the random function $d^2(x,y(t))$, and is related to curvature on the manifold \parencite{pennec2017hessian}. Assuming this quantity is bounded allows us to bound the parallel transport distance in terms of the geodesic distance, as shown in Lemma \ref{lemma:technical_1}.

\noindent \textbf{Remarks on Assumption \ref{assumption:distributional}:}\\
\noindent Items \ref{item:assumption:distributional1}-\ref{item:assumption:distributional4} of Assumption \ref{assumption:distributional} are used to facilitate the application of our multivariate CCA results in Section~\ref{thm:main_theorem_multivariate_case} to the sparse-functional setting considered here. Specifically, Items \ref{item:assumption:distributional3}-\ref{item:assumption:distributional4} are used in particular to get fast-rate bounds, which match the root-n estimation rate of the functional quantities. We note that in item \ref{item:assumption:distributional3}, we do not require that $\theta_k$ and $b_k$ have the same sparsity structure, but only that they are both $s$-sparse.
Item \ref{item:assumption:distributional4} is a generalization of the standard restricted-eigenvalue condition in Lasso theory \parencite{hastie2015statistical} and is equivalent to the one proposed in \textcite{gaynanova2020prediction}. Item \ref{item:assumption:distributional5} is a weak assumption about the boundedness of the variance of $y(t)$ on the manifold and along with Assumption \ref{assumption:manifold} is necessary to show that $\Ex{ d\bl \hat{\mu}(t),\mu(t) \br^2}$ is root-$n$ consistent.

\noindent \textbf{Remarks on Assumption \ref{assumption:rate}:}\\
\noindent Item \ref{item:assumption:rate1} of Assumption \ref{assumption:rate}, that the correlation with the high-dimensional data is captured in a finite-dimensional subspace of the functional data, is necessary to even define the canonical directions for functional data, as stated in the main manuscript. The condition that $d \leq p$ formalizes our asymmetrical treatment of the data, and allows for the CCA problem to become a sparse regression problem. $d \geq \dcorr$ is necessary to ensure we capture all of the components of $\Logy$ which are correlated with $X$. Items \ref{item:assumption:rate2} - \ref{item:assumption:rate5} are mainly used to simplify the theorem statements by allowing us to bound norms of estimated quantities using the corresponding population quantities. In particular, Item \ref{item:assumption:distributional2} is used in conjunction with Lemma \ref{lemma:conditions_for_assumptions_to_hold} to show that $\twonorm{\hat{\Sigma}_Z} \lesssim \twonorm{\sigmay}$. Item \ref{item:assumption:distributional3} allows us to only make a group restricted eigenvalue assumption on $\sigmax^{1/2}$ rather than the data matrix $\frac{1}{\sqrt{N}}\mathbb{X}$ (see Lemma \ref{lemma:probabilistic_group_restricted_eigenvalue_condition}). Item \ref{item:assumption:distributional4} states that the variances $\omega_j$ of the principal scores $Y_j$ should not shrink too quickly as $N$ and $d$ grow. Note that if $d$ is assumed constant, the condition reduces to $s = o(N)$. This is used to establish the simplifying bounds given in Lemmas \ref{lemma:conditions_for_assumptions_to_hold} and \ref{lemma:gamma_and_gamma_tilde_equivalent}. Item \ref{item:assumption:distributional5} is used to replace $\hat{\gamma}_k$ and $\tilde{\gamma}_k$ with the population canonical correlations $\gamma_k$ (see Lemma \ref{lemma:gamma_and_gamma_tilde_equivalent} and the discussion in the proof of Theorem \ref{thm:multivariate_canonical_vector_bound_probabilistic_slow}). Items \ref{item:assumption:rate6} and \ref{item:assumption:rate7} are technical conditions. These mainly arise from the complexity of the setting considered here. The assumption $\Ex{\phidif^4}^{1/2} \lesssim \Ex{\phidif^2}$ could be replaced with a boundness assumption on $\munorm{\Log_{\mu}y_i}$, or alternatively, by adopting a sample splitting strategy to estimate $\hat{\mu}$, the principal components $\{\hat{\phi}_j\}$, and to carry out CCA. This can be seen from the second term of equation (\ref{eq:Z_minus_Y_explanation2}), where the absence of one of these assumptions requires that we use the Cauchy-Schwarz inequality, introducing fourth moments. Note that if we were to assume that $\hat{\mu} = \mu$, then $\Ex{\munorm{\Log_{\hat \mu} y_i \thetagamma \Log_{\mu}y_i}^4}^{1/2} = 0$, immediately satisfying the condition in item 7.

\noindent \textbf{Remarks on Assumption \ref{assumption:minor}:}\\
\noindent Items \ref{item:assumption:minor1}-\ref{item:assumption:minor3} are not critical and only serve the purpose of simplifying the theorem statements. Item \ref{item:assumption:minor5} is introduced to account for the sign ambiguity of the CCA solutions.

\subsection{Proofs for results in Section \ref{appendix:proof_of_canonical_function_section}} \label{sec:apdx:proofs_for_results_in_canonical_function_proof}
\textbf{Proof of Lemma \ref{lemma:start_of_proof}:}\\We define $\tilde{\psi}_k = \sum_{j=1}^d \hat{\phi}_j a_{kj}$. For ease of notation, we drop the $k$ in writing $\hat \psi_k$, $\psi_k$ and $\tilde{\psi}_k$, $\eta_k$, etc.
We have
\begin{equation} \label{eq:can_fun_start}
    \norm{\hat \psi \thetagamma \psi}_{\mu}^2 = \munorm{\Gamma_{\hat{\mu},\mu}\hat{\psi} - \psi}^2 \leq 2\munorm{\gammamuhat \hat{\psi} - \gammamuhat \tilde{\psi}}^2 + 2 \munorm{\gammamuhat \psitil - \psi}^2.
\end{equation}
The first term in equation \eqref{eq:can_fun_start} is
\begin{equation*}
    2\munorm{\gammamuhat \bl \hat{\psi} - \tilde{\psi} \br}^2 = \muhatnorm{\Gamma_{\mu,\hat{\mu}}\bl \gammamuhat \bl \hat{\psi} - \tilde{\psi} \br \br}^2 = 2\muhatnorm{\psihat - \psitil}^2.
\end{equation*}
Define $\psibar$ as $\sum_{j=1}^d \hat{\phi}_j \eta_j$.
Then the second term in equation \eqref{eq:can_fun_start} is
\begin{equation*}
    \leq 4 \munorm{\gammamuhat \psitil - \gammamuhat \psibar}^2 + 4 \munorm{\gammamuhat \psibar - \psi}^2.
\end{equation*}
Therefore,
\begin{equation}\label{eq:can_fun_main}
     \norm{\hat \psi \thetagamma \psi}_{\mu}^2 \lesssim \muhatnorm{\psihat - \psitil}^2 + \muhatnorm{\psitil - \psibar}^2 + \munorm{\psibar \thetagamma \psi}^2.
\end{equation}
The first term in equation \eqref{eq:can_fun_main} is
\begin{equation}
    \muhatnorm{\psihat - \psitil}^2 = \muhatnorm{\sum_{j=1}^d\hat{\phi}_j \hat{\eta}_j - \sum_{j=1}^d a_j \hat{\phi}_j}^2 = \muhatnorm{\sum_{j=1}^d\hat{\phi}_j \bl \hat{\eta}_j - a_j \br}^2 = \twonorm{\hat{\eta} - a}^2,
\end{equation}
where in the third equality we have used that the $\hat{\phi}_j$ are orthonormal in $\Jmuhat$.
Similarly, the second term is
\begin{equation}
    \muhatnorm{\psitil - \psibar}^2 = \twonorm{a - \eta}^2.
\end{equation}
By the triangle inequality, the third term is
\begin{equation}
     \munorm{\psibar \thetagamma \psi}^2 \leq \bl \infnorm{\eta}\sum_{j=1}^d \munorm{\hat{\phi}_j \thetagamma \phi_j} \br^2,
\end{equation}
completing the proof. \hfill \qedsymbol\\

\noindent \textbf{Proof of Lemma \ref{lemma:A_minus_C_Frobenious}:}\\
For ease of notation, we write $X_S$ as $X$ throughout the proof of the lemma. From the definition of $A$ and $C$, we have
\begin{equation}
    \Fnorm{A-C} = \Fnorm{\Sigma_{X}^{-1/2}\Sigma_{{X}Z}\sigmaz^{-1/2} - \Sigma_{X}^{-1/2}\Sigma_{{X}Y}\sigmay^{-1/2}} = \Fnorm{\Sigma_{X}^{-1/2}\bl \Ex{XZ^{\T}}\sigmaz^{-1/2} - \Ex{XY^{\T}}\sigmay^{-1/2}\br}.
\end{equation}
From the linearity of expectation, this is
\begin{equation}
     = \Fnorm{\Ex{\Sigma_{X}^{-1/2}X\bl Z^{\T}\sigmaz^{-1/2} - Y^{\T}\sigmay^{-1/2}\br}} = \Fnorm{\Ex{\Sigma_{X}^{-1/2}X\bl \sigmaz^{-1/2}Z - \sigmay^{-1/2}Y\br^{\T}}}.
\end{equation}
By Theorem 2.6.7 of \textcite{hsing2015theoretical}, using the Frobenious norm of an outer product, and the Cauchy-Schwarz inequality, we have
\begin{align}
      \Fnorm{A-C} &\leq \Ex{\Fnorm{\Sigma_{X}^{-1/2}X\bl \sigmaz^{-1/2}Z - \sigmay^{-1/2}Y\br^{\T}}}\\
     &= \Ex{\twonorm{\Sigma_{X}^{-1/2}X}\twonorm{ \sigmaz^{-1/2}Z - \sigmay^{-1/2}Y}}\\
     &\leq \Ex{\twonorm{\Sigma_{X}^{-1/2}X}^2}^{1/2}\Ex{\twonorm{ \sigmaz^{-1/2}Z - \sigmay^{-1/2}Y}^2}^{1/2}.
\end{align}
We have that $\Ex{\twonorm{\Sigma_{X}^{-1/2}X}^2} = \Ex{\operatorname{tr}\bl\sigmax^{-1}XX^{\T}\br} = \operatorname{tr}\bl I_s\br = s$. Using this along with identity \ref{item:AaBb} from Section \ref{subsec:identities} to upper bound $\Ex{\twonorm{ \sigmaz^{-1/2}Z - \sigmay^{-1/2}Y}^2}^{1/2}$ completes the proof.\\

\noindent \textbf{Proof of Lemma \ref{lemma:minus_square_root_difference_ZY}:}\\
The first statement follows from the second statement, by identity \ref{item:MVT_minus_square_root_matrices} from Section \ref{subsec:identities} and Lemma \ref{lemma:minus_square_root_difference_ZY}. To show the second statement, we begin with
\begin{equation}
    \twonorm{\sigmay - \sigmaz} = \twonorm{\Ex{YY^{\T}} - \Ex{ZZ^{\T}}} = \twonorm{\Ex{YY^{\T} - ZZ^{\T}}},
\end{equation}
and using Theorem 2.6.7 of \textcite{hsing2015theoretical}, this is
\begin{equation}
    \leq \Ex{\twonorm{YY^{\T} - ZZ^{\T}}} = \Ex{\twonorm{YY^{\T} - ZY^{\T} + ZY^{\T}-ZZ^{\T}}} = \Ex{\twonorm{Z\bl Z-  Y\br^{\T} + \bl Z - Y\br Y^{\T}}}.
\end{equation}
From this, the triangle inequality, and using the two-norm of an outer product, we have
\begin{equation}
    \twonorm{\sigmay - \sigmaz} \leq \Ex{\twonorm{Z\bl Z-  Y\br^{\T}}} + \Ex{\twonorm{ \bl Z - Y\br Y^{\T}}} = \Ex{\twonorm{Z}\twonorm{Z-Y}} + \Ex{\twonorm{Z-Y}\twonorm{Y}}
\end{equation}
By the Cauchy-Schwarz inequality, the right hand side is
\begin{equation}
    = \Ex{\bl \twonorm{Z} + \twonorm{Y}\br \twonorm{Z-Y}} \leq \Ex{\bl\twonorm{Z} + \twonorm{Y}\br^2}^{1/2} \Ex{\twonorm{Z-Y}^2}^{1/2}
\end{equation}
The second statement of the lemma follows, and the proof is complete. \hfill \qedsymbol\\

\noindent \textbf{Proof of Lemma \ref{lemma:z_minus_y}:}\\
Define $W \in \rone^d$ as the random vector with $W_j \equiv \innerdouble{\Log_{\mu}y_1}{\gammamuhat \hat{\phi}_j}_{\mu}$ for $j=1 , \ldots d$. Then
\begin{equation}
    \twonorm{Z-Y}^2 \leq 2\twonorm{Z-W}^2 + 2\twonorm{W-Y}^2.
\end{equation}
We have
\begin{align}
    Z_j - W_j &= \innerdouble{\Log_{\hat{\mu}}y_1}{\hat{\phi}_j}_{\hat{\mu}} - \innerdouble{\Log_{\mu}y_1}{\gammamuhat \hat{\phi}_j}_{\mu}\\
    & = \innerdouble{\Log_{\hat{\mu}}y_1}{\hat{\phi}_j}_{\hat{\mu}} -\innerdouble{\Gamma_{\mu,\hat{\mu}}\Log_{\mu}y_1}{\hat{\phi}_j}_{\hat{\mu}}\\
    & = \innerdouble{\Log_{\hat{\mu}}y_1-\Gamma_{\mu,\hat{\mu}}\Log_{\mu}y_1}{\hat{\phi}_j}_{\hat{\mu}},
\end{align}
and therefore, by the Cauchy-Schwarz inequality, and because the $\hat{\phi}_j$ are orthonormal along $\hat{\mu}$,
\begin{align}
    \left | Z_j - W_j \right | &\leq \norm{\Log_{\hat{\mu}}y_1-\Gamma_{\mu,\hat{\mu}}\Log_{\mu}y_1}_{\hat{\mu}}\norm{\hat{\phi}_j}_{\hat{\mu}}\\
    & = \muhatnorm{\Log_{\mu} y_1 \thetagamma \Log_{\hat{\mu}}y_1}\\
    & = \munorm{\Log_{\hat \mu} y_1 \thetagamma \Log_{\mu}y_1},
\end{align}
where in the last equality we have used that $\norm{U \thetagamma V}_{\mu} = \norm{V \thetagamma U}_{\hat{\mu}}$ for $U \in \Jmuhat$ and $V \in \Jmu$. We also have
\begin{align}
    W_j - Y_j &= \innerdouble{\Log_{\mu}y_1}{\gammamuhat \hat{\phi}_j}_{\mu} - \innerdouble{\Log_{\mu}y_1}{\phi_j}_{\mu}\\
    & = \innerdouble{\Log_{\mu}y_1}{ \hat{\phi}_j \thetagamma \phi_j}_{\mu},
\end{align}
so that again by the Cauchy-Schwarz inequality,
\begin{equation}
    \left | W_j - Y_j \right | \leq \munorm{\Log_{\mu}y_1}\phidif.
\end{equation}
Thus,
\begin{align}
    \twonorm{Z-Y}^2 & \leq 2\sum_{j=1}^d \left | Z_j - W_j \right |^2 + \left | W_j - Y_j \right |^2 \\
    & \leq 2\sum_{j=1}^d \munorm{\Log_{\hat \mu} y_1 \thetagamma \Log_{\mu}y_1}^2 + \munorm{\Log_{\mu}y_1}^2\phidif^2,
\end{align}
from which the statement of the lemma follows, and the proof is complete. \hfill \qedsymbol\\

\noindent \textbf{Proof of Lemma \ref{lemma:phi_dif}:}\\
From \textcite{lin2019intrinsic} (page 3551), $\gammamuhat \hat{\phi}_j$ are the eigenvectors of $\Phi \hat{\C}$. By definition, the $\phi_j$ are the eigenvectors of $\C$. Thus, we can use the Davis-Kahan Theorem for Hilbert spaces \parencite{jirak2020perturbation} to obtain that
\begin{equation}
    \gammamuhat \hat{\phi}_j - \phi_j \leq 2\sqrt{2}\operatorname{max}\bl \bl \omega_{j-1} - \omega_j\br^{-1},\bl \omega_j-\omega_{j+1}\br^{-1}\br \norm{\C - \Phi \hat{\C}}_{\op},
\end{equation}
where $\omega_0 \equiv \infty$ and $\norm{\C}_{\op} \equiv \supp{\munorm{\C U}}{U \in \Jmu, \munorm{U} = 1}$.
Adding and subtracting $\hat{\C}_{\mu}$ in $\norm{\C - \Phi \hat{\C}}_{\op}$ and using identity \ref{item:norm_squared_triangle_inequality} from Section \ref{subsec:identities}, the proof is complete. \hfill \qedsymbol\\

\noindent \textbf{Proof of Lemma \ref{lemma:operator_expectation_bounds}:}\\
The first statement is equivalent to Lemma 5.2 of \textcite{cardot1999functional}.
To show the second statement, we begin with the definition of $\hat{\C}$ and Proposition 2 item 5 of \textcite{lin2019intrinsic}, from which we deduce that
\begin{equation}
    \Phi \hat{\C} = \frac{1}{N}\sum_{i=1}^N\bl \gammamuhat \Log_{\hat{\mu}}y_i\br \otimes \bl \gammamuhat \Log_{\hat{\mu}}y_i\br.
\end{equation}
This implies
\begin{equation}
    \hat{\C}_{\mu} - \Phi \hat{\C} = \frac{1}{N}\sum_{i=1}^N a_i \otimes a_i - b_i \otimes b_i,
\end{equation}
where we denote $a_i \equiv \Log_{\mu}y_i$ and $b_i \equiv \gammamuhat \Log_{\hat{\mu}} y_i$.
It is straightforward to show that $a \otimes a - b \otimes b = a \otimes \bl a-b \br + \bl a - b \br \otimes \bl b-a \br + \bl a-b\br \otimes a$, where $a,b \in \Jmu$. From Theorem 3.4.7. of \textcite{hsing2015theoretical}, $\norm{a \otimes b}_{\op} = \norm{a}_{\mu}\norm{b}_{\mu}$, and therefore,
\begin{equation}
    \norm{a \otimes a - b \otimes b}_{\op} \leq 2\munorm{a}\munorm{a-b} + \munorm{a-b}^2.
\end{equation}
Then,
\begin{equation}
    \norm{\hat{\C}_{\mu} - \Phi \hat{\C}_{\op}} \leq \frac{1}{N} \sum_{i=1}^N 2\munorm{\Log_{\mu}y_i}\munorm{\Log_{\hat{\mu}}y_i \thetagamma \Log_{\mu} y_i} + \munorm{\Log_{\hat{\mu}}y_i \thetagamma \Log_{\mu} y_i}^2,
\end{equation}
so that
\begin{equation} \label{eq:operator_difference_eq}
    \norm{\hat{\C}_{\mu} - \Phi \hat{\C}_{\op}}^2 \lesssim  \bl \frac{1}{N} \sum_{i=1}^N 2\munorm{\Log_{\mu}y_i}\munorm{\Log_{\hat{\mu}}y_i \thetagamma \Log_{\mu} y_i} \br^2 + \bl \frac{1}{N} \sum_{i=1}^N \munorm{\Log_{\hat{\mu}}y_i \thetagamma \Log_{\mu} y_i}^2 \br^2.
\end{equation}
It is straightforward to show that, for any i.i.d.~random variables $W_i$ with finite variance we have
\begin{equation}
    \Ex{\bl \frac{1}{N}\sum_{i=1}^N W_i \br^2} \leq \Ex{W_1^2},
\end{equation}
where the $W_i$ are not required to have mean $0$.
Taking the expectation in equation \eqref{eq:operator_difference_eq}, applying this last result on each term, using the Cauchy-Schwarz inequality, and subsequently using the assumption stated in the Lemma concerning $\munorm{\Log_{\hat{\mu}}y_i \thetagamma \Log_{\mu} y_i}$, the proof is complete. \hfill \qedsymbol\\

\noindent \textbf{Proof of Lemma \ref{lemma:technical_1}:}\\
We start with a mean value theorem result for smooth functions $f$ from $[0,t_1]$ into a normed vector space $V$, where $t_1 \in \rone$. By Theorem 1.1.1. of \textcite{hormander2015analysis}, we have
\begin{equation} \label{eq:MVT_vector_valued_functions}
    \norm{f(t_1) - f(0)} = \left |b-a \right | \supp{f'(c)}{c \in [0,t_1]}.
\end{equation}
We set
\begin{equation}
    f(t) = \Gamma_{\gamma(t),\gamma(0)} U\bl \gamma(t)\br - U\bl \gamma(0) \br,
\end{equation}
where $U$ is a smooth vector field on $\M$, $U: \M \ra T\M$, and $\gamma(t)$ is the minimizing geodesic between two points $x,y \in \M$ ($\gamma(0) = x, \gamma\bl (x,y)\br = y$. Letting $t_1 = d(x,y)$, then $\gamma:[0,t_1] \ra \M$, and $f:[0,t_1] \ra T_x\M$. Letting $\norm{W}_x$ denote the norm of $W \in T_x\M$, and using $f(t)$ in equation \eqref{eq:MVT_vector_valued_functions}, we have
\begin{equation}
    \norm{\Gamma_{y,x}U(y) - U(x)} \leq d(x,y) \supp{\norm{f'(c)}}{c \in [0,d(x,y)]}.
\end{equation}
We can determine $f'(c)$:
\begin{align}
    f'(c) &= \lim_{t \ra 0^+}\frac{f(c+t) - f(c)}{t}\\
    & = \lim_{t \ra 0^+}\frac{\Gamma_{\gamma(c+t),x}U\bl \gamma(c+t) \br - U(x) - \Gamma_{\gamma(c),x}U\bl \gamma(c) \br + U(x)}{t}\\
    & = \lim_{t \ra 0^+}\frac{\Gamma_{\gamma(c+t),x}U\bl \gamma(c+t) \br - \Gamma_{\gamma(c),x}U\bl \gamma(c) \br}{t}.
\end{align}
Using that $\Gamma_{z,x}U(z) = \Gamma_{y,x}\bl \Gamma_{z,y} U(z) \br$, where $x = x$, $y = \gamma(c)$, and $z = \gamma(c+t)$, we have
\begin{align}
    f'(c) &=  \lim_{t \ra 0^+}\frac{\Gamma_{\gamma(c),x} \left [ \Gamma_{\gamma(c+t),\gamma(c)}U\bl \gamma(c+t) \br - U\bl \gamma(c) \br \right ]}{t}\\
    & = \Gamma_{\gamma(c),x} \left [\lim_{t \ra 0^+}\frac{ \Gamma_{\gamma(c+t),\gamma(c)}U\bl \gamma(c+t) \br - U\bl \gamma(c) \br }{t} \right ]\\
    & = \Gamma_{\gamma(c),x} \left [\nabla_{\gamma'(c)}U \right ].
\end{align}
Since for any smooth vector field $W$, $\norm{\Gamma_{y,x}\bl W \br}_x = \norm{W}_y$, the proof is complete. \hfill \qedsymbol\\

\noindent \textbf{Proof of Lemma \ref{lemma:technical_2}:}\\
We apply Lemma \ref{lemma:technical_1} to the vector $V_t(p) \equiv \Log_p y_1(t)$ for fixed $t$. We have that (equation (25) of \textcite{kendall2011limit})
\begin{equation}
    V_t(p) = \operatorname{grad}\bl -\frac{1}{2}d\bl p,y_1(t) \br^2 \br = \operatorname{grad}\bl f_t \br (p).
\end{equation}
Then, by the definition of the Riemannian Hessian $H_t$ of $f_t$, for a smooth curve $\gamma$ on $\M$ at time $c$,
\begin{equation}
    \norm{\nabla_{\gamma'(c)} V_t\bl \gamma(c) \br}_{\gamma(c)} = \norm{\nabla_{\gamma'(c)} \operatorname{grad}\bl f_t \br (p)}_{\gamma(c)} = \norm{H_t\bl \gamma(c) \br \bl \gamma'(c) \br }_{\gamma(c)}.
\end{equation}
Using Lemma \ref{lemma:technical_1}, we choose $x = \mu(t)$ and $y = \hat{\mu}(t)$, and we denote $\gamma_t(s)$ as the minimizing geodesic between $ \mu(t)$ and $\hat{\mu}(t)$:
\begin{align}
    \norm{\Gamma_{\hat{\mu}(t),\mu(t)} \Log_{\hat{\mu(t)}}y_1(t) - \Log_{\mu(t)}y_1(t)} &\leq d\bl \hat{\mu}(t),\mu(t) \br \supp{\norm{H_t\bl \gamma_t(c) \br \bl \gamma_t'(c) \br }_{\gamma_t(c)}}{c \in \left [0, d \bl \hat{\mu}(t), \mu(t) \br \right]}\\
    & \lesssim d\bl \hat{\mu}(t),\mu(t) \br 
\end{align}
for all $t \in \Tsc$ with probability one, where in the last inequality we have used Assumption \ref{item:hessian_assumption_bound} in the Lemma statement along with the fact that, since $\gamma_t(s)$ is a minimizing geodesic, $\norm{\gamma'(s)}_{\gamma(s)} = 1$ for all $s$. Then,
\begin{align}
    \Ex{\norm{\Log_{\hat{\mu}}y_1 \thetagamma \Log_{\mu}y_1}_{\mu}^2 }&=  
    \Ex{\int_{\Tsc} \norm{\Gamma_{\hat{\mu}(t),\mu(t)} \Log_{\hat{\mu(t)}}y_1(t) - \Log_{\mu(t)}y_1(t)}^2 dt}\\
    & \lesssim \Ex{\int_{\Tsc}  d\bl \hat{\mu}(t),\mu(t) \br^2 dt} .\label{eq:exp_int_bound}
\end{align}
By Tonelli's theorem,
\begin{equation}
    \Ex{\int_{\Tsc}  d\bl \hat{\mu}(t),\mu(t) \br^2 dt} = \int_{\Tsc} \Ex{  d\bl \hat{\mu}(t),\mu(t) \br^2} dt.
\end{equation}
Next, we will apply Corollary 4 of \textcite{schotz2019convergence} to bound $\Ex{  d\bl \hat{\mu}(t),\mu(t) \br^2}$. To do so, we need the following definitions related to the metric entropy of geodesic balls in $\M$. Let $B_{\delta}(a) \equiv \left \{ x \in \M : d(x,a) \leq \delta \right \}$ be the ball of radius $r$ centered at $a$ on $\M$, and $N(B,r)\equiv \operatorname{min}\bl k \in \mathbb{N} | \exists q_1, \ldots q_k \in \M : B \subseteq \bigcup_{j=1}^k B_r(q_j) \br$ be the covering number of a set $B$ using radius $r$. Then, the entropy assumption in the statement of Corollary 4 of \textcite{schotz2019convergence}, that there exists $0< \beta <1$ such that, for all $\delta,r>0$, $\log \bl N\bl B_{\delta}\bl\mu(t)\br,r \br \br^{1/2} \lesssim \bl \frac{\delta}{r}\br^\beta$, is satisfied (\textcite{ahidar2020convergence} Example 2.3).

Now additionally using Assumption \ref{item:hadamard_assumption} made in the statement of the Lemma, we can apply Corollary 4 of \textcite{schotz2019convergence} with $\epsilon = 1$ to obtain
\begin{equation}
    \Ex{ d\bl \hat{\mu}(t),\mu(t) \br^2} \lesssim \Ex{d\bl y_1(t),y_2(t) \br^3}^{2/3}\frac{1}{N}
\end{equation}
for all $t \in \Tsc$.

Using Assumption \ref{item:expectation_finite_assumption} made in the statement of the Lemma, we have
\begin{equation} \label{eq:frechet_mean_expectation_bound}
    \Ex{ d\bl \hat{\mu}(t),\mu(t) \br^2} \lesssim \frac{1}{N},
\end{equation}
which we can combine with equation \eqref{eq:exp_int_bound} to establish
\begin{equation}
    \Ex{\norm{\Log_{\hat{\mu}}y_1 \thetagamma \Log_{\mu}y_1}_{\mu}^2 } \lesssim \int_{\Tsc} \frac{1}{N} dt = \frac{\tau}{N},
\end{equation}
completing the proof. \hfill \qedsymbol\\

\noindent \textbf{Proof of Lemma \ref{lemma:conditions_for_assumptions_to_hold}:}\\
Note that $\Ex{\twonorm{Y}^2} = \Ex{Y^{\T}Y} = \Ex{\operatorname{tr}\bl YY^{\T}\br} = \operatorname{tr}\bl \Ex{YY^{\T}}\br = \operatorname{tr}\bl \sigmay \br$. Then, the first two statements of the Lemma follow from observing that $\sigmay$ is diagonal with entries $\omega_j$, the eigenvalues of the covariance operator of $\Logy$, by the results of Lemma \ref{closure_image_facts_thm}.

To show $\Ex{\twonorm{Z}^2} \lesssim \Ex{\twonorm{Y}^2}$, we begin with the triangle inequality:
\begin{equation}
    \Ex{\twonorm{Z}^2} \lesssim \Ex{\twonorm{Z-Y}^2} + \Ex{\twonorm{Y}^2}.
\end{equation}
Using Lemma \ref{lemma:z_minus_y_expectation_bound} and $\Ex{\twonorm{Y}^2} = \sum_{j=1}^d\omega_j$, we observe for $\Ex{\twonorm{Z}^2} \lesssim  \Ex{\twonorm{Y}^2}$ to hold, it is necessary that $\frac{d}{N}\underset{j=1,\ldots d}{\operatorname{max}}\bl \frac{1}{ \omega_j-\omega_{j+1}}\br^2\bl \sum_{j=1}^d\br^{-1} \lesssim 1$. This follows from assumption \eqref{eq:lemma_assumption} provided in the Lemma, because under the other assumptions provided in the Lemma, both $\bl \sum_{j=1}^d\omega_j\br/{\omega_d^2}$ and $\sum_{j=1}^d\omega_j$ are greater than or equal to 1. From this, we also deduce that $\Ex{\twonorm{Z-Y}^2} = o(1)$.

To show that $\twonorm{\sigmaz^{-1}} \lesssim \twonorm{\sigmay^{-1}}$, a more involved argument is required. We again begin with the triangle inequality:
\begin{align}
    \twonorm{\sigmaz^{-1}}^2 &\lesssim \twonorm{\sigmaz^{-1} - \sigmay^{-1}}^2 + \twonorm{\sigmay^{-1}}^2\\
    & \leq  \twonorm{\sigmaz^{-1}}^2\twonorm{\sigmaz - \sigmay}^2  \twonorm{\sigmay^{-1}}^2 + \twonorm{\sigmay^{-1}}^2.
\end{align}
This implies
\begin{align}
    \twonorm{\sigmaz^{-1}}^2 - \twonorm{\sigmaz^{-1}}^2\twonorm{\sigmaz - \sigmay}^2  \twonorm{\sigmay^{-1}}^2 \lesssim \twonorm{\sigmay^{-1}}^2,
\end{align}
so that
\begin{align}
    \twonorm{\sigmaz^{-1}}^2\bl 1 -\twonorm{\sigmaz - \sigmay}^2  \twonorm{\sigmay^{-1}}^2\br \leq \twonorm{\sigmay^{-1}}^2,
\end{align}
and therefore, if $\twonorm{\sigmaz - \sigmay}^2  \twonorm{\sigmay^{-1}}^2 = o(1)$ held, then we would have $\twonorm{\sigmaz^{-1}} \lesssim \twonorm{\sigmay^{-1}}$ and the proof would be complete. Thus, to show $\twonorm{\sigmaz^{-1}} \lesssim \twonorm{\sigmay^{-1}}$, it suffices to show that $\frac{d}{N}\frac{\sum_{j=1}^d\omega_j}{\omega_d^2}\underset{j=1,\ldots d}{\operatorname{max}}\bl \frac{1}{ \omega_j-\omega_{j+1}}\br^2 = o(1)$ implies $\twonorm{\sigmaz - \sigmay}^2  \twonorm{\sigmay^{-1}}^2 = o(1)$. From the second item of Lemma \ref{lemma:minus_square_root_difference_ZY} and from $\Ex{\twonorm{Z}^2} \lesssim \Ex{\twonorm{Y}^2}$ which has already been shown, we have
\begin{align}
    \twonorm{\sigmaz - \sigmay}^2 &\leq \operatorname{max}\bl \E \left [ \twonorm{Z}^2 \right ],\E \left [ \twonorm{Y}^2 \right] \br \E \left [\twonorm{Z-Y}^2 \right]\\
    & \lesssim \E \left [\twonorm{Z-Y}^2 \right]\sum_{j=1}^d\omega_j.
\end{align}
Thus, for $\twonorm{\sigmaz - \sigmay}^2  \twonorm{\sigmay^{-1}}^2 = o(1)$ to hold, it suffices to show that $\frac{1}{\omega_d^2}\E \left [\twonorm{Z-Y}^2 \right]\sum_{j=1}^d\omega_j = o(1)$. From Lemma \ref{lemma:z_minus_y_expectation_bound}, we have made exactly the assumption so that $\frac{1}{\omega_d^2}\E \left [\twonorm{Z-Y}^2 \right]\sum_{j=1}^d\omega_j = o(1)$ holds, completing the proof that $\twonorm{\sigmaz^{-1}} \lesssim \twonorm{\sigmay^{-1}}$.

To show the final statement of the Lemma, that $\twonorm{\sigmaz} \lesssim \twonorm{\sigmay}$, we can make an argument similar to the one used to show $\twonorm{\sigmaz^{-1}} \lesssim \twonorm{\sigmay^{-1}}$. From this, it suffices to show that $\frac{d}{N}\frac{\sum_{j=1}^d\omega_j}{\omega_1^2}\underset{j=1,\ldots d}{\operatorname{max}}\bl \frac{1}{ \omega_j-\omega_{j+1}}\br^2 = o(1)$. Since $\omega_d \leq \omega_1$, $\frac{d}{N}\frac{\sum_{j=1}^d\omega_j}{\omega_1^2}\underset{j=1,\ldots d}{\operatorname{max}}\bl \frac{1}{ \omega_j-\omega_{j+1}}\br^2 = o(1)$ holds under our assumptions and the proof is complete. \hfill \qedsymbol\\

\noindent \textbf{Proof of Lemma \ref{lemma:gamma_and_gamma_tilde_equivalent}:}\\
To bound $|\gamma_k - \tilde{\gamma}_k|$, we can use Weyl's inequality (\textcite{bhatia2013matrix} Corollary III.2.6.), because $\gamma_k$ and $\tilde{\gamma}_k$ are the $k$ eigenvalues of the matrices $|C|$ and $|A|$ respectively \parencite{uurtio2018tutorial}:
\begin{equation}
|\gamma_k - \tilde{\gamma}_k| \leq \twonorm{\left | A \right | - \left | C \right |} \leq \Fnorm{\left | A \right | - \left | C \right |}\lesssim \Fnorm{A-C},
\end{equation}
where in the second inequality we have used identity \ref{item:bhatia_VII_39} from Section \ref{subsec:identities}. Combining Lemmas
\ref{lemma:z_minus_y_expectation_bound}, \ref{lemma:conditions_for_assumptions_to_hold}, and \ref{lemma:A_minus_C_bound}, we observe that the stated assumption in the Lemma implies $|\gamma_k - \tilde{\gamma}_k| \lesssim \min \{\gamma_k,\tilde{\gamma}_k\}$, establishing the stated conclusion using the assumption that $\gamma_k$ and $\tilde{\gamma}_k$ are bounded from below. That $\gamma_k^2$ and $\tilde{\gamma}_k^2$ are asymptotically equivalent follows from the same argument, in addition to the function $f(x) = x^2$ being Lipschitz continuous on the interval $[0,1]$.

\section{Asymmetric Sparse-Functional CCA: Proof of Theorem \ref{thm:consistency_of_scores_main}} \label{appendix:consistency_of_scores}
In this section we prove Theorem \ref{thm:consistency_of_scores_main}, which shows that our sample estimates of the canonical variables, derived from our proposed estimates of the canonical vectors, grow asymptotically close to the population canonical variable solutions of the infinite-dimensional CCA problem in Theorem \ref{theorem_problem}. We denote by $\bl U_k,V_k \br$ the $k$th canonical variable solution pair to the infinite-dimensional problem in Theorem \ref{theorem_problem}. We denote by $\curl \gamma_k^*\curr_{k=1}^p$ the canonical correlations that are attained by these solution pairs. We denote with $K = \operatorname{max} \left \{i \in \{ 1, \ldots p\}: \gamma_i^* > 0 \right \}$ the number of nontrivial canonical vectors in this problem, and to simplify the notation, we use the convention ${\gamma_0^*}^2=\infty$ and ${\gamma_{K+1}^*}^2=-\infty$. We often refer to canonical variables as canonical scores or simply scores, which are distinct from the principal scores derived from the principal component expansion of $\chi_1$. The difference will be clear from context, but in general, `scores' with no preceding modifier refers to the canonical scores.

Since we are showing that two random variables are close, we need to specify the probability spaces and random variables of interest. In this section we suppose that we observe realizations of $\bl X_i,y_i \br_{i=1}^N$ which is used in the Sparse-Functional Asymmetric CCA (Algorithm \ref{alg:asymmetric_sparse_fun_cca}). We assume these are i.i.d samples that all live on the probability space $\bl \Omega,\calF,\mathbb{P} \br$. $d$ denotes the number of principal components we choose to approximate $\Logy$ by. We obtain estimates of $\hat \mu$, $\left \{ \hat \phi_j \right \}_{j=1}^d$, $\hat B$, $\curl \hat \omega_j \curr_{j=1}^d$, $ \curl \hat \gamma_j \curr_{j=1}^d$, $\sigmayhat, \sigmaxhat$, $\hat H = \left[\hat \eta_1, \ldots, \hat \eta_d \right]$, $\hat{T} = \left[ \hat \theta_1, \ldots, \hat \theta_d\right]$, and $\hat{\psi}_k = \sum_{j=1}^d \hat{\eta}_{kj} \hat{\phi}_j$ for $k = 1, \ldots d$.

We will define the scores from the sample procedure as follows. Observing a new and independent data point $\bl \xtest,\ytest \br$ from the same distribution as the sample, which also lives on $\bl \Omega,\calF,\mathbb{P} \br$, we define
\begin{equation}
    \bl \hat U_k, \hat V_k \br \equiv \bl \innerdouble{\Log_{\hat \mu}\ytest}{\hat \psi_k}_{\hat \mu}, \hat \theta_k^{\T}\xtest\br,
\end{equation}
which are the canonical scores we would obtain from the new data point and our sample canonical vector estimates.
To be precise, we denote by $\bl U_k,V_k \br$ the $k$th canonical variable solution pair to the infinite-dimensional problem in Theorem \ref{theorem_problem}, between $\chi_1 = \Log_{\mu}\ytest$ and $\chi_2 = \xtest$, for $k=1 ,\ldots K$. We would like to show that $\hat U_k$ becomes close to $U_k$ and $\hat V_k$ becomes close to $V_k$ in an asymptotic sense.

Throughout the proof, we make the same assumptions in Section \ref{sec:final_rates_sparse_functional}, with the important exception that, we no longer make a finite-dimensional correlation structure Assumption \ref{assumption_cor} between $X$ and $y$. Thus, we allow for all principal scores of the functional data $y$ to potentially be correlated with the components of $X$.

The main idea of the proof of the bounds is similar to that of the canonical vectors, and indeed we rely on the results of the previous supplementary sections. We define several intermediate scores in order to show that the sample score and the infinite-dimensional score are close together.

The first intermediate scores come from the population $d$-dimensional canonical correlation problem described in Theorem \ref{theorem_problem_finite} and equation \eqref{eq:finite_dim_score_problem}, which we denote as $\bl U_k^{(d)},V_k^{(d)} \br$. By Theorem \ref{theorem_problem_finite}, these can equivalently be derived from the canonical vector solutions given by the following CCA problem:
\begin{align} \label{eq:canonical_pair_test}
        (\eta_1,\theta_1) &= \argmax{\operatorname{Corr}^2\bl \eta^{\T}\Ytest,\theta^{\T}\xtest \br}{\eta \in \rone^d, \theta \in \rone^p, \Var(\eta^{\T}\Ytest) = \Var(\theta^{\T}\xtest) = 1},\\
        (\eta_k,\theta_k) &= \argmax{\operatorname{Corr}^2\bl \eta^{\T}\Ytest,\theta^{\T}\xtest \br}{\substack{\eta \in \rone^d, \theta \in \rone^p, \Var(\eta^{\T}\Ytest) = \Var(\theta^{\T}\xtest) = 1\\ \Cov \left ( \eta^{\T}\Ytest, \eta_i^{\T}\Ytest\right)=0, i=1, \ldots, k \\
    \Cov \left ( \theta^{\T}\xtest, \theta_i^{\T}\xtest\right)=0, i=1, \ldots, k}}, \qquad k = 2, \ldots K
    \end{align}
Here, $\Ytest$ is the $d$-dimensional random vector such that $\bl \Ytest \br_j= \innerdouble{\Log_{\mu}\ytest}{\phi_j}_{\mu}$, the $j$th principal score associated with the population principal component $\phi_j$ of $\Log_{\mu}\ytest$, for $j = 1, \ldots d$. By definition, $\bl U_k^{(d)},V_k^{(d)} \br = \bl \eta_k^{\T}\Ytest,\theta_k^{\T}\xtest \br$. We refer to $\curl \gamma_k\curr_{k=1}^K$ as the canonical correlations attained in this problem.

The second intermediate scores are derived from the population canonical correlation problem derived from the `estimates' of $Y$ that we make when using $\hat \mu$ and $\hat \phi_j$:
\begin{align}
        (a_1,b_1) &= \argmax{\operatorname{Corr}^2\bl a^{\T}Z,b^{\T}X \br}{a \in \rone^d, b \in \rone^p, \Var(a^{\T}Z) = \Var(b^{\T}X) = 1}, \label{eq:CCA_ztest_1}\\
        \label{eq:CCA_ztest_k}(a_k,b_k) &= \argmax{\operatorname{Corr}^2\bl a^{\T}Z,b^{\T}X \br}{\substack{a \in \rone^d, b \in \rone^p, \Var(a^{\T}Z) = \Var(b^{\T}X) = 1\\ \Cov \left ( a^{\T}Z, a_i^{\T}Z\right)=0, i=1, \ldots, k \\
    \Cov \left ( b^{\T}X, b_i^{\T}X\right)=0, i=1, \ldots, k}}, \qquad k = 2, \ldots K.
    \end{align}
Here, $Z$ is defined as
\begin{equation}
    \bl Z\br_j \equiv \innerdouble{\Log_{\hat \mu} y_1}{\hat \phi_j}_{\hat \mu}.
\end{equation}
We let $\tilde{\gamma}_1^2 \ldots \tilde{\gamma}_K^2$ denote the squared canonical correlations attained by the pairs $\bl a_1, b_1\br, \ldots \bl a_K, b_K\br$. We also define the random variable $\ztest \in \rone^d$ so that
\begin{equation}
    \bl \ztest\br_j \equiv \innerdouble{\Log_{\hat \mu} \ytest}{\hat \phi_j}_{\hat \mu}.
\end{equation}
Note that the distribution of $\ztest$ is not equal to that of $Z$, since $\ytest$ is independent of $\hat \mu$ and the $\hat \phi_j$, while $y_1$ is not.

We define the secondary intermediate scores as $ U_k^{Z,1} \equiv a_k^{\T}\Ytest$, $U_k^{Z,2} \equiv a_k^{\T}\ztest$, and $V_k^{Z}\equiv b_k^{\T}\xtest$. The additional score for $U_k$ arises due to the complexity of estimating $\hat \mu$ and $\hat \phi_j$, whereas the proof for the bound on $V_k$ is slightly simpler.

Following \textcite{hsing2015theoretical} Theorem 10.2.3, our goal is to show a probabilistic bound on $\E \left [ \bl U_k - \hat U_k\br^2 \left. \right \vert \bl X_i,y_i\br_{i=1}^N \right ]$ and $\E \left [ \bl V_k - \hat V_k\br^2 \left. \right \vert \bl X_i,y_i\br_{i=1}^N \right ]$ as the sample size $N$, the number of principal components we select $d$, and the dimension of the high-dimensional data $p$ go to infinite. We choose this notion of error, where we condition on the sample, because we can derive a result which is comparable to our results for the canonical vectors while also integrating out the randomness of $\bl\xtest,\ytest\br$.

We begin with the proof of the bound for $V_k$ since it is slightly simpler than that of $U_k$. For convenience throughout the proofs, we drop the $k$ in the notation of the scores and canonical vectors. For example, we write $U_k,U_k^{Z,1},\eta_k$ as $U,U^{Z,1},\eta$.
\subsection{Proof of bound for high-dimensional score}
The proof strategy is to decompose the quantity of interest $\bl V - \hat V\br^2$ into three parts:
\begin{equation}
    \bl V - \hat V\br^2 \lesssim {\bl V - V^{(d)} \br^2} + {\bl V^{(d)} - V^Z \br^2} + {\bl V^Z - \hat V \br^2}
\end{equation}
so that
\begin{align}
    \E \left [ \bl V - \hat V\br^2 \left. \right \vert \bl X_i,y_i\br_{i=1}^N \right ] & \lesssim \E \left [ {\bl V - V^{(d)} \br^2} + {\bl V^{(d)} - V^Z \br^2} + {\bl V^Z - \hat V \br^2} \left. \right \vert \bl X_i,y_i\br_{i=1}^N \right ] \\
    & \equiv \text{Term I } + \text{Term II } + \text{Term III}
\end{align}
\textit{Term I:}
\begin{equation}
    \text{Term I} = \E \left [ {\bl V - V^{(d)} \br^2} \left. \right \vert \bl X_i,y_i\br_{i=1}^N \right ] = {\bl V - V^{(d)} \br^2}.
\end{equation}
since $V$ and $V^{(d)}$ are independent of the sample. Then, Theorem \ref{thm:finite_infinite_score} gives the probabilistic bound
\begin{equation}
    \text{Term I} = O_P\bl\frac{{\gamma_1^*}^2 \norm{\mathscr{C}_{12} - \mathscr{C}_{12}^{(d)}}^2}{\min_{j \neq k} \left |{\gamma_k^*}^2 - {\gamma_j^*}^2 \right |} \br,
\end{equation}
since bounds in expectation imply probabilistic bounds. See Section \ref{sec:finite_infinite_CCA_theory} for definitions of the operator $\mathscr{C}_{12}$ and its principal component approximation $\mathscr{C}_{12}^{(d)}$.\\
\textit{Term II:}
\begin{equation}
    \text{Term II} = \E \left [ \bl V^{(d)} - V^Z \br^2 \left. \right \vert \bl X_i,y_i\br_{i=1}^N \right ].
\end{equation}
Let the support $S \subseteq \{ 1, \ldots p \}$ represent the indices of non-zero elements of $\theta$ or $b$, with cardinality $|S| \leq 2s$. Denote $\xtests \in \rone^{|S|}$ as the random vector consisting of only the entries $\{\bl \xtest \br_j:j \in S\}$. Similarly, we define $\theta_{S}$ and $b_{S}$ from $\theta$ and $b$ respectively, as well as $\Sigma_{\xtest,S}$ from $\Sigma_{\xtest}$, following Section \ref{appendix:proof_of_canonical_function_section}.

By definition,
\begin{align}
    V^{(d)} - V^Z &= \theta^{\T}\xtest - b^{\T}\xtest\\
    &= \theta_S^{\T}\xtests - b^{\T}_S\xtests\\
    &= \bl \theta_S - b_S \br^{\T}\xtests.
\end{align}
Therefore,
\begin{align}
    \text{Term II} &= \E \left [ \bl \theta_S - b_S \br^{\T}\xtests\xtests^{\T} \bl \theta_S - b_S \br  \left. \right \vert \bl X_i,y_i\br_{i=1}^N \right ]\\ &= \bl \theta_S - b_S \br^{\T} \Sigma_{X,S} \bl \theta_S - b_S \br\\
    &  = \twonorm{\Sigma_{X,S}^{1/2}\bl \theta_S - b_S \br}^2.
\end{align}
We can bound this by appealing to equation \eqref{eq:b_tilde_theta_tilde_term} and Lemma \ref{lemma:theta_minus_b_bound}, so that we can remove the factor of $\twonorm{\Sigma_{X,S}^{-1/2}}^2$. We obtain
\begin{equation}
    \text{Term II} = O_P\bl \frac{ s \twonorm{\sigmay^{-1}}^3 \Ex{\twonorm{Y}^2}^2}{{\operatorname{min}\bl \gamma_{k-1}-\gamma_k,\gamma_k - \gamma_{k+1}\br^2}} \frac{\tau d}{N}\Ex{\munorm{\Log_{\mu}y_1}^4}^{3/2} \underset{j=1,\ldots d}{\operatorname{max}}\bl \frac{1}{ \omega_j-\omega_{j+1}}\br^2 \br.
\end{equation}
\textit{Term III:}\\
\begin{equation}
    \text{Term III} = \E \left [ {\bl V^Z - \hat V \br^2}  \left. \right \vert \bl X_i,y_i\br_{i=1}^N \right ].
\end{equation}
By definition,
\begin{align}
    V^Z - \hat V &= b^{\T}\xtest - \hat \theta^{\T}\xtest\\
    &= \bl b-\hat \theta \br^{\T}\xtest.
\end{align}
Similarly to Term II, we obtain
\begin{equation}
    \text{Term III} = \twonorm{\sigmax^{1/2}\bl \hat \theta - b \br }^2.
\end{equation}
We can appeal to equation \eqref{eq:thm_C_2_theta_bound} following the same argument used to bound $\theta-\hat \theta$ in the proof of Theorem \ref{thm:multivariate_canonical_vector_bound_probabilistic_fast}, but where we place a $\sigmax^{1/2}$ in front of the relevant terms throughout to obtain
\begin{equation}
    \twonorm{\sigmax^{1/2}\bl \hat \theta - b \br} \lesssim \bl \frac{\twonorm{\sigmax^{1/2}\theta}}{\tilde\gamma^2} + \frac{\twonorm{\sigmax ^{1/2}\tilde B}}{\tilde\gamma {\operatorname{min}\bl\tilde\gamma_{k-1}^2-\tilde\gamma_k^2,\tilde\gamma_k^2-\tilde\gamma_{k+1}^2\br} } \br \twonorm{\tilde B^{\T}\sigmax \tilde B - \bhat^{\T}\sigmaxhat\bhat} + \frac{\twonorm{\sigmax^{1/2}\bl \tilde B-\bhat\br}}{\tilde\gamma}.
\end{equation}
Here, $\tilde B = \sigmax^{-1}\Sigma_{XZ}\sigmaz^{-1/2}$. Noting that 
\begin{equation}\sigmax^{1/2}\tilde B = \sigmax^{-1/2}\Sigma_{XZ}\Sigma_Z^{-1/2}
\end{equation}
so $\twonorm{\sigmax^{1/2}\tilde B} = \tilde{\gamma}_1$ and $\twonorm{\sigmax^{1/2}\theta} = 1$, we have
\begin{equation}
    \text{Term III}\lesssim  \frac{1}{ {\operatorname{min}\bl\tilde\gamma_{k-1}^2-\tilde\gamma_k^2,\tilde\gamma_k^2-\tilde\gamma_{k+1}^2\br^2} } \twonorm{\tilde B^{\T}\sigmax \tilde B - \bhat^{\T}\sigmaxhat\bhat}^2 + {\tilde\gamma^{-2}}\twonorm{\sigmax^{1/2}\bl \tilde B-\bhat\br}^2.
\end{equation}
Now we bound the two relevant terms on the right-hand-side. By equation \eqref{eq:sigmax_one_half_B_bound}, Lemma \ref{lemma:other_terms}, Remark \ref{remark:lambda_0_prob_bound_subgauss_bound}, and Lemma \ref{lemma:lambda_0_final_bound}, we have
\begin{equation}
    \Fnorm{\sigmax^{1/2}(\bhat-\tilde B)}^2 = O_P \bl \kappa_Xs \twoinfnorm{\sigmax} \frac{d\log p}{N} \br.
\end{equation}
By Corollary \ref{cor:simplified_operator_norm_bound},
\begin{equation}
     \twonorm{\tilde B^{\T}\sigmax \tilde B - \bhat^{\T}\sigmaxhat\bhat}^2 = O_P\bl \tilde \gamma_1^2 \twoinfnorm{\sigmax}s\kappa  \frac{d\log p}{N}  \br.
\end{equation}
Putting the last three equations together,
\begin{align}
    \text{Term III} &= O_P\bl\frac{\tilde \gamma_1^2 \twoinfnorm{\sigmax}s\kappa  }{ {\operatorname{min}\bl\tilde\gamma_{k-1}^2-\tilde\gamma_k^2,\tilde\gamma_k^2-\tilde\gamma_{k+1}^2\br^2} }\frac{d\log p}{N}  + \tilde \gamma^{-2} \kappa_Xs \twoinfnorm{\sigmax} \frac{d\log p}{N}\br\\
    & = O_P\bl \frac{\twoinfnorm{\sigmax}s\kappa  }{ {\operatorname{min}\bl\tilde\gamma_{k-1}^2-\tilde\gamma_k^2,\tilde\gamma_k^2-\tilde\gamma_{k+1}^2\br^2} } \frac{d\log p}{N} \br.
\end{align}
\textit{Final bound for} $V_k$:\\
Now we can combine our bounds for terms I, II, and III to obtain
\begin{align}
    \E \left [ \bl V - \hat V\br^2 \left. \right \vert \bl X_i,y_i\br_{i=1}^N \right ] &=O_P\bl\frac{{\gamma_1^*}^2 \norm{\mathscr{C}_{12} - \mathscr{C}_{12}^{(d)}}^2}{\min_{j \neq k} \left |{\gamma_k^*}^2 - {\gamma_j^*}^2 \right |} \br \\
    &+ O_P\bl \frac{ s \twonorm{\sigmay^{-1}}^3 \Ex{\twonorm{Y}^2}^2}{{\operatorname{min}\bl \gamma_{k-1}-\gamma_k,\gamma_k - \gamma_{k+1}\br^2}} \frac{\tau d}{N}\Ex{\munorm{\Log_{\mu}y_1}^4}^{3/2} \underset{j=1,\ldots d}{\operatorname{max}}\bl \frac{1}{ \omega_j-\omega_{j+1}}\br^2 \br\\
    & +  O_P\bl \frac{\twoinfnorm{\sigmax}s\kappa  }{ {\operatorname{min}\bl\tilde\gamma_{k-1}^2-\tilde\gamma_k^2,\tilde\gamma_k^2-\tilde\gamma_{k+1}^2\br^2} } \frac{d\log p}{N} \br.
\end{align}
Simplifying this, we obtain
\begin{align}
    \E \left [ \bl V_k - \hat V_k\br^2 \left. \right \vert \bl X_i,y_i\br_{i=1}^N \right ] &=O_P\bl\frac{{\gamma_1^*}^2 \norm{\mathscr{C}_{12} - \mathscr{C}_{12}^{(d)}}^2}{\min_{j \neq k} \left |{\gamma_k^*}^2 - {\gamma_j^*}^2 \right |} \br \\
    &+ O_P\bl \frac{ \twoinfnorm{\sigmax} \tau s\kappa \omega_d^{-3} \bl \sum_{j=1}^d\omega_j\br^2\Ex{\munorm{\Log_{\mu}y_1}^4}^{3/2}}{\min_{j \neq k} \min \curl \left |{\gamma_k}^2 - {\gamma_j}^2 \right |, \left|{\gamma_k} - {\gamma_j} \right | \curr^2} \frac{ d\log p}{N} \underset{j=1,\ldots d}{\operatorname{max}}\bl \frac{1}{ \omega_j-\omega_{j+1}}\br^2 \br.
\end{align}
\subsection{Proof of bound for Riemannian-functional score}
The proof strategy is to decompose the quantity of interest $\bl U - \hat U\br^2$ into four parts:
\begin{equation}
    \bl U - \hat U\br^2 \lesssim {\bl U - U^{(d)} \br^2} + {\bl U^{(d)} - U^{Z,1} \br^2} +{\bl U^{Z,1} - U^{Z,2} \br^2} + {\bl U^{Z,2} - \hat U \br^2}
\end{equation}
so that
\begin{align}
    \E \left [ \bl U - \hat U\br^2 \left. \right \vert \bl X_i,y_i\br_{i=1}^N \right ] & \lesssim \E \left [ {\bl U - U^{(d)} \br^2} + {\bl U^{(d)} - U^{Z,1} \br^2} +{\bl U^{Z,1} - U^{Z,2} \br^2} + {\bl U^{Z,2} - \hat U \br^2} \left. \right \vert \bl X_i,y_i\br_{i=1}^N \right ] \\
    & \equiv \text{Term I } + \text{Term II } + \text{Term III} + \text{Term IV}
\end{align}
\textit{Term I:}\\
Following the same argument used in the proof for the bound on $V$, Theorem \ref{thm:finite_infinite_score} gives the probabilistic bound
\begin{equation}
    \text{Term I} = O_P\bl\frac{{\gamma_1^*}^2 \norm{\mathscr{C}_{12} - \mathscr{C}_{12}^{(d)}}^2}{\min_{j \neq k} \left |{\gamma_k^*}^2 - {\gamma_j^*}^2 \right |} \br.
\end{equation}
\textit{Term II:}\\
\begin{equation}
    \text{Term II} = \E \left [ \bl U^{(d)} - U^{Z,1} \br^2 \left. \right \vert \bl X_i,y_i\br_{i=1}^N \right ].
\end{equation}
By definition,
\begin{align}
    U^{(d)} - U^Z &= \eta^{\T}\Ytest - a^{\T}\Ytest\\
    &= \bl \eta - a \br^{\T}\Ytest,
\end{align}
so that
\begin{align}
    \text{Term II} &=  \E \left [ \bl \eta - a \br^{\T}\Ytest\Ytest^{\T}\bl \eta - a \br \left. \right \vert \bl X_i,y_i\br_{i=1}^N \right ] \\
    &= \twonorm{\sigmay^{1/2}\bl \eta-a\br}^2.\\
    &= \twonorm{ \tilde\eta-\sigmay^{1/2}a}^2\\
    &= \twonorm{\tilde \eta - \sigmay^{1/2}\sigmaz^{-1/2}\tilde a}^2 \\
    &\lesssim  \twonorm{\tilde \eta - \tilde a}^2 + \twonorm{\bl I - \sigmay^{1/2}\sigmaz^{-1/2} \br\tilde a}^2
\end{align}
The first term on the right hand side shares the same bound as Term II in the proof for the bound for $V$, using Lemma \ref{lemma:theta_minus_b_bound} and equation \eqref{eq:a_minus_eta_tilde_bound}:
\begin{equation}
    \twonorm{\tilde \eta - \tilde a}^2 = O_P\bl \frac{ s \twonorm{\sigmay^{-1}}^3 \Ex{\twonorm{Y}^2}^2}{{\operatorname{min}\bl \gamma_{k-1}-\gamma_k,\gamma_k - \gamma_{k+1}\br^2}} \frac{\tau d}{N}\Ex{\munorm{\Log_{\mu}y_1}^4}^{3/2} \underset{j=1,\ldots d}{\operatorname{max}}\bl \frac{1}{ \omega_j-\omega_{j+1}}\br^2 \br.
\end{equation}
To bound $\twonorm{\bl I - \sigmay^{1/2}\sigmaz^{-1/2} \br\tilde a}^2$, we use $\twonorm{\tilde a} = 1$, Lemma \ref{lemma:minus_square_root_difference_ZY}, and
\begin{align}
    \twonorm{\bl I - \sigmay^{1/2}\sigmaz^{-1/2} \br\tilde a}^2  & \leq \twonorm{\sigmay^{1/2}\bl \sigmay^{-1/2} - \sigmaz^{-1/2}\br}^2\\
    &\lesssim \twonorm{\sigmay}\E \left [\twonorm{Z-Y}^2 \right] \twonorm{\sigmay^{-1}}^3 \E \left [ \twonorm{Y}^2 \right].
\end{align}
By Lemma \ref{lemma:z_minus_y_expectation_bound}, we then have
\begin{equation}
    \twonorm{\bl I - \sigmay^{1/2}\sigmaz^{-1/2} \br\tilde a}^2 \lesssim \twonorm{\sigmay} \twonorm{\sigmay^{-1}}^3 \E \left [ \twonorm{Y}^2 \right]\frac{\tau d}{N}\Ex{\munorm{\Log_{\mu}y_1}^4}^{3/2} \underset{j=1,\ldots d}{\operatorname{max}}\bl \frac{1}{ \omega_j-\omega_{j+1}}\br^2 .
\end{equation}
Combining these bounds and simplifying, we finally obtain
\begin{align}
    \text{Term II}& = O_P \bl \frac{\omega_1 \omega_d^{-3}\bl \sum_{j=1}^d\omega_j\br^2 }{{\operatorname{min}\bl \gamma_{k-1}-\gamma_k,\gamma_k - \gamma_{k+1}\br^2}} \frac{\tau sd}{N}\Ex{\munorm{\Log_{\mu}y_1}^4}^{3/2} \underset{j=1,\ldots d}{\operatorname{max}}\bl \frac{1}{ \omega_j-\omega_{j+1}}\br^2\br.
\end{align}
\textit{Term III:}\\
\begin{equation}
    \text{Term III} = \E \left [ \bl U^{Z,1} - U^{Z,2} \br^2 \left. \right \vert \bl X_i,y_i\br_{i=1}^N \right ].
\end{equation}
By definition,
\begin{align}
    U^{Z,1} - U^{Z,2} &= a^{\T}\Ytest - a^{\T}\ztest\\
    & = a^{\T}\bl \Ytest - \ztest\br,
\end{align}
so that by the Cauchy-Schwarz inequality and Lemma \ref{lemma:conditions_for_assumptions_to_hold},
\begin{align}
    \bl U^{Z,1} - U^{Z,2}\br^2 & \leq \twonorm{a}^2\twonorm{ \Ytest - \ztest}^2\\
    &\leq \twonorm{\sigmaz^{-1/2}}^2\twonorm{ \Ytest - \ztest}^2\\
    & \lesssim \twonorm{\sigmay^{-1}}\twonorm{ \Ytest - \ztest}^2.
\end{align}
Despite the fact that here we have $\ztest$, we can appeal to Lemma \ref{lemma:z_minus_y_expectation_bound} to bound $\twonorm{ \Ytest - \ztest}^2$, since replacing $y_1$ with $\ytest$ does not change the proof. Using Lemma \ref{lemma:z_minus_y_expectation_bound} and the law of total expectation we obtain
\begin{align}
    \text{Term III} &\lesssim \twonorm{\sigmay^{-1}}\E \left [ \twonorm{ \Ytest - \ztest}^2 \left. \right \vert \bl X_i,y_i\br_{i=1}^N \right ]\\
    &= O_P\bl \omega_d^{-1}\frac{\tau d}{N}\Ex{\munorm{\Log_{\mu}y_1}^4}^{3/2} \underset{j=1,\ldots d}{\operatorname{max}}\bl \frac{1}{ \omega_j-\omega_{j+1}}\br^2 \br.
\end{align}
\textit{Term IV:}\\
\begin{equation}
    \text{Term IV} = \E \left [ \bl U^{Z,2} - \hat U \br^2 \left. \right \vert \bl X_i,y_i\br_{i=1}^N \right ].
\end{equation}
By definition,
\begin{align}
     U^{Z,2} - \hat U &= a^{\T}\ztest - \hat\eta^{\T}\ztest\\
     & = \bl a - \hat \eta \br^{\T} \ztest,
\end{align}
so that similarly to Term III we obtain
\begin{equation}
    \text{Term IV} \lesssim  \twonorm{a - \hat \eta}^2\E \left [ \twonorm{\ztest}^2\left. \right \vert \bl X_i,y_i\br_{i=1}^N \right ].
\end{equation}
While $\Ex{\twonorm{\ztest}^2} \neq \Ex{\twonorm{Z}^2} $, we can use the same argument we used to show $\Ex{\twonorm{Z}^2} \lesssim \Ex{\twonorm{Y}^2} $ to show that $\Ex{\twonorm{\ztest}^2} \lesssim \Ex{\twonorm{Y}^2}  = \sum_{j=1}^d\omega_j$ (Lemma \ref{lemma:conditions_for_assumptions_to_hold}). Then by the law of total expectation and Theorem \ref{thm:multivariate_canonical_vector_bound_probabilistic_fast} applied to $Z$ and $X$, we have
\begin{align}
    \text{Term IV} & = O_P\bl  \frac{d\log (p)}{N}\twonorm{\sigmaz^{-1}}\operatorname{max}\left \{ \frac{\tilde \gamma_1^2\twoinfnorm{\sigmax} s\kappa}{{\operatorname{min}\bl\tilde\gamma_{k-1}^2-\tilde\gamma_k^2,\tilde\gamma_k^2-\tilde\gamma_{k+1}^2\br^2}}, \twonorm{\sigmaz}\twonorm{\sigmaz^{-1}} \right \} \br \cdot O_P\bl \sum_{j=1}^d\omega_j\br\\
    & = O_P \bl \frac{d\log (p)}{N}\bl \sum_{j=1}^d\omega_j\br\omega_1\omega_d^{-2} \frac{\gamma_1^2\twoinfnorm{\sigmax} s\kappa}{{\operatorname{min}\bl\gamma_{k-1}^2-\gamma_k^2,\gamma_k^2-\gamma_{k+1}^2\br^2}}  \br.
\end{align}
\textit{Final bound for} $U_k$:\\
Now we can combine our bounds for terms I, II, III and IV to obtain
\begin{align}
    \E \left [ \bl U - \hat U\br^2 \left. \right \vert \bl X_i,y_i\br_{i=1}^N \right ] &=O_P\bl\frac{{\gamma_1^*}^2 \norm{\mathscr{C}_{12} - \mathscr{C}_{12}^{(d)}}^2}{\min_{j \neq k} \left |{\gamma_k^*}^2 - {\gamma_j^*}^2 \right |} \br  \\
    & + O_P \bl \frac{\omega_1 \omega_d^{-3}\bl \sum_{j=1}^d\omega_j\br^2 }{{\operatorname{min}\bl \gamma_{k-1}-\gamma_k,\gamma_k - \gamma_{k+1}\br^2}} \frac{\tau sd}{N}\Ex{\munorm{\Log_{\mu}y_1}^4}^{3/2} \underset{j=1,\ldots d}{\operatorname{max}}\bl \frac{1}{ \omega_j-\omega_{j+1}}\br^2\br\\
    & + O_P\bl \omega_d^{-1}\frac{\tau d}{N}\Ex{\munorm{\Log_{\mu}y_1}^4}^{3/2} \underset{j=1,\ldots d}{\operatorname{max}}\bl \frac{1}{ \omega_j-\omega_{j+1}}\br^2 \br\\
    & + O_P \bl \frac{d\log (p)}{N}\bl \sum_{j=1}^d\omega_j\br\omega_1\omega_d^{-2} \frac{\gamma_1^2\twoinfnorm{\sigmax} s\kappa}{{\operatorname{min}\bl\gamma_{k-1}^2-\gamma_k^2,\gamma_k^2-\gamma_{k+1}^2\br^2}}  \br
\end{align}
Simplifying this gives
\begin{align}
    \E \left [ \bl U_k - \hat U_k\br^2 \left. \right \vert \bl X_i,y_i\br_{i=1}^N \right ] &=O_P\bl\frac{{\gamma_1^*}^2 \norm{\mathscr{C}_{12} - \mathscr{C}_{12}^{(d)}}^2}{\min_{j \neq k} \left |{\gamma_k^*}^2 - {\gamma_j^*}^2 \right |} \br \\
    &+ O_P\bl \frac{ \twoinfnorm{\sigmax} \tau s\kappa \omega_1 \omega_d^{-3} \bl \sum_{j=1}^d\omega_j\br^2\Ex{\munorm{\Log_{\mu}y_1}^4}^{3/2}}{\min_{j \neq k} \min \curl \left |{\gamma_k}^2 - {\gamma_j}^2 \right |, \left|{\gamma_k} - {\gamma_j} \right | \curr^2} \br\\
    &\cdot O_P\bl \frac{ d\log p}{N} \underset{j=1,\ldots d}{\operatorname{max}}\bl \frac{1}{ \omega_j-\omega_{j+1}}\br^2 \br.
\end{align}
We note that the second term on the right-hand side is nearly identical to the bound on $\E \left [ \bl V - \hat V\br^2 \left. \right \vert \bl X_i,y_i\br_{i=1}^N \right ]$, except here we have an additional $\omega_1$ term. Therefore, we may concisely write the final bound on the score errors for both $U$ and $V$ together using the bound for $U$, since we have already assumed for convenience that $\omega_1 \geq 1$.

Before stating the final bounds, we state a final lemma which allows us to use the infinite-dimensional canonical correlations.
\begin{lemma}
    For every $k = 1, 2, \ldots K$, $\gamma_k^*$ is asymptotically equivalent to $\gamma_k$ as $d$ goes to infinite.
\end{lemma}
The proof follows directly from Theorem 4.2.8 of \textcite{hsing2015theoretical} applied to $\mathscr{C}_{12}\mathscr{C}_{21}$ and $\mathscr{C}_{12}^{(d)}\mathscr{C}_{21}^{(d)}$ which gives that
\begin{equation} \label{eq:correlation_bound_operators}
    \sup_{k\geq1} \left |{\gamma_k^*}^2 - \gamma_k^2  \right | \leq \norm{\mathscr{C}_{12}\mathscr{C}_{21} - \mathscr{C}_{12}^{(d)}\mathscr{C}_{21}^{(d)}}.
\end{equation}
For bounded operators $A$ and $B$ with adjoints $A^*$ and $B^*$, we have
\begin{equation}
    AA^* - BB^* = (A-B)A^* - B(A^*-B^*),
\end{equation}
so that
\begin{equation}
    \norm{AA^* - BB^*} \leq \bl \norm{A} + \norm{B} \br \norm{A-B}.
\end{equation}
Applying this inequality with $A = \mathscr{C}_{12}$ and $B = \mathscr{C}_{12}^{(d)}$, the quantity on the right-hand side of equation \eqref{eq:correlation_bound_operators} converges to $0$ as $d$ grows to infinity since
\begin{equation} \label{eq:C_12_HS_norm}
    \norm{\mathscr{C}_{12}-\mathscr{C}_{12}^{(d)}} \leq \norm{\mathscr{C}_{12}-\mathscr{C}_{12}^{(d)}}_{\text{HS}},
\end{equation}
which decays to $0$ as $d$ grows since $\mathscr{C}_{12}$ being finite-dimensional means is it necessarily Hilbert-Schmidt.

We finally state the error bound on the convergence of the estimated canonical variables to the population infinite-dimensional canonical variables.
\begin{theorem}[Convergence of canonical variables] \label{thm:consistency_of_scores}
Let $(U_k,V_k)$ be the solution pair to the problem in Theorem \ref{theorem_problem} between $\chi_1 = \Log_{\mu}\ytest$ and $\chi_2 = \xtest$, for $k=1,\ldots K$. Let $\bl \hat U_k, \hat V_k \br \equiv \bl \innerdouble{\Log_{\hat \mu}\ytest}{\hat \psi_k}_{\hat \mu}, \inner{\xtest}{\hat \theta_k}\br$, where $\hat \theta_k$, $\hat \eta_k$ and $\hat \mu$ have been estimated via Algorithm \ref{alg:asymmetric_sparse_fun_cca}.
Then, under Assumptions \ref{assumption:manifold}-\ref{assumption:minor}, but without assuming a finite-dimensional correlation structure on $X$ and $y$ in Assumption \ref{assumption:rate}, we have
\begin{align}
     &\max \left \{ \E \left [ \bl U_k - \hat U_k\br^2 \left. \right \vert \bl X_i,y_i\br_{i=1}^N \right ], \E \left [ \bl V_k - \hat V_k\br^2 \left. \right \vert \bl X_i,y_i\br_{i=1}^N \right ] \right \}
     \\
    &=O_P\bl\frac{{\gamma_1^*}^2 \norm{\mathscr{C}_{12} - \mathscr{C}_{12}^{(d)}}^2}{\min_{j \neq k} \left |{\gamma_k^*}^2 - {\gamma_j^*}^2 \right |} + \frac{ ds\log p}{N}\frac{ \tau\twoinfnorm{\sigmax}\kappa  \Ex{\munorm{\Log_{\mu}y_1}^4}^{3/2}}{\min_{j \neq k} \min \curl \left |{\gamma_k^*}^2 - {\gamma_j^*}^2 \right |, \left|{\gamma_k^*} - {\gamma_j^*} \right | \curr^2}  \frac{\omega_1  \bl \sum_{j=1}^d\omega_j\br^2}{\omega_d^{3}\underset{j=1,\ldots d}{\operatorname{min}}\bl { \omega_j-\omega_{j+1}}\br^{2}} \br.
\end{align}
    
\end{theorem}
\section{Additional identities and inequalities}
\label{subsec:identities}
In the proofs, we use several identities and inequalities involving matrices. For definitions of the various matrix operations used below we refer to Section \ref{appendix:notation}. In the following, $A$ and $B$ denote matrices for which the specified matrix multiplications are valid.
\begin{enumerate}
    \item \label{item:first} For $x \in \rone^p$, $\infnorm{x} \leq \twonorm{x} \leq \onenorm{x}$.
    \item $\twonorm{x} \leq \sqrt{p}\infnorm{x}$, $\onenorm{x} \leq \sqrt{p}\twonorm{x}$, and $\onenorm{x} \leq p \infnorm{x}$.

    \item \label{item:fourth} $\maxnorm{A} \leq \twoinfnorm{A} \leq \twonorm{A} \leq \Fnorm{A} \leq \onetwo{A}$

    \item \label{item:induced_norms} For induced norms, $\norm{AB}_{\beta,\alpha} \leq \norm{A}_{\gamma,\alpha}\norm{B}_{\beta,\gamma}$ 
     (\textcite{trefethen2022numerical} equation (3.14)). In particular, $\norm{AB}_2 = \norm{AB}_{2,2} \leq \norm{A}_{\infty,2}\norm{B}_{2,\infty}$.
     \item \label{item:twonorm_bound_2} For any matrix $A$, $\twonorm{A} = \twonorm{A^{\T}A}^{1/2}$.
     \item \label{item:third} In addition to its definition as the largest singular value, $\twonorm{A}$ is the norm induced by $\twonorm{\cdot}$ and $\twonorm{\cdot}$.
     \item In addition to its definition as the norm induced by the $\norm{\cdot}_{2}$ and $\norm{\cdot}_{\infty}$ norms, $\twoinfnorm{A} = \operatorname{max}_i(\twonorm{A_i})$, where $A_i$ is the $i$th row of $A$. (\textcite{cape2019two} Proposition 6.1.)
     
    \item $\twoinfnorm{AB} \leq \twoinfnorm{A}\twonorm{B}$. (\textcite{cape2019two} Proposition 6.5.)
    \item For $B \in \rone^{p \times d}$, $\twoinfnorm{B} \leq \sqrt{d} \maxnorm{B}$.
    \item \label{item:inftwonorm} For the induced norm $\norm{\cdot}_{\infty,2}$, $\norm{B^{\T}}_{\infty,2} \leq \onetwo{B}$.
    
    \item \label{item:twonorm_bound} If $A \in \rone^{d\times p}$, and $B \in \rone^{p \times d}$, then $\twonorm{AB} \leq \twoinfnorm{A^{\T}}\onetwo{B}$. Additionally, $\twonorm{AB} \leq \onetwo{A^{\T}}\twoinfnorm{B}$.
    \item \label{item:one_two_norm_inequality}$\onetwo{AB} \leq \onetwo{A}\twonorm{B}$.
    \item \label{item:MVT_square_root_matrices} If $A,B \in \rone^{d \times d}$ are positive definite, then $\twonorm{A^{1/2} - B^{1/2}} \leq \frac{1}{2}\operatorname{max}\bl \twonorm{A^{-1}}, \twonorm{B^{-1}} \br^{1/2} \twonorm{A - B}$.
    \item \label{item:MVT_minus_square_root_matrices} If $A,B \in \rone^{d \times d}$ are positive definite, then $\twonorm{A^{-1/2} - B^{-1/2}} \leq \frac{1}{2}\operatorname{max}\bl \twonorm{A^{-1}}, \twonorm{B^{-1}} \br^{3/2} \twonorm{A - B}$.
    \item \label{item:product_SPD_matrices} For any positive definite matrices $A,B$ of the same size, the product $AB$ is diagonalizable with positive eigenvalues. Additionally, $AB$ has the same eigenvalues as $(AB^2A)^{1/2}$. In particular, $\twonorm{AB-I} =\twonorm{(AB^2A)^{1/2}-I}$. Note that $AB$ may not be symmetric, and therefore is not necessarily positive definite.
    \item \label{item:difference_in_inverses_bound} For positive real numbers $x$ and $y$, $\left | x^{-1} - y^{-1} \right | \leq {\operatorname{min}(x,y)}^{-3}\left | x^2 - y^2 \right |$.
    \item \label{item:AaBb}For $A,B \in \rone^{d \times d}$, $a,b \in \rone^d$, $\twonorm{Aa-Bb}^2 \lesssim \twonorm{A-B}^2\twonorm{a}^2 + \twonorm{B}^2\twonorm{a-b}^2$
    \item \label{item:norm_squared_triangle_inequality} For any norm $\norm{\cdot}$, $\norm{a+b}^2 \leq 3 \bl \norm{a}^2 + \norm{b}^2\br$.
    \item \label{item:bhatia_VII_39} For matrices $A,B$ of the same dimensions, $\Fnorm{\left | A \right | - \left | B \right |} \leq \sqrt{2}\Fnorm{A-B}$, where $\left |A \right | \equiv \bl A^{\T}A \br^{1/2}$. (\textcite{bhatia2013matrix} equation VII.39)

\end{enumerate}

\begin{proof}
    We prove the non-standard or non-straightforward results.\\
    \textit{Proof of \ref{item:inftwonorm}}:\\
    To show $\norm{B^{\T}}_{\infty,2} \leq \onetwo{B}$, we use the definition:
    \begin{equation}
        \norm{B^{\T}}_{\infty,2} \equiv \supp{\twonorm{B^{\T}x}}{\infnorm{x} = 1}.
    \end{equation}
    Without loss of generality, let $B$ be an element of $\rone^{p \times d}$.
    $\twonorm{B^{\T}x} = \twonorm{\sum_{i=1}^p b_i x_i}$ where $b_i$ is the $i$th row of $B$, and $x_i$ is the $i$th entry of $x \in \rone^p$. For $x \in \rone^p$ with $\infnorm{x}=1$, we have
    \begin{equation}
         \twonorm{\sum_{i=1}^p b_i x_i} \leq \sum_{i=1}^p |x_i| \twonorm{b_i} \leq \sum_{i=1}^p \twonorm{b_i} = \onetwo{B},
    \end{equation}
    using the triangle inequality and because $\infnorm{x}=1$. This completes the proof.\\
    \textit{Proof of \ref{item:twonorm_bound}}:\\
    To show $\twonorm{AB} \leq \twoinfnorm{A^{\T}}\onetwo{B}$, we begin by using item \ref{item:induced_norms} to obtain
    \begin{equation}
        \twonorm{AB} = \twonorm{B^{\T}A^{\T}} \leq \norm{B^{\T}}_{\infty,2}\norm{A^{\T}}_{2,\infty}.
    \end{equation}
    Using item \ref{item:inftwonorm} we have
    \begin{equation}
         \norm{B^{\T}}_{\infty,2}\norm{A^{\T}}_{2,\infty} \leq \onetwo{B} \norm{A^{\T}}_{2,\infty},
    \end{equation}
    completing the proof of the first statement. To show the second statement, we proceed similarly but apply item \ref{item:induced_norms} to $\twonorm{AB}$ instead of $\twonorm{B^{\T}A^{\T}}$.\\
    \textit{Proof of \ref{item:one_two_norm_inequality}:}\\
    To show $\onetwo{AB} \leq \onetwo{A}\twonorm{B}$, we begin with the definition. Without loss of generality, $A \in \rone^{p \times q}$ and $B \in \rone^{p \times r}$. We have
    \begin{equation}
        \onetwo{AB} = \sum_{i=1}^p\twonorm{\bl AB \br_i} = \sum_{i=1}^p\twonorm{A_i^{\T} B} \leq  \sum_{i=1}^p\twonorm{a_i}\twonorm{B} = \onetwo{A}\twonorm{B},
    \end{equation}
    where $\bl C \br_i$ denotes the $i$th row of a matrix $C$ and we have used item \ref{item:third}.\\
    \textit{Proof of \ref{item:MVT_square_root_matrices}:}\\
    The statement follows from equation X.46 of \textcite{bhatia2013matrix} by choosing $r=1/2$ and since the $2$-norm is unitarily invariant.\\
    \textit{Proof of \ref{item:MVT_minus_square_root_matrices}:}\\
    We begin with the equality $A^{-1} - B^{-1} = A^{-1}\bl B - A \br B^{-1}$, which holds for any square invertible matrices $A$ and $B$ of the same size. This implies
    \begin{equation}
        \twonorm{A^{-1} - B^{-1}} \leq \twonorm{A^{-1}} \twonorm{A-B} \twonorm{ B^{-1}}.
    \end{equation}
     We apply this to the matrices $A^{1/2}$ and $B^{1/2}$ to obtain
    \begin{equation}
        \twonorm{A^{-1/2} - B^{-1/2}} \leq \twonorm{A^{-1/2}} \twonorm{A^{1/2}-B^{1/2}} \twonorm{B^{-1/2}}.
    \end{equation}
    Using item \ref{item:MVT_square_root_matrices} on $\twonorm{A^{1/2}-B^{1/2}}$ in the last inequality, we deduce that
    \begin{align}
        \twonorm{A^{-1/2} - B^{-1/2}} & \leq \frac{1}{2}\operatorname{max}\bl \twonorm{A^{-1}}, \twonorm{B^{-1}} \br^{1/2} \twonorm{A - B} \twonorm{A^{-1/2}} \twonorm{B^{-1/2}}\\
        & = \frac{1}{2}\twonorm{A^{-1/2}} \twonorm{B^{-1/2}}\operatorname{max}\bl \twonorm{A^{-1/2}}, \twonorm{B^{-1/2}} \br \twonorm{A - B} 
    \end{align}
    One of $\twonorm{A^{-1/2}}$ and $\twonorm{B^{-1/2}}$ is larger, and in either case, the statement to be proven holds. \\
    \textit{Proof of \ref{item:product_SPD_matrices}}:\\
    The first claim, that $AB$ is diagonalizable with positive eigenvalues, follows from Proposition 6.1 of \textcite{serre2010matrices}.~To show that $AB$ has the same eigenvalues as $(AB^2A)^{1/2}$, let $C = AB$, so that $(AB^2A)^{1/2} = \bl CC^{\T}\br^{1/2}$. Letting $U\Sigma V^{\T} = C$ be a singular value decomposition of $C$, we have
    \begin{equation}
         \bl CC^{\T}\br^{1/2} = \bl U \Sigma V^{\T}V \Sigma U^{\T} \br^{1/2} = \bl U \Sigma^2 U^{\T}\br^{1/2} = U \Sigma U^{\T},
    \end{equation}
    where in the last line we have used that $C$ has positive singular values from the first claim.\\
    The last claim follows from the previous claim, since for a diagonalizable matrix $C$ where $C = WDW^{-1}$ with $D$ diagonal, we have
    \begin{equation}
        C-I = WDW^{-1} - WIW^{-1} = W(D-I)W^{-1},
    \end{equation} so that $C-I$ has eigenvalues equal to those of $C$ minus $1$.\\
    \textit{Proof of \ref{item:difference_in_inverses_bound}}:\\
    Letting $f(x) = x^{-1/2}$, the mean value theorem implies
    \begin{equation}
        \left | f(x) - f(y) \right | \leq \operatorname{min}\bl x, y\br^{-3/2} \left | x - y \right |.
    \end{equation}
    Plugging in $x^2$ for $x$ and $y^2$ for $y$, we obtain the result.\\
    \textit{Proof of \ref{item:AaBb}}:\\
    The statement follows from adding and subtracting $Ba$ and the inequality$\twonorm{a+b}^2 \leq 2\twonorm{a}^2 + 2\twonorm{b}^2$.\\
    \textit{Proof of \ref{item:norm_squared_triangle_inequality}}:\\
    From the triangle inequality, we have
    \begin{equation}
        \twonorm{a+b}^2 \leq \bl \twonorm{a} + \twonorm{b} \br^2 = \twonorm{a}^2 + 2 \twonorm{a}\twonorm{b} + \twonorm{b}^2
    \end{equation}
    One of $\twonorm{a}$ or $\twonorm{b}$ is larger than the other, and in either case, we deduce that $\twonorm{a}^2 + 2 \twonorm{a}\twonorm{b} + \twonorm{b} \leq 3\bl \twonorm{a}^2 + \twonorm{b}^2\br$, completing the proof.
\end{proof}

\section{Intrinsic RFPCA algorithm}
\label{sec:apdx:irfpca}
For completeness, in this section, we outline the main steps of the Intrinsic RFPCA algorithm. Let $M$ be the dimension of $\M$, $d$ the number of principal components to compute, $N$ the number of observations, and $L$ the number of time-steps.
\begin{enumerate}
    \item Estimate $\mu: \Tsc \to \M$, the functional Frech\'et mean of $y$, by computing the Frech\'et mean of $\{y_i(t_l)\}_i$ separately for each point in $\{t_l\}_{l=1,\ldots,L}$ where the functional data are observed.
    \item Compute the linear representations $\Log_{\hat{\mu}}y_i \in \Jmuhat$, for $i = 1,\ldots N$. In practice, this can be done by computing $\Log_{\hat{\mu}(t_l)}y_i(t_l) \in T_{\hat{\mu}(t_l)}\M$ for every $t_l$ by using the $\Log$ map on $\M$.
    
    \item  Let $E(x)$ be an orthonormal frame for the tangent space centered at $x \in \M$ (see Section~\ref{sec:orthonormal_frame} for an example). An orthonormal frame is a collection of tangent bases $E(x)= \{E_1(x), \ldots E_M(x) \}$ for $\txm$, with $x\in \M$, which varies smoothly with $x$, i.e. for each $k= 1, \ldots M$, $E_k(x)$ is a smooth map from $\M$ to $T\M$. For any fixed $x \in \M$, the functions $\{E_k(x)\}$ are orthonormal with respect to the inner product on $T_x\M$, that is, $\inner{\cdot}{\cdot}_x$.
    
    Compute the functional `coefficients' $\hat{Z}_i: \Tsc \to \rone^M$ of the expansion of $\Log_{\hat{\mu}}y_i: \Tsc \to T\M$ relative to $E$, for each $i = 1, \ldots N$. In practice, $\hat{Z}_i(t_l) \in \rone^M$ is computed separately for every $l = 1, \ldots L$. The $k$th entry of $\hat{Z}_i(t_l) \in \rone^M$ is computed as $\inner{\Log_{\hat{\mu}(t_l)}y_i(t_l)}{E_k(\hat{\mu}(t_l)}_{\hat{\mu}(t_l)}$ for $k = 1, \ldots M$. The resulting $\{\hat{Z}_i\}$ are estimates of the realizations of a real vector-valued random process $Z: \Tsc \rightarrow \rone^M$, with $k$th component $Z_k: \Tsc \to \rone$ given by $Z_k(t) = \inner{\Log_{\mu(t)}y(t)}{E_k(\mu(t)}_{\mu(t)}$.
    
    The process $Z$ is a real vector-valued process with the same principal scores as the process $\Logy$, and the $j$th principal component of $Z$, $\pi_j: \Tsc \rightarrow \rone^M$, is related to the $j$th principal component of $\Logy$, $\phi_j \in \Jmu$, via $\phi_j = \sum_{k=1}^M \pi_{jk} E_k$. Here, $\pi_{jk}$ denotes the $k$th entry of $\pi_j$ for $k=1, \ldots M$.
    \item Apply Multivariate Functional Principal Component Analysis (MFPCA) \parencite{happ2018multivariate} to the functions $\{\hat{Z}_i\}$ to estimate $d$ principal component functions $\hat{\pi}_{j}$ of the functional coefficients.

    Estimate the principal component functions $\{\phi_j\}$ of the $\Log$ representations of the functional data as $\hat{\phi}_j = \sum_{k=1}^M \hat{\pi}_{jk} E_k (\hat{\mu})$, for $j = 1, \ldots d$. Then, estimate the associated scores as $\hat{Y}_{ij} = \frac{1}{L}\sum_{l=1}^L Z_i(t_l)^{\T} \hat{\pi}_j(t_l)$, for $j=1, \ldots d$ and $i = 1, \ldots N$. The score estimates do not depend on the choice of the orthonormal frame $E$ (Proposition 5, item 2 of \textcite{lin2019intrinsic}). Compute variance estimates $\hat{\omega}_j = \Var(\hat Y_{ij})$.
    
    \item Return the estimated scores $\{\hat{Y}_{ij}\}$, variances $\{\hat{\omega}_j\}$ and principal component functions $\{\hat{\phi}_j\}$.
\end{enumerate}

\subsection{An orthonormal frame for the manifold of SPD matrices}\label{sec:orthonormal_frame}
Here we provide an explicit construction of an orthonormal frame in the setting where $\M$ is the manifold of $\rone^{m \times m}$ symmetric positive matrices equipped with the affine invariant metric, as in our application setting. The maps $\Log$ and $\Exp$ maps are defined as $\operatorname{Log}_F(G) = F^{1/2}\operatorname{log}\bl F^{-1/2}GF^{-1/2} \br F^{1/2}$ and $\Exp_F(W) = F^{1/2}\operatorname{exp}\bl F^{-1/2}WF^{-1/2} \br F^{1/2}$. Moreover, the inner product at $F \in \M$ between $W,Z \in T_F \M$ is defined as $\inner{W}{Z}_{\M} = \operatorname{tr}\bl F^{-1}WF^{-1}Z\br$. In this setting, we can use the result from Section 3.3.3.3.~of \textcite{pennec2019riemannian}, which provides an explicit construction of an orthonormal frame $E(F)$ for the tangent bundle, evaluated at an arbitrary $F \in \M$. This can be defined as
\begin{equation}
\label{eq:orthonormal_frame}
E_{i j}\bl F \br= \begin{cases}\bl F^{1/2}e_i\br \bl F^{1/2}e_i\br^{\T} & (1 \leq i=j \leq m)\\ 
\frac{1}{\sqrt{2}}\left(\bl F^{1/2}e_i\br \bl F^{1/2}e_j \br^{\T}+\bl F^{1/2}e_j\br \bl F^{1/2}e_i\br^{\T}\right) & (1 \leq i<j \leq m),\end{cases}
\end{equation}
where $e_i$ denotes the $i$th standard unit vector in $\rone^m$, and $F \in \rone^{m \times m}$. For a fixed $F \in \M$, there are $M = m(m + 1)/2$ unique $E_{ij}\bl F \br$, where $M$ is the dimension of $\M$.
We also note that the iterative algorithm proposed in \textcite{cheng2016recursive} can be used to estimate the Fr\'echet mean $\mu$, and is detailed in equations (13) and (14) of \textcite{cheng2016recursive}.

\setcounter{maxnames}{10}
\printbibliography

@book{pennec2019riemannian,
  title={{Riemannian Geometric Statistics in Medical Image Analysis}},
  author={Pennec, Xavier and Sommer, Stefan and Fletcher, Tom},
  year={2019},
  publisher={Academic Press},
}

@article{fletcher2007riemannian,
  title={Riemannian geometry for the statistical analysis of diffusion tensor data},
  author={Fletcher, P Thomas and Joshi, Sarang},
  journal={Signal Processing},
  volume={87},
  number={2},
  pages={250--262},
  year={2007},
  publisher={Elsevier}
}

@article{cupidon_properties_2008,
	title = {Some properties of canonical correlations and variates in infinite dimensions},
	volume = {99},
	issn = {0047259X},
	url = {https://linkinghub.elsevier.com/retrieve/pii/S0047259X07000978},
	doi = {10.1016/j.jmva.2007.07.007},
	language = {en},
	number = {6},
	urldate = {2023-07-23},
	journal = {Journal of Multivariate Analysis},
	author = {Cupidon, J. and Eubank, R. and Gilliam, D. and Ruymgaart, F.},
	month = jul,
	year = {2008},
	pages = {1083--1104},
}

@article{barachant2013classification,
  title = {Classification of covariance matrices using a {{Riemannian-based}} kernel for {{BCI}} applications},
  author = {Barachant, Alexandre and Bonnet, St{\'e}phane and Congedo, Marco and Jutten, Christian},
  year = {2013},
  month = jul,
  journal = {Neurocomputing},
  series = {Advances in Artificial Neural Networks, Machine Learning, and Computational Intelligence},
  volume = {112},
  pages = {172--178},
  issn = {0925-2312},
  doi = {10.1016/j.neucom.2012.12.039},
  urldate = {2024-07-16}
}

@article{bhattacharya2003large,
  title = {Large Sample Theory of Intrinsic and Extrinsic Sample Means on Manifolds. {{I}}},
  author = {Bhattacharya, Rabi and Patrangenaru, Vic},
  year = {2003},
  month = feb,
  journal = {Annals of Statistics},
  volume = {31},
  number = {1},
  pages = {1--29},
  issn = {00905364},
  doi = {10.1214/aos/1046294456}
}

@article{cai2016estimating,
  title = {Estimating Structured High-Dimensional Covariance and Precision Matrices: {{Optimal}} Rates and Adaptive Estimation},
  shorttitle = {Estimating Structured High-Dimensional Covariance and Precision Matrices},
  author = {Cai, T. Tony and Ren, Zhao and Zhou, Harrison H.},
  year = {2016},
  month = jan,
  journal = {Electronic Journal of Statistics},
  volume = {10},
  number = {1},
  pages = {1--59},
  publisher = {{Institute of Mathematical Statistics and Bernoulli Society}},
  issn = {1935-7524, 1935-7524},
  doi = {10.1214/15-EJS1081},
  urldate = {2023-11-01}
}

@article{cardot1999functional,
  title = {Functional Linear Model},
  author = {Cardot, Herv{\'e} and Ferraty, Fr{\'e}d{\'e}ric and Sarda, Pascal},
  year = {1999},
  month = oct,
  journal = {Statistics and Probability Letters},
  volume = {45},
  number = {1},
  pages = {11--22},
  issn = {01677152},
  doi = {10.1016/S0167-7152(99)00036-X}
}

@misc{carmichael2020learning,
  title = {Learning {{sparsity}} and {{block diagonal structure}} in {{multi-view mixture models}}},
  author = {Carmichael, Iain},
  year = {2020},
  eprint = {2012.15313},
  archiveprefix = {arXiv}
}

@misc{chen2013sparse,
  title = {Sparse {{CCA}} via {{precision adjusted iterative thresholding}}},
  author = {Chen, Mengjie and Gao, Chao and Ren, Zhao and Zhou, Harrison H.},
  year = {2013},
  eprint = {1311.6186},
  urldate = {2020-02-13},
  archiveprefix = {arXiv}
}

@article{cho2022tangent,
  title = {Tangent Functional Canonical Correlation Analysis for Densities and Shapes, with Applications to Multimodal Imaging Data},
  author = {Cho, Min Ho and Kurtek, Sebastian and Bharath, Karthik},
  year = {2022},
  month = may,
  journal = {Journal of Multivariate Analysis},
  volume = {189},
  pages = {104870},
  issn = {0047259X},
  doi = {10.1016/j.jmva.2021.104870},
  urldate = {2024-07-17},
  langid = {english}
}

@article{dai2017optimal,
  title = {Optimal {{Bayes}} Classifiers for Functional Data and Density Ratios},
  author = {Dai, Xiongtao and M{\"u}ller, Hans Georg and Yao, Fang},
  year = {2017},
  journal = {Biometrika},
  volume = {104},
  number = {3},
  eprint = {1605.03707},
  pages = {545--560},
  issn = {14643510},
  doi = {10.1093/biomet/asx024},
  archiveprefix = {arXiv}
}

@article{dai2018principal,
  title = {Principal Component Analysis for Functional Data on {R}iemannian Manifolds and Spheres},
  author = {Dai, Xiongtao and M{\"u}ller, Hans Georg},
  year = {2018},
  month = dec,
  journal = {Annals of Statistics},
  volume = {46},
  number = {6B},
  eprint = {1705.06226},
  pages = {3334--3361},
  issn = {00905364},
  doi = {10.1214/17-AOS1660},
  archiveprefix = {arXiv}
}

@article{dryden2009noneuclidean,
  title = {Non-{{Euclidean}} Statistics for Covariance Matrices, with Applications to Diffusion Tensor Imaging},
  author = {Dryden, Ian L. and Koloydenko, Alexey and Zhou, Diwei},
  year = {2009},
  month = sep,
  journal = {Annals of Applied Statistics},
  volume = {3},
  number = {3},
  eprint = {0910.1656v1},
  pages = {1102--1123},
  issn = {19326157},
  doi = {10.1214/09-AOAS249},
  archiveprefix = {arXiv}
}

@article{dubey2020functional,
  title = {Functional models for time-varying random objects},
  author = {Dubey, Paromita and M{\"u}ller, Hans Georg},
  year = {2020},
  journal = {Journal of the Royal Statistical Society. Series B: Statistical Methodology},
  volume = {82},
  number = {2},
  eprint = {1907.10829},
  pages = {275--327},
  issn = {14679868},
  doi = {10.1111/rssb.12337},
  archiveprefix = {arXiv}
}

@article{dubey2021modeling,
  title = {Modeling {{time-varying random objects}} and {{dynamic networks}}},
  author = {Dubey, Paromita and M{\"u}ller, Hans Georg},
  year = {2021},
  journal = {Journal of the American Statistical Association},
  eprint = {2104.04628},
  issn = {1537274X},
  doi = {10.1080/01621459.2021.1917416},
  archiveprefix = {arXiv}
}

@article{feng2018anglebased,
  title = {Angle-Based Joint and Individual Variation Explained},
  author = {Feng, Qing and Jiang, Meilei and Hannig, Jan and Marron, J. S.},
  year = {2018},
  month = jul,
  journal = {Journal of Multivariate Analysis},
  volume = {166},
  eprint = {1704.02060},
  pages = {241--265},
  publisher = {Academic Press Inc.},
  issn = {10957243},
  doi = {10.1016/j.jmva.2018.03.008},
  urldate = {2020-07-13},
  archiveprefix = {arXiv}
}

@article{gao2017sparse,
  title = {Sparse {{CCA}}: {{Adaptive}} Estimation and Computational Barriers},
  author = {Gao, Chao and Ma, Zongming and Zhou, Harrison H.},
  year = {2017},
  month = oct,
  journal = {Annals of Statistics},
  volume = {45},
  number = {5},
  eprint = {1409.8565},
  pages = {2074--2101},
  issn = {00905364},
  doi = {10.1214/16-AOS1519},
  urldate = {2020-02-13},
  archiveprefix = {arXiv}
}

@article{gaynanova2016simultaneous,
  title = {Simultaneous {{sparse estimation}} of {{canonical vectors}} in the p {$\geq$} {{N setting}}},
  author = {Gaynanova, Irina and Booth, James G. and Wells, Martin T.},
  year = {2016},
  journal = {Journal of the American Statistical Association},
  volume = {111},
  number = {514},
  pages = {696--706},
  issn = {1537274X},
  doi = {10.1080/01621459.2015.1034318}
}

@article{gaynanova2020prediction,
  title = {Prediction and Estimation Consistency of Sparse Multi-Class Penalized Optimal Scoring},
  author = {Gaynanova, Irina},
  year = {2020},
  journal = {Bernoulli},
  volume = {26},
  number = {1},
  eprint = {1809.04669},
  pages = {286--322},
  issn = {13507265},
  doi = {10.3150/19-BEJ1126},
  urldate = {2020-12-25},
  archiveprefix = {arXiv}
}

@article{ghodrati2022distributionondistribution,
  title = {Distribution-on-Distribution Regression via Optimal Transport Maps},
  author = {Ghodrati, Laya and Panaretos, Victor M},
  year = {2022},
  month = dec,
  journal = {Biometrika},
  volume = {109},
  number = {4},
  pages = {957--974},
  issn = {1464-3510},
  doi = {10.1093/biomet/asac005},
  urldate = {2024-07-16}
}

@article{ghosal2023frechet,
  title = {Fr{\'e}chet Single Index Models for Object Response Regression},
  author = {Ghosal, Aritra and Meiring, Wendy and Petersen, Alexander},
  year = {2023},
  month = jan,
  journal = {Electronic Journal of Statistics},
  volume = {17},
  number = {1},
  issn = {1935-7524},
  doi = {10.1214/23-EJS2120},
  urldate = {2024-07-17},
  langid = {english}
}

@article{glasser2013minimal,
  title = {The Minimal Preprocessing Pipelines for the {{Human Connectome Project}}},
  author = {Glasser, Matthew F. and Sotiropoulos, Stamatios N. and Wilson, J. Anthony and Coalson, Timothy S. and Fischl, Bruce and Andersson, Jesper L. and Xu, Junqian and Jbabdi, Saad and Webster, Matthew and Polimeni, Jonathan R. and Van Essen, David C. and Jenkinson, Mark},
  year = {2013},
  month = oct,
  journal = {NeuroImage},
  volume = {80},
  pages = {105--124},
  publisher = {Elsevier Inc.},
  issn = {10538119},
  doi = {10.1016/j.neuroimage.2013.04.127},
  isbn = {1053-8119},
  pmid = {23668970}
}

@article{glasser2016multimodal,
  title = {A Multi-Modal Parcellation of Human Cerebral Cortex},
  author = {Glasser, Matthew F. and Coalson, Timothy S. and Robinson, Emma C. and Hacker, Carl D. and Harwell, John and Yacoub, Essa and Ugurbil, Kamil and Andersson, Jesper and Beckmann, Christian F. and Jenkinson, Mark and Smith, Stephen M. and Van Essen, David C.},
  year = {2016},
  journal = {Nature},
  volume = {536},
  number = {7615},
  pages = {171--178},
  publisher = {Nature Publishing Group},
  issn = {14764687},
  doi = {10.1038/nature18933},
  isbn = {0008-5472 (Print){\textbackslash}r0008-5472 (Linking)},
  pmid = {27437579}
}

@article{happ2018multivariate,
  title = {Multivariate {{functional principal component analysis}} for {{data observed}} on {{different}} ({{dimensional}}) {{domains}}},
  author = {Happ, Clara and Greven, Sonja},
  year = {2018},
  journal = {Journal of the American Statistical Association},
  volume = {113},
  number = {522},
  eprint = {1509.02029},
  pages = {649--659},
  issn = {1537274X},
  doi = {10.1080/01621459.2016.1273115},
  urldate = {2021-08-18},
  archiveprefix = {arXiv}
}

@article{he2010functional,
  title = {Functional Linear Regression via Canonical Analysis},
  author = {He, Guozhong and M{\"u}ller, Hans-Georg and Wang, Jane-Ling and Yang, Wenjing},
  year = {2010},
  month = aug,
  journal = {Bernoulli},
  volume = {16},
  number = {3},
  issn = {1350-7265},
  doi = {10.3150/09-BEJ228},
  urldate = {2023-11-01},
  langid = {english}
}

@article{hotelling1936relations,
  title = {Relations {{between two sets}} of {{variates}}},
  author = {Hotelling, H.},
  year = {1936},
  month = dec,
  journal = {Biometrika},
  volume = {28},
  number = {3-4},
  pages = {321--377},
  issn = {0006-3444},
  doi = {10.1093/biomet/28.3-4.321}
}

@article{huang2015functional,
  title = {Functional Partial Canonical Correlation},
  author = {Huang, Qing and Renaut, Rosemary},
  year = {2015},
  month = may,
  journal = {Bernoulli},
  volume = {21},
  number = {2},
  eprint = {1506.00414},
  primaryclass = {math, stat},
  issn = {1350-7265},
  doi = {10.3150/14-BEJ597},
  urldate = {2023-04-28},
  archiveprefix = {arXiv}
}

@article{hutchison2013dynamic,
  title = {Dynamic Functional Connectivity: {{Promise}}, Issues, and Interpretations},
  author = {Hutchison, R. Matthew and Womelsdorf, Thilo and Allen, Elena A. and Bandettini, Peter A. and Calhoun, Vince D. and Corbetta, Maurizio and Della Penna, Stefania and Duyn, Jeff H. and Glover, Gary H. and {Gonzalez-Castillo}, Javier and Handwerker, Daniel A. and Keilholz, Shella and Kiviniemi, Vesa and Leopold, David A. and {de Pasquale}, Francesco and Sporns, Olaf and Walter, Martin and Chang, Catie},
  year = {2013},
  journal = {NeuroImage},
  volume = {80},
  pages = {360--378},
  publisher = {Elsevier Inc.},
  issn = {10538119},
  doi = {10.1016/j.neuroimage.2013.05.079},
  isbn = {1095-9572 (Electronic){\textbackslash}r1053-8119 (Linking)},
  pmid = {23707587}
}

@article{kereta2021estimating,
  title = {Estimating Covariance and Precision Matrices along Subspaces},
  author = {Kereta, {\v Z}eljko and Klock, Timo},
  year = {2021},
  month = jan,
  journal = {Electronic Journal of Statistics},
  volume = {15},
  number = {1},
  issn = {1935-7524},
  doi = {10.1214/20-EJS1782},
  urldate = {2024-07-17},
  langid = {english}
}

@misc{kessler2023computational,
  title = {Computational {{inference}} for {{directions}} in {{canonical correlation analysis}}},
  author = {Kessler, Daniel and Levina, Elizaveta},
  year = {2023},
  number = {arXiv:2308.11218},
  eprint = {2308.11218},
  primaryclass = {stat},
  publisher = {arXiv},
  urldate = {2023-08-23},
  archiveprefix = {arXiv}
}

@incollection{kim2014canonical,
  title = {Canonical {{correlation analysis}} on {{Riemannian manifolds}} and {{its applications}}},
  booktitle = {Computer {{Vision}} -- {{ECCV}} 2014},
  author = {Kim, Hyunwoo J. and Adluru, Nagesh and Bendlin, Barbara B. and Johnson, Sterling C. and Vemuri, Baba C. and Singh, Vikas},
  editor = {Fleet, David and Pajdla, Tomas and Schiele, Bernt and Tuytelaars, Tinne},
  year = {2014},
  volume = {8690},
  pages = {251--267},
  publisher = {Springer International Publishing},
  address = {Cham},
  doi = {10.1007/978-3-319-10605-2_17},
  urldate = {2024-04-01},
  copyright = {http://www.springer.com/tdm},
  isbn = {978-3-319-10604-5 978-3-319-10605-2},
  langid = {english}
}

@book{kokoszka2017introduction,
  title = {Introduction to {{Functional Data Analysis}}},
  author = {Kokoszka, Piotr and Reimherr, Matthew},
  year = {2017},
  month = sep,
  journal = {Introduction to Functional Data Analysis},
  publisher = {{Chapman and Hall/CRC}},
  address = {New York, New York, USA},
  doi = {10.1201/9781315117416},
  isbn = {978-1-315-11741-6}
}

@article{liegeois2019resting,
  title = {Resting Brain Dynamics at Different Timescales Capture Distinct Aspects of Human Behavior},
  author = {Li{\'e}geois, Rapha{\"e}l and Li, Jingwei and Kong, Ru and Orban, Csaba and Van De Ville, Dimitri and Ge, Tian and Sabuncu, Mert R. and Yeo, B. T.Thomas},
  year = {2019},
  journal = {Nature Communications},
  volume = {10},
  number = {1},
  issn = {20411723},
  doi = {10.1038/s41467-019-10317-7},
  urldate = {2021-06-11},
  pmid = {31127095}
}

@article{lin2013group,
  title = {Group Sparse Canonical Correlation Analysis for Genomic Data Integration},
  author = {Lin, Dongdong and Zhang, Jigang and Li, Jingyao and Calhoun, Vince D and Deng, Hong-Wen and Wang, Yu-Ping},
  year = {2013},
  month = dec,
  journal = {BMC Bioinformatics},
  volume = {14},
  number = {1},
  pages = {245},
  issn = {1471-2105},
  doi = {10.1186/1471-2105-14-245},
  urldate = {2024-04-01},
  langid = {english}
}

@article{lin2019intrinsic,
  title = {Intrinsic {{Riemannian}} Functional Data Analysis},
  author = {Lin, Zhenhua and Yao, Fang},
  year = {2019},
  month = dec,
  journal = {The Annals of Statistics},
  volume = {47},
  number = {6},
  issn = {0090-5364},
  doi = {10.1214/18-AOS1787},
  urldate = {2023-02-08}
}

@article{lin2020mapping,
  title = {Mapping Brain--Behavior Networks Using Functional and Structural Connectome Fingerprinting in the {{HCP}} Dataset},
  author = {Lin, Ying-Chia and Baete, Steven H and Wang, Xiuyuan and Boada, Fernando E},
  year = {2020},
  journal = {Brain and Behavior},
  volume = {10},
  number = {6},
  pages = {e01647},
  issn = {2162-3279},
  doi = {10.1002/brb3.1647},
  urldate = {2024-07-30},
  langid = {english}
}

@misc{liu2021noneuclidean,
  title = {Non-{{Euclidean analysis}} of {{joint variations}} in {{multi-object shapes}}},
  author = {Liu, Zhiyuan and Schulz, J{\"o}rn and Taheri, Mohsen and Styner, Martin and Damon, James and Pizer, Stephen and Marron, J. S.},
  year = {2021},
  month = sep,
  number = {arXiv:2109.02230},
  eprint = {2109.02230},
  primaryclass = {cs, stat},
  publisher = {arXiv},
  urldate = {2024-03-29},
  archiveprefix = {arXiv}
}

@article{liu2022improved,
  title = {Improved {{interpretability}} of {{brain-behavior CCA with domain-driven dimension reduction}}},
  author = {Liu, Zhangdaihong and Whitaker, Kirstie J. and Smith, Stephen M. and Nichols, Thomas E.},
  year = {2022},
  month = jun,
  journal = {Frontiers in Neuroscience},
  volume = {16},
  pages = {851827},
  issn = {1662-453X},
  doi = {10.3389/fnins.2022.851827},
  urldate = {2023-09-18},
  langid = {english}
}

@article{lock2013joint,
  title = {Joint and Individual Variation Explained ({{JIVE}}) for Integrated Analysis of Multiple Data Types},
  author = {Lock, Eric F. and Hoadley, Katherine A. and Marron, J. S. and Nobel, Andrew B.},
  year = {2013},
  month = mar,
  journal = {Annals of Applied Statistics},
  volume = {7},
  number = {1},
  eprint = {1102.4110},
  pages = {523--542},
  issn = {19326157},
  doi = {10.1214/12-AOAS597},
  archiveprefix = {arXiv},
  pmid = {23745156}
}

@book{marron2021object,
  title = {Object {{Oriented Data Analysis}}},
  author = {Marron, J.S. and Dryden, Ian L.},
  year = {2021},
  edition = {1},
  publisher = {{Chapman and Hall/CRC}},
  address = {Boca Raton},
  doi = {10.1201/9781351189675},
  urldate = {2023-03-17},
  isbn = {978-1-351-18967-5},
  langid = {english}
}

@article{masarotto2019procrustes,
  title = {Procrustes Metrics on Covariance Operators and Optimal Transportation of {{Gaussian}} Processes},
  author = {Masarotto, Valentina and Panaretos, Victor M. and Zemel, Yoav},
  year = {2019},
  month = may,
  journal = {Sankhya: The Indian Journal of Statistics},
  volume = {81A},
  eprint = {1801.01990},
  pages = {172--213},
  publisher = {Sankhya A},
  issn = {09727671},
  doi = {10.1007/s13171-018-0130-1},
  archiveprefix = {arXiv}
}

@article{mckeague2022significance,
  title = {Significance Testing for Canonical Correlation Analysis in High Dimensions},
  author = {McKeague, Ian W and Zhang, Xin},
  year = {2022},
  month = nov,
  journal = {Biometrika},
  volume = {109},
  number = {4},
  pages = {1067--1083},
  issn = {0006-3444, 1464-3510},
  doi = {10.1093/biomet/asab059},
  urldate = {2023-09-04},
  langid = {english}
}

@article{murden2022interpretive,
  title = {Interpretive {{JIVE}}: {{Connections}} with {{CCA}} and an Application to Brain Connectivity},
  shorttitle = {Interpretive {{JIVE}}},
  author = {Murden, Raphiel J. and Zhang, Zhengwu and Guo, Ying and Risk, Benjamin B.},
  year = {2022},
  journal = {Frontiers in Neuroscience},
  volume = {16},
  issn = {1662-453X},
  urldate = {2023-08-21}
}

@article{pervaiz2020optimising,
  title = {Optimising Network Modelling Methods for {{fMRI}}},
  author = {Pervaiz, Usama and Vidaurre, Diego and Woolrich, Mark W. and Smith, Stephen M.},
  year = {2020},
  month = may,
  journal = {NeuroImage},
  volume = {211},
  pages = {116604},
  issn = {10538119},
  doi = {10.1016/j.neuroimage.2020.116604},
  urldate = {2022-11-07},
  langid = {english}
}

@article{petersen2019frechet,
  title = {Fr{\'e}chet Regression for Random Objects with {{Euclidean}} Predictors},
  author = {Petersen, Alexander and M{\"u}ller, Hans Georg},
  year = {2019},
  month = apr,
  journal = {Annals of Statistics},
  volume = {47},
  number = {2},
  eprint = {1608.03012},
  pages = {691--719},
  issn = {00905364},
  doi = {10.1214/17-AOS1624},
  archiveprefix = {arXiv}
}

@article{petersen2022modeling,
  title = {Modeling {{probability density functions}} as {{data objects}}},
  author = {Petersen, Alexander and Zhang, Chao and Kokoszka, Piotr},
  year = {2022},
  journal = {Econometrics and Statistics},
  volume = {21},
  pages = {159--178},
  issn = {24523062},
  doi = {10.1016/j.ecosta.2021.04.004},
  urldate = {2022-07-19}
}

@article{pigoli2014distances,
  title = {Distances and Inference for Covariance Operators},
  author = {Pigoli, Davide and Aston, John A.D. and Dryden, Ian L. and Secchi, Piercesare},
  year = {2014},
  journal = {Biometrika},
  volume = {101},
  number = {2},
  pages = {409--422},
  issn = {14643510},
  doi = {10.1093/biomet/asu008}
}

@article{power2011functional,
  title = {Functional {{network organization}} of the {{human brain}}},
  author = {Power, Jonathan D. and Cohen, Alexander L. and Nelson, Steven M. and Wig, Gagan S. and Barnes, Kelly Anne and Church, Jessica A. and Vogel, Alecia C. and Laumann, Timothy O. and Miezin, Fran M. and Schlaggar, Bradley L. and Petersen, Steven E.},
  year = {2011},
  journal = {Neuron},
  volume = {72},
  number = {4},
  pages = {665--678},
  publisher = {Elsevier Inc.},
  issn = {08966273},
  doi = {10.1016/j.neuron.2011.09.006},
  pmid = {22099467}
}

@book{ramsay2015functional,
  title = {Functional {{Data Analysis}}},
  author = {Ramsay, James O. and Silverman, Bernard W.},
  year = {2015},
  journal = {International Encyclopedia of the Social \& Behavioral Sciences: Second Edition},
  publisher = {Springer-Verlag},
  address = {New York},
  doi = {10.1016/B978-0-08-097086-8.42046-5},
  isbn = {978-0-08-097087-5}
}

@article{shao2022intrinsic,
  title = {Intrinsic {{Riemannian}} Functional Data Analysis for Sparse Longitudinal Observations},
  author = {Shao, Lingxuan and Lin, Zhenhua and Yao, Fang},
  year = {2022},
  month = jun,
  journal = {The Annals of Statistics},
  volume = {50},
  number = {3},
  issn = {0090-5364},
  doi = {10.1214/22-AOS2172},
  urldate = {2024-04-01},
  langid = {english}
}

@article{shin2015canonical,
  title = {Canonical Correlation Analysis for Irregularly and Sparsely Observed Functional Data},
  author = {Shin, Hyejin and Lee, Seokho},
  year = {2015},
  month = feb,
  journal = {Journal of Multivariate Analysis},
  volume = {134},
  pages = {1--18},
  issn = {0047259X},
  doi = {10.1016/j.jmva.2014.10.001},
  urldate = {2023-11-01},
  langid = {english}
}

@article{shu2020dcca,
  title = {D-{{CCA}}: {{A decomposition-based canonical correlation analysis}} for {{high-dimensional datasets}}},
  author = {Shu, Hai and Wang, Xiao and Zhu, Hongtu},
  year = {2020},
  month = dec,
  journal = {Journal of the American Statistical Association},
  volume = {115},
  number = {529},
  eprint = {2001.02856},
  pages = {292--306},
  issn = {1537274X},
  doi = {10.1080/01621459.2018.1543599},
  archiveprefix = {arXiv}
}

@article{smith2013restingstate,
  title = {Resting-State {{fMRI}} in the {{Human Connectome Project}}},
  author = {Smith, Stephen M. and Beckmann, Christian F. and Andersson, Jesper and Auerbach, Edward J. and Bijsterbosch, Janine and Douaud, Gwena{\"e}lle and Duff, Eugene and Feinberg, David A. and Griffanti, Ludovica and Harms, Michael P. and Kelly, Michael and Laumann, Timothy and Miller, Karla L. and Moeller, Steen and Petersen, Steve and Power, Jonathan and {Salimi-Khorshidi}, Gholamreza and Snyder, Abraham Z. and Vu, An T. and Woolrich, Mark W. and Xu, Junqian and Yacoub, Essa and U{\v g}urbil, Kamil and Van Essen, David C. and Glasser, Matthew F.},
  year = {2013},
  month = oct,
  journal = {NeuroImage},
  volume = {80},
  pages = {144--168},
  issn = {10538119},
  doi = {10.1016/j.neuroimage.2013.05.039},
  urldate = {2019-11-08},
  pmid = {23702415}
}

@article{smith2015positivenegative,
  title = {A Positive-Negative Mode of Population Covariation Links Brain Connectivity, Demographics and Behavior},
  author = {Smith, Stephen M. and Nichols, Thomas E. and Vidaurre, Diego and Winkler, Anderson M. and Behrens, Timothy E.J. and Glasser, Matthew F. and Ugurbil, Kamil and Barch, Deanna M. and Van Essen, David C. and Miller, Karla L.},
  year = {2015},
  journal = {Nature Neuroscience},
  volume = {18},
  number = {11},
  pages = {1565--1567},
  publisher = {Nature Publishing Group},
  issn = {15461726},
  doi = {10.1038/nn.4125},
  isbn = {1546-1726 (Electronic){\textbackslash}r1097-6256 (Linking)},
  pmid = {26414616}
}

@article{stocker2023functional,
  title = {Functional {{additive models}} on {{manifolds}} of {{planar shapes}} and {{forms}}},
  author = {St{\"o}cker, Almond and Steyer, Lisa and Greven, Sonja},
  year = {2023},
  month = oct,
  journal = {Journal of Computational and Graphical Statistics},
  volume = {32},
  number = {4},
  pages = {1600--1612},
  issn = {1061-8600, 1537-2715},
  doi = {10.1080/10618600.2023.2175687},
  urldate = {2024-07-17},
  langid = {english}
}

@article{uurtio2018tutorial,
  title = {A {{tutorial}} on {{canonical correlation methods}}},
  author = {Uurtio, Viivi and Monteiro, Jo{\~a}o M. and Kandola, Jaz and {Shawe-Taylor}, John and {Fernandez-Reyes}, Delmiro and Rousu, Juho},
  year = {2018},
  month = nov,
  journal = {ACM Computing Surveys},
  volume = {50},
  number = {6},
  pages = {1--33},
  issn = {0360-0300, 1557-7341},
  doi = {10.1145/3136624},
  urldate = {2024-04-01},
  langid = {english}
}

@article{vanessen2012human,
  title = {The {{Human Connectome Project}}: {{A}} data acquisition perspective},
  author = {Van Essen, D. C. and Ugurbil, K. and Auerbach, E. and Barch, D. and Behrens, T. E.J. and Bucholz, R. and Chang, A. and Chen, L. and Corbetta, M. and Curtiss, S. W. and Della Penna, S. and Feinberg, D. and Glasser, M. F. and Harel, N. and Heath, A. C. and {Larson-Prior}, L. and Marcus, D. and Michalareas, G. and Moeller, S. and Oostenveld, R. and Petersen, S. E. and Prior, F. and Schlaggar, B. L. and Smith, S. M. and Snyder, A. Z. and Xu, J. and Yacoub, E.},
  year = {2012},
  month = oct,
  journal = {NeuroImage},
  volume = {62},
  number = {4},
  pages = {2222--2231},
  issn = {10538119},
  doi = {10.1016/j.neuroimage.2012.02.018},
  pmid = {22366334}
}

@article{wang2020finding,
  title = {Finding the Needle in a High-Dimensional Haystack: {{Canonical}} Correlation Analysis for Neuroscientists},
  shorttitle = {Finding the Needle in a High-Dimensional Haystack},
  author = {Wang, Hao-Ting and Smallwood, Jonathan and {Mourao-Miranda}, Janaina and Xia, Cedric Huchuan and Satterthwaite, Theodore D. and Bassett, Danielle S. and Bzdok, Danilo},
  year = {2020},
  month = aug,
  journal = {NeuroImage},
  volume = {216},
  pages = {116745},
  issn = {10538119},
  doi = {10.1016/j.neuroimage.2020.116745},
  urldate = {2024-07-17},
  langid = {english}
}

@article{wang2021eigenvectorbased,
  title = {Eigenvector-Based Sparse Canonical Correlation Analysis: {{Fast}} Computation for Estimation of Multiple Canonical Vectors},
  author = {Wang, Wenjia and Zhou, Yi Hui},
  year = {2021},
  journal = {Journal of Multivariate Analysis},
  volume = {185},
  eprint = {2004.10231},
  issn = {10957243},
  doi = {10.1016/j.jmva.2021.104781},
  urldate = {2020-04-23},
  archiveprefix = {arXiv}
}

@article{witten2009penalized,
  title = {A Penalized Matrix Decomposition, with Applications to Sparse Principal Components and Canonical Correlation Analysis},
  author = {Witten, Daniela M. and Tibshirani, Robert and Hastie, Trevor},
  year = {2009},
  journal = {Biostatistics},
  volume = {10},
  number = {3},
  pages = {515--534},
  issn = {14654644},
  doi = {10.1093/biostatistics/kxp008},
  urldate = {2021-03-24},
  pmid = {19377034}
}

@article{xia2018linked,
  title = {Linked Dimensions of Psychopathology and Connectivity in Functional Brain Networks},
  author = {Xia, Cedric Huchuan and Ma, Zongming and Ciric, Rastko and Gu, Shi and Betzel, Richard F. and Kaczkurkin, Antonia N. and Calkins, Monica E. and Cook, Philip A. and {Garc{\'i}a de la Garza}, Angel and Vandekar, Simon N. and Cui, Zaixu and Moore, Tyler M. and Roalf, David R. and Ruparel, Kosha and Wolf, Daniel H. and Davatzikos, Christos and Gur, Ruben C. and Gur, Raquel E. and Shinohara, Russell T. and Bassett, Danielle S. and Satterthwaite, Theodore D.},
  year = {2018},
  journal = {Nature Communications},
  volume = {9},
  number = {1},
  issn = {20411723},
  doi = {10.1038/s41467-018-05317-y},
  urldate = {2020-05-06},
  pmid = {30068943}
}

@article{yang2015independence,
  title = {Independence {{test}} for {{high dimensional data based}} on {{regularized canonical correlation coefficients}}},
  author = {Yang, Yanrong and Pan, Guangming},
  year = {2015},
  journal = {The Annals of Statistics},
  volume = {43},
  number = {2},
  eprint = {43556524},
  eprinttype = {jstor},
  pages = {467--500},
  publisher = {Institute of Mathematical Statistics},
  issn = {0090-5364},
  urldate = {2023-03-16}
}

@article{yang2021survey,
  title = {A {{survey}} on {{canonical correlation analysis}}},
  author = {Yang, Xinghao and Liu, Weifeng and Liu, Wei and Tao, Dacheng},
  year = {2021},
  month = jun,
  journal = {IEEE Transactions on Knowledge and Data Engineering},
  volume = {33},
  number = {6},
  pages = {2349--2368},
  issn = {1041-4347, 1558-2191, 2326-3865},
  doi = {10.1109/TKDE.2019.2958342},
  urldate = {2024-04-01},
  copyright = {https://ieeexplore.ieee.org/Xplorehelp/downloads/license-information/IEEE.html},
  langid = {english}
}

@article{yao2005functional,
  title = {Functional Data Analysis for Sparse Longitudinal Data},
  author = {Yao, Fang and M{\"u}ller, Hans Georg and Wang, Jane Ling},
  year = {2005},
  month = jun,
  journal = {Journal of the American Statistical Association},
  volume = {100},
  number = {470},
  pages = {577--590},
  issn = {01621459},
  doi = {10.1198/016214504000001745},
  isbn = {0162-1459}
}

@article{yoon2020sparse,
  title = {Sparse Semiparametric Canonical Correlation Analysis for Data of Mixed Types},
  author = {Yoon, Grace and Carroll, Raymond J. and Gaynanova, Irina},
  year = {2020},
  journal = {Biometrika},
  volume = {107},
  number = {3},
  eprint = {1807.05274},
  pages = {609--625},
  issn = {14643510},
  doi = {10.1093/biomet/asaa007},
  urldate = {2020-12-25},
  archiveprefix = {arXiv}
}

@article{yu2015useful,
  title = {A Useful Variant of the {{Davis-Kahan}} Theorem for Statisticians},
  author = {Yu, Y. and Wang, T. and Samworth, R. J.},
  year = {2015},
  journal = {Biometrika},
  volume = {102},
  number = {2},
  eprint = {1405.0680},
  pages = {315--323},
  issn = {14643510},
  doi = {10.1093/biomet/asv008},
  urldate = {2020-02-12},
  archiveprefix = {arXiv}
}

@article{yuan2022doublematched,
  title = {Double-{{matched matrix decomposition}} for {{multi-view data}}},
  author = {Yuan, Dongbang and Gaynanova, Irina},
  year = {2022},
  month = oct,
  journal = {Journal of Computational and Graphical Statistics},
  volume = {31},
  number = {4},
  pages = {1114--1126},
  issn = {1061-8600, 1537-2715},
  doi = {10.1080/10618600.2022.2067860},
  urldate = {2024-07-17},
  langid = {english}
}

@article{zhang2020mixedeffect,
  title = {Mixed-{{effect time-varying network model}} and {{application}} in {{brain connectivity analysis}}},
  author = {Zhang, Jingfei and Sun, Will Wei and Li, Lexin},
  year = {2020},
  journal = {Journal of the American Statistical Association},
  volume = {115},
  number = {532},
  pages = {2022--2036},
  doi = {10.1080/01621459.2019.1677242},
  urldate = {2021-12-07}
}

@article{zhao2021covariate,
  title = {Covariate {{assisted principal}} Regression for Covariance Matrix Outcomes},
  author = {Zhao, Yi and Wang, Bingkai and Mostofsky, Stewart H and Caffo, Brian S and Luo, Xi},
  year = {2021},
  month = jul,
  journal = {Biostatistics},
  volume = {22},
  number = {3},
  pages = {629--645},
  issn = {1465-4644, 1468-4357},
  doi = {10.1093/biostatistics/kxz057},
  urldate = {2024-04-01},
  copyright = {https://academic.oup.com/journals/pages/open\_access/funder\_policies/chorus/standard\_publication\_model},
  langid = {english}
}

@article{zhou2021dynamic,
  title = {Network regression with graph laplacians},
  author = {Zhou, Yidong and Müller, Hans-Georg},
  date = {2022},
  journal = {Journal of Machine Learning Research},
  volume = {23},
  number = {320},
  pages = {1--41},
  url = {http://jmlr.org/papers/v23/22-0681.html}
}

@article{zhu2023statistical,
  title = {Statistical {{learning methods}} for {{neuroimaging data analysis}} with {{applications}}},
  author = {Zhu, Hongtu and Li, Tengfei and Zhao, Bingxin},
  year = {2023},
  month = aug,
  journal = {Annual Review of Biomedical Data Science},
  volume = {6},
  number = {1},
  pages = {73--104},
  issn = {2574-3414, 2574-3414},
  doi = {10.1146/annurev-biodatasci-020722-100353},
  urldate = {2024-02-22},
  langid = {english}
}

@article{zhuang2020technical,
  title = {A Technical Review of Canonical Correlation Analysis for Neuroscience Applications},
  author = {Zhuang, Xiaowei and Yang, Zhengshi and Cordes, Dietmar},
  year = {2020},
  month = jun,
  journal = {Human Brain Mapping},
  volume = {41},
  number = {13},
  pages = {3807--3833},
  issn = {1065-9471},
  doi = {10.1002/hbm.25090},
  urldate = {2024-04-01},
  pmcid = {PMC7416047},
  pmid = {32592530}
}

@article{zou2006sparse,
  title = {Sparse Principal Component Analysis},
  author = {Zou, Hui and Hastie, Trevor and Tibshirani, Robert},
  year = {2006},
  month = jun,
  journal = {Journal of Computational and Graphical Statistics},
  volume = {15},
  number = {2},
  eprint = {1205.0121v2},
  pages = {265--286},
  issn = {10618600},
  doi = {10.1198/106186006X113430},
  archiveprefix = {arXiv},
  isbn = {106186006X},
  pmid = {21811560}
}

@book{Lee2012,
author="Lee, John M.",
title="Smooth {Manifolds}",
bookTitle="Introduction to Smooth Manifolds",
year="2012",
publisher="Springer New York",
address="New York, NY",
isbn="978-1-4419-9982-5",
doi="10.1007/978-1-4419-9982-5_1",
url="https://doi.org/10.1007/978-1-4419-9982-5_1"
}

@book{lee2018introduction,
  title={{Introduction to Riemannian Manifolds}},
  author={Lee, John M},
  volume={2},
  year={2018},
  publisher={Springer}
}

@book{hsing2015theoretical,
  title={{Theoretical Foundations of Functional Data Analysis, With An Introduction to Linear Operators}},
  author={Hsing, Tailen and Eubank, Randall},
  volume={997},
  year={2015},
  publisher={John Wiley \& Sons}
}

@article{jirak2020perturbation,
  title={Perturbation bounds for eigenspaces under a relative gap condition},
  author={Jirak, Moritz and Wahl, Martin},
  journal={Proceedings of the American Mathematical Society},
  volume={148},
  number={2},
  pages={479--494},
  year={2020}
}

@book{hormander2015analysis,
  title={The {A}nalysis of {L}inear {P}artial {D}ifferential {O}perators {I}: {D}istribution {T}heory and {F}ourier {A}nalysis},
  author={H{\"o}rmander, Lars},
  year={2015},
  publisher={Springer}
}

@article{schotz2019convergence,
  title={Convergence rates for the generalized {F}r{\'e}chet mean via the quadruple inequality},
  author={Sch{\"o}tz, Christof},
  year={2019}
}

@book{vershynin2018high,
  title={{High-Dimensional Probability: An Introduction with Applications in Data Science}},
  author={Vershynin, Roman},
  volume={47},
  year={2018},
  publisher={Cambridge university press}
}

@incollection{cheng2016recursive,
  title={Recursive computation of the {F}r{\'e}chet mean on non-positively curved Riemannian manifolds with applications},
  author={Cheng, Guang and Ho, Jeffrey and Salehian, Hesamoddin and Vemuri, Baba C},
  booktitle={Riemannian Computing in Computer Vision},
  pages={21--43},
  year={2016},
  publisher={Springer}
}

@Article{glmnet,
    title = {Regularization Paths for Generalized Linear Models via
      Coordinate Descent},
    author = {Jerome Friedman and Robert Tibshirani and Trevor Hastie},
    journal = {Journal of Statistical Software},
    year = {2010},
    volume = {33},
    number = {1},
    pages = {1--22},
    doi = {10.18637/jss.v033.i01},
  }

@book{hastie2015statistical,
  title={{Statistical Learning with Sparsity: the Lasso and Generalizations}},
  author={Hastie, Trevor and Tibshirani, Robert and Wainwright, Martin},
  year={2015},
  publisher={CRC press}
}

@article{krzysko2013canonical,
  title={Canonical correlation analysis for functional data},
  author={Krzy{\'s}ko, Miros{\l}aw and Waszak, {\L}ukasz},
  journal={Biometrical Letters},
  volume={50},
  number={2},
  pages={95--105},
  year={2013}
}

@book{trefethen2022numerical,
  title={{Numerical Linear Algebra}},
  author={Trefethen, Lloyd N and Bau, David},
  volume={181},
  year={2022},
  publisher={Siam}
}

@article{cape2019two,
  title={THE TWO-TO-INFINITY NORM AND SINGULAR SUBSPACE GEOMETRY WITH APPLICATIONS TO HIGH-DIMENSIONAL STATISTICS},
  author={Cape, Joshua and Tang, Minh and Priebe, Carey E},
  journal={The Annals of Statistics},
  volume={47},
  number={5},
  pages={2405--2439},
  year={2019},
  publisher={JSTOR}
}

@book{bhatia2013matrix,
  title={{M}atrix {A}nalysis},
  author={Bhatia, Rajendra},
  volume={169},
  year={2013},
  publisher={Springer Science \& Business Media}
}

@book{serre2010matrices,
  title={{M}atrices: {T}heory and {A}pplications},
  author={Serre, Denis},
  year={2010},
  publisher={Springer}
}

@article{kendall2011limit,
  title={Limit theorems for empirical Fr{\'e}chet means of independent and non-identically distributed manifold-valued random variables},
  author={Kendall, Wilfrid S and Le, Huiling},
  journal={Brazilian Journal of Probability and Statistics},
  volume={25},
  number={3},
  pages={323--352},
  year={2011}
}

@article{pennec2017hessian,
  title={Hessian of the {R}iemannian squared distance},
  author={Pennec, Xavier},
  journal={Preprint},
  year={2017}
}

@article{ahidar2020convergence,
  title={Convergence rates for empirical barycenters in metric spaces: curvature, convexity and extendable geodesics},
  author={Ahidar-Coutrix, Adil and Le Gouic, Thibaut and Paris, Quentin},
  journal={Probability Theory and Related Fields},
  volume={177},
  number={1},
  pages={323--368},
  year={2020},
  publisher={Springer}
}

@article{boulesteix2007partial,
  title={Partial least squares: a versatile tool for the analysis of high-dimensional genomic data},
  author={Boulesteix, Anne-Laure and Strimmer, Korbinian},
  journal={Briefings in bioinformatics},
  volume={8},
  number={1},
  pages={32--44},
  year={2007},
  publisher={Oxford University Press}
}

@article{charkaborty2022testing,
  title={Testing for the rank of a covariance operator},
  author={Charkaborty, Anirvan and Panaretos, Victor M},
  journal={The Annals of Statistics},
  volume={50},
  number={6},
  pages={3510--3537},
  year={2022},
  publisher={Institute of Mathematical Statistics}
}

@article{zhou2023functional,
  title={Functional linear regression for discretely observed data: from ideal to reality},
  author={Zhou, Hang and Yao, Fang and Zhang, Huiming},
  journal={Biometrika},
  volume={110},
  number={2},
  pages={381--393},
  year={2023},
  publisher={Oxford University Press}
}
\end{refsection}

\end{document}